\documentclass[reqno,12pt,oneside]{report}% right-side equation numbering, 12 point font, print one-sided 

\usepackage{booktabs}
\usepackage{CJKutf8}
\usepackage{rac}         % Use Rackham thesis style file
\usepackage[toc,page]{appendix}
\makeatletter
\let\old@chapter\@chapter% Store \@chapter
\usepackage[ruled]{algorithm2e}
\let\@chapter\old@chapter% Restore \@chapter
\makeatother
\usepackage{setspace}
\usepackage[intlimits]{amsmath} % Puts the limits of integrals on top and bottom
\usepackage{indentfirst}
\usepackage{amsxtra}     % Use various AMS packages
\usepackage{amsthm}
\usepackage{amssymb}
\usepackage{amsfonts}
\usepackage{bbm}
\usepackage{bigints}
\usepackage{bm}
\usepackage{enumitem}
\usepackage[yyyymmdd]{datetime}

\usepackage{fixmath}
\usepackage{float}
\usepackage{mathtools}
\usepackage{multirow}
\usepackage{multicol}
\usepackage[frozencache,cachedir=.]{minted}
\usepackage{textcomp}
\usepackage{gensymb}
\usepackage{graphicx}    % Add some packages for figures. Read epslatex.pdf on ctan.tug.org
\usepackage{epstopdf}
\AppendGraphicsExtensions{.tif}
\PassOptionsToPackage{hyphens}{url}
\usepackage[hidelinks,pagebackref]{hyperref}
\usepackage{bookmark}
\usepackage{longtable}
\usepackage{listings}
\usepackage{rotating}
\usepackage{tikz}
\usetikzlibrary{shapes.gates.logic.US,trees,positioning,arrows}
\usepackage{color}
\usepackage{pdflscape}
\usepackage{tikz}
\usetikzlibrary{shapes}
\usepackage{epsfig}
\usepackage{xcolor}
\usepackage{scalerel,stackengine}

\usepackage{notoccite} %avoid numbering citations from ListOfFigure

\stackMath
\newcommand\reallywidehat[1]{%
\savestack{\tmpbox}{\stretchto{%
  \scaleto{%
    \scalerel*[\widthof{\ensuremath{#1}}]{\kern-.6pt\bigwedge\kern-.6pt}%
    {\rule[-\textheight/2]{1ex}{\textheight}}%WIDTH-LIMITED BIG WEDGE
  }{\textheight}% 
}{0.5ex}}%
\stackon[1pt]{#1}{\tmpbox}%
}

\usepackage{subcaption}
\usepackage{verbatim}
\usepackage[sort&compress,numbers]{natbib}
\usepackage[printonlyused]{acronym} % For the List of Abbreviations. Read acronym.pdf on ctan.tug.org
\usepackage{setspace}    % Allows you to specify the line spacing
\theoremstyle{plain}

\theoremstyle{definition}

\theoremstyle{remark}

\numberwithin{theorem}{chapter}     % Numbers theorems "x.y" where x
\makeatletter
\def\cleardoublepage{\clearpage\if@twoside \ifodd\c@page\else
\hbox{}
\thispagestyle{empty}
\newpage
\if@twocolumn\hbox{}\newpage\fi\fi\fi}
\makeatother 

\begin{document}
\bibliographystyle{unsrt}    % Set the bibliography style. agu04, plain, alpha, etc.
% \bibliography{xampl}

% Title page as required by Rackham dissertation guidelines
\titlepage{Electron Acceleration and Radiation Generation from Relativistic Laser-Plasma Interactions at High Repetition-Rate}{Jinpu Lin}{Doctor of Philosophy}
{Nuclear Engineering and Radiological Sciences}{2021}
{Professor Karl M. Krushelnick, Chair \\
Professor Igor Jovanovic \\
Research Scientist John Nees \\
Professor Theodore Norris \\
Professor Alexander G.R. Thomas}

% Begin the front matter as required by Rackham dissertation guidelines
\initializefrontsections

% Optional Frontispiece
\frontispiece{

\vskip 0.16\textheight
\begin{center}
    \includegraphics[width=.95\textwidth]{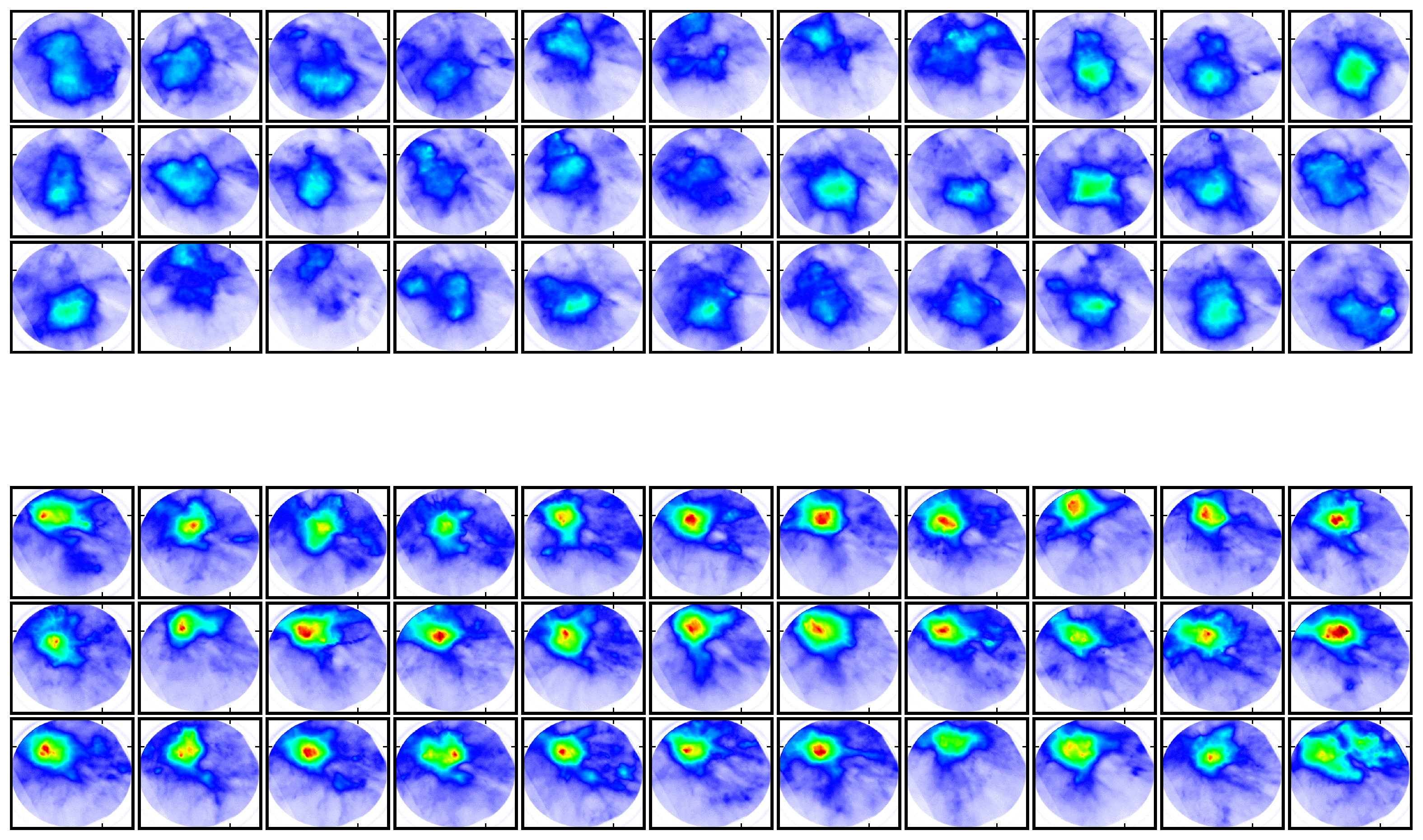} \\
    Optimizing beam profile of electrons in a laser-wakefield accelerator.
\end{center}

}

\newpage
\thispagestyle{empty}
\begin{center}

\hspace{0pt}
\vfill
Jinpu Lin

linjinp@umich.edu

ORCID iD: 0000-0003-1251-0457

\copyright Jinpu Lin 2021

All Rights Reserved
\vfill
\hspace{0pt}

\end{center}

% % Optional, but recommended, Copyright page
% \copyrightpage{Jinpu Lin}{linjinp@umich.edu}{0000-0003-1251-0457}

% Page numbering. If you don't include a frontispiece or copyright page, you'll need to change this for two-sided printing.
\makeatletter
\if@twoside \setcounter{page}{4} \else \setcounter{page}{1} \fi
\makeatother
 
% Optional Dedication page
\dedicationpage{
\begin{center}
\textbf{Dedication}
\end{center}
\begin{CJK*}{UTF8}{gbsn}
To my grandfather, 胡安.
\end{CJK*}
}

% Optional Acknowledgements page
\startacknowledgementspage
This dissertation would not have been possible without the support from many, many people. I especially thank my advisor Prof. Karl Krushelnick for his guidance and patience, and for encouraging me to explore the exciting topics in this field. Karl has provided me all the opportunities to travel around the world for experiments, conferences, and workshops. His work ethnic and passion are inspiring, and I have learned a lot from him. I am extremely grateful to Prof. Alec Thomas for introducing me to the group when I was an undergrad. Alec has guided me through my Ph.D. career, and I really appreciate his time and help in physics discussion and simulation, his precious comments and critical thinking for my paper writing, and the journal club he organized. I would like to thank Research Scientist John Nees for training me in the lab and answering all my naive questions with patience. I admire John's enthusiasm and knowledge in optics, and I would not have gone so far without his mentorship. Thank you to Prof. Igor Jovanovic for the valuable advice on OPA and on radiation detection, and all the insightful comments in the group meetings. Igor is not only an amazing researcher but also a fantastic graduate chair, and his charisma has motivated me to pursue a career path in academia. I would like to thank Assistant Research Scientist Yong Ma and Milos Burger for their generous support, both academically and personally. Thank you Yong for guiding me on the setting up the simulations and data analysis, and thank you Milos for the hands-on instructions in fixing the laser. I really appreciate the valuable advice on my career development from both of you!

Many thanks to my fellow graduate students, especially Paul Campbell for being a fantastic colleague and friend throughout my Ph.D. career (and for introducing me to the soccer team), Brandon Russell for all the fun in the conference trips and lunch breaks (and for going through the material in Jackson with me all the way to the second chapter), Amina Hussein for the supportive conversations and for passing her knowledge of the challenging classes to me, Peter Kordell for sharing his experience when I needed encouragement and for guiding me through the optical system design with all the patience, Keegan Behm and Thomas Batson for welcoming me to the group and being supportive in my first days, and Jesus Hinojosa, Hongmei Tang, and Brendan Stassel for being genial officemates. It has been a great pleasure to work with my fellow lab mates Patrick Skrodzki, Xuan Xiao, Jon Murphy, Nick Peskosky, Lauren Finney, Hao Huang, and Jungmoo Hah. I am grateful and happy to spend the lab days with them and to learn from them. Thank you to my collaborators Nick Beier and Tam Nguyen at UC Irvine for helping me set up the HHG experiment and for the friendship since then, to Daniel Woodbury and Robert Schwartz at the University of Maryland for hosting the LWFA experiment and taking me to lunches, to Mark Mathis for helping me with the genetic algorithm coding and deformable mirror setup, and to Qian Qian for the contribution to the machine learning project. I also would like to thank Andre Antonie, Jason Cardarelli, Ryan Sandberg, Stephen Dilorio, Mario Balcazar, Nick Ernst, and all members in the HFS group for making it exciting to discuss physics.

To my friends and family - I couldn't have got thus far without you! Thank you to Sunming Qin 
% \begin{CJK*}{UTF8}{gbsn}
% 萌萌
% \end{CJK*}
for sharing every experience on defense preparation, and to Weiwei Jiang and Weishu Wu for the joyful conversation that kept me positive during the pandemic lockdown. I do not know how to express my gratitude towards my parents for their unconditional love and endless support. Thank you for everything. Claire, thank you for going through all of this with me, and I can't wait for what comes next.
\label{Acknowledgements}

% Optional Preface page
%\startprefacepage
%\input{Preface}
%\label{Preface}

% Table of contents, list of figures, etc.
\tableofcontents     % Required
\listoffigures       % Required if there is more than one figure
\listoftables        % Required if there is more than one table
\listofappendix    % Required if there is more than one appendix
\listofabbreviations % Optional. Abbreviations should be stored in a file named abbr.tex

% \addcontentsline{toc}{chapter}{LIST OF PUBLICATIONS}
% \chapter*{\centering \large\selectfont{LIST OF PUBLICATIONS}}
% \input{listOfPublications.tex}
% \label{listOfPublications}

% Optional in-dissertation Abstract Page
\startabstractpage

This dissertation explores the interaction between high-intensity lasers and plasmas to accelerate electrons and produce radiation via experimental and computational efforts. The laser pulses used in this dissertation have ultrashort duration ($<100$ fs), near-infrared to mid-infrared wavelength (0.8 $\mu m$, 2 $\mu m$, or 3.9 $\mu m$), millijoules of energy, and high repetition rates (480 Hz or 20 Hz). The plasma sources applied are from solid-density targets (overdense) or gaseous targets (underdense). With the high-repetition-rate capability, statistical methods are employed to optimize certain aspect of the experiments and to interpret the physics.

In the solid target experiments, electron acceleration in the form of attosecond bunches and radiation generation via \acf{HHG} and via characteristic x-ray emission are presented. In the gas target experiments, electron acceleration via \acf{LWFA} is demonstrated with the help of statistical methods, including genetic algorithms and supervised learning methods. 

MeV-level attosecond electron bunches from the interactions between ultrashort pulses (30 fs, 0.8 $\mu m$, 12 mJ) and solid targets (fused silica and copper) are investigated through similarities between experimental and simulated electron energy spectra. The bunch duration and temporal structure are measured in \acf{PIC} simulations. The experimental observation of such bunches occurs mainly in the specular reflection direction when focusing the laser pulse onto a sub-wavelength boundary of thick overdense plasmas at grazing incidence. To isolate a single electron bunch, simulations using single cycle laser pulses are performed. Particle tracking is applied to analyze the effects of carrier-envelope phase, preplasma density profile, laser intensity, and the focal spot size.

Surface \acs{HHG} and corresponding phenomena are studied using femtosecond mid-infrared laser pulses (2 $\mu m$, 1.6 mJ, 67 fs) interacting with solid targets (fused silica and silicon). Experimental measurements of the \acs{HHG} spectra and the beam divergence are reported. The power-law scaling of harmonic efficiency vs. harmonic order is examined. The intensity of horizontally-polarized harmonics and vertically-polarized harmonics are measured when the driving laser pulses are polarized in horizontal, vertical, left-circular, and right-circular directions. The scaling of the third harmonic efficiency vs. laser intensity is also investigated.

Characteristic x-ray emission from laser-solid interactions are presented. Laser pulses with various wavelengths and pulse energies are used to interact with overdense plasma of various preplasma profiles from a molybdenum target. The study is performed both experimentally with hundreds of thousands of laser shots, and computationally with \acs{PIC} simulations scanning over the 4-dimensional parameter space consisting of laser wavelength, pulse energy, preplasma profile, and x-ray emission properties.

Statistical methods are used to improve the focus of laser beams in high numerical aperture systems as an efficient route to increasing intensities in the ultrafast regime to relativistic levels. A method that optimizes the focus of a high-power laser without attenuation is demonstrated experimentally using near-infrared (0.8 $\mu m$) and mid-infrared (2 $\mu m$) laser pulses, where the second harmonic generation at full intensity in a low-pressure gas provides a figure of merit for optimizing the shape of a deformable mirror via a genetic algorithm. Nonlinear and thermal aberrations are corrected, and aberrations caused by filters are avoided.

Coherent control of the dynamics of laser-wakefield acceleration driven by ultrashort ($\sim 100$ fs) mid-infrared ($\sim 3.9~\mu$m) laser pulses is demonstrated, where plasma densities up to $3\times 10^{19}cm^{-3}$ (or $40\%$ of the critical density at $\lambda=3.9 \mu m$) are used. MeV-level, collimated electron beams with non-thermal, peaked energy spectra are generated. Optimization of electron beam qualities, including the total charge, energy spectra, beam pointing, and stability, is realized through adaptive control of the laser wavefront using a deformable mirror and a genetic algorithm. The improvement in the electron beam quality is explained by \acs{PIC} simulations using the optimal wavefront.

Applications of machine learning techniques in relativistic laser-plasma experiments are explored beyond optimization purposes. With trained supervised learning models, the beam charge of electrons produced in a laser wakefield accelerator is predicted given the laser wavefront change caused by a deformable mirror. Feature importance analysis on the trained models shows that specific aberrations in the laser wavefront are favored in generating higher beam charges. The predictive models enable operations beyond merely searching for an optimal beam charge. The quality of the measured data is characterized, and anomaly detection is demonstrated. The model robustness against measurement errors is examined by applying a range of virtual measurement error bars to the experimental data. This work demonstrates a route to machine learning applications in the highly nonlinear problem of relativistic laser-plasma interaction for in-depth data analysis to assist physics interpretation.
\label{Abstract}

\startthechapters 

%% Chapters %%
\chapter{Introduction}
\label{chap:Intro}
\section{Motivation}
\begin{figure}[ht]
\centering
\includegraphics[width=0.95\columnwidth
% , height=0.6\columnwidth
]{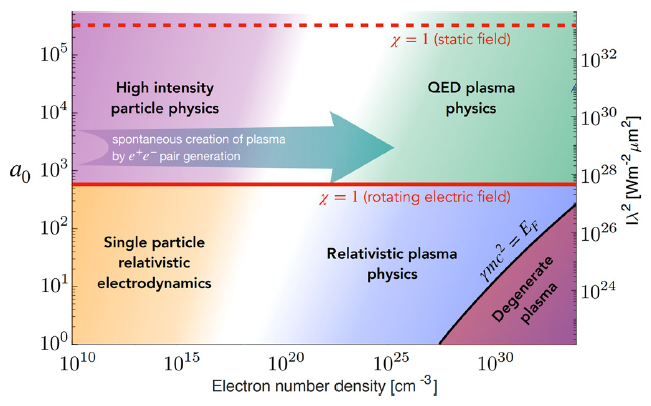}
\caption{Strong field physics regimes classified by laser-plasma conditions. Adapted from Ref. \cite{zhang2020relativistic} with permission. Note that the laser intensity is shown in units of $W \cdot m^{-2}$ rather than $W \cdot cm^{-2}$.}
\label{FigIntroZhang}
\end{figure}

The development of laser technology has enabled new fields of physics, such as nonlinear optics. In the past few decades, reaching relativistic laser intensities has been realized through the invention of the \acs{CPA} technique \cite{strickland1985compression}, for which Donna Strickland and Gerard Mourou were awarded the Nobel Prize in Physics in 2018. At such laser intensities, joules of energy can be compressed spatially into a few microns (a millionth of a meter) and temporally into tens of femtoseconds (a millionth of a billionth of a second), creating ultra-intense electric and magnetic fields in the laboratory. The fields are so strong that they can easily pull away electrons from their atomic orbits and produce plasmas. While most of the universe consists of plasmas, interactions between strong laser fields and plasmas have opened a new field of study. Fig. \ref{FigIntroZhang} demonstrates different regimes of strong-field physics as a function of the plasma density and either the field strength $a_0$ or the laser intensity, plotted by Zhang \textit{et al.} \cite{zhang2020relativistic}. At lower plasma densities, the collective behavior of plasmas is suppressed, and thus the dynamics of particles are more of interest. The laser-plasma conditions used in this dissertation work fall into the bottom right region on Fig. \ref{FigIntroZhang}, where high energy particle beams and bright radiation sources are generated and can serve as sources for many other fields of physics. The \acf{SFQED} \cite{greiner1985quantum} regime can be accessed with even high laser intensities, in which electrons-positrons pairs can be produced. These produced pairs change the plasma density in the region, and also change the charge distribution and thus reshape the electric fields, adding complexity to the interactions between the fields and the plasmas. Although the required laser intensity for \acf{QED} plasma physics is beyond current capabilities and are mostly studied theoretically \cite{di2012extremely, ridgers2012dense, bulanov2013electromagnetic, gonoskov2015extended, blackburn2019reaching, zhang2020relativistic}, studies in this regime can be related to many astrophysical phenomena \cite{goldreich1969pulsar, remington2006experimental, ruffini2010electron}. 

This dissertation explores particle acceleration and radiation generation via relativistic laser-plasma interactions. Accelerating particles to approach the speed of light is important for a variety of scientific and technological applications. These high-energy particles can not only be used for particle colliders but also radiate and produce bright ultrashort bursts of x-rays to probe dynamics inside an atom or a molecule. To date, the most powerful particle accelerators are the \acf{LINACs}, and the most powerful coherent radiation sources are \acf{XFELs}.

\begin{figure}[ht]
\centering
\includegraphics[width=0.95\columnwidth
% , height=0.6\columnwidth
]{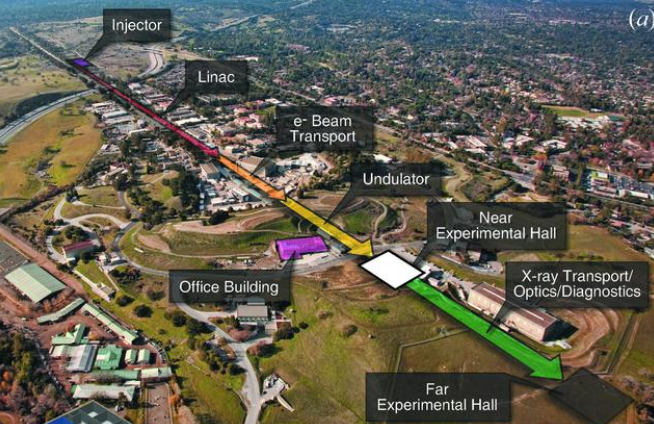}
\caption{Areal view of the \acs{LCLS} in Stanford. Adapted from Ref. \cite{white2015linac} with permission.}
\label{FigIntroSLAC}
\end{figure}

Although these facilities offer particle and radiation sources with unprecedented qualities, they are far from affordable. Fig. \ref{FigIntroSLAC} shows an example of the \acf{LCLS} in Stanford, which utilizes the 20 GeV electron beams from the Stanford Linear Accelerator \cite{white2015linac} and is thus 3 km long. The \acs{XFELs} worldwide are compared and listed in Ref. \cite{XFELsAll}. The next generation of particle accelerators (\acs{LINACs}) and radiation sources (\acs{XFELs}) will be tens of kilometers long and will cost billions of dollars. 

The field of relativistic laser-plasma interactions provides possible alternatives at a substantially smaller size and cost. Advanced accelerators based on the concepts of \ac{LWFA} and \ac{PWFA} utilize the mechanism of "wakefields" to generate accelerating gradients up to 100 GeV/m, which are several orders of magnitude greater than those produced in conventional accelerators. Electron beams with quasi-monoenergetic peaks up to 7.8 GeV have been demonstrated in \acs{LWFA} in 2019 \cite{gonsalves2019petawatt}, and the current energy level for \acs{PWFA} is on the order of 10 GeV \cite{cakir2019brief}, showing that wakefield accelerators are promising candidates. Although both \acs{LWFA} and \acs{PWFA} fall in the category of relativistic plasma physics, \acs{LWFA} is directly driven by high-intensity lasers and will be studied in this dissertation work. \acs{LWFA} was first proposed by Tajima and Dawson \cite{tajima1979laser} in 1979. Since then, extensive studies have been performed with analytical models and computer simulations \cite{pukhov2002laser, lu2007generating, esarey2009physics} as well as with experimental results \cite{malka2002electron, mangles2004monoenergetic, geddes2004high, faure2004laser, leemans2006gev, gonsalves2019petawatt} to understand its mechanisms. However, there still remain issues with the electron beam pointing, stability, and energy spread, making them difficult to serve as relativistic particle sources for many applications. Therefore, precise control of the produced electron beams has become the critical question to address in the next decade.

Compact bright coherent radiation sources with ultrashort pulse duration and ultrahigh photon energy are also expected from relativistic laser-plasma interactions. There are several mechanisms that can generate such radiation, including betatron radiation in \acs{LWFA}, compact \acf{FELs} based on laser-plasma accelerators, and surface \acf{HHG} from overdense plasma targets. In the highly nonlinear regime of \acs{LWFA}, relativistic electrons in the strong wakefield oscillate and radiate betatron x-rays \cite{rousse2004production, kneip2010bright, albert2016applications}. On the other side, the electron bunches produced in \acs{LWFA} have short bunch duration ($\sim$femtoseconds) and high current ($\sim$kA), which makes it possible to reduce the size of the undulator of \acf{FELs} \cite{gruner2007design, nakajima2008towards, couprie2018towards}. Although a few experimental demonstrations of undulator radiation have been presented \cite{schlenvoigt2008compact, fuchs2009laser, anania2014ultrashort, andre2018control, maier2020water}, realizing \acs{LWFA}-based \acs{FELs} remains a challenging task due to the quality of the electron beams produced by current \acf{LWFAs}. Another approach to compact radiation sources is surface \acs{HHG}. When ultra-intense lasers interact with overdense plasmas, the critical surface of the plasma oscillates as it is periodically driven by the laser fields and so radiates photons at wavelengths of harmonics of the driving laser. \acs{HHG} as a promising ultrashort x-ray source has been studied extensively \cite{dromey2006high, teubner2009high, nomura2009attosecond, dollar2013scaling} and reviewed in a recent article \cite{edwards2020x}. Furthermore, the harmonics produced in such processes are compressed both in time (attoseconds) and in space (x-ray wavelengths), resulting in several orders of magnitude increase in peak intensity according to a prediction in Ref. \cite{quere2021reflecting}.

\section{High repetition-rate laser-plasma experiments and statistical methods}
\label{sec:IntroHigh}

While \acs{CPA} technology has enabled unprecedented laser intensities, the highest energy laser facilities usually operate in single-shot mode, namely firing a few shots a day. On the other side, development of laser systems with higher repetition rates but lower peak power is always of fundamental interest, particularly for applications. Lasers that output $\sim$ TW power ($\sim 20$ mJ energy), $\sim30$ fs duration light at $\sim$ kHz repetition-rate have been widely embraced in several laboratories worldwide, such as the \acf{Lambda-cubed} at the \acf{CUOS} at the University of Michigan. More recently, there have been rapid developments in lasers with higher output energy ($>$ tens of Joules) at repetition rates $\sim10$ Hz \cite{nakamura2017diagnostics, sistrunk2017all, pilar2018characterization, roso2018high, jourdain2021l4n}. Accordingly, plasma targets and diagnostics tools that work at high repetition rates are also drawing increasing attention \cite{prencipe2017targets, zaffino2018preparation, george2019high, musgrave2019gallium, henares2019development, chagovets2021automation}. 

With the high repetition-rate operation capabilities in both the laser systems and the plasma targets, extensive studies have been conducted for relativistic laser-plasma interactions at high repetition rates, including interactions with gas targets \cite{he2013high, salehi2017mev, guenot2017relativistic, faure2018review, gotzfried2018research, salehi2019high, maier2020decoding, rovige2021optimization}, with liquid targets \cite{hah2016high, feister2017relativistic, morrison2018mev, hah2018characterization, becker2019characterization}, and with solid targets \cite{mordovanakis2009quasimonoenergetic, easter2010thesis, bocoum2016anticorrelated, giulietti2017d, zaim2019few, dover2020demonstration}. For relativistic laser-plasma experiments, having high repetition rates yields many advantages. First, applications of the particle and radiation sources from laser-plasma interactions usually demand more than single-shot mode, and most applications do not require record high beam energies. Secondly, accumulating a large number of measurements can be beneficial for events that happen at a small probability, such as \acf{SFQED} phenomena. For potential vacuum polarization experiments with high energy, high repetition rate lasers in the future, the challenges in designing high repetition rate targets automatically go away since no target other than a vacuum is needed. Lastly, having higher repetition rates allows the use of statistical methods to assist the experiments, such as machine learning.

Being one of the most impactful technological advances of the decade, \ac{ML} has not found many inspiring applications in laser-plasma interactions in the last few years. The development of high-repetition-rate laser facilities that can deliver at least thousands of shots a day enables the applications of statistical methods. Traditional statistical method applications are mainly via genetic algorithms \cite{nayuki2005production, he2015coherent, englesbe2016control, hah2017enhancement, lin2018focus, noaman2018controlling, streeter2018temporal, lefebvre2018phase, dann2019laser, lin2019adaptive, smith2020optimizing,   finney2021filament}, which have been effective for optimization purposes but can produce results that are difficult to interpret. Instead, \acs{ML} methods can generate predictive models that reveal more information in the dataset to understand the physical processes and provide control over complex parameter space and enable anomaly detection.

Researchers in the \acf{ICF} community have adopted \acs{ML} as an effective tool and led the broader laser-plasma community. Humbird \textit{et al.} \cite{humbird2018deep} have developed  the \acs{DJINN} algorithm based on neural networks and random forest to train and model data from \acs{ICF} experiments. Various supervised learning regression algorithms have been applied to assist \acs{ICF} data analysis \cite{gaffney2019making, gopalaswamy2019tripled, humbird2019parameter, hatfield2019using, hsu2020analysis, ruby2021high}. Since \acs{ICF} has limitations in the laser and target repetition rate and computer simulations are used often to help, transfer learning has been applied to reduce the bias of computer simulations using just a few experimental data \cite{humbird2019transfer, kustowski2019transfer}.

\acs{ML} is also drawing increasing attention in the wakefield accelerator community in the last year or two. Automation and control of \acs{LWFA} have been demonstrated using Bayesian optimization where six controllable parameters were tuned \cite{shalloo2020automation}. Applications beyond optimization purpose, such as feature analysis and anomaly detection, have been explored and will be discussed by employing multiple supervised learning architectures in this thesis. In \acs{PWFA}, \ac{RL} has been applied to trajectory optimization in the \ac{AWAKE} experiment, showing the ability to optimize within just a few iterations after training for 300 iterations \cite{kain2020sample}.

Researchers in other branches of laser-plasma interactions have adopted \acs{ML} as well. For example, Gonoskov \textit{et al.} \cite{gonoskov2019employing} have employed neural networks to resolve theoretical and experimental difficulties in high-order-harmonic spectra. Neural networks and deep learning also find extensive applications in \acf{LIBS} \cite{li2020laser, ewusi2020automatic, castorena2021deep}. 

In this thesis, we apply both genetic algorithms and supervised learning methods to experiments performed at high repetition rate laser facilities. Details of the results will be presented in Chap. \ref{chap:ML}, with a primary focus on \acs{LWFA} which also discusses laser focus optimization. Statistical methods together with high repetition rate experimental capabilities pave the way towards precise control of laser-plasma experiments, aiming to produce relativistic particles and bright radiation at smaller sizes and more affordable sizes than those of traditional sources.

% \section{Mid-infrared laser-plasma interactions}
% longer wavelength: larger ponderomotive force, less collisional absorption

\section{Dissertation outline}
The content of the dissertation is organized as follows:
% \begin{enumerate}
\begin{itemize}
\item Chap. \ref{chap:Theory} reviews the theoretical background on laser-plasma interactions at relativistic intensities. Topics include ultrashort pulse amplification theories, basic ionization mechanisms, and high-intensity laser interacting with single atoms, overdense plasmas, and underdense plasmas.

\item Chap. \ref{chap:Exp} describes the laser system, plasma target preparation, and diagnostics tools used in this dissertation. Chap. \ref{chap:Exp} also discusses the simulation methods and statistical methods assisting the experiments.

\item Chap. \ref{chap:Solid} covers three experiments where relativistic intensity laser pulses interact with solid-density plasmas. Sec. \ref{sec:HHG} discusses surface high-order harmonic generation using a 2.05 $\mu m$ laser, and part of the material has been published in \cite{beier2019relativistic}. Sec. \ref{sec:Atto} studies MeV-level attosecond electron bunches when a laser pulse is at grazing incidence onto the target, which has been published in \cite{lin2020towards}. Sec. \ref{sec:Xray} investigates characteristic x-ray emissions with a parametric study, including governing laser-plasma parameters such as laser wavelength, laser pulse energy, and preplasma profile.

\item Chap. \ref{chap:ML} covers three projects where statistical methods are applied to relativistic laser-plasma experiments at high repetition rates. Sec. \ref{sec:FocusOpt} presents focus optimization at relativistic intensity with high numerical aperture and adaptive optics, which has been published in \cite{lin2018focus}. Sec. \ref{sec:MIRLWFA} demonstrates closed-loop optimization of laser-wakefield acceleration driven by ultrashort ($\sim$100 fs) mid-infrared ($\sim$ 3.9$\mu m$) laser pulses, which has been published in \cite{lin2019adaptive}. Sec. \ref{sec:MLLWFA} explores the applications of machine learning techniques in relativistic laser-plasma experiments beyond optimization purposes.

\item Chap. \ref{chap:Conclusion} concludes and summarizes the work done in this dissertation and provides an outlook on potential future work.

% \end{enumerate}
\end{itemize}

Since this thesis covers a variety of topics in relativistic laser-plasma interactions, detailed introductions of each topic will be given at the beginning of the corresponding chapter and section.

\chapter{Physical Theories}
\label{chap:Theory}
In this chapter, the fundamental theory involved in short-pulse laser-plasma interactions will be reviewed. The primary references for this chapter include the textbook by Andrew Weiner \cite{weiner2011ultrafast} on ultrafast lasers, the textbook by Paul Bellan on plasma physics \cite{bellan2008fundamentals}, the textbooks by Paul Gibbon \cite{gibbon2005short} and by William Kruer \cite{kruer2019physics} on laser-plasma interactions, the review article by Esarey \textit{et al.} \cite{esarey2009physics} on laser-wakefield accelerators, and the PhD dissertations by Zhaohan He \cite{he2014laser}, by Peter Kordell \cite{kordell2019collisionless}, by Paul Campbell \cite{campbell2019laboratory}, and by Amina Hussein \cite{hussein2019laser}.

\section{Ultrashort pulse amplification}
\label{sec:theoryLaser}
Reaching relativistic laser intensities ($>10^{18} W\cdot cm^{-2}$) and above is necessary to study relativistic plasma physics and other strong-field physics phenomena, as is illustrated in Fig. \ref{FigIntroZhang}. A typical mode-locked oscillator generates $\sim100$ fs, $\sim$nJ pulses, which corresponds to maximum intensities $\sim10^{12} W\cdot cm^{-2}$. Therefore, amplification is needed. This section will introduce the concepts of broadband (femtosecond) amplifiers from a theoretical point of view.

\begin{figure}[H]
\centering
\includegraphics[width=0.8\columnwidth]{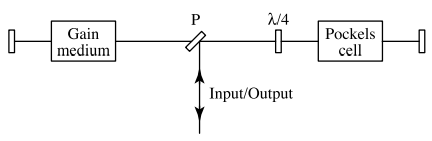}
\caption{A sketch of a generic regenerative amplifier. P: thin-film polarizer; $\lambda/4$: quarter-wave plate. Adapted from Ref. \cite{weiner2011ultrafast} with permission.}
\label{FigTheoryRegen}
\end{figure}

The basic principle of femtosecond pulse amplification is not different from that of narrowband amplifiers: extract energy from gain media. However, the gain material for femtosecond pulses must possess a broad optical bandwidth. Solid-state materials such as Ti:Sapphire with bandwidth $\sim10^{14}$ Hz thus dominates femtosecond amplifier technology. There are two common configurations of the amplifiers: the regenerative amplifier scheme and the multipass amplifier scheme. In the regenerative amplifier scheme the pulse to be amplified is trapped inside a laser resonator using a polarization gating approach, as is shown in Fig. \ref{FigTheoryRegen}. Light in the resonator passes repeatedly through a thin-film polarizer, a quarter-wave plate, and an electro-optic Pockels cell. The Pockels cell acts like another quarter-wave plate when a voltage step ($\sim kV$) is applied to it, enabling polarization switch in the regenerative amplifier cavity so that the polarizer can either reflect or transmit the beam.

\begin{figure}[ht]
\centering
\includegraphics[width=0.8\columnwidth]{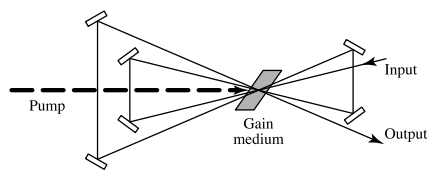}
\caption{A sketch of a generic multipass amplifier. Adapted from Ref. \cite{weiner2011ultrafast} with permission.}
\label{FigTheoryMultipass}
\end{figure}

The multipass configuration is relatively straightforward. As is illustrated in Fig. \ref{FigTheoryMultipass}, a series of mirrors is used to pass the beam through the amplifier crystal sequentially at different angles. Thus, the different passes are separated geometrically. Geometric complexity typically limits the number of passes to eight or so in femtosecond amplifiers. 

The main difference between these two configurations lies in the beam separation. In regenerative amplifiers, the beams are separated using the time-gating of the polarization, while in multipass amplifiers, the beams are separated spatially from different angles. Both schemes are commonly used in today's high-power laser facilities. One would think that ultrashort pulses can be amplified to extreme intensities by adding amplifiers to the system. However, beams in the amplifiers are subject to fundamental laws that limit the accessible peak power, such as nonlinear beam propagation.

\subsection{Nonlinear beam propagation}
As an intense laser beam propagates in a nonlinear index material, it accumulates phase shift. This can be characterized by the B-integral:

\begin{equation}
    B = \frac{2\pi}{\lambda} \int n_2 I(z) dz
    \label{EqTheoryBintegral}
\end{equation}

where $I(z)$ is the intensity along the propagation axis, z is the position in beam propagation direction, and $n_2$ is the nonlinear refractive index defined as follows:

\begin{equation}
    n = n_0 + n_2\cdot I
    \label{EqTheoryN2}
\end{equation}

where $n_2\cdot I$ is the change in refractive index caused by an intense laser beam. In high-power laser systems, $B<1$ is usually required to avoid serious phase shift accumulation. The direct consequence of having a large B-integral is self-focusing, where the Kerr effect builds up a nonuniform phase delay decreases radially. This is also called "Kerr lensing" as it behaves as a lens. As a result, the laser beam is focused to a smaller radius as it propagates, leading to an increased intensity which can exceed the damage threshold of the amplifier gain medium. Furthermore, self-focusing can amplify small modulations in the spatial intensity profile of the beam and result in instabilities. The requirement on the B-integral imposes a fundamental limit on the maximum intensity achievable in femtosecond amplifiers, especially in regenerative amplifiers with longer material path lengths.

\subsection{\acf{CPA}}

\begin{figure}[H]
\centering
\includegraphics[width=0.95\columnwidth]{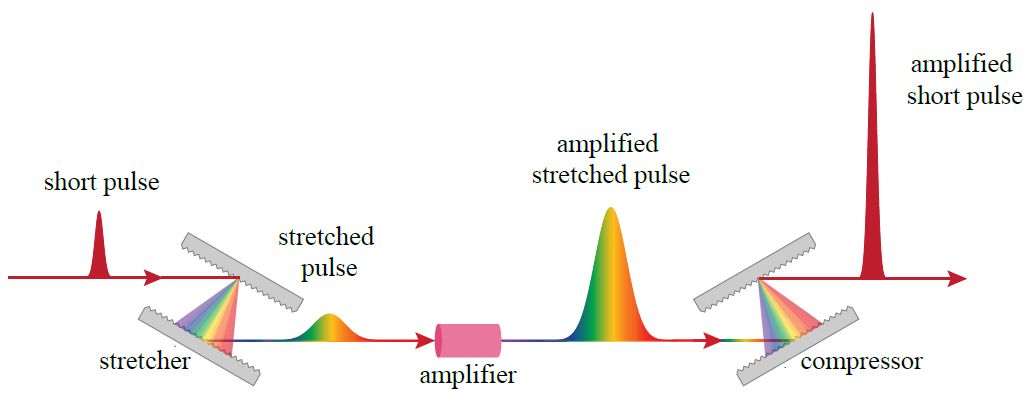}
\caption{Chirped pulse amplification concept. Adapted from Ref. \cite{campbell2019laboratory} with permission.}
\label{FigTheoryCPA}
\end{figure}

The limitation on laser intensity can be overcome by the chirped pulse amplification technique, as is illustrated in Fig. \ref{FigTheoryCPA}. The input pulse is passed through a first dispersive system (stretcher), usually a pair of diffraction gratings, leading to a highly chirped, temporally stretched pulse. The stretched pulse preserves the bandwidth of the input pulse but lowers the peak power by the stretching factor. The stretched pulse is then brought into the amplifiers introduced previously. The peak intensity of the chirped pulse is reduced to the extent that damage in the amplifiers is no longer a concern. A second dispersive system (compressor), usually a pair of diffraction gratings with the sign of dispersion opposite that in the stretching stage, is used to compress the amplified pulse to ultrashort pulse duration. With proper dispersion balance in the compressor, bandwidth-limited pulses with no chirp can be achieved. 

There are a few practical challenges regarding the \acs{CPA} challenges. Since the pulse goes through a dispersive medium such as the amplifier gain material in the middle stage, it experiences extra dispersion. This extra dispersion also needs to be compensated in the compressor. However, it is challenging to balance the low-order dispersion and the high-order dispersion at the same time. Besides, the spatial beam profile in the gratings requires great care and caution. Because optical frequencies separate within the grating pair, spatial phase errors can translate into spectral phase errors. Possible sources of such errors include spatial clipping at the edge of optics and aberrations in the focusing optics in the stretcher.

\begin{figure}[ht]
\centering
\includegraphics[width=0.9\columnwidth]{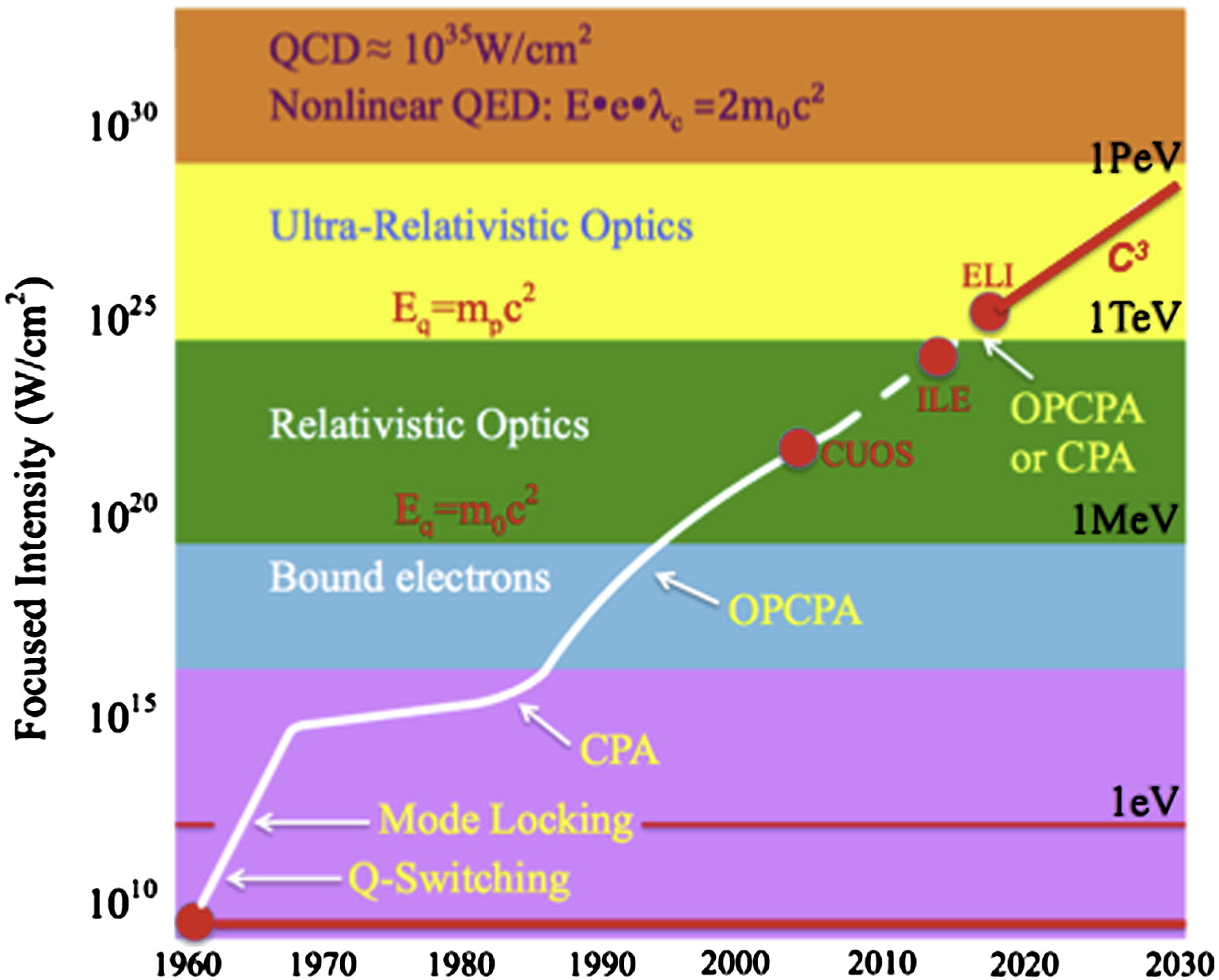}
\caption{Laser focus intensity vs. years. Adapted from Ref. \cite{mourou2012exawatt} with permission.}
\label{FigTheoryMourou}
\end{figure}

Utilizing the \acs{CPA} technique, laser intensities above $10^{15} W\cdot cm^{-2}$ have become accessible, as is shown in Fig. \ref{FigTheoryMourou}. A record intensity at $2\times10^{22} W\cdot cm^{-2}$ has been demonstrated \cite{yanovsky2008ultra} at \acs{CUOS} in 2008, and more recently at $5.5\times10^{22} W\cdot cm^{-2}$ at CoReLS in South Korea \cite{yoon2019achieving}.

\subsection{\acf{OPA}}
% Weiner page 245
The primary reference for this subsection is the detailed review article by Manzoni \textit{et al.} in Ref. \cite{manzoni2016design}.
The amplification methods discussed above are all based on population inversion in the laser gain media. A primary disadvantage of these amplified pulses is fixed wavelength operation with moderate tunability, constrained by the narrow wavelength range of emission in population inversion. However, having frequency tunability is of increasing interest to the high field science community, especially for studying wavelength-sensitive phenomena and for pump-probe experiments. \acf{OPA} is the most common method to enable frequency tuning in high-intensity femtosecond lasers. It utilizes the second-order nonlinear optical effect and provides amplification over a broad wavelength range from visible to mid-infrared. 
A pump beam at high intensity and high frequency $\omega_p$ overlaps spatially and temporally with a signal beam at low intensity and low frequency $\omega_s$ in a nonlinear crystal, resulting in energy transfer from the pump beam to the signal beam. To meet the conservation of energy, a third idler beam is generated at frequency $\omega_i$:
\begin{equation}
    \omega_p = \omega_s + \omega_i
    \label{EqTheoryOPAenergy}
\end{equation}

The intensities of the signal and idler beams depend on the efficiency in the \acs{OPA} process. To have the highest efficiency, the phase-matching condition, is required:
\begin{equation}
    \vec{k_p} = \vec{k_s} + \vec{k_i}
    \label{EqTheoryOPAphaseMatch}
\end{equation}
where $\vec{k}$ is the wave vector and Eq. \ref{EqTheoryOPAphaseMatch} can be regarded as the conservation of momentum.

In addition to transferring energy from a fixed frequency pump beam to a tunable frequency signal beam, the \acs{OPA} also reduces the nanosecond prepulse caused by the \acf{ASE}, leading to a higher contrast ratio than it has in \acs{CPA} systems. It is because the \acs{OPA} is pumped by femtosecond lasers, and the amplification happens only when it is pumped. On the contrary, the contrast ratio in \acs{CPA} systems is usually lower because the nanosecond pump lasers in the amplifiers amplify the spontaneous emission together with the main pulse. Having a high contrast ratio is crucial to many laser-plasma experiments with solid targets, in which the preplasmas caused by the prepulses play a significant role. It is desired to have a controllable prepulse to govern the preplasma profile in such experiments.

Because of the merits of \acs{OPA} mentioned above, it is integrated with the \acs{CPA} concept for \acf{OPCPA}. Instead of the Ti:Sapphire crystals used in most femtosecond \acs{CPA} systems, \acs{OPCPA} systems are usually pumped with the energetic picosecond pulses generated by Nd-doped or Yb-doped crystals. To match the long pulse duration of the pump, the seed is stretched to a similar duration to achieve efficient energy extraction in the amplification stage. It is then compressed to a near-bandwidth-limited pulse, analogous to the \acs{CPA} scheme. \acs{OPCPA} produces few-cycle pulses pumped by energetic pump beams, and is therefore regarded as one of the most promising routes towards extreme laser intensities.
% do i need this section here?

\section{Ionization mechanisms}
\label{sec:theoryIonization}
% Karl's notes. Gibbon Sec 2.
In the previous section, we have introduced methods to obtain high-intensity lasers. Plasma sources, on the other side, are also crucial for laser-plasma interactions. This section will introduce the basic ionization mechanisms that produce free electrons and charged ions as plasma sources.

Ionization happens when an electron receives an external kick to escape from the nucleus. An intuitive way to picture this is the Bohr model. The distance between an electron and the nucleus in a hydrogen atom is $a_B=5.3\times10^{-11}$ m. From Coulomb's law, the electric field that the electron experiences due to the nucleus is:
\begin{equation}
    E_a = \frac{e}{4\pi\epsilon_0 a_B^2} = 5.1\times10^9 V/m
    \label{EqTheoryBohrField}
\end{equation}
This electric field corresponds to the so called atomic intensity:
\begin{equation}
    I = \frac{1}{\mu_0} |\langle E\times B\rangle | = \frac{\epsilon_0 c}{2} E_a^2 = 3.5\times10^{16} W\cdot cm^{-2}
    \label{EqTheoryBohrIntensity}
\end{equation}
where $\epsilon_0$ and $\mu_0$ are the vacuum permittivity and permeability, respectively. That being said, ionization will happen as long as the peak laser intensity exceed $3.5\times10^{16} W\cdot cm^{-2}$. In fact, this is a simplified model, and ionization can occur through more complicated mechanisms.

\subsection{Photon-ionization}
The most straightforward mechanism is photo-ionization, also described as the well-known photoelectric effect. Photo-ionization happens when the photon energy exceeds the ionization potential of an electron, which is the amount of energy required to remove the electron from an atom. The ionization potential is the lowest for ionizing an electron from a neutral atom, and increases for ionization an electron from more positively-charged ions. For a ground-state electron in a hydrogen atom, the ionization potential is 13.6 eV, corresponding to a photon wavelength of $91$ nm. It is well below the wavelength of the high-intensity lasers described in the previous section, such as $\sim800$ nm from Ti:Sapphire lasers and $1064$ nm from Nd:YAG lasers. Therefore, photo-ionization is not likely to happen in most laser-plasma experiments at relativistic intensities.

However, photo-ionization is the dominant ionization mechanism for \acf{XFELs} interacting with atoms. The photon energy of typical \acs{XFELs} is between 100 eV and a few tens of keV, which exceeds the ionization potential of most ground state electrons but matches the ionization potential of inner-shell electrons. When interacting with \acs{XFELs}, an atom will be left in a transient state after losing an inner-shell electron due to photoionization. Although the transient state usually stabilizes in the form of radiation (characteristic x-ray emission, to be discussed in Sec. \ref{sec:Xray}), it can also be followed by the emission of another electron, called an Auger electron, to fill the inner-shell vacancy.

% \subsection{Multi-Photon Ionization}
Although the photon energy of $\sim1\;\mu m$ lasers can not exceed typical ionization potentials, the sum of multiple photons can. This phenomenon is called the \acf{MPI}, where two or more photons are absorbed simultaneously to provide enough energy to ionize an electron. The rate of \acs{MPI} can be estimated using perturbation theory as:
\begin{equation}
    \Gamma_n = \sigma_n\cdot I_L^n
    \label{EqTheoryMPI}
\end{equation}
where n is the number of photons, $\sigma_n$ is the cross-section which decreases with n, and $I_L$ is the laser intensity. This process can occurs at laser intensities $>10^{10} W\cdot cm^{-2}$. The rate of \acs{MPI} can be further increased when resonance absorption is achieved to an excited intermediate state so that another photon can ionize the atom. This process is called the \acf{REMPI}.

The ionized electron can pick up extra energy from the photons through \acf{ATI}:
\begin{equation}
    E_f = (n+s)\hbar\omega - I_p
    \label{EqTheoryATI}
\end{equation}
where s is the number of excess photons, $E_{ion}$ is the ionization potential of the electron, and $E_{f}$ is the final kinetic energy of the electron.

\subsection{Tunnel Ionization}
As the laser intensity increases to above $10^{14} W\cdot cm^{-2}$ and gets close to the atomic intensity defined in Eq. \ref{EqTheoryBohrIntensity}, the electric field of the laser pulse is within an order of magnitude of the Coulomb field on the electron given by Eq. \ref{EqTheoryBohrField}. Consequently, the perturbation theory assumed in \acs{MPI} is violated and the atomic binding energy is disturbed, allowing the electron to tunnel through the disturbed potential.

Tunneling ionization and \acs{MPI} can be distinguished using the Keldysh parameter \cite{keldysh1965ionization}:
\begin{equation}
    \gamma_K = \frac{Time\;to\; tunnel\; out}{Period\; of\; laser\; field} \sim \sqrt{\frac{I_p}{U_p}}
    \label{EqTheoryKeldysh}
\end{equation}
where $I_p$ is the ionization potential, and $U_p$ is the ponderomotive potential to be derived in Eq. \ref{EqTheorysinglePondero5}. Tunneling ionization dominates when $\gamma_K<1$, while \acs{MPI} dominates when $\gamma_K>1$. It has to be pointed out that the ponderomotive potential scales with $(I\cdot\lambda)^2$, thus tunneling ionization favors larger laser fields and longer laser wavelengths. The threshold intensity to have a larger rate in tunneling ionization than in \acs{MPI} can be approximated as:
\begin{equation}
    I_L [W\cdot cm^{-2}] > 5.4\times10^{12} \frac{I_p [eV]}{(\lambda [\mu m])^2}
    \label{EqTheoryTunnelThreshold}
\end{equation}

\subsection{Direct field ionization}
% aka, barrier suppression ionization
Direct field ionization, also known as barrier suppression ionization, happens when the laser intensity increases to a point where the laser field completely dominates over the potential barrier. In this case, the potential that the electron experiences is a superposition of the Coulomb potential and a laser electric field $E_L$:
\begin{equation}
    V(x) = \frac{-Ze^2}{4\pi\epsilon_0 x}-\frac{eE_L}{4\pi\epsilon_0}
    \label{EqTheoryDFI}
\end{equation}
where x is the position in space. The maximum value of V(x) occurs where $x_{max}=(ZE_L/e)^{1/2}$ by setting $\frac{dV(x)}{dx}=0$. If this maximum potential exceeds the ionization potential of the electron ($V_{max}=I_p$), direct field ionization takes over. Therefore, the threshold laser field is obtained:
\begin{equation}
    E_{L,th} = 4\pi^2\epsilon_0^2\cdot \frac{I_p^2}{Ze^3}
    \label{EqTheoryDFIelectric}
\end{equation}
The threshold intensity, called appearance intensity, is:
\begin{equation}
    I_{app} = 8\pi^4\epsilon_0^5\cdot\frac{c I_p^4}{Z^2 e^6} \sim 4\times10^9\frac{(I_p[eV])^4}{Z^2} [W\cdot cm^{-2}]
    \label{EqTheoryDFIintensity}
\end{equation}
For the ground-state electron in a hydrogen atom whose ionization potential is 13.6 eV, the appearance intensity is calculated to be $1.4\times10^{14} W\cdot cm^{-2}$. This intensity is two orders of magnitude lower than the atomic intensity from Eq. \ref{EqTheoryBohrIntensity}.

\subsection{Collisional ionization}

As the laser field ionizes electrons from atoms, the emitted electrons can collide with ions and ionize them to release more electrons, causing an avalanche effect. This is known as the collisional ionization or avalanche ionization. The rate of collisions for a single particle with a field of target particles with density $n_T$ is:
\begin{equation}
    r(v,n) = n_T\cdot\sigma(v)\cdot v
    \label{EqTheoryCollisionRate}
\end{equation}
where $\sigma$ is the cross section of the process and v is the particle velocity. For example, the collision rate for K-shell electrons is $n_T\cdot\sigma_K(v)\cdot v$ given the cross-section for K-shell ionization $\sigma_K(v)$. To get the collisional ionization rate, we need to sum over possible collisional events in all shells for a total cross-section $\sigma_{total}=\sum_i N_i\sigma_i(v)$:
\begin{equation}
    r_{col}(v,n) = n_T\cdot \sum_i N_i\sigma_i(v) \cdot v
    \label{EqTheoryCollisionTotal}
\end{equation}
where $\sigma_i$ is the cross-section for the $i_{th}$ shell and $N_i$ is the number of electrons in that shell.
% The electron-ion collision frequency is approximated to be:
% \begin{equation}
%     \nu_{e,i} [s^{-1}] \simeq 3\times 10^{-6}\; ln(\Lambda)\; Z\; n_e[cm^{-3}]\; (\theta[eV])^{-3/2}
%     \label{EqTheoryCollisionIonize}
% \end{equation}
% where $ln(\Lambda)$ is the Coulomb logarithm and $\theta$ is the electron temperature. For hot plasmas, electron-ion collision happens on a lower rate.

% Free electrons can also collide with neutrals if the plasma is weakly ionized. Unlike the collision between charged particles governed by the Coulomb force, electron-neutral is governed by short-range forces. In this case, the neutral can be regarded as a hard body and the cross-section is simply defined by its atomic radius. The electron-neutral collision frequency is therefore:
% \begin{equation}
%     \nu_{e,n} = n_e\cdot\sigma\cdot v
%     \label{EqTheoryCollisionIonizeNeu}
% \end{equation}
% where $\sigma$ is the cross-section and v is the velocity. 
 
Note that the collision rates scale with the electron density. For a high target density or a long pulse duration, there will be enough collisional events such that collisional ionization dominates over the laser ionization mechanisms. Collisional ionization at solid densities can be complicated to model and sometimes statistical approximations are used, and it remains an ongoing research field. It also worth emphasizing that the collisional ionization process is separate from electron-ion collision processes in plasma physics for thermalization of the charged particle population.

\section{Laser interacting with single electrons}
\label{sec:theorySingleElectron}
When an electron is ionized by intense femtosecond lasers, it is automatically subject to the strong laser field. In this section, we will picture its dynamics in a simplified plane wave as well as the time-averaged effects caused by a tightly-focused short pulse.
\subsection{Single electron motion in an electromagnetic plane wave}
% use LinSciRep 
Consider a free electron in an electromagnetic wave, its equation of motion is given by the Lorentz equation:
\begin{equation}
    \frac{d\mathbf{p}}{dt} = -e(\mathbf{E}+\mathbf{v}\times\mathbf{B})
    \label{EqTheorysingleEOM}
\end{equation}
where $\mathbf{p}$ is the electron momentum and $\mathbf{v}$ is the electron velocity, and the electric field $\mathbf{E}$ and the magnetic field $\mathbf{B}$ can be described by the vector potential of the wave $\mathbf{A}$:
\begin{equation}
    \mathbf{E} = -\frac{\partial \mathbf{A}}{\partial t},\;\;\;\;\; \mathbf{B} = \nabla\times\mathbf{A}
    \label{EqTheorysingleEB}
\end{equation}
For a linear polarized plane wave propagating along the $\hat{z}$ axis, the vector potential is:
\begin{equation}
    \mathbf{A}(z,t) = A_0 sin(k z-\omega t) \hat{x}
    \label{EqTheorysingleA}
\end{equation}
The vector potential can be normalized by considering the non-relativistic regime where $\mathbf{p}=m\mathbf{v}$ and $\mathbf{E}\gg\mathbf{v}\times\mathbf{B}$. Assuming the electron is initially at rest, the equation of motion in Eq. \ref{EqTheorysingleEOM} can be easily integrated to obtain:
\begin{equation}
    \mathbf{v_{os}} = \frac{-ie\mathbf{E}}{m\omega},\;\;\;\;\; \mathbf{a}=\frac{\mathbf{v_{os}}}{c}=\frac{e\mathbf{E}}{m\omega c}
    \label{EqTheorysingleQuiver}
\end{equation}
where $\mathbf{v_{os}}$ is called the quiver velocity and $\mathbf{a}$ is the normalized vector potential. Recall Eq. \ref{EqTheoryBohrIntensity} that the laser intensity can be calculated from the electric field as $I=\frac{\epsilon_0 c}{2} E^2$, the normalized vector potential can be expressed in terms of the laser intensity:
\begin{equation}
    a^2 = \langle\frac{eE}{m\omega c}\rangle^2 = \frac{e^2 I\lambda^2}{2\pi^2\epsilon m^2c^5}
    \label{EqTheorysingleVec}
\end{equation}
The peak normalized vector potential, $a_0$, is defined at the peak laser intensity after plugging in the scientific constants into Eq. \ref{EqTheorysingleVec}:
\begin{equation}
    a_0 = \sqrt{\frac{I_{peak}[W cm^{-2}](\lambda[\mu m])^2}{1.37\times10^{18}}}
    \label{EqTheorysinglea0}
\end{equation}

When $a_0>1$, the quiver motion of the electron becomes relativistic and the laser-plasma interaction is said to be relativistic. In the relativistic regime, however, the Lorentz factor $\gamma$ needs to be taken into account so that the electron momentum becomes $\mathbf{p}=\gamma m\mathbf{v}$. By definition, the Lorentz factor is related with momentum in the following identity:
\begin{equation}
    \gamma^2-(p/mc)^2=1
    \label{EqTheorysingleGamma}
\end{equation}
To solve the equation of motion Eq. \ref{EqTheorysingleEOM}, the energy equation is also needed:
\begin{equation}
    \frac{d}{dt}\gamma mc^2 = -e \mathbf{v} \cdot \mathbf{E}
    \label{EqTheorysingleEnergy}
\end{equation}

% For simplicity, we will set m=c=1 in the next few paragraphs. 
Combining  Eq. \ref{EqTheorysingleEB} and Eq. \ref{EqTheorysingleEnergy} with Eq. \ref{EqTheorysingleEOM}, we can derive the momentum change of an electron in a two-dimensional plane wave:
\begin{equation}
    p_{x}-p_{x 0}=amc,\;\;\;\;\; p_{z}-p_{z 0}=\gamma-\gamma_{0}=\frac{a^{2}+2a\cdot p_{x0}}{2\left(\gamma_{0}-p_{z 0}\right)}mc
    \label{EqTheorysingleMomentum}
\end{equation}
where $p_{x}$ is the transverse momentum and $p_{z}$ is the longitudinal momentum. If the electron is initially at rest, Eq. \ref{EqTheorysingleMomentum} reduces to:
\begin{equation}
    p_{x}=amc,\;\;\;\;\; p_{z}=(a^2/2) mc
    \label{EqTheorysingleMomentum0}
\end{equation}

It has to be emphasized that an electron does not pick up any energy or momentum over a laser cycle in a plane wave if it is in phase with the laser field. This is shown mathematically in Eq. \ref{EqTheorysingleMomentum} and Eq. \ref{EqTheorysingleMomentum0} since $\mathbf{a}(\phi)=\mathbf{a}(\phi+2\pi)$.

\subsection{Ponderomotive force}
\begin{figure}[H]
\centering
\includegraphics[height=0.8\textheight]{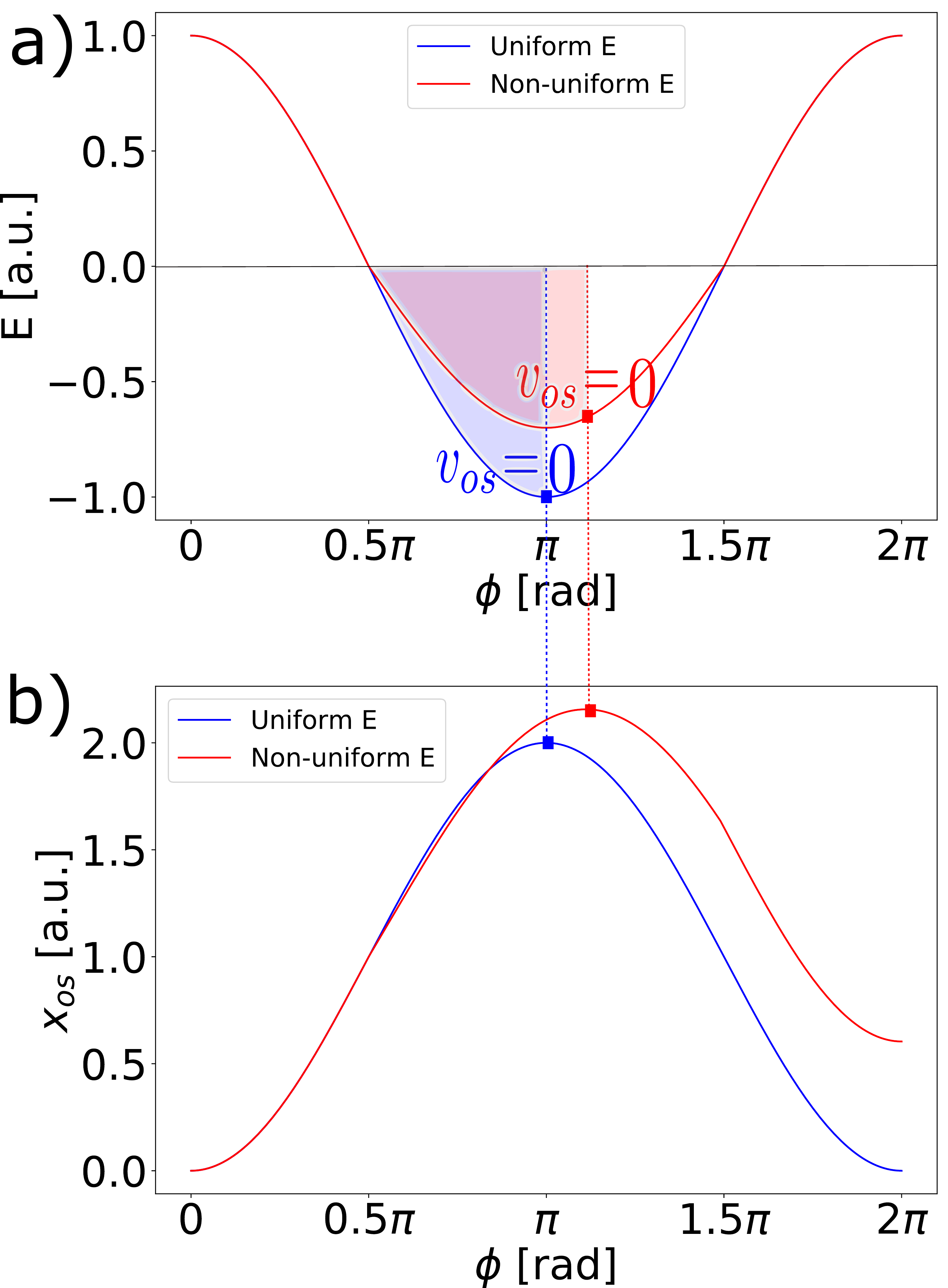}
\caption{Electric field (a) and oscillation amplitude (b) vs. laser phase in uniform (blue) and non-uniform (red) electric fields.}
\label{FigTheoryPonderomotive}
\end{figure}
In the previous subsection where a plane wave is assumed, the perpendicular motion of an electron can be described by the quiver oscillation. Integrating the oscillation velocity in Eq. \ref{EqTheorysingleQuiver} gives the quiver amplitude:
\begin{equation}
    \mathbf{x_{os}} = \frac{-e\mathbf{E}}{m\omega^2} + \mathbf{x_{0}}
    \label{EqTheorysingleQuiverAmp}
\end{equation}
The electron always set back to the starting position $\mathbf{x_{0}}$ after a laser period. However, this is no longer the case if the electric field strength is not uniform in space. Fig. \ref{FigTheoryPonderomotive} sketches the oscillation trajectory of an electron in a uniform electric field and in a non-uniform electric field. In the latter case, the electric field (in red) remains the same as the uniform field when $x_{os}<1$ but is set lower when $x_{os}>1$. As a result, the quiver velocity does not reduce to 0 at $\phi=\pi$ but at a later phase, such that the shaded area in red equals the shaded area in blue. Note that the phase where $v_{os}=0$ corresponds to the turning point where the tangent of the oscillation trajectory is 0, as is shown in Fig. \ref{FigTheoryPonderomotive}. Entering the region of a weaker electric field, the electron moves an extra distance in $\hat{x}$ before turning back. A similar phenomenon occurs in the second half of the laser period, and the red trajectory does not set back to 0 as the blue trajectory does by $\phi=2\pi$. Therefore, the electron drifts towards the region of a weaker electric field over a laser period. This time-averaged effect due to spatial gradient in the electric field strength is known as the ponderomotive force.

The ponderomotive force plays a key role in relativistic laser-plasma interactions because focusing a short-pulse laser for relativistic intensities always leads to an intensity gradient in both space and time. Over many laser cycles, electrons get expelled and eventually move out of the region of highest intensity.

The mathematical expression for the ponderomotive force can be derived by Taylor expanding the equation of motion (Eq. \ref{EqTheorysingleEOM}) in  $\hat{x}$ about the $0^{th}$ order quiver motion \cite{kordell2019collisionless}:
\begin{equation}
    \mathbf{F} = \frac{d\mathbf{p}}{dt} = -e((\mathbf{x_{os}}\cdot\nabla)\mathbf{E}+\mathbf{v_{os}}\times \mathbf{B})
    \label{EqTheorysinglePondero}
\end{equation}

Plugging in Eq. \ref{EqTheorysingleQuiver} and Eq. \ref{EqTheorysingleQuiverAmp}, Eq. \ref{EqTheorysinglePondero} becomes:
\begin{equation}
    \mathbf{F} = -\frac{e^2}{m\omega^2}((\mathbf{E}\cdot\nabla)\mathbf{E}+\mathbf{E}\times(\nabla\times\mathbf{E})) = -\frac{e^2}{2m\omega^2}\nabla(E^2)
    \label{EqTheorysinglePondero2}
\end{equation}
Take into account the Lorentz factor $\gamma$ for relativistic motion and time average Eq. \ref{EqTheorysinglePondero2}, the ponderomotive force is \cite{quesnel1998theory}:
\begin{equation}
    \mathbf{F_p} = \langle F\rangle = -\frac{e^2}{2m\langle\gamma\rangle\omega^2}\nabla\langle E^2\rangle
    \label{EqTheorysinglePondero3}
\end{equation}
where $\langle \rangle$ denotes the average over a laser period, and the time-averaged Lorentz factor $\langle\gamma\rangle=\sqrt{1+\langle p_x\rangle/m^2c^2+\langle p_z\rangle/m^2c^2}$ from Eq. \ref{EqTheorysingleGamma}. In the weakly relativistic regime, a fair assumption to make is that the quiver motion dominates the electron momentum:
\begin{equation}
    \langle\gamma\rangle \simeq \sqrt{1+\langle p_x\rangle/m^2c^2} = \sqrt{1+\langle a\rangle^2}
    \label{EqTheorysinglegamma2}
\end{equation}
using the relation in Eq. \ref{EqTheorysingleMomentum0}. Note that $\langle a\rangle^2=a_0^2$ for circularly polarized laser pulses and $\langle a\rangle^2=a_0^2/2$ for linearly polarized laser pulses.
A more frequently used expression of the ponderomotive force is obtained by replacing $\mathbf{E}$ in Eq. \ref{EqTheorysinglePondero3} with $\mathbf{a}$ using Eq. \ref{EqTheorysingleQuiver}:
\begin{equation}
    \mathbf{F_p} = -\frac{e^2}{2m\langle\gamma\rangle\omega^2} \cdot (\frac{mc\omega}{e})^2 \nabla\langle a^2\rangle = \frac{mc^2}{2\langle\gamma\rangle} \nabla\langle a^2\rangle
    \label{EqTheorysinglePondero4}
\end{equation}
The ponderomotive potential, defined as $\mathbf{F}=-\nabla U$, is therefore:
\begin{equation}
    U_p = mc^2(\langle\gamma\rangle-1)
    \label{EqTheorysinglePondero5}
\end{equation}

\section{Laser interacts with overdense plasmas}
\label{sec:theoryOverdense}
\subsection{Plasma density profile}
The density of the plasmas near a solid target is characterized by a scale length $L_s$, and the plasma density profile is usually approximated by an exponential function:

\begin{equation}
    n_e = n_s\cdot \exp(-z/L_s),\;\;\;\;\;
    L_s = n_e\cdot \left\vert\frac{dn_e}{dz}\right\vert^{-1}
    \label{EqTheoryDensity}
\end{equation}
where $n_s$ is the solid density, $n_e$ is the electron density, and $\hat{z}$ axis is in the target normal direction that perpendicular to the solid surface. The density of the plasmas determines the plasmas oscillation frequency:
\begin{equation}
    \omega_{pe}^2 = \frac{n_e e^2}{m_e\epsilon_0}
    \label{EqTheoryFreq}
\end{equation}
The critical density $n_{cr}$ is defined as the plasma density where the plasma frequency matches the laser frequency:
\begin{equation}
    n_{cr} = \frac{m_e\epsilon_0\omega_L^2}{e^2}
    \label{EqTheoryCrit}
\end{equation}
Plasmas with densities higher than the critical density are defined as overdense plasmas, and plasmas with densities lower than the critical density are defined as underdense plasmas. The surface where the plasma density equals the critical density is called the critical surface. At the critical density, the plasmas become opaque for an electromagnetic wave. Namely, a laser pulse can not propagate into overdense plasmas as it will get reflected at the critical surface.

\begin{figure}[ht]
\centering
\includegraphics[width=0.95\columnwidth]{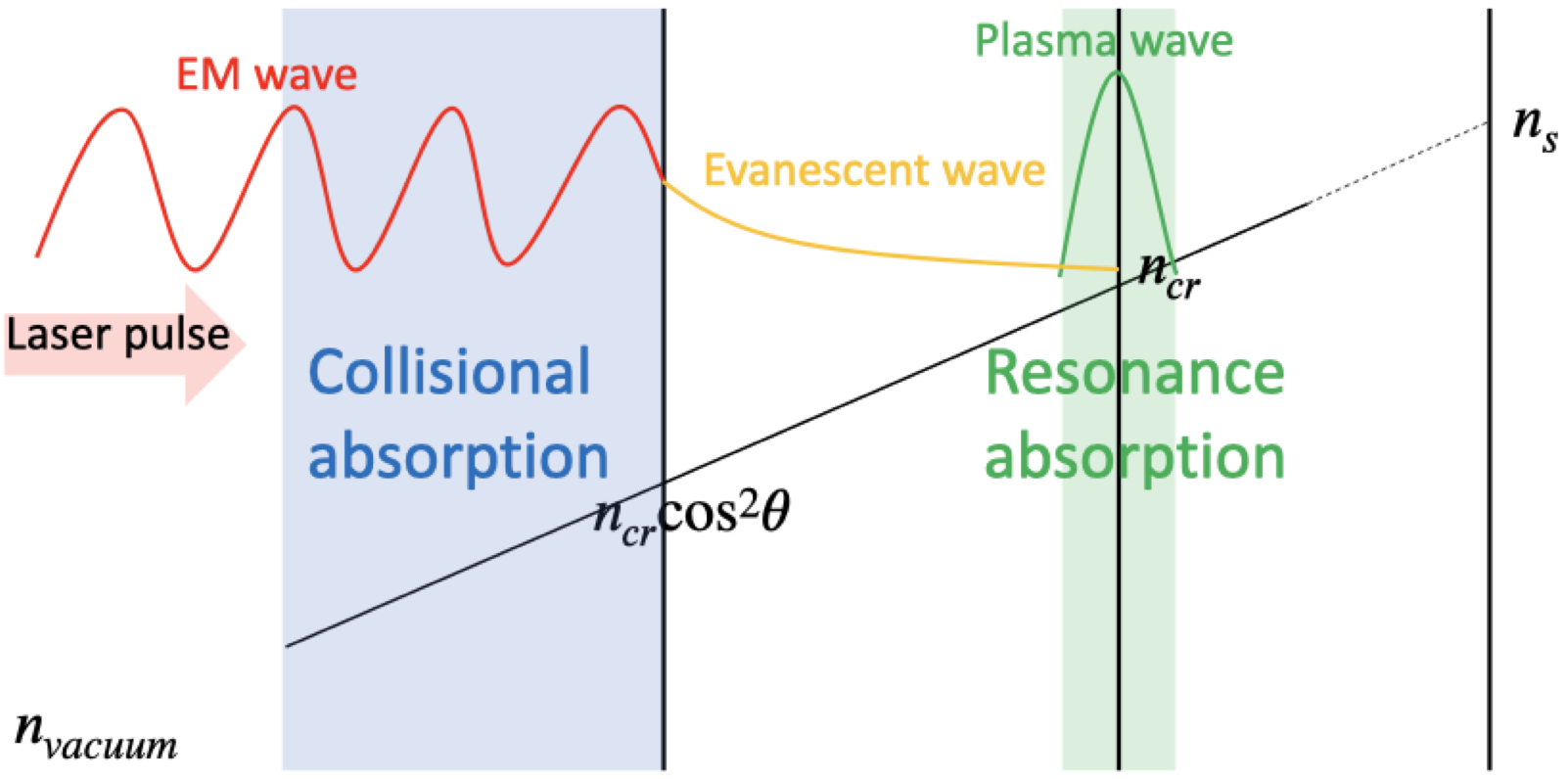}
\caption{Illustration of physical processes along the plasma density gradient.}
\label{FigTheorySolid}
\end{figure}

However, the effective critical surface can be located at even lower plasma densities. Consider a P-polarized interaction at oblique incidence with incident angle $\theta$ defined from the target normal. Note that in laser-plasma interactions, a "P-polarized interaction" refers to an interaction where the electric field of the laser pulse is parallel to the plane of incidence. In contrast, an "S-polarized interaction" refers to an interaction where the electric field of the laser pulse is parallel to the plane of the wave vector and the target normal. Instead of the critical surface, the laser pulse gets reflected at an "effective" critical surfaces where $n_e=n_{cr}cos^2\theta$. Beyond this effective critical surface, the energy of the laser electromagnetic wave is transferred to the plasma via an evanescent wave: 
\begin{equation}
    E_{evanescent}\sim E_0 \exp(-z/L_{skin}),\;\;\;\;\;
    L_{skin}=\frac{c}{\omega_{pe}} \sqrt{\frac{1}{1-\frac{\omega^2}{\omega^2_{pe}}cos^2\theta}}
    \label{EqTheorySkin}
\end{equation}
where $L_{skin}$ is the decay length, or the collisionless skin depth. A schematic of the concepts discussed above is shown in Fig. \ref{FigTheorySolid}.

It has to be pointed out that the plasma density profile can be modified during the interaction due to the ponderomotive force introduced in Sec. \ref{sec:theorySingleElectron}. As electrons are moved away from the region of highest intensity by the ponderomotive force, density steepening occurs around this region. In relativistic laser-solid interactions, the laser pulse is usually focused onto the critical surface. Therefore, density steepening with a decreased plasma density scale length around the critical surface is often expected.

\subsection{Laser absorption mechanisms}
In this section, we will go through the major absorption mechanisms when laser interacts with overdense plasmas. In short, the laser energy is primarily coupled into plasma waves via resonance absorption, to thermal electrons via collisional absorption, and to hot electrons via $J\times B$ heating. A detailed review on the absorption of short-pulse lasers by solids is available in Ref. \cite{wilks1997absorption} by Wilks and Kruer.

\subsubsection{Resonance absorption}
As the evanescent laser wave penetrates the overdense plasma, it drives the electrons to oscillate around. Meanwhile, the ions barely move because of their heavier mass, setting up a charge separation between the electrons and ions. The charge separation leads to an electrostatic field oscillating correspondingly. At the critical surface where the plasma oscillation frequency matches the laser frequency, electron plasma waves are resonantly excited. During this process, part of the laser energy is transferred to the plasma waves. This phenomenon that happens around the critical surface is called resonance absorption, as is illustrated in Fig. \ref{FigTheorySolid}.

In the same geometry of a P-polarization and an angle of incidence $\theta$, the fraction of laser energy that absorbed by the plasma waves is:
\begin{equation}
    f_{resonance} \approx \Phi^2(\tau)/2 
    \label{EqTheoryResAbsFrac}
\end{equation}
where the parameter $\Phi(\tau)$ is determined by the angle of incidence and the density scale length:
\begin{equation}
    \Phi(\tau)\approx 2.3\tau \exp{\left[-\frac{2}{3}\tau^3\right]}, \;\;\; \;\; \tau=(\omega_L L_s/c)^{1/3} 
    \label{EqTheoryResAbsFrac2}
\end{equation}
The optimal condition for resonance absorption occurs at:
\begin{equation}
    \theta_{max} \approx \arcsin{\left[0.8\left(\frac{c}{\omega_L L_s}\right)^{1/3}\right]}
    \label{EqTheoryResAbsAngle}
\end{equation}
where $\omega_L$ is the laser frequency.

\subsubsection{Vacuum heating}
Resonance absorption occurs when the plasma density is moderately developed, meaning that the density scale length is comparable to or larger than the quiver amplitude of the electrons defined in Eq. \ref{EqTheorysingleQuiverAmp}. It agrees with the schematic in Fig. \ref{FigTheorySolid} where the resonance absorption region in green scans across only a small portion along the density gradient.

However, vacuum heating, also known as Brunel absorption, takes over resonance absorption at a very sharp plasma density gradient ($L_s<0.1\lambda_L$). In such conditions, the quiver amplitude of the electrons becomes larger than the plasma density scale length, and the green region in  Fig. \ref{FigTheorySolid} would have expanded all the way to the left at much lower densities approaching vacuum. In other words, the plasma density is so sharp that the electrons driven by the laser field are oscillating across not the resonance region around the critical surface but instead the whole region between the solid density and vacuum. Although the laser field can only reach the skin depth defined in Eq. \ref{EqTheorySkin}, electrons are now able to penetrate deeper into areas of higher densities due to the relatively larger quiver amplitude compared to the density scale length. Since the collision rate increases with plasma density, the kinetic energy of the quiver electrons is absorbed by the plasmas in the region of high density via collision.

The fraction of laser energy absorbed in vacuum heating can be obtained by solving an equation system:
\begin{equation}
    \begin{cases}
    f_{vacuum} = \frac{1}{\pi a_{0}} f\left[\left(1+f^{2} a_{0}^{2} \sin ^{2} \theta\right)^{1 / 2}-1\right] \frac{\sin \theta}{\cos \theta} \\
    f = 1+\sqrt{1-f_{vacuum}}
    \end{cases}
    \label{EqTheoryVacHeat}
\end{equation}
where $a_0$ is the peak normalized vector potential and $\theta$ is the angle of incidence. The field amplification factor f in the second line in Eq. \ref{EqTheoryVacHeat} is defined as: 
\begin{equation}
    f=\frac{driving\; \mathbf{E}\; field}{incident\; \mathbf{E}\; field + reflected\; \mathbf{E}\; field}
    \label{EqTheoryVacHeat2}
\end{equation}
The first line in Eq. \ref{EqTheoryVacHeat} can be further simplified at the low-intensity limit and the high-intensity limit:
\begin{equation}
    f_{vacuum} = 
    \begin{cases} 
    \frac{a_0}{2\pi}\frac{sin^3\theta}{cos\theta} f^3, & \mbox{if } a_0\ll 1 \\ 
    \frac{1}{\pi}\frac{sin^3\theta}{cos\theta} f^2, & \mbox{if } a_0\gg 1 
    \end{cases}
    \label{EqTheoryVacHeat3}
\end{equation}

\subsubsection{J \texorpdfstring{$\times$}{X} B heating}
$\mathbf{J}\times \mathbf{B}$ heating is a relativistic effect that happens when the interaction is so strong that the electrons oscillate at nearly the speed of light and the $\mathbf{v}\times \mathbf{B}$ component in the Lorentz equation (Eq. \ref{EqTheorysingleEOM}) is no longer negligible. In $\mathbf{J}\times \mathbf{B}$ heating, electrons are directly accelerated into the high density plasma by the laser field, similar to the mechanism of vacuum heating. However, the oscillations governed by the $\mathbf{v}\times \mathbf{B}$ component happens at twice the laser frequency since both $\mathbf{v}$ and $\mathbf{B}$ are at laser frequency. The longitudinal force caused by a linearly polarized laser pulse is given by:
\begin{equation}
    f_z = -\frac{m}{4} \frac{\partial}{\partial z}v^2_{os}(z)\;(1-cos(2\omega_L t))
    \label{EqTheoryJXB}
\end{equation}
The first term is the DC ponderomotive force which modifies the plasma density profile, while the second AC term $cos(2\omega_L t)$ heats the electrons. $\mathbf{J}\times \mathbf{B}$ heating is most efficient in normal incidence geometry, but does not work for circularly polarized laser pulses.

\subsubsection{Collisional absorption}
While all three absorption mechanisms discussed above are collisionless, the collision between particles also plays a vital role in energy transfer, especially in \acf{ICF} processes. Collisional absorption, also known as Inverse Bremsstrahlung, happens before the laser pulse arrives at the critical surface, as is illustrated in Fig. \ref{FigTheorySolid}. Unlike the resonance absorption mechanisms, no extra plasma waves are present in collisional absorption and the laser field is the only field that matters. Recall Eq. \ref{EqTheorysingleMomentum} and Eq. \ref{EqTheorysingleMomentum0}, an electron does not pick up any energy or momentum over a laser cycle in a plane wave if it is in phase with the laser field. However, if the electron hits an ion during its oscillation period, it gets dephased from the laser field and gains energy. During these collisional events, a portion of the laser energy is absorbed by the electrons.

The fraction of energy transferred in collisional absorption depends in detail on the density profile of the plasma. For an S-polarized interaction with an exponential plasma density profile, the absorption coefficient is:
\begin{equation}
    f_{collisional} =  1-exp\left[-\frac{8}{3} \frac{\nu_{e,i} L}{c} cos^3\theta\right]
    \label{EqTheoryColAbsFrac}
\end{equation}

where $\nu_{e,i}$ is the electron-ion collisional frequency. For a linear density profile, Eq. \ref{EqTheoryColAbsFrac} becomes:
\begin{equation}
    f_{collisional} =  1-exp\left[-\frac{32}{15} \frac{\nu_{e,i} L}{c} cos^5\theta\right]
    \label{EqTheoryColAbsFrac2}
\end{equation}
Note that this calculated absorption coefficient can change during the interaction. It is because the electron-ion collisional frequency $\nu_{e,i}$ depends on the electron density and temperature, which are constantly varying as the interaction evolves. An empirical scaling law from experimental results states:
\begin{equation}
    f_{collisional} \propto\frac{Z^{3/2}\;\tau_L^{0.6}}{I_L^{0.4}\;\lambda_L^2 } 
    \label{EqTheoryColAbsFrac3}
\end{equation}
where Z is the number of electrons in the ion, $\tau_L$ is the duration of the incident laser pulse, $I_L$ is the laser intensity, and $\lambda_L$ is the laser wavelength.

\section{Laser-driven electron acceleration in underdense plasmas}
\label{sec:theoryUnderdense}
% cite papers: amina, jessica shaw
% cite esarey paper
% not my main focus - don't spend too much time on it! write ~2 page, and the complete it after defense
% use humboldt proposal
Laser-driven plasma-based accelerators were first proposed by Tajima and Dawson \cite{tajima1979laser}. When an intense laser pulse propagates into an underdense plasma, the ponderomotive force, as is introduced in Sec. \ref{sec:theorySingleElectron}, expels the plasma electrons from the driving laser pulse. As the electrons are displaced, a cavity of ions is formed behind the driver beam and co-propagates. The expelled electrons and the stationary ions then set up a charge separation, generating large-amplitude plasma waves (wakefields) in "bubble" structures. Electrons that trapped in the plasma waves expedience a longitudinal field exceeding GeV/m and are thus accelerated to relativistic velocities. 
\begin{figure}[ht]
\centering
\includegraphics[width=0.9\columnwidth]{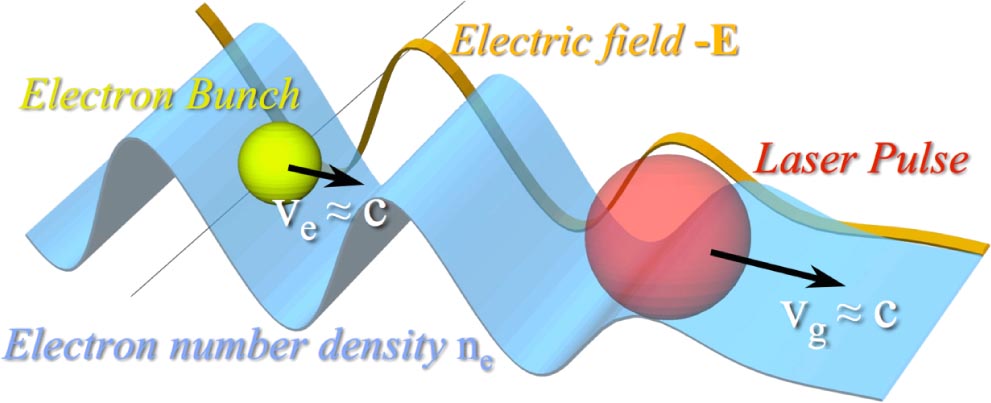}
\caption{\acs{LWFA} principle. Electrons are trapped by the plasma wave and accelerated by the longitudinal field, moving behind the laser pulse. Adapted from Ref. \cite{albert2014laser} with permission.}
\label{FigTheoryLWFAprinciple}
\end{figure}

The wakefield is driven most efficiently when the laser pulse length is on the order of the plasma period: $L\sim\lambda_p$, where the pulse length is simply the pulse duration times the group velocity. If the laser pulse length is long compared to the plasma wavelength, different electron acceleration mechanisms come to play, including the \acf{SM-LWFA} and the \acf{DLA}.

% self-modulated? copy MLLWFA intro here? 
To operate in the \acs{SM-LWFA} regime, the pulse power must be larger than the critical power ($P>P_{cr}$) and the pulse length must be longer than the plasma wavelength ($L > \lambda_p$). The critical power required for relativistic optical guiding is defined as \cite{sprangle1987relativistic}:
\begin{equation}
\label{EqTheoryUnderCriticalPower}
     P_{cr}=17.4\; n_{cr}/n_e [GW]
\end{equation}
In most cases where the laser power can not be increased easily, the critical power can be reached with either high plasma density or long laser wavelength ($n_{cr}\propto \lambda_L^{-2}$ from Eq. \ref{EqTheoryCrit}), according to Eq. \ref{EqTheoryUnderCriticalPower}. Increasing the plasma density also effectively decreases the plasma wavelength (Eq. \ref{EqTheoryFreq}), making it easier to achieve the $L > \lambda_p$ requirement. In \acs{SM-LWFA}, the laser power is so high ($P>P_{cr}$) that the pulse self-focuses, and the pulse is so long ($L > \lambda_p$) that it breaks up into a train of short pulses, where each of these short pulses has a pulse length matching the plasma wavelength. The plasma wave produces periodic regions of enhanced focusing and diffraction and modulates the laser pulse. Multiple "buckets" are formed as the pulse breaks into shorter pulses, and electrons are accelerated in these "buckets". Electron beams accelerated via \acs{SM-LWFA} usually have high beam charges but lack monoenergetic energy spectra.

To operate in the \acs{DLA} regime, an ultra-relativistic laser intensity is desired besides a long pulse duration (usually picosecond pulses). In \acs{DLA}, the magnetic field of such an intense pulse becomes so strong that the $\mathbf{v}\times\textbf{B}$ force dominates over other longitudinal forces, directly accelerating the electrons in the laser propagation direction. Instead of the bubble structures in \acs{LWFA}, electrons are accelerated in a quasi-static channel, which is formed because the ponderomotive force becomes strong enough to expel almost all the electrons.
% references : jessica shaw? luoise? feleci albert?

% cite papers: amina, jessica shaw
% cite esarey paper

\chapter{Experimental Methods}
\label{chap:Exp}
\section{Laser systems}
\label{sec:ExplaserSystem}
\subsection{The \acf{Lambda-cubed}}
Most of the experiments were set up and performed in the \acf{Lambda-cubed} at the \acf{CUOS} at the University of Michigan. The laser system was designed to study distinctive physical effects in the relativistic $\lambda^3$ regime \cite{nees2006distinctive}. The intensity of a short-pulse laser is constrained by the diffraction-limited focal spot size ($\sim\lambda^2$) and the single-cycle pulse duration ($\sim c/\lambda$), giving name to the \acs{Lambda-cubed} where the intensity $I\sim c/\lambda^3$. The laser system works at 480 Hz, and the output beam is at $\sim$800 nm wavelength, $\sim$35 fs pulse duration, and with up to $\sim$20 mJ energy. The \acs{ASE} contrast is $\sim10^8$ on the picosecond to nanosecond time scale. 
\begin{figure}[ht]
\centering
\includegraphics[width=0.95\columnwidth]{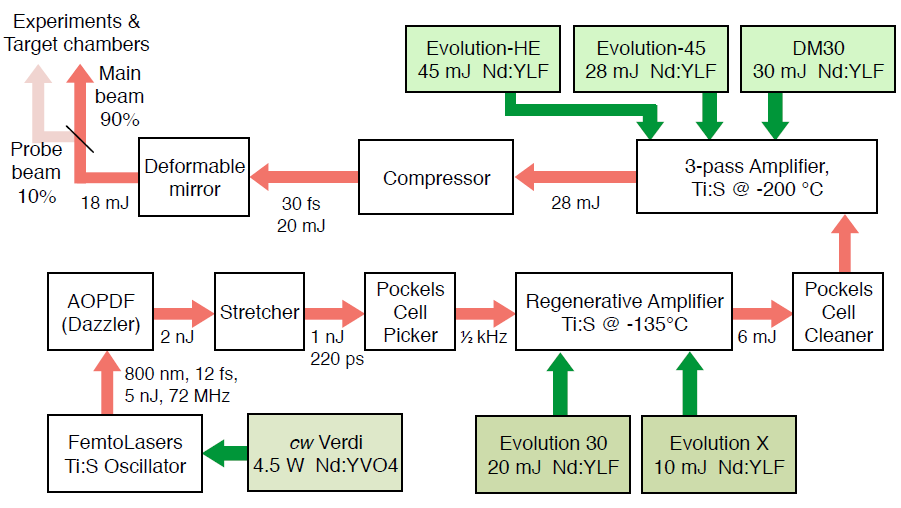}
\caption{Schematic of the \acs{Lambda-cubed}. Adapted from Ref. \cite{he2014laser} with permission.}
\label{FigExpLambda3}
\end{figure}

The \acs{Lambda-cubed} is seeded by pulses with 30 nm spectral bandwidth at $\sim80$ MHz from a Ti:Sapphire oscillator (Coherent MICRA 5W), in which the cavity mirror displacement is electronically controlled to enable auto mode-lock. Pulse shaping is realized through a dazzler, or an \acf{AOPDF}, to program the spectral phase and amplitude of the pulses. The \acs{Lambda-cubed} utilizes the \acs{CPA} technique to amplify the seed. After passing through a grating pair, the pulses are stretched to 40 ps. A regenerative amplifier with two pump lasers followed by a three-pass amplifier gives up to 24 mJ of energy before compression. The concepts of these amplification techniques have been introduced in Sec. \ref{sec:theoryLaser}. The amplified beam is re-compressed using a pair of transmission grating down to $\sim35$ fs and $\leq20$ mJ. The grating pair sits on a multidimensional translation stage, allowing $\mu m$ resolution adjustments in directions perpendicular to and parallel to the grating surfaces. Rotation of the grating pair with regard to the uncompressed beam is also available on a routine basis. A sketch of the \acs{Lambda-cubed} in 2014 \cite{he2014laser} is shown in Fig. \ref{FigExpLambda3} for reference.

\subsection{The mid-infrared \acs{OPA} at \acs{CUOS}}
The produced 800 nm beam from the \acs{Lambda-cubed} serves as the pump beam for an \acf{OPA}, generating a 2 $\mu m$ \acf{MIR} signal beam and a 1.3 $\mu m$ idler beam. \acs{OPA} is an amplification technique which enables frequency tuning, and its working principle has been introduced in Sec. \ref{sec:theoryLaser}.

\begin{figure}[ht]
\centering
\includegraphics[width=0.76\columnwidth]{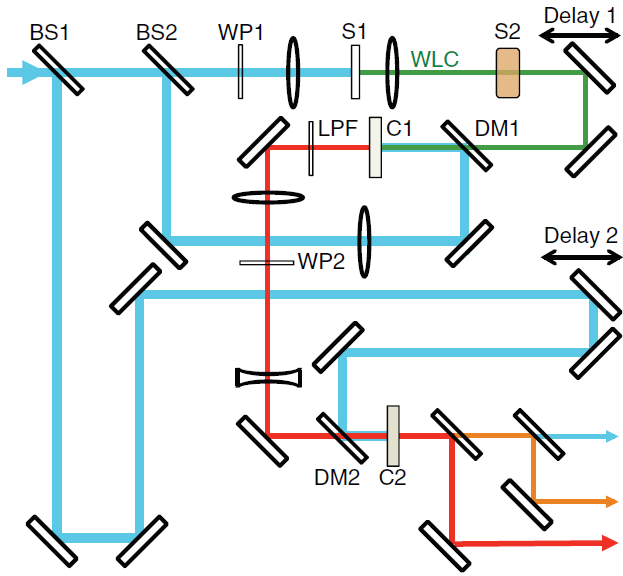}
\caption{Schematic of the \acs{OPA}. BS, beam splitter; WP, $\lambda$/2 wave plate; DM, dichroic mirror; S1, sapphire plate; S2, ZnSe plate; LPF, long-pass filter; C1, type I BBO crystal; and C2, type II BBO crystal. Adapted from Ref. \cite{xu2014nondegenerate} with permission.}
\label{FigExpOPA}
\end{figure}

The \acs{OPA} used in this work was designed by Xu \textit{et al.} \cite{xu2014nondegenerate} at Pennsylvania State University. The beam paths and the optical components in the \acs{OPA} are illustrated in Fig. \ref{FigExpOPA}. The 800 nm pump beam is in blue, the 2.05 $\mu m$ signal beam is in red, the 1.31 $\mu m$ idler beam is in orange, and the \acf{WLC} is in green. The 800 nm pump beam comes from the top left and splits into three paths after passing through two beam splitters. The beam in the first path with $\sim1\;\mu J$ energy is focused onto a 3-mm-thick sapphire plate to generate \acs{WLC} and then stretched by a 10-mm-thick $Z_nS_e$ plate. The stretched \acs{WLC} is then directed to the first BBO crystal for amplification, phase-matched with the beam in the second path reflected off BS2 with $\sim800\;\mu J$ energy. The idler from the first amplification stage is at 2 $\mu m$, which serves as the seed for the second amplification stage. The beam in the last path reflected off BS1 carrying most of the pump energy ($>12$ mJ) overlaps with the 2 $\mu m$ on the second BBO crystal. In both amplification stages, a tunable time-delay is allowed in the 800 nm pump beams to select the desired portion of the stretched \acs{WLC}. After amplification, the signal, the idler, and the residual pump are separated using dichroic mirrors.

\begin{figure}[H]
\centering
\includegraphics[width=0.56\columnwidth]{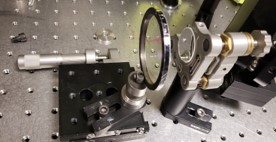}
\caption{Tunable prepulse from a nitrocellulose pellicle on a translation stage.}
\label{FigExpPrepulse}
\end{figure}

The 2 $\mu m$ beam coming out of the \acs{OPA} is then directed to a vacuum chamber for laser-plasma experiments. Since the dichroic mirrors in the \acs{OPA} can not separate colors with $100\%$ efficiency, a 1.65 $\mu m$ long-pass filter (Andover 1.65ILP-25) is used to let through only the 2 $\mu m$ signal with $\sim80\%$ beam energy remaining. A 2 $\mu m$ thick nitrocellulose pellicle (National Photocolor) is placed on a translation stage in the beam path to pick up a prepulse with $\sim8\%$ of the main pulse energy. The translation stage locates right in front of a flat mirror where the beam arrives from near normal incidence. A picture of the prepulse delay stage is shown in Fig. \ref{FigExpPrepulse}. The prepulse delay can be adjusted between 0 - 187 ps by tuning a translation stage, and the prepulse intensity is $\sim$two orders of magnitude lower than the main pulse intensity because of a much larger focal spot. 

A $5\times$ reflective telescope consisting of a f=-200 mm convex mirror and a f=1000 mm concave mirror is built to expand the beam diameter to 50 mm. Increasing the beam diameter allows the use of two-inch optics, such as a two-inch deformable for more actuators and thus better wavefront control, and two-inch \acf{OAP}s for larger f-number and thus higher focal intensity.

\subsection{The mid-infrared \acs{OPCPA} at the University of Maryland}
A compact \acs{OPCPA} laser system was used for the \acs{LWFA} experiment to be discussed in Sec. \ref{sec:MIRLWFA}, which was conducted at the Intense Laser Matter Interactions group at the University of Maryland. The \acs{OPCPA} system operates at a central wavelength of 3.9 $\mu m$ with the tail-to-tail spectrum extending over 0.6 $\mu m$. It delivers $25\pm1$ mJ, 87 fs pulses ($<$7 optical cycles) at a repetition of 20 Hz. The system is seeded by Yb:KGW modelocked oscillator and pumped by a Yb:CaF$_2$ \acs{CPA}. The parametric amplification is performed in three \acs{OPA} stages using KTP cyrstals. Detailed design of the system is available in Ref. \cite{andriukaitis201190}.

% \section{Laser characterization}
% \label{sec:OpticalCharact}
% \input{Chap3Exp/LaserCharact/LaserCharact.tex}

\section{Plasma targets}
\label{sec:ExpTargetPrep}
% don't put experimental setup sketches here!
Most of the experiments (except for the \acs{LWFA} experiments) were performed in a vacuum chamber with 58 cm inner-diameter at \acs{CUOS}. The chamber is pumped by an oil-free scroll vacuum pump (SCROLLVAC SC 15D) to tens of mTorr pressure, and the pressure is measured with a convection gauge (KJL300807). A two-inch diameter, 5 mm thick fused-silica window is used to transmit the laser beam into the chamber at normal incidence. A $\sim5$mm-thick solid target is mounted on a rotary stage inside the chamber, which allows both rotation and radial movement. It enables the laser pulses to continuously interact with undamaged target areas, making the most use of the high-repetition-rate capability provided by the \acs{Lambda-cubed}. The rotary stage sits on another translation stage with 2 $\mu m$ minimum controllable step moving along the laser propagation direction, allowing the solid target to walk within the Rayleigh range ($\sim10\;\mu m$) of the laser focal spot. The rotary stage and the translation stages are controlled by motors outside the chamber such that the target can be moved without breaking the vacuum condition.

A major concern regarding rotary stages is the inevitable target wobbling towards and away from the laser focus as it rotates, leading to interactions at lower intensities in a portion of a rotation period. Therefore, the period wobbling needs to be constrained within the Rayleigh range of the laser focal spot. We achieve this using the following tools and routine. A dial indicator measures the pressure it feels on a surface and translates it into a distance reading, and the mount stage has three pins to push the target's rear surface from positions of the same radius but different angles. Before the experiments, we apply displacement to the target rear surface from three different touching points while measuring the maximum wobbling distance during a rotation period with the dial indicator, until the reading of the dial indicator is within its resolution. The resolution of the dial indicator is 2 $\mu m$, which is much smaller than the Rayleigh range ($\sim10\;\mu m$) of a 2 $\mu m$-wavelength laser beam at f/1.3 focus.

\subsection{Solid targets for surface \acs{HHG}}
Silicon and glass (fused silica) targets are used for the surface \acs{HHG} experiment to be discussed in Sec. \ref{sec:HHG}. The fused silica targets are 100 mm in diameter and 5 mm thick. The silicon targets are made by attaching silicon wafers to the glass targets. The silicon wafers (University Wafer ID 589) are double-side polished, 100 mm in diameter, and 500 $\mu m$ thick. Unlike the $\sim\mu m$ thick targets used for other laser-solid interaction experiments, such as ion acceleration, the targets used here are much thicker than the skin depth of the laser pulses. As a result, the laser pulses do not penetrate through the targets, even with relativistic effects taken into account.

Since the surface \acs{HHG} process favors a short plasma density gradient in general, no external preplasma is introduced in this experiment. The contrast between the main pulse intensity and the prepulse intensity is $\sim10^8$ owing to the short pulse duration of the pump in the \acs{OPA}, as is discussed in Sec. \ref{sec:theoryLaser}. However, some degree of flexibility in the preplasma density profile is available by switching the target material. The work function of silicon is 4.8 eV, while the work function of fused silica is 5.0 eV \cite{fomenko2012handbook}. Taking advantage of the different work functions, different damage thresholds (and thus different preplasma density profiles) are obtained when using silicon targets versus fused silica targets.

\subsection{Solid targets for characteristic x-ray emission}
A molybdenum target is used for the characteristic x-ray generation experiment to be discussed in Sec. \ref{sec:Xray}. The molybdenum target is 100 mm in diameter and $\sim9.5$ mm thick. Laser ablation on the molybdenum target produces heavy debris that can damage the \acf{OAP}, which is $\sim75$ mm away from the target surface plane. Therefore, a 2 $\mu m$ thick nitrocellulose pellicle (National Photocolor) is placed in between the target and the \acs{OAP} without blocking the beam path. The pellicle is usually burned in less than a minute of consecutive operation, which corresponds to $\sim30,000$ shots. Vacuum conditions have to be broken to open the chamber lid and replace the pellicle, even if the target surface has not been fully used. Note that a copper disk was also used, but much more debris was produced than that from the molybdenum target, making it difficult to collect the dataset for our parametric study ($\sim700,000$ shots in total). After taking laser shots all over the target surface, the target is reused by polishing the damaged surface with silicon carbide sandpapers (mesh numbers range from 120 to 600, in units of $\frac{\#\;of\;particles}{inch^2}$) and a precision polishing machine (Malvern Instrument Multipol 2).

The target preplasma profile is studied as a governing parameter in the characteristic x-ray emission process. The preplasma density gradient is controlled by the time delay between an external prepulse and the main pulse arrival. A tunable prepulse with a time delay of 0 - 187 ps after the main pulse, as is discussed in Sec. \ref{sec:ExplaserSystem}, is available. It yields a tunable preplasma density profile with density scale length $0.1\lambda<L_s<5.5\lambda$.

\subsection{Gas targets for \acs{LWFA}}
Hydrogen gas was used as the target for the \acs{LWFA} experiment to be discussed in Sec. \ref{sec:MIRLWFA}. The diameter of the orifice nozzle is 150 $\mu m$, and the plasma number density is up to $3\times 10^{19}cm^{-3}$ without cryogenic cooling. Since the \acs{LWFA} experiment in Sec. \ref{sec:MIRLWFA} was performed with mid-infrared laser pulses at laser wavelength $\lambda_L=3.9\;\mu m$, this plasma density corresponds to $40\%$ of the critical density. The jet was mounted onto a 3-D translation stage to adjust the position of the laser focus throughout the hydrogen gas target.
% maryland

Argon gas is used as the target for the \acs{LWFA} experiment to be discussed in Sec. \ref{sec:MLLWFA}. The free-flowing argon gas is from a capillary with inner-diameter of 100 $\mu m$, where the optimal backing pressure is in the range of $21\sim23$ psi. The plasma density profile is roughly a Gaussian with a 120 $\mu m$ FWHM, and the peak density reaches $6.5\times10^{18} cm^{-3}$. The laser focus is generally placed on the down-ramp of the target density profile.

\section{Laser-plasma diagnostics}
\label{sec:ExpLPDiag}
% define \acf{CCD} here
\subsection{Mid-infrared laser diagnostics}
The 2 $\mu m$ laser pulses are diagnosed differently from the commonly-used near-infrared laser pulses at 800 nm wavelength. Since the 2 $\mu m$ pulses are not visible, the alignment is performed using \acs{MIR} liquid crystal detector cards (Thorlabs VRC6S) taking advantage of the thermal effects, or using regular paper cards with infrared viewers (Photographic Equipment). The spectrum of the 2 $\mu m$ pulses is measured with an infrared spectrometer (Ocean Optics Nirquest). Alternatively, it is diagnosed with a grating spectrometer (Horiba iHR550) and an InSb point detector (Hamamatsu P4631-03). The power of the 2 $\mu m$ pulses is measured with an integrated power meter (Coherent PowerMax PM30). In some circumstances where a faster response than that of a thermal power meter is needed, such as tuning the \acs{OPA} delay stages for maximal output energy, a PbS fixed gain detector (Thorlabs PDA30G) with 250 $\mu s$ rising time is used. The PbS detector has a much smaller damage threshold than the maximal energy of the 2 $\mu m$ pulses, and thus neutral density filters are placed in front of the detector for protection. Therefore, in a daily routine, the PbS detector with a short response time is first used to maximize the \acs{OPA} output energy, and then the power meter is used to measure the absolute energy in the beam.

The duration of the 2 $\mu m$ pulses is measured using an intensity autocorrelator. In the autocorrelator, a 2 $\mu m$ pulse is split into two pulses, and a delay stage is used to tune the time delay between them. The two pulses are focused onto a nonlinear crystal and overlap with each other spatially. The time delay is adjusted so that the two pulses also overlap temporally, \acf{SHG} occurs. A \acf{CCD} camera (Mightex CGE-B013-U) is used to capture the \acs{SHG} signal at 1 $\mu m$ wavelength, and the signal position on the camera changes as the time delay changes. By tuning the time delay and recording the displacement on the delay stage, the pixel size is translated to real distance units, and the pulse duration is inferred after dividing by the speed of light. Note that the intensity width is multiplied by a deconvolution factor of $\sqrt{2}$ for the pulse duration, assuming a Gaussian shape. For more accurate measurements without this assumption in pulse shape, a diffraction grating work at 2 $\mu m$ wavelength is required.

\subsection{Surface \acs{HHG} spectral diagnostics}

The surface \acs{HHG} spectra were measured using a Thorlabs CCS200 spectrometer, whose detection range covers 200 nm - 1020 nm. In the \acs{HHG} experiments driven by the 2 $\mu m$ pulses from the \acs{OPA}, which will be described in Sec. \ref{sec:HHG}, $2^{nd}$, $3^{rd}$, $4^{th}$, $5^{th}$, and $6^{th}$ harmonics were observed on the spectrometer. Since the spectrometer has a low grating efficiency approaching the edge of the detection range, an intensity calibration was necessary. We used an Ocean Optics DH2000 Deuterium-Tungsten Halogen lamp with a known emission spectrum to calibrate the intensity readings of the Thorlabs CCS200 spectrometer.

\begin{figure}[H]
\centering
\includegraphics[width=0.98\textwidth, height=0.25\textheight]{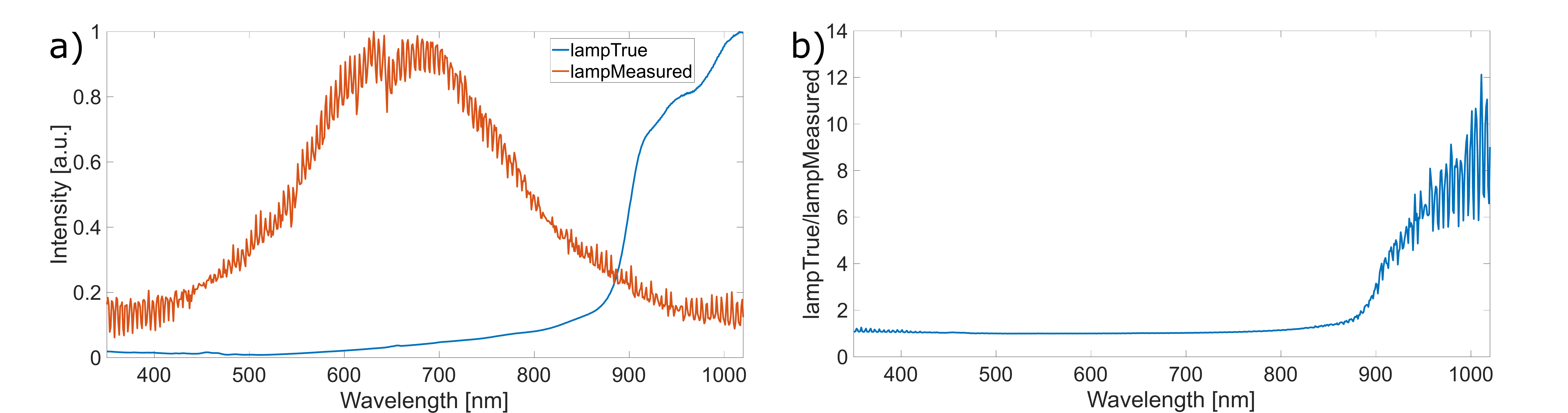}
\caption{Spectrometer calibration using the lamp source. (a): The blue curve is the emission spectrum of the lamp provided by the manufacture; the orange curve is the spectrum measured with the Thorlabs CCS200 spectrometer. (b): Ratio between the known spectral intensity and the measured spectral intensity.}
\label{ExpFigSpectrometerCalibration}
\end{figure}

Fig. \ref{ExpFigSpectrometerCalibration} shows the process of calibrating the spectrometer with the lamp source. The emission spectrum of the lamp source was measured and is plotted in orange in Fig. \ref{ExpFigSpectrometerCalibration}a, while the spectrum provided by the manufacture is plotted in blue. The ratio between the known spectral intensity and the measured spectral intensity is plotted in Fig. \ref{ExpFigSpectrometerCalibration}b. The grating efficiency of the spectrometer drops drastically at the mid-infrared range, which leads to the increase in the intensity ratio in  Fig. \ref{ExpFigSpectrometerCalibration}b. For reference, the MFG \acs{CCD} detector on the spectrometer has a $\sim5\%$ responsivity at 1000 nm. The low responsivity results in the amplification of noise around that wavelength, shown in  Fig. \ref{ExpFigSpectrometerCalibration}b.

\begin{figure}[H]
\centering
\includegraphics[width=0.98\textwidth, height=0.25\textheight]{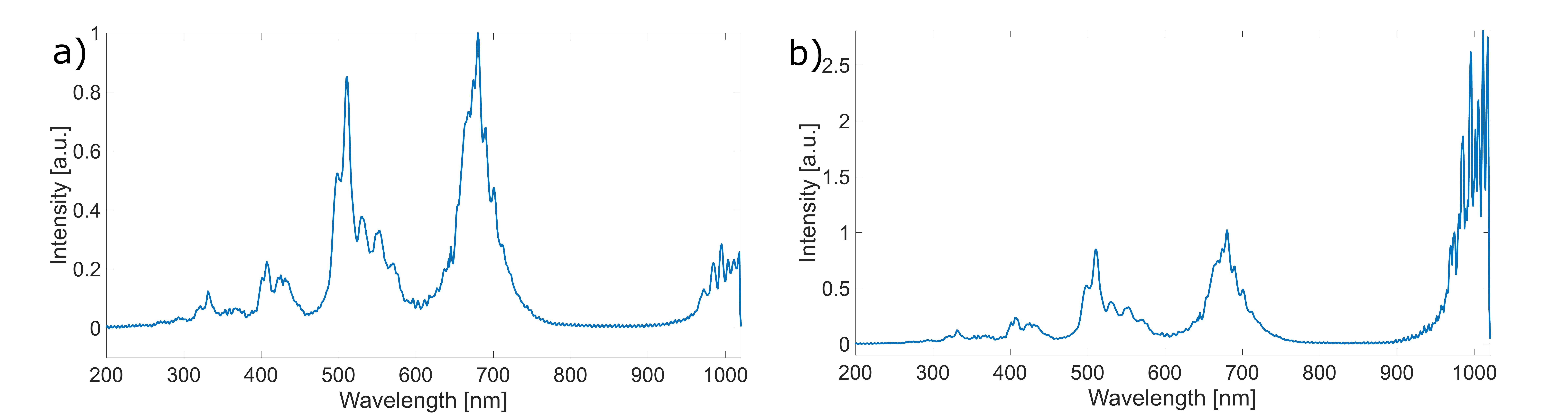}
\caption{\acs{HHG} spectrum before (a) and after (b) correction.}
\label{ExpFigSpectrumCorrection}
\end{figure}

Since the intensity ratio is close to 1 below 900 nm, most of the measured harmonics were not affected by the calibration, except for the second harmonic at $\sim$1000 nm. An example of the measured harmonic spectrum before and after correction is shown in Fig. \ref{ExpFigSpectrumCorrection}. Although the peak for the second harmonic at 1000 nm is not ideal due to the noise amplification as well as the measurement cut-off at 1020 nm, it is clear that the second harmonic intensity is higher than the higher-order harmonics after the spectrum correction.

\subsection{X-ray spectral diagnostics}

The characteristic x-ray emission experiment, which is to be discussed in detail in Sec. \ref{sec:Xray}, was primarily diagnosed with an Amptek XR-100 Cadmium Telluride (CdTe) single-photon detector for x-ray spectral measurement. The detector area is $5\;mm\times5\;mm$, while a 100 $\mu m$ thick beryllium window is placed in front of the CdTe diode. The detector has 1024 channels and the detection range extends beyond 100 keV, which is much higher than the characteristic emission line of molybdenum ($k_\alpha=17$ keV, and $k_\beta=19$ keV). A fast threshold was set to eliminate the low energy noise at channel number 55, which is calibrated to be $\sim6$ keV in the calibration process to be discussed below. The rising time was set to be 3.2 $\mu s$. The detector was operated at a voltage of 950 V and a temperature of 211 K.

\begin{figure}[ht]
\centering
\includegraphics[width=0.98\textwidth, height=0.3\textheight]{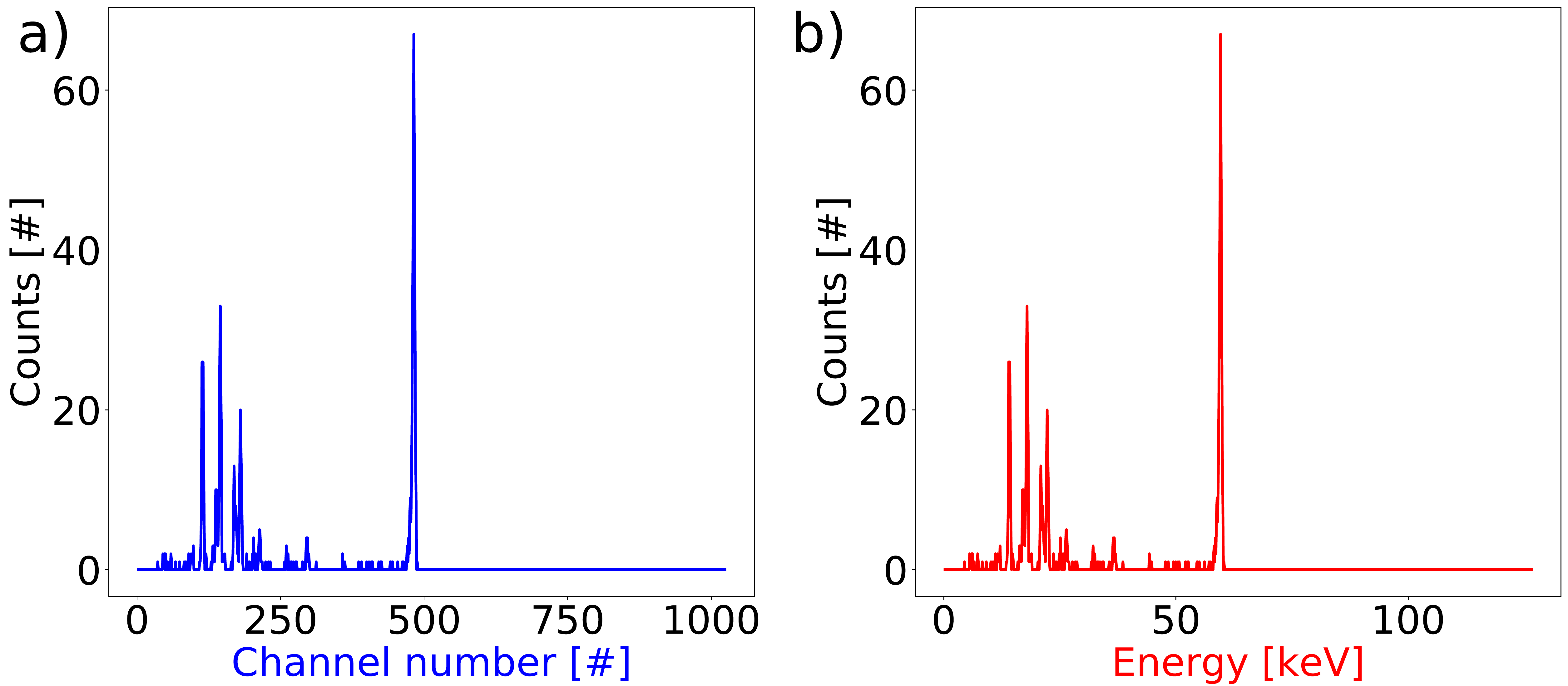}
\caption{Raw x-ray spectrum (a) and calibrated spectrum (b) of an \acs{am-241} source.}
\label{FigExpAm241}
\end{figure}

The detection efficiency through the 100 $\mu m$ thick beryllium window is well-calibrated by the manufacture. Therefore, unlike the y-calibration in the previous subsection for the Thorlabs CCS200 spectrometer, an x-calibration is to be conducted here to convert the channel numbers in the Amptek detector to real energy units. An \acf{am-241} source with distinct peaks in its known spectrum is used for this calibration. Fig. \ref{FigExpAm241}a shows the raw spectrum recorded for the \acs{am-241} source, integrated over 20 seconds. The channel numbers of the two strongest peaks in Fig. \ref{FigExpAm241}a are 113 and 481. In the known \acs{am-241} emission spectrum, the energy of these two peaks are 13.95 keV and 59.54 keV, respectively. Consequently, a linear transformation is performed to convert the x=axis from channel number to energy, as is shown in Fig. \ref{FigExpAm241}b. The calibrated units are used in the measurement of characteristic x-ray emissions and bremsstrahlung radiation in Sec. \ref{sec:Xray}.

\section{Computational modeling: \acs{PIC} simulations}
\label{sec:ExpPIC}
Computational tools can give access to regimes where the experimental capabilities can not readily reach. Using computer simulations can also avoid all the challenges in experimental diagnostics. The \acf{PIC} method is widely used in plasma physics for simulating collisionless plasma kinetics. Since the plasma number density in relativistic laser-plasma interactions is usually beyond $10^{19}cm^{-3}$, it would be too computationally expensive to simulate the real particles. Instead, macroparticles representing a collection of real particles are used. The motion of the macroparticles are solved iteratively on a grid in the following loop:
\begin{enumerate}
\item Interpolate the electric fields and magnetic fields on the grids to the position of the macroparticles.

\item Particle pusher: update the momentum and position of the particles using the equation of motion \ref{EqTheorysingleEOM}.

\item Calculate the currents and charge densities from the updated distribution of the macroparticles.

\item Field solver: with the currents and charge densities, solve the electric and magnetic fields using Maxwell's equation.

\end{enumerate}

In setting up \acs{PIC} simulations, the grid size and the time step should be carefully chosen. The grid size must resolve the relevant length scales, such as the Debye length and the laser wavelength. The time step must be smaller than the shortest time scale of any significant physical process. Moreover, for the Courant criterion must be satisfied for numerical stability: $\Delta t^2<\frac{1}{\Delta_x^2}+\frac{1}{\Delta_y^2}+\frac{1}{\Delta_z^2}$. Even smaller grid sizes and time steps are required for certain circumstances, such as interactions at high plasma densities or studying attosecond phenomena.

In this thesis, \acs{PIC} simulations were mainly performed using the OSIRIS 4.X framework in 2D3v Cartesian geometry. OSIRIS is a massively parallel \acs{PIC} code for modeling relativistic laser-plasma interactions developed by the OSIRIS Consortium, consisting of UCLA and IST (Lisbon, Portugal) \cite{OsirisRef, OsirisRef2}. Parallel computing was performed utilizing the Great Lakes (Flux) high-power computing cluster at the University of Michigan.

\section{Statistical methods for adaptive optical systems}
\label{sec:ExpAdaptiveOS}
\subsection{Adaptive optical systems}
% see alex englesbe 2021 paper
An \acf{AOS} is a closed-loop system consisting of an adaptive optic, a measurement device, and a controller. An example of an \acs{AOS} is illustrated in Fig. \ref{MIRLWFAsetUp} in Sec. \ref{sec:MIRLWFA}. \acs{AOS}es are first designed for terrestrial telescopes to correct the wavefront distortion caused by the Earth's atmosphere \cite{beckers1993adaptive}. For ultrafast lasers at relativistic intensities, wavefront aberrations due to the focusing optics, such as \acs{OAP}s, are often inevitable, being the main obstacle from having diffraction-limited focal spots at extreme intensities. With the help of \acs{AOS}es and machine learning algorithms, we can not only correct wavefront aberrations in laser pulses but also control the wavefront for other physical processes in relativistic laser-plasma interactions.

Such active feedback loops are also widely used in laser pulse shaping using dazzlers (\acs{AOPDF}s). For instance, Dann \textit{et al.} adjusted the spectral phase of a driving laser pulse to optimize the properties of the electron beam from a laser-wakefield accelerator \cite{dann2019laser}.

A commonly used adaptive optic is the \acf{DM}. Its mirror surface consists of many sub-regions, and each sub-region is connected to a separate programmable actuator. By controlling the displacement of each actuator, the mirror surface can be pushed and pulled at different regions to achieve any desired deformation. As a result, the wavefront of the laser beam reflecting off the \acs{DM} can be manipulated.

\subsection{Genetic algorithms for optimization}
\begin{figure}[ht]
\centering
\includegraphics[width=0.9\columnwidth]{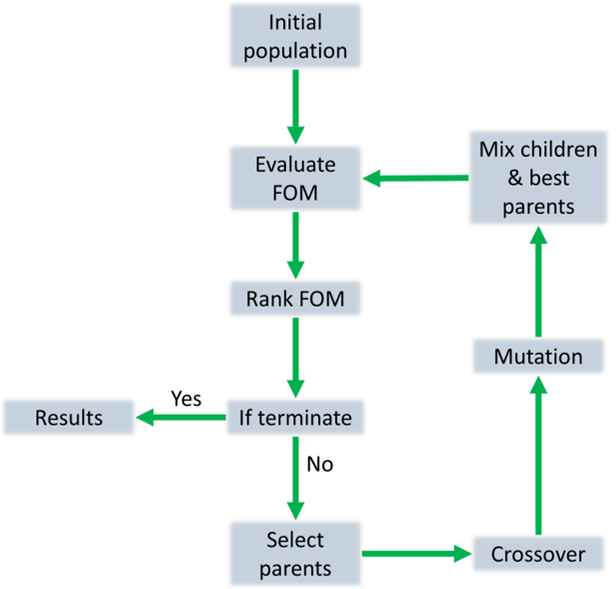}
\caption{Working principle of the genetic algorithm.}
\label{FigExpGA}
\end{figure}

% see alex englesbe 2021 paper
Optimization in an \acs{AOS} is often executed with the genetic algorithm. The \acf{GA}, also called the evolutionary algorithm, utilizes biologically inspired logic to repeatedly make semi-random searches till reaching convergence \cite{holland1992adaptation}. Given a sufficient number of iterations, the \acs{GA} will find the global maximum. In practice, a local maximum close to the global maximum is usually obtained due to the limited operation time of the experiments. There are two common ways to define the solution space of the \acs{GA} regarding the genetic representation when implementing a \acs{GA} with an \acs{AOS} and a \acs{DM}. In this thesis work, we define the basis set upon the actuators' voltages on the \acs{DM}. Alternatively, one can employ the first several orders in the Zernike polynomial.

Fig. \ref{FigExpGA} demonstrates the workflow of the genetic algorithm. The \acs{GA} starts from randomly initializing the population in genetic representation (actuators' voltages or Zernike coefficients), as is discussed above. In either case, each individual gene represents the wavefront change caused by the deformable mirror. A laser pulse with such a wavefront is then taken into the system, and a measurement is made. For example, in the experiment optimizing a laser-wakefield accelerator (Fig. \ref{MIRLWFAsetUp} in Sec. \ref{sec:MIRLWFA}), the measurement is made on the produced electron beam on a LANEX screen using a \acs{CCD} camera. The measurement is evaluated mathematically to obtain a \acf{FOM}, which could be the electron beam total charge, the electron beam profile, or any other properties of the electron beam. The \acs{FOM}s of all individuals are ranked and the best ones are selected as the parents for the next generation. 

\begin{figure}[ht]
\centering
\includegraphics[width=0.9\columnwidth]{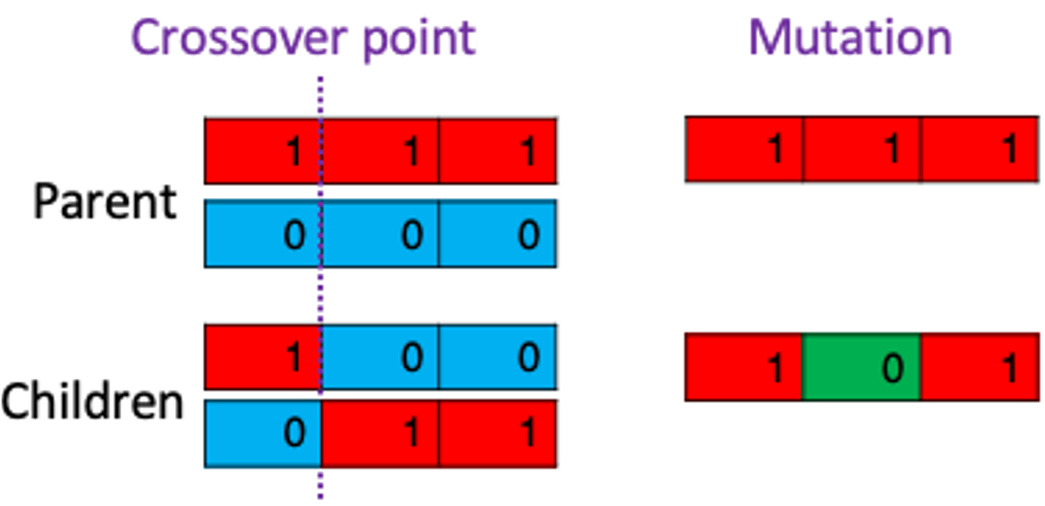}
\caption{Illustration of crossover and mutation operations in the genetic algorithm.}
\label{FigExpGAoperations}
\end{figure}

The parents are used to produce children through crossover or mutation, as is illustrated in Fig. \ref{FigExpGAoperations}. For a parent gene [1, 1, 1] and a parent gene [0, 0, 0], setting the crossover point to the beginning of the second point would yield two children genes [1, 0, 0] and [0, 1, 1]. Crossover broadens the sampling parameter space, but it also evolves towards a local maximum. Therefore, the mutation is introduced to extend towards the global maximum. The mutation operator takes a single parent gene [1, 1, 1] to produce a child gene, for example [1, 0, 1], at a specified mutation probability. With the crossover and mutation operators, a pool of children genes is generated. To avoid cases where all children are inferior to their parents, the best parents are maintained and mixed with the generated children to form the gene pool for the next generation. The wavefront of each individual in the mixed gene pool will be taken to the system to evaluate its \acs{FOM}. The cycle repeats until reaching some termination condition.

In this thesis, the genetic algorithm is used to optimize the laser focus and the electron beams from \acs{LWFA}, as is to be discussed in detail in Sec. \ref{sec:FocusOpt} and Sec. \ref{sec:MIRLWFA}, respectively.

\subsection{Supervised learning methods for modeling}
% machine learning vs. genetic algorithms - ted norris course project
Although the genetic algorithm has been an effective tool in high repetition rate laser labs, it has its intrinsic weaknesses. In the selection procedure, a \acs{GA} keeps a portion of the individuals with the highest \acs{FOM} in each generation, say the best $10\%$, and make them serve as the “parent” in the next generation. Equivalently, $90\%$ of the data recorded are abandoned in a \acs{GA} in every iteration. The waste of information can be avoided using machine learning methods. In supervised learning, which will be discussed in detail in the following paragraphs, all data are saved and no selection procedure is required at all since the goal is modeling rather than optimization. In reinforcement learning which does real-time optimization, experiences in previous steps can be stored as a replay buffer and used in future steps. Nevertheless, machine learning techniques are working with a ten-fold larger “effective dataset” than that the genetic algorithms have, and understandably draw more complete conclusions. It also explains the fact that \acs{GA}s are difficult to find smooth convergence since using only the newest measurements can lead to overfitting.

Another weakness of the \acs{GA} is that it oftentimes provides little information other than a local optimum, making it difficult to perform physical interpretation. Therefore, we use supervised learning methods to generate predictive models to go beyond optimization.

% copy those methods over. need this many pages
Four supervised learning regression methods are used in Sec. \ref{sec:MLLWFA} to predict the electron beam charges based on laser wavefront changes. Supervised learning is a branch of machine learning, which learns a function that maps an input (feature) to an output (label) based on example input-output pairs in a training sample. In each of the following supervised learning methods, the model is trained on the training dataset recursively until it can accurately predict the labels using the features. The model performance is then characterized by the test dataset. 

\subsubsection{Random forest}
The \acf{RF} regressor\cite{liaw2002classification} is a popular bagged algorithm for high-dimensional and nonlinear regression. It is based on the concept of a decision tree, which splits the dataset along some dimensions and recursively divides the space into regions with similar labels. Being the most popular bagging (Bootstrap Aggregation) ensemble algorithm, random forest samples, with replacement at uniform probability, the original dataset $D$ into $m$ datasets ($D_1, D_2, ..., D_m$) with the same size as $D$. For instance, if the original dataset contains three samples $D=[a,b,c]$, then $D_1$ could be $[a,c,c]$. For each dataset $D_j$ in the forest, we train a full decision tree by splitting the data to $k<n$ dimensions. Only k features are to be considered when looking for the best split. Since the trees become much more different as they select different features, we have to increase the number of trees and average over individual regressors. This bagging process helps reduce variance effectively. Denote $h_{D_i}$ as the regressor for the dataset $D_i$, then the bagged regressor $\hat{h}$ is expressed as \cite{kilian}:
\begin{equation}
% h_i=\frac{1}{m}\sum_{i=1}^m h_i(x)
\hat{h}=\frac{1}{m}\sum_{i=1}^m h_{D_i}
\label{EqExpRF}
\end{equation}
% The \acf{RF} regressor\cite{liaw2002classification} is a popular bagged algorithm for high-dimensional and nonlinear regression. It is based on the concept of a decision tree, which splits the dataset along some dimensions and recursively divides the space into regions with similar labels. Being the most popular bagging (Bootstrap Aggregation) ensemble algorithm, random forest samples, with replacement at uniform probability, the original dataset $D$ into $m$ datasets ($D_1, D_2, ..., D_m$) with the same size as $D$. For instance, if the original dataset contains three samples $D=[a,b,c]$, then $D_1$ could be $[a,c,c]$. For each dataset $D_j$ in the forest, we train a full decision tree by splitting the data to $k<n$ dimensions. Only k features are to be considered when looking for the best split. Since the trees become much more different as they select different features, we have to increase the number of trees and average over individual regressors. This bagging process helps reduce variance effectively. Denote $h_i$ as the regressor of the $i_{th}$ tree, then the bagged regressor is \cite{kilian}:
% \begin{equation}
% % \label{rfEq}
% h_i=\frac{1}{m}\sum_{i=1}^m h_i(x)
% \label{EqExpRF}
% \end{equation}

In this study, we utilize the \textit{Sklearn.ensemble.RandomForestRegressor} library in Scikit-learn \cite{pedregosa2011scikit} to implement the algorithm. The hyper-parameters to tune are the number of trees, the maximum depth in a tree, and the maximum number of features when splitting. 

\subsubsection{Deep neural network}
A \acf{DNN} is a feed-forward artificial neural network with multiple hidden layers. The goal of a feed-forward network is to approximate some function $f$ \cite{goodfellow2016deep}. A typical one hidden layer can be mathematically described with weight $w$ and bias $w_0$ by Eq. \ref{EqExpNN1}:
\begin{equation}
\label{EqExpNN1}
y = f(w_0+w_1^Tx)
\end{equation}

According to the Universal Approximation Theorem, a feed-forward network with even one hidden layer 
% \qian{(I think should be 1 hidden layer)} 
can approximate any continuous function from one finite-dimensional space to another under some conditions \cite{hornik1991approximation}. Although practically it may lead to an infeasibly large layer and fail to generalize, \acs{DNN}s are powerful function approximators when appropriately learned. Compared to shallow models, \acs{DNN}s usually can extract better features and learn more effectively. 

In this work, we build a fully-connected five-layer DNN using the \textit{Tensorflow.Keras} library \cite{chollet2015keras} based on Google’s deep learning software TensorFlow \cite{abadi2016tensorflow}. When constructing the network, we use the rectified linear unit (ReLU) function and the Sigmoid function as the activation functions for different layers.
The activation function for the first, the fourth, and the fifth layer is the rectified linear unit (ReLU) function, and that for the second and the third layer is the Sigmoid function. 
The cost function is the mean squared error loss governed by the Adam optimizer \cite{kingma2014adam} to update the network weights. A $L_2$ norm regularization is added to the loss function to reduce overfitting. The main tuning parameters are the number of layers, the number of neurons in each layer, the epoch size, and the initialization of the weight matrix.

\subsubsection{Deep jointly-informed neural networks}
The \acf{DJINN} is a machine learning algorithm that constructs and initializes the deep feed-forward neural networks based on decision trees. It was developed by Humbird \textit{et al.}\cite{humbird2018deep}, and it has shown success in training ICF datasets \cite{humbird2019parameter, gaffney2019making, hsu2020analysis} as well as standard regression datasets such as Boston housing prices, California housing prices, and diabetes disease progression \cite{humbird2018deep}. The algorithm starts by constructing a decision tree or an ensemble of trees, where the number of trees will be the number of networks in the later stage. It then maps the decision trees to deep neural networks by taking the decision paths as guidance for the network architecture and weight initialization. The networks are trained with backpropagation using TensorFlow \cite{abadi2016tensorflow}. In the network architecture, the activation function is the rectified linear unit and the cost function is governed by Adam optimizer \cite{kingma2014adam}. Without optimizing the architecture of the neural networks, \acs{DJINN} displays comparable performance to optimized architectures at a significantly lower computational cost using their datasets.

The \acs{DJINN} regression source code is accessible at the LLNL/\acs{DJINN} github directory. The main tuning parameters for this study are the maximum depth of trees and the number of trees (nets) in ensemble.
% \abigail{please complete}
% \jinpu{Thanks, will do!}

\subsubsection{Gaussian process}
The \acf{GP} is a non-parametric Bayesian algorithm for supervised learning problems \cite{rasmussen2003gaussian}. While most machine learning algorithms fit the dataset into a model function with weight parameters and use that function to make predictions, Bayesian methods avoid the intermediate step and make predictions directly from the dataset. This is achieved by integrating all possible weight functions in the universe \cite{kilian}: 

\begin{equation}
\label{EqExpGP}
P(y|x,D)=\int _{w} P(y|x,w)\cdot P(w|D)\; dw
\end{equation}
where P is the prediction, w is the weight matrix, and D is the dataset. In Gaussian process regression, we assume that the data can be fit by some model with weight function w and a Gaussian distributed noise $\epsilon$: $y=f(x)=w^{T}x+\epsilon$. Thus the term $P(y|x,w)$ in Eq. \ref{EqExpGP} is a Gaussian distribution. The second term on the right-hand-side can also be proved to be Gaussian using the Bayes' rule:

\begin{equation}
\label{EqExpBayes}
P(w|D)=\frac{P(D|w)\;P(w)}{P(D)}
\end{equation}

where $P(w)$ is the prior distribution, $P(w|D)$ is the posterior distribution after D is considered, and $P(D)$ is a normalization. By choosing a Gaussian prior, Bayes' rule leads to a Gaussian posterior $P(w|D)$. Marginalizing out w in Eq. \ref{EqExpGP}, we know $P(y|x,D)$ is also in a Gaussian distribution. 
% In other words, we are able to make predictions with some understandable distribution despite that we do not learn a specific model. It also yields merit in GP: it gives a distribution of prediction values instead of only one number.
The Gaussian process has a closed-form posterior distribution that can be used to quantify the uncertainy of the estimate (posterior mode) through a confidence interval.

Another advantage of Gaussian process regression is that we can specify prior information about the shape of the model by selecting certain kernel functions. As is introduced previously, the probability function of the prediction can be expressed by a Gaussian distribution: $P(y|x,D)=N(\mu, \Sigma)$, where $\mu$ is the mean and $\Sigma$ is the covariance matrix. The covariance matrix can be kernelized using selected kernel functions. In this project, we implement the algorithm using the $Sklearn.gaussian\_process$ library in Scikit-learn \cite{pedregosa2011scikit} with a combination of Matern kernel and Rational Quadratic kernel. The hyper-parameters to tune are the smoothness, the length-scale, and the scale mixture parameter in the kernels.
% see alex englesbe 2021 paper

\chapter{Laser-solid Interactions at Relativistic Intensities}
\label{chap:Solid}
\section{Introduction}
\label{sec:SolidIntro}
The interaction mechanisms between laser pulses and solid targets can be complicated. The textbook by Paul Gibbon \cite{gibbon2005short} covers different physics models across several orders of magnitude in laser intensity and in laser pulse duration. In this dissertation work, we discuss the interactions that happen at about the relativistic intensity (a0$\sim$1) with ultrashort laser pulses ($\sim$30 fs). Such an interaction is one of the most efficient ways to couple energy to solid-density plasmas, as almost all the laser energy are absorbed by the electrons while ions gain energy from the collision with electrons in a longer time frame. Due to their heavy mass, ions can usually be regarded as immobile in relativistic short-pulse laser-solid interactions.

Many studies in this regime are performed with plasma mirrors. Plasma mirrors are used to clean the laser pulse contrast in high-intensity laser facilities, as they are transmissive at low intensities and reflective at higher intensities when the pulse ionizes the material to form a plasma surface. Plasma mirrors not only cleans the laser contrast, they are also efficient in light manipulation due to their ultrafast and nonlinear response to light field. In such interaction processes when plasma mirrors are ionized, a sheath of plasmas is formed, whose size is $\sim \mu m$ and dynamic response is $<$picoseconds. Relativistic electron bunches form at the surface of the solid, moving at nearly the speed of light and having attosecond bunch duration. Light with broad spectra peaked at various harmonics of the laser beam frequency is also generated, known as the \acf{HHG}. More information about the development of plasma mirrors and their characteristics can be found in the following literature \cite{thaury2007plasma, vincenti2014optical, bocoum2015spatial, wilson2018development}.

When laser interacts with metallic targets with larger atomic numbers, these targets apparently do not act as plasma mirrors to improve laser contrast as they are not optically transmissive in any case. However, physics processes that generate relativistic electron bunches still occur. Moreover, characteristic x-rays can be emitted from metallic targets, whose spectra peak at some specific energies. Unlike \acs{HHG} where the wavelength of the produced x-rays depends on the driver laser, the wavelength of the characteristic x-ray emission due to electron impact is determined by the material properties.

In this section, we will report results for surface \acs{HHG} in Sec. \ref{sec:HHG}, attosecond electron bunch generation in Sec. \ref{sec:Atto}, and characteristic x-ray emission in Sec. \ref{sec:Xray}.
\newpage

\section[]{High-order harmonic generation using mid-infrared laser pulses\footnotemark}
\label{sec:HHG}
\footnotetext{Part of this section co-authored with Beier, N., Nguyen, T., Nees, J., Krushelnick, K., and Dollar, F. (2019): Relativistic short- pulse high harmonic generation at 1.3 and 2.1 $\mu$m wavelengths. New Journal of Physics, 21(4), 043052.}
\subsection{Introduction}
\ac{HHG} through relativistic laser-plasma interactions
is a promising source for isolated attosecond high brightness pulses in the soft and hard x-ray regimes \cite{naumova2004relativistic, tsakiris2006route, teubner2009high, nomura2009attosecond, mikhailova2012isolated, kallala2020techniques}. Compared to x-ray tubes and other small bright continuum sources, laser-based harmonic sources can generate shorter ($\sim$femtosecond or shorter) pulses. \acf{XFELs} and synchrotrons can provide ultrashort pulses but require huge facilities. Laser-based \acs{HHG} sources are compact in size, and recent study on the power-law scaling shows that \acs{HHG} can be competitive with \acs{XFELs} for peak power at few-keV photon energies \cite{edwards2020x}. Understanding \acs{HHG} can benefit the advancement of attosecond science, which enables applications in the water window \cite{gibbon1996harmonic, ren2018attosecond} as well as in probing electron dynamics on the nanometer/attosecond scale \cite{krausz2009attosecond, boutu2012ultrafast}. \acs{HHG} is also associated with the potential laser intensity boost towards the Schwinger limit for studying strong field \ac{QED} effects. A relativistic oscillating plasma mirror can compress the pulse in time down to the attosecond range, and can shorten the wavelength by producing harmonics and thus decrease the focal spot size, achieving several orders of magnitude increase in the laser intensity \cite{quere2021reflecting}.
% It can also provide the huge intensities
% required for probing the non-linear quantum electrodynamics properties of the vacuum. Refocusing harmonics to $I>10^{29}w cm^{-2}$from an incident laser intensity of $I>10^{22}w cm^{-2}$
% can approach the Schwinger limit for electron-positron pair production in vacuum (electric field $10^{16}V cm^{-1}$). 
The \acs{HHG} process occurs when focusing ultrashort intense laser pulses to some media, including underdense plasmas produced by gas targets \cite{krausz2009attosecond, ren2018attosecond}, overdense plasmas formed at solid surfaces \cite{teubner2009high, gibbon1996harmonic}, and even bulk crystals \cite{ghimire2011observation, vampa2015linking}. The generation of harmonics in the gas medium is well understood, but there are limitations in the conversion efficiency due to the limitations in driving laser intensity. On the other hand, \acs{HHG} using solid targets, or plasma mirrors, can be driven at relativistic laser intensities and is believed to generate high brightness extreme-ultraviolet and x-ray radiation with improving conversion efficiency. Relativistic \acs{HHG} has been studied across various regimes and geometries both theoretically \cite{bulanov1994interaction, von1996high, gibbon1996harmonic, lichters1996short, naumova2004relativistic, gordienko2004relativistic, baeva2006theory, zhang2015enhancement, edwards2016waveform, edwards2016multipass, lecz2018enhancement, xu2019effect}, and experimentally \cite{dromey2006high, tarasevitch2007transition, easter2010high, heissler2012few, dollar2013scaling, kahaly2013direct, easter2013angular, cantono2018extreme, mitrofanov2018high, beier2019relativistic}. 

Since the harmonic cutoff and bandwidth broadening scales with the laser wavelength, there has been growing interest in using longer wavelength lasers to drive \acs{HHG} \cite{mitrofanov2018high, beier2019relativistic}. Other parameters in relativistic laser-plasma interactions, such as the ponderomotive potential and the critical density, also scale with wavelength. In this work, we investigate \acs{HHG} and corresponding phenomena using femtosecond mid-infrared laser pulses interacting with solid targets. We report experimental measurements of \acs{HHG} spectra and beam divergence using $2\mu m$ laser pulses. The power-law scaling of harmonic efficiency vs. harmonic order is examined. We show the intensity of horizontally-polarized harmonics and vertically-polarized harmonics when the driving laser pulses are polarized in horizontal, vertical, left-circular, and right-circular directions. The scaling of the third harmonic efficiency vs. laser intensity is also investigated.
% we also show the scaling with laser intensity? Shall I include it if it's not surely correct?? How do explain the drop in THG signal? Wrong measurements? Why not skip it?

\subsection{Surface \acs{HHG} mechanisms}
% this should be in the theory chapter? or not? maybe not. discuss lwfa/hhg/k-alpha in their own sections
The underlying physics of ultrafast laser-solid \acs{HHG} can be described by two mechanisms at two distinct regimes: the \acf{CWE} model dominates at sub-relativistic intensities ($a_0<1$) and short scale length ($\lambda/50<L<\lambda/15$) \cite{quere2006coherent}, while the \acf{ROM} model dominates at relativistic intensities ($a_0>1$) \cite{von1996high} and sub-wavelength scale length. Note that the optimal scale length for relativistic \acs{HHG} remains an open question. Dollar \textit{et al.} \cite{dollar2013scaling} found that the optimal scale length is $L\sim\lambda/5$ with experiments at $a_0=30$, and Kahaly \textit{et al.} \cite{kahaly2013direct} investigated the scale length dependence experimentally at $a_0=4.9$. Recently, Edwards \textit{et al.} \cite{edwards2020x} examined the \acs{HHG} efficiency at various intensities ($1<a_0<1000$) in simulation, which shows that the optimal gradient is dependent on the driving laser intensity regarding the target density.
% For example, \acs{ROM} is the dominant mechanism at both $a_0=2$ and $a_0=100$, but the latter would favor a sharper plasma gradient.

In \acs{CWE} at sub-relativistic intensities, the phase-matching between the driving laser electromagnetic field and the plasma oscillation (induced in the wake of energetic electron bunches) leads to the light emission of attosecond pulses lasting for one plasma oscillation period in each laser cycle. The attosecond pulse train interferes constructively and destructively to give even and odd harmonics \cite{quere2006coherent}. In \acs{ROM} at above relativistic intensities, the critical surface driven by the laser acts as a mirror to reflect the incident radiation. The periodic oscillation of the critical surface leads to alternate compression and stretching of the reflected laser based on the Doppler effect. The theoretical model with mathematical expressions can be found in \cite{lichters1996short}. At extreme intensities where $a_0\gg 1$, a theory called the coherent synchrotron emission model was proposed \cite{mikhailova2012isolated} to explain \acs{HHG} in the basic laws of synchrotron radiation. Under the combined force of laser and ionic potential, hot electrons from relativistic laser-solid interactions are accelerated in a different direction from their velocities, resulting in synchrotron-like trajectories. This synchrotron radiation gives rise to coherent radiation with a high-frequency cutoff.

Even and odd order harmonics can have different polarization depending on the polarization of the driving laser pulses. P-polarized pulses generate horizontally polarized harmonics, while S-polarized pulses generate even harmonics in horizontal polarization and odd harmonics in vertical polarization, according to the selection rules of the \acs{ROM} model \cite{lichters1996short}. On the other side, the \acs{CWE} model predicts \acs{HHG} signals only when the driving laser pulses are P-polarized \cite{quere2006coherent}. Circular polarization pulses have also been studied, and are found to cause a deflection in the angle of emission of the harmonics \cite{easter2013angular}.

\subsection{Experimental setup}
\begin{figure}[ht]
\centering
\includegraphics[height=0.25\textheight]{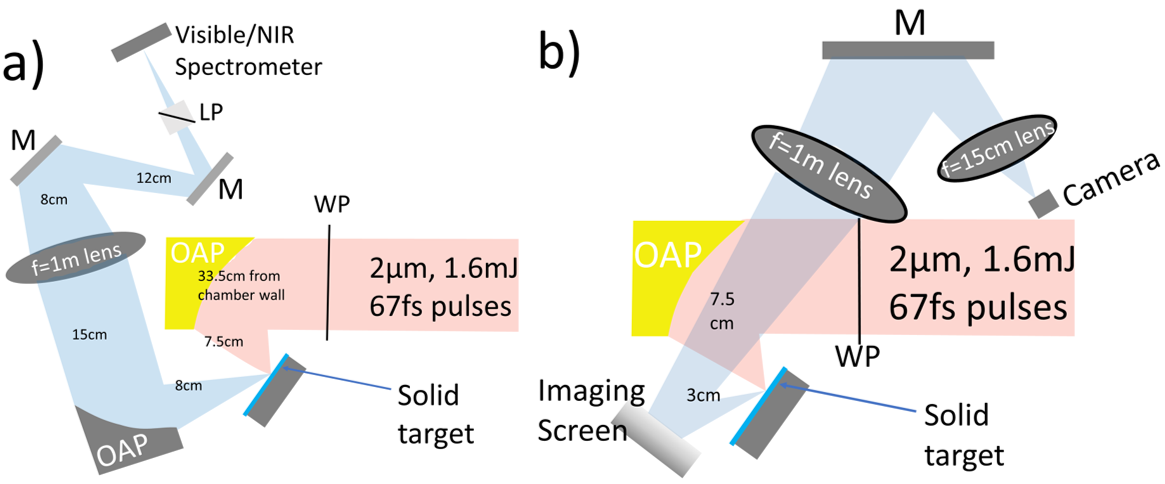}
\caption{Schematic of the experimental setup to measure the harmonic spectra (a) and the harmonic divergence (b). OAP: off-axis paraboloid; M: flat mirror; WP: wave plate; LP: linear polarizer; solid target: silicon or fused silica target.}
\label{HHGSetup}
\end{figure}
Fig. \ref{HHGSetup} demonstrates the setup for the \acs{HHG} experiment using 2 $\mu m$, 1.6 mJ, 67 fs laser pulses from the \acs{OPA} in \acs{Lambda-cubed}. Details with regard to the laser system and the \acs{OPA} have been introduced in Chap. \ref{sec:ExplaserSystem}. The output polarization from the \acs{OPA} was controlled with a half wave plate to provide pulses in horizontal, vertical, or circular polarization.  The laser pulses were focused by a gold-coated f/1.3 off-axis paraboloid onto the target at $\sim45^{\circ}$ incidence angle. The focal spot size (FWHM) was $5\mu m$, resulting in a peak intensity $I=\frac{0.94\cdot E_p}{\tau_p\cdot\pi\cdot w^2/2}=8.7\times 10^{16}\;W/cm^2$ and normalized vector potential $a_0\sim$ 0.5. The 2 $\mu m$ focal spot was optimized through a genetic algorithm and a deformable mirror measuring the second harmonic produced from plasma formation in the rarefied gas. Details of the focus optimization method are described in Section. \ref{sec:FocusOpt} or in \cite{lin2018focus}. The laser was focused through a thin pellicle to protect the \acs{OAP} from target debris. A rotary target stage was used to provide degrees of freedom in rotation as well as movement radially and longitudinally. The experiments were performed using a 5 mm thick fused silica target with or without 500 $\mu m$ thick silicon wafers attached to it. In Fig. \ref{HHGSetup}a, the generated harmonics were collected by a silver-coated f/2 off-axis paraboloid and a f=1m lens onto a Thorlabs CCS200 spectrometer (detection range 200nm - 1020nm). A Zinc Calcite linear polarizer was inserted to select harmonics in horizontal or vertical polarization. In Fig. \ref{HHGSetup}b, a diffusing screen was placed 2 cm in the
specular direction from the target to intercept the harmonics. The screen was imaged using a set of fused-silica lenses onto a Mightex CGE B013 U \acs{CCD} camera (active imager size 6.25mm $\times$ 5.01mm). Bandpass filters centered at $671\pm10$ nm (Edmund Optics 65-717), $500\pm10$ nm (Edmund Optics 65-694), and $400\pm50$ nm (Edmund Optics 65-741) were used to select the $3^{rd}$, $4^{th}$, and $5^{th}$ harmonic, respectively. Divergence measurements were integrated from 5 to 500 shots. Shot to shot fluctuations of $\sim15\%$ intensity were present in the divergence measurements due to stage instability.

\subsection{Results}
\begin{figure}[ht]
\centering
\includegraphics[height=0.3\textheight]{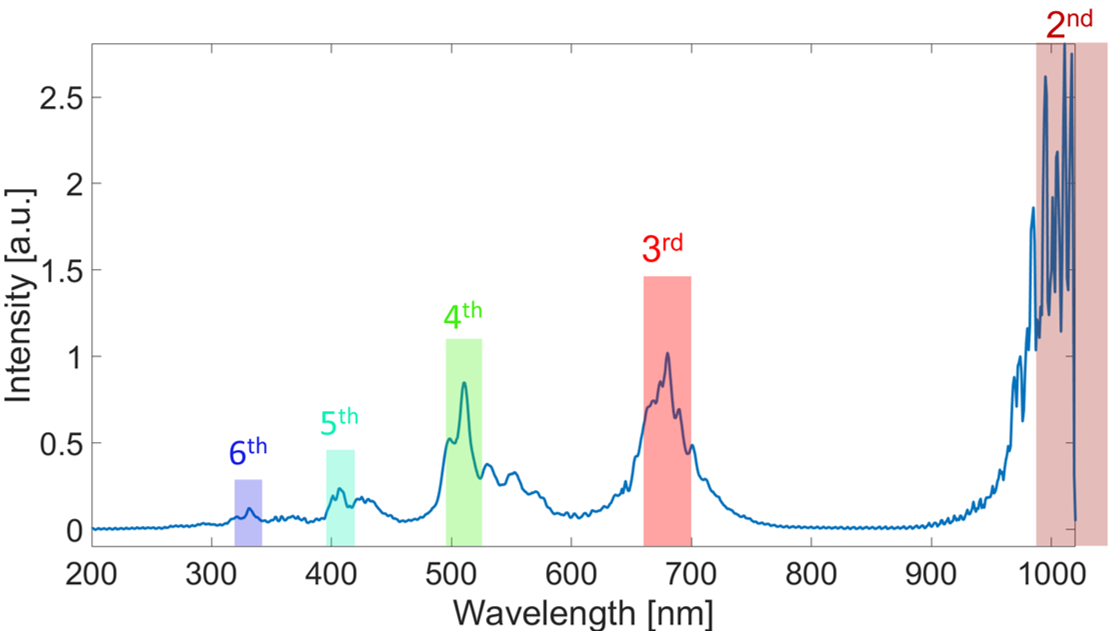}
\caption{A spectrum of horizontally polarized harmonics created by P-polarized iteractions with a silicon target. Orders of harmonics are labeled above the shaded areas, respectively. The spectrum was integrated over 10 shots.}
\label{HHGfig2}
\end{figure}

Fig. \ref{HHGfig2} shows an example of the measured harmonic spectrum from a silicon target driven by P-polarized pulses. The peak at the second harmonic is not complete due to the sharp cutoff at 1020 nm in the detection range of the spectrometer, and is noisy due to the noise amplification at low responsivity around this wavelength. Since the spectrometer has a low grating efficiency approaching the edge of the detection range, an intensity calibration is performed to correct the spectrum. Details of the calibration process have been explained in Sec. \ref{sec:ExpLPDiag}. The $2^{nd}$ through $6^{th}$ harmonics are observed and labeled within the colored area.

\begin{figure}[ht]
\centering
\includegraphics[height=0.3\textheight]{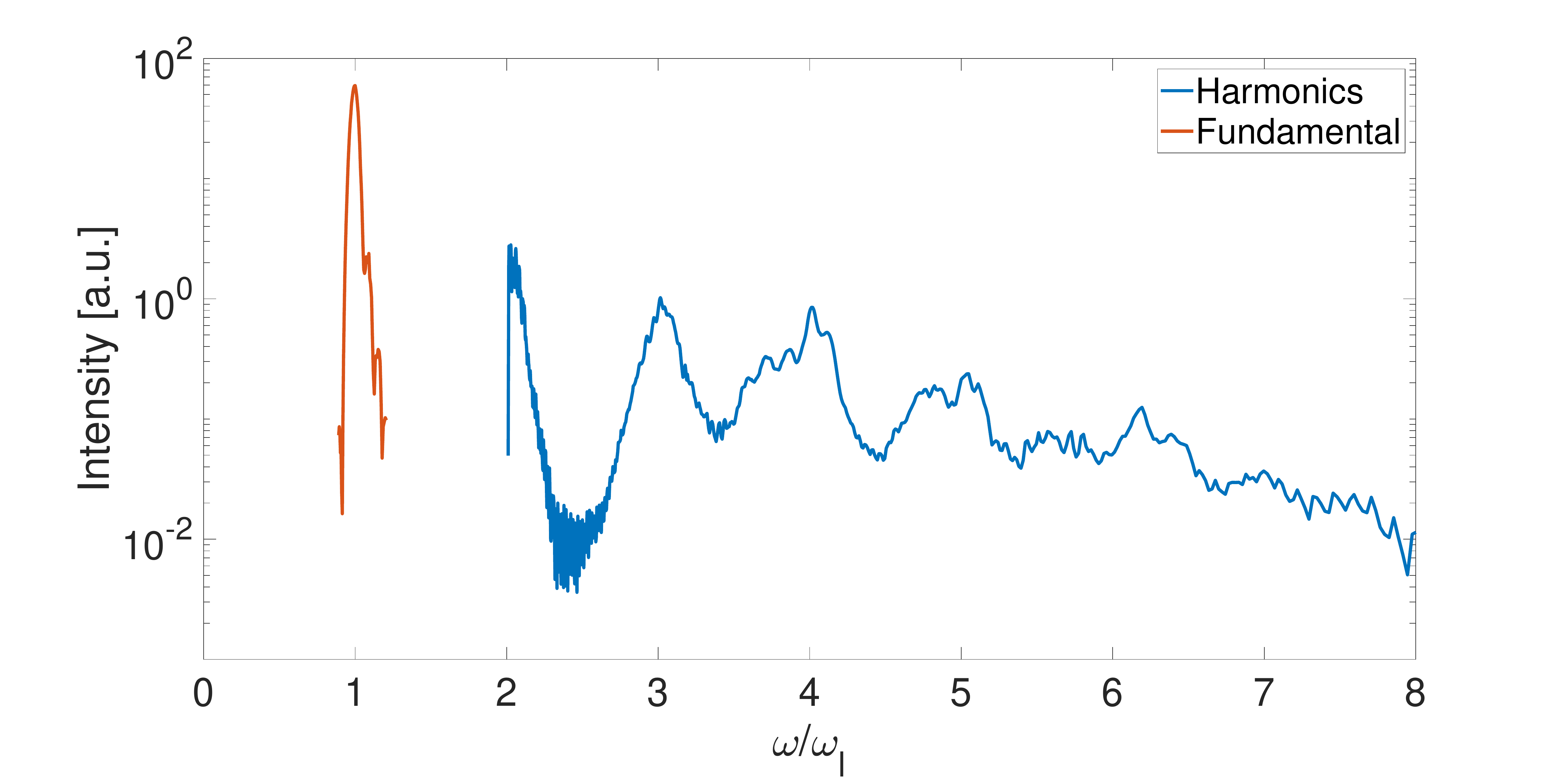}
\caption{The spectrum in Fig. \ref{HHGfig2} in frequency domain. The 2 $\mu m$ fundamental is plotted in orange while the harmonics are plotted in blue. Intensities of the two spectra are not calibrated thus not comparable.}
\label{HHGfigFreq}
\end{figure}

The spectrum of the 2 $\mu m$ driving laser was also measured using a Horiba iHR550 grating spectrometer and a Hamamatsu InSb P4631-03 point detector. The fundamental spectrum is plotted together with the harmonics in Fig. \ref{HHGfigFreq} in the frequency domain. Since the measured intensities from two distinct detectors were not calibrated to real units, the y-axis is normalized to arbitrary units and cannot be compared directly. It is observed that the high-frequency part of the fundamental' spectrum is modulated, and this modulation is also observed on the harmonics' spectra. Moreover, the modulation is noticeably broadened and amplified at higher-order harmonics. It should be pointed out that extra modulation can occur as the harmonics penetrate through plasmas, leading to more unique features in each harmonic's spectrum.

% Table generated by Excel2LaTeX from sheet 'Sheet2'
\begin{table}[ht]
  \centering
    \begin{tabular}{|c|c|c|c|c|c|c|}\hline
      Order & fundamental & $2^{nd}$   & $3^{rd}$ & $4^{th}$ & $5^{th}$ & $6^{th}$ \\\hline
    Wavelength [nm] & 2050$\pm$65 & $>$1017 & 680$\pm$36 & 511$\pm$24 & 407$\pm$17 & 331$\pm$14 \\\hline
    $\lambda_l/n$ [nm] &   2050   & 1025 & 683   & 512  & 410  & 342 \\\hline
    Intensity [a.u.] &    /   & $\sim$2.809 & 1.022   & 0.849  & 0.237  & 0.125 \\\hline
    $\Delta\omega/\omega$ & 0.058  & $>$0.04  & 0.055  & 0.048  & 0.054  & 0.056 \\\hline
    \end{tabular}%
  \caption{Features of the harmonics shown in Fig. \ref{HHGfig2} and Fig. \ref{HHGfigFreq}. $\lambda_l$=2050 nm is the fundamental wavelength and n is the harmonic order.}
  \label{tab:hhgScaling}%
\end{table}%

Tab. \ref{tab:hhgScaling} illustrates the central wavelength, spectral width, intensity, and bandwidth at \acf{FWHM} of each harmonic. Since the spectrum at the second harmonic is noisy and incomplete, the numbers listed in its column are approximations based on the best that can be observed. The first two rows show the central wavelengths of the measured harmonics and the central wavelengths calculated from the fundamental wavelengths at 2050 nm. The central wavelengths of the $3^{nd}$ through $5^{th}$ harmonics are within $\pm3$ nm from the calculated values, while the $6^{th}$ harmonic is noticeably red-shifted.
The third row shows the peak intensity of each harmonic. The intensity of the fundamental is not listed for comparison since it was measured with a different detector and was not calibrated to real units. Although the absolute harmonic efficiencies are not available, normalizing the intensities to arbitrary units provides adequate information to study the scaling laws of the harmonic efficiencies. Details of the power-law scaling are discussed in the next paragraph. The last row lists the relative \acs{FWHM} bandwidth of each harmonic normalized to the harmonic frequency. Although the $\Delta\omega/\omega$ of the $4^{th}$ harmonic is relatively smaller, most of the observed harmonics inherited the bandwidth from the spectrum of the fundamental.

\begin{figure}[H]
\centering
\includegraphics[height=0.4\textheight]{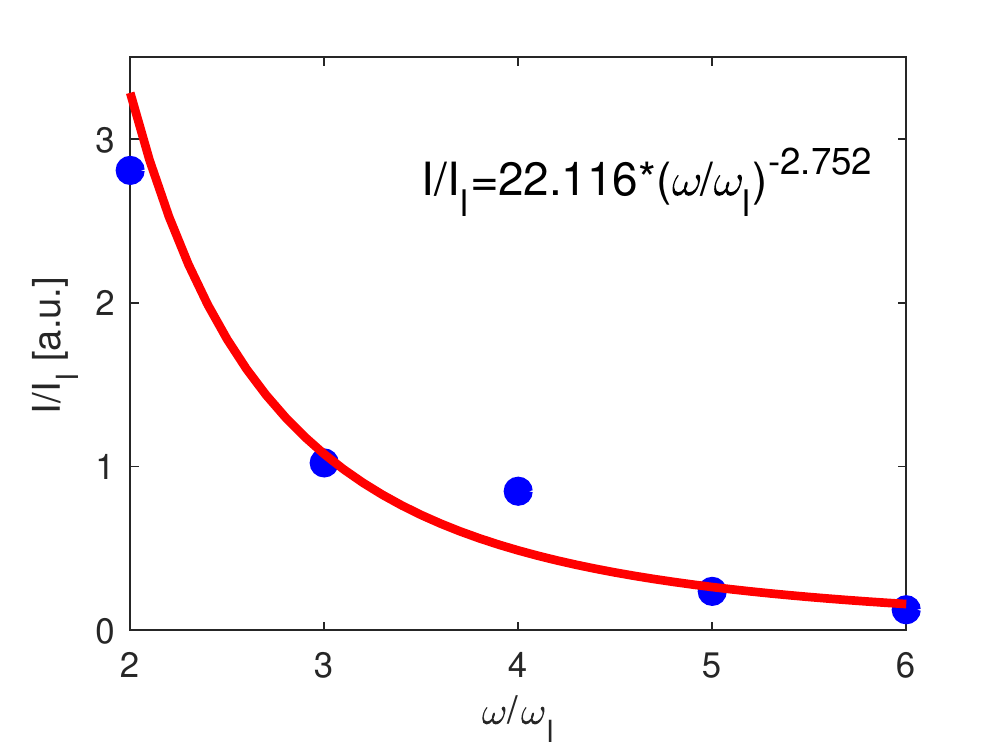}
\caption{Conversion efficiency vs. harmonic order. The solid curve is a power-law fit.}
\label{HHGfigPowerLaw}
\end{figure}

An important question to address in \acs{HHG} studies is the scaling law of harmonic efficiency, which is often expressed in a power function: $I(\omega)\propto (\omega/\omega_L)^p$ up to a cutoff frequency. Knowing the value of the coefficient p is of great significance because it determines the value of \acs{HHG} applications as x-ray sources. Compared to other x-ray sources like \acs{XFELs}, PW-class lasers may have advantages in source brightness if large p-values can be achieved. Dollar \textit{et al.} \cite{dollar2013scaling} experimentally studied the power-law scaling and observed $p\sim-4.5$, and  Easter in his thesis \cite{easter2010thesis} reported similar values of $-5.5<p<-3.4$ across a range of intensities where $5\times10^{17}W\mu m^2cm^{-2}<I\lambda^2<1\times10^{19}W\mu m^2cm^{-2}$. Analytic studies using the \acs{ROM} model suggests a p=-8/3 limit up to a \acs{HHG} cutoff $\propto\gamma^3$ \cite{baeva2006theory}, where $\gamma$ is the Lorentz factor, and such values have been observed in experiments \cite{dromey2006high, jahn2019towards}. Smaller values approaching the theoretically predicted -4/3 or -6/5 limit \cite{an2010enhanced} have also been reported experimentally \cite{dromey2012coherent}. A complete review of the power-law scaling in relativistic \acs{HHG}, including numerical simulations over a large parameter space, is provided by Edwards and Mikhailova in Ref. \cite{edwards2020x}. It is shown that the value of p is dependent on the similarity parameter $S=\frac{n_e/n_{cr}}{a_0}$, where $n_e$ is the electron density and $n_{cr}$ is the critical density. The similarity theory for relativistic laser-plasma interactions is proposed by Gordienko \textit{et al.} \cite{gordienko2005scalings}. For $1/S<0.1$, the power-law coefficient $p\leq-8/3$. To have p-values close to the predicted -4/3 limit, the laser-plasma condition favors $0.3<1/S<0.5$. In our case, fitting the harmonic intensities in Tab. \ref{tab:hhgScaling} using a power function gives p=-2.752 with regression determination $R^2=0.93$, as is shown in Fig. \ref{HHGfigPowerLaw}. This number is close to the predicted value of -8/3=-2.67 in theory \cite{baeva2006theory}. Since the experiment was conducted at just shy of the relativistic intensity ($a_0\sim0.5$), it is not surprising that our p-value is closer to the -8/3 prediction than the -4/3 prediction. 

It has to be pointed out that the reported results above are analyzed using one spectrum integrated over ten shots. Among all the $\sim150$ spectra recorded on that shot day, $\sim40$ show similar spectral features and have $-3<p<-2.7$. The periodic wobbling of the target on the rotary stage should be responsible for the other spectra, as many were taken when the target moved out of focus.

\subsubsection{\acs{HHG} divergence}
\begin{figure}[H]
\centering
\includegraphics[height=0.3\textheight]{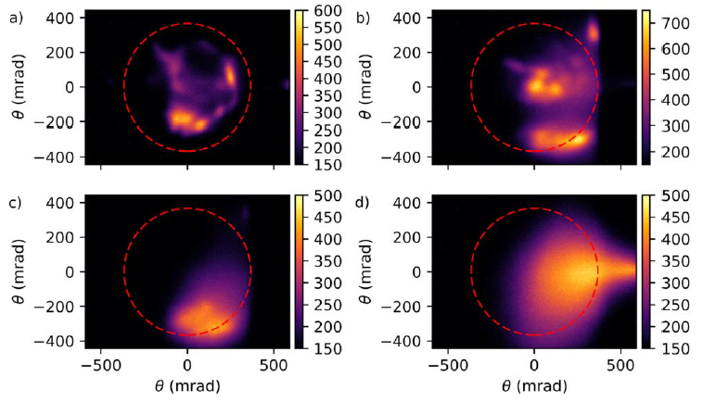}
\caption{Divergence of the harmonics: (a): $3w_l$ from silicon target; (b): $3w_l$ from fused silica target; (c): $4w_l$ from fused silica target; (d): $3w_l$ from fused silica target. Dashed circles represent the divergence of the fundamental defined by the f/1.3 focusing geometry. The figure has been published in Ref. \cite{beier2019relativistic}.}
\label{HHGfigDivergence}
\end{figure}
Fig. \ref{HHGfigDivergence} shows the measured divergence of harmonics from p-polarized interactions with silicon and fused silica targets. 
The images were taken after passing through bandpass filters for each harmonic wavelength, as is described previously, and structure for lower
order harmonics are observed. As seen in Fig. \ref{HHGfigDivergence}a–c, it appears as though the low order harmonics are generated in beamlets, which are not colinear with the specular direction. Broadband plasma recombination emission was detected during the divergence measurements at the $5^{th}$ harmonic, which should be responsible for the profile in Fig. \ref{HHGfigDivergence}d. Note that measurements in Fig. \ref{HHGfigDivergence}b–d were performed with fused silica targets rather than silicon wafer targets reported in Fig. \ref{HHGfig2}a and Fig. \ref{HHGfigFreq}. The $5^{th}$ harmonic was never observed from fused silica targets interactions, which could be explained by the diffuse plasma discharge observed in Fig. \ref{HHGfigDivergence}d.

\subsubsection{\acs{HHG} polarization dependence}
\begin{figure}[H]
\centering
\includegraphics[width=0.95\textwidth]{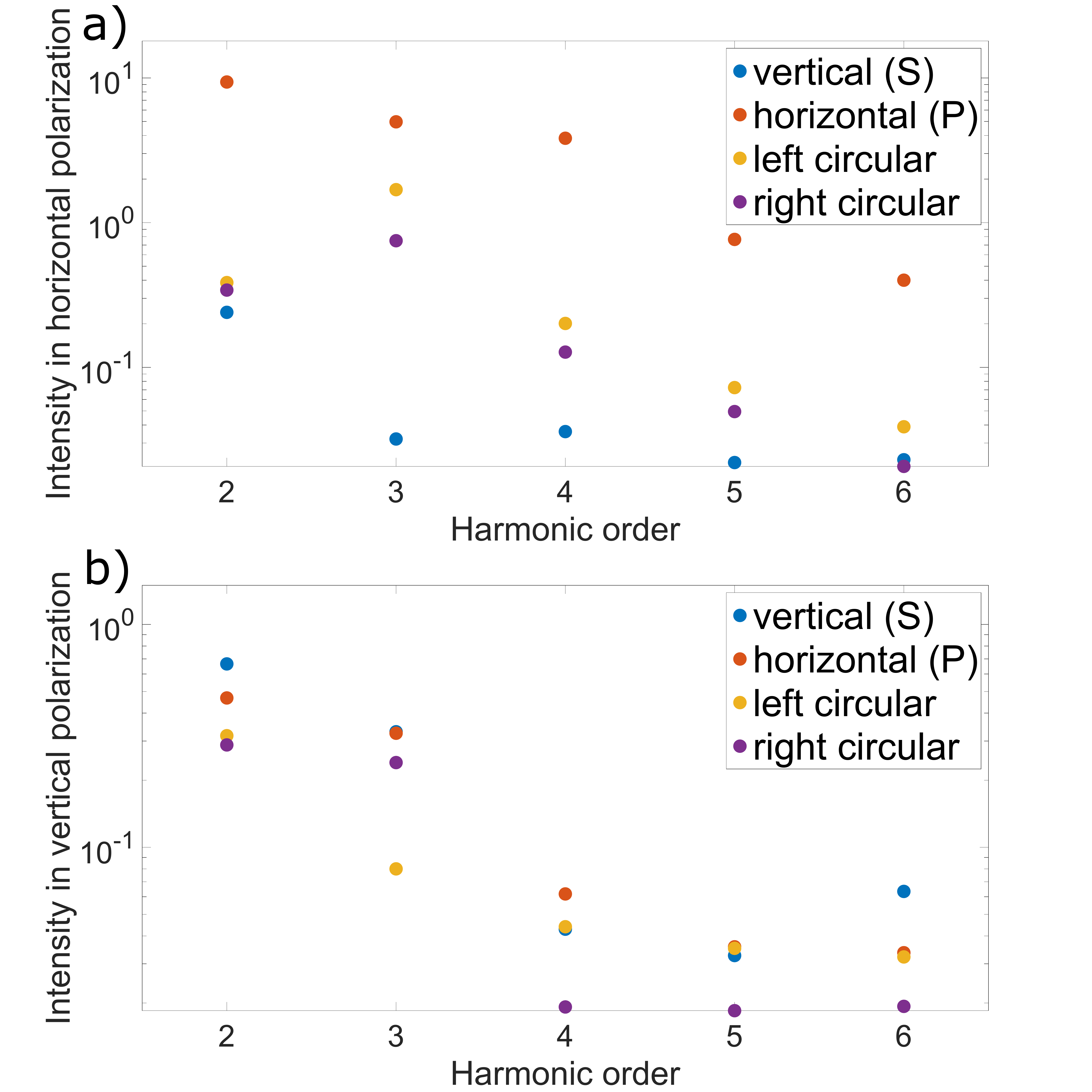}
\caption{Intensity of (a): horizontally-polarized harmonics, and (b) vertically-polarized harmonics. The driving laser pulses were polarized in horizontal (orange), vertical (blue), left circular (yellow), and right circular (pink) directions. Each of the data point was integrated over 5 shots.}
\label{HHGfigPol}
\end{figure}

The polarization dependence of the harmonics from silicon targets was investigated by placing a wave plate in the driving beam path and a linear polarizer in the harmonic diagnostic path. The experimental setup is illustrated in Fig. \ref{HHGSetup}a. The 2$\mu m$ driving pulse was polarized in the linear directions using a half wave plate (CWO-2050-02-08-R10) or in the circular directions using both the half wave plate and a quarter wave plate (WPQ10M-2020). The driving laser experienced $\sim10\%$ decrease in pulse energy when passing through a wave plate. Note that a laser pulse is called "left hand polarized" when the electric field rotates in the counter-clockwise direction, and "right hand polarized" when the electric field rotates in the clockwise direction. Fig. \ref{HHGfigPol}a shows the intensities of the horizontally-polarized part of the second to sixth harmonics. It is observed that the orange dots are an order of magnitude higher than others for most harmonics, which corresponds to horizontally-polarized driving laser pulses, or namely P-polarized interactions. On the other hand, the blue dots from vertically polarized laser pulses or S-polarized interactions are lower than those from any other polarization. Compared to P-polarized interactions, S-polarized interactions generate horizontally-polarized harmonics at a negligible efficiency (two orders of magnitude lower). This phenomenon agrees with the polarization selection rules predicted by the \acs{CWE} model \cite{quere2006coherent}, as is tabulated in Tab. \ref{tab:hhgSelect}. In \acs{ROM}, horizontally-polarized even order harmonics are expected when the driving interaction is S-polarized \cite{lichters1996short}, which is not observed in Fig. \ref{HHGfigPol}a. However, it has to be pointed out that the electric fields in our experiments differ from the plane wave approximation used in theories because of the high numerical aperture focus.

\begin{table}[ht]
  \centering
    \begin{tabular}{|c|c|c|c|}
    \toprule
    \multirow{2}[4]{*}{Driving laser} & \multicolumn{2}{c|}{Harmonics (ROM)} & \multirow{2}[4]{*}{Harmonics (CWE)} \\
\cmidrule{2-3}          & Odd   & Even  &  \\
    \midrule
    Horizontal & Horizontal & Horizontal & Yes  \\
    \midrule
    Vertical & Vertical & Horizontal & No \\
    \midrule
    Circular & Both  & Both  & / \\
    \bottomrule
    \end{tabular}%
  \caption{Polarization dependence of \acs{ROM} \cite{lichters1996short} and \acs{CWE} \cite{quere2006coherent} harmonics at normal incidence.}
  \label{tab:hhgSelect}%
\end{table}%

Circularly-polarized laser pulses were also employed to generate horizontally-polarized harmonics, and the efficiency was lower than that for P-polarized interactions but higher than that for S-polarized interactions. This observation is in line with the experimental results from Easter \textit{et al.} \cite{easter2013angular}, which compares harmonic spectra from circularly-polarized pulses to those from horizontally-polarized pulses. Pulses in left-hand circular polarization (yellow dots) and in right-hand circular polarization (pink dots) have an observable difference in harmonic efficiency, as is shown in Fig. \ref{HHGfigPol}a. However, it should be pointed out that circularly-polarized pulses can deflect the emitting harmonics at a small angle \cite{easter2013angular}.

Fig. \ref{HHGfigPol}b demonstrates the intensities of the vertically-polarized part of the second to sixth harmonics. In general, harmonic intensities in Fig. \ref{HHGfigPol}b is an order of magnitude lower than those in Fig. \ref{HHGfigPol}a. Even and odd order harmonics display no distinct behavior as is predicted by \acs{ROM}. However, it is noticeable that circularly-polarized laser pulses are less efficient in generating vertically-polarized harmonics than linearly-polarized laser pulses.

\subsubsection{Scaling with laser intensity}

\begin{figure}[H]
\centering
\includegraphics[width=0.95\textwidth]{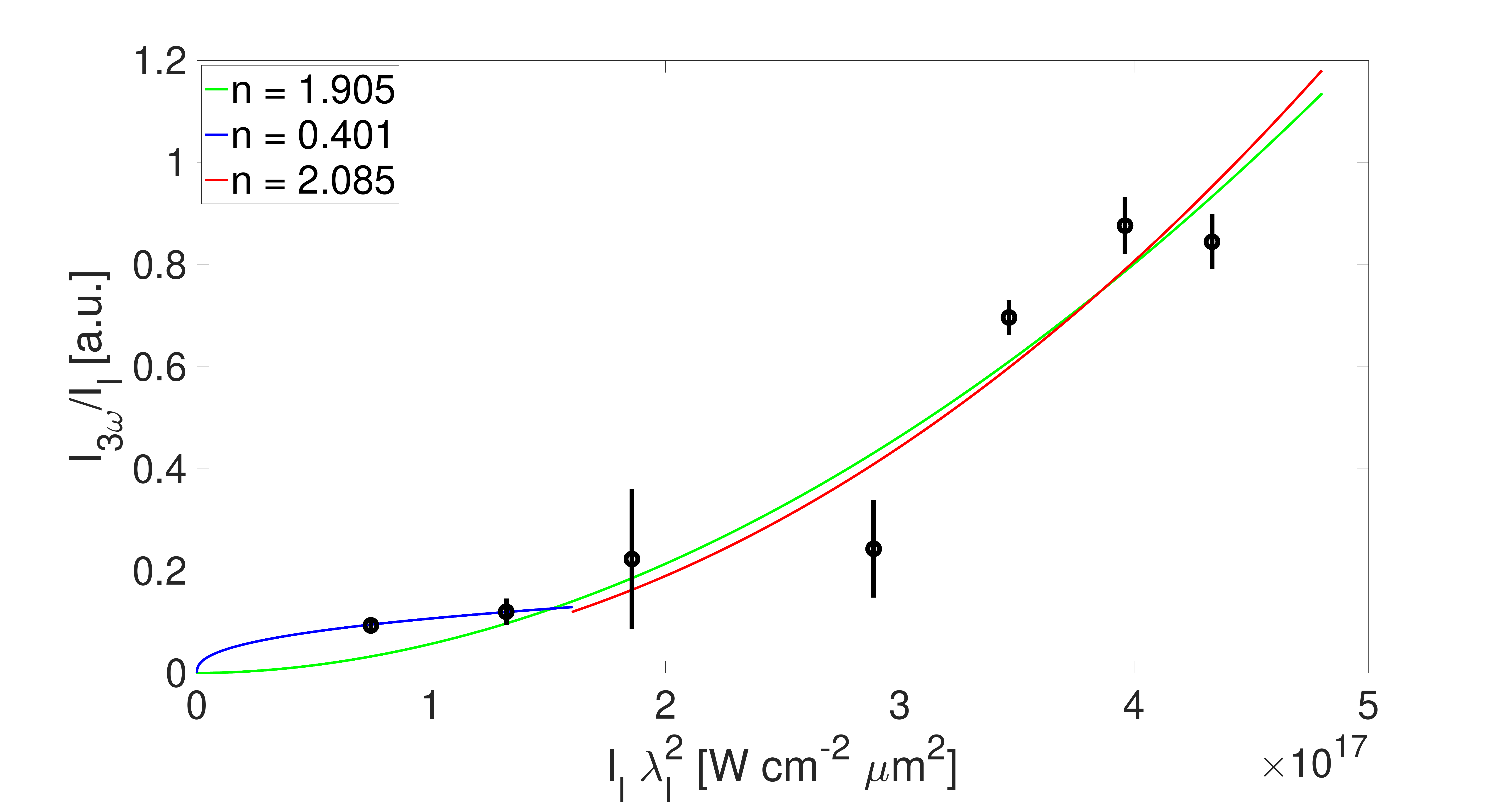}
\caption{Intensity scaling of the third harmonic vs. laser intensity. $\sim$Three measurements were made at each laser intensity to provide the errorbar, while each of the measurement was integrated over $25\sim100$ shots. The green, blue, and red curves are fitted from different portion of the intensity range.}
\label{HHGfigIntensity}
\end{figure}

The efficiencies of the third harmonic from silicon targets were plotted as the blue dots in Fig. \ref{HHGfigIntensity} over a range of laser intensities within $3.7\times10^{16}\;W cm^{-2} \mu m^2<I_l\lambda_l^2<4.3\times10^{17}\;W cm^{-2} \mu m^2$. Fitting all the measurements into a power scaling as a function of the laser intensity $\propto I_l^n$ yields n=1.905 in the green line. The intensity scaling can be very different in different \acs{HHG} models. Quere \textit{et al.} \cite{quere2006coherent} experimentally measured the \acs{CWE} harmonics within $4\times10^{15}\;W cm^{-2} \mu m^2<I_l\lambda_l^2<2\times10^{16}\;W cm^{-2} \mu m^2$ and showed $n\sim0.4$. Gibbon \textit{et al.} \cite{gibbon1996harmonic} predicted a highly nonlinear behavior ($n=2\sim2.5$) in the \acs{ROM} model with simulation results within $2\times10^{17}\;W cm^{-2} \mu m^2<I_l\lambda_l^2<1\times10^{19}\;W cm^{-2} \mu m^2$. Since our experiment was performed in the transition regime of these two intensities ranges, it is worth splitting the green curve in Fig. \ref{HHGfigIntensity} into two parts. The blue line covers $3.7\times10^{16}\;W cm^{-2} \mu m^2<I_l\lambda_l^2<1.3\times10^{17}\;W cm^{-2} \mu m^2$, and its scaling parameter n=0.401 matches the \acs{CWE} observations very well. The red line covers $1.9\times10^{16}\;W cm^{-2} \mu m^2<I_l\lambda_l^2<4.3\times10^{17}\;W cm^{-2} \mu m^2$, and is fitted into a larger scaling parameter at n=2.085. As the laser intensity increases, relativistic effects become stronger and the \acs{ROM} becomes more dominant, which leads to a more nonlinear intensity scaling.

\subsection{Discussion}
In summary, we have demonstrated harmonic spectra and harmonic divergence measured experimentally when 2 $\mu m$ laser pulses interacted with silicon and glass targets. The experiments were performed at a weak relativistic regime ($a_0=0.5$), and the harmonic generation mechanism sits in the transition regime between the \acs{ROM} dominant regime and the \acs{CWE} dominant regime. The harmonic efficiency scales with harmonic order in a power law as $I(\omega)\propto (\omega/\omega_L)^{-2.752}$, which is close to the frequently quoted value of -8/3 predicted by the \acs{ROM} model. We have also investigated the third harmonic efficiency vs. laser intensity in a power function $I_{3\omega}\propto I_L^n$, which shows an increasingly nonlinear scaling as the laser intensity ramps up. We have studied the intensity of harmonics polarized in horizontal and vertical directions, when the driving laser pulses are polarized in horizontal, vertical, left-circular, and right-circular directions. For linearly polarized driving pulses, the results do not show the distinct feature in even and odd harmonics predicted by the unique selection rule of the \acs{ROM} model. For circularly-polarized laser pulses, both even and odd harmonics were observed, which is in line with the \acs{ROM} selection rule. In terms of harmonic efficiency, generating horizontally-polarized harmonics with P-polarized interactions is favored. 

As a potential source for bright attosecond x-rays, a major challenge in practical applications of relativistic \acs{HHG} is that the harmonics tend to emit as periodic pulse trains rather than single pulses. There have been extensive studies to isolate single attosecond pulses from the pulse train, utilizing either spatial separation or temporal separation. The former can be achieved by introducing an angular dispersion in the laser beam to allow different frequency components of the laser pulse to disperse along the perpendicular axis and leads to the spatial chirp. As a result, the laser cycle period varies along the perpendicular axis, leading to a rotating wave vector. This attosecond lighthouse effect has been presented experimentally by Wheeler \textit{et al.} \cite{wheeler2012attosecond}. The latter can potentially be achieved by the polarization gating technique, which has been proposed and studied with \acs{PIC} simulations \cite{rykovanov2008intense, yeung2015noncollinear, chen2018isolated}. The concept is based on the fact that \acs{HHG} using elliptically polarized pulses at normal incidence has significantly lower efficiency than that of using linearly polarized pulses. By focusing a left hand circularly polarized pulse and a right hand circularly polarized pulse onto a plasma mirror, the two electric fields rotating in opposite directions overlap at focus. A linear gate lasting a laser cycle is thus formed, and the harmonics generated during this gate period are no longer suppressed.

Another future application of \acs{HHG} is to push the limit of the peak intensity that a laser pulse can ever achieve. The peak laser intensity $I\sim\frac{E}{\tau\cdot A}$, where E is the pulse energy, $\tau$ is the pulse duration, and A is the focal spot size. While numerous progress has been achieved towards higher laser pulse energy and shorter pulse duration, the smallest possible focal spot size is constrained by the nature of light, namely the diffraction limit. However, \acs{HHG} provides an alternative to achieving a smaller focal spot size by converting an optical wavelength to its harmonic wavelength. Quere \textit{et al.} \cite{quere2021reflecting} predict an intensity gain up to $10^6$ for a 4-PW laser system. However, there are challenges to overcome before it can be realized experimentally, including but not limited to controlling the plasma mirror curvature and measuring extreme intensities.

\clearpage
\section[]{Towards isolated attosecond electron bunches\footnotemark}
\label{sec:Atto}
\footnotetext{This section co-authored with Batson, T., Nees, J., Thomas, A. G. R., and Krushelnick, K. (2020): Towards isolated attosecond electron bunches using ultrashort-pulse laser-solid interactions. Scientific reports, 10(1), 1-11.}
Driving high-intensity ultrashort laser pulses into overdense plasmas tends to produce electron beams with higher charge number at lower peak energy compared to those from underdense plasmas in \acf{LWFA}. This interaction between relativistic laser pulses and solids has been studied in the past decades in different regimes of laser pulse characteristics and plasma conditions. When the pulse duration is ultrashort ($\sim$ 30fs) and the plasma density scale-length is short compared to laser wavelength ($L_s=n_e\cdot(\frac{dn}{de})^{-1}<0.1\lambda$), the incident radiation drives a periodic motion of the critical surface to stretch and compress the light reflected off the surface within each cycle. This is known as the relativistic oscillating mirror model, which leads to the generation of attosecond light sources through \acf{HHG} \cite{bulanov1994interaction, lichters1996short, baeva2006theory, heissler2012few, dollar2013scaling} as well as ejection of electrons from the surface. The reflected laser field then accelerates these electrons to relativistic energy through the \acf{VLA} mechanism \cite{popov2009vacuum, chopineau2019identification}, usually resulting in a ring-shaped electron beam structure \cite{thevenet2016vacuum, tsymbalov2019well}. If the plasma scale-length is longer than the laser wavelength, the interaction mainly happens near the critical density and the acceleration mechanism becomes extremely complex, including ponderomotive acceleration \cite{zhang2004emission}, surface quasistatic fields \cite{li2006observation} and direct laser acceleration. Varying the scale-length from $<0.1\lambda$ to $5.5\lambda$, an optimal condition for generating quasi-monoenergetic electrons is found at $L_s=0.5\lambda$ using relativistic high repetition rate laser system with abundant statistics \cite{mordovanakis2009quasimonoenergetic}. Scaling laws of electron temperature have also been studied \cite{mordovanakis2010temperature}. A theory for the electron acceleration mechanism in this regime is the standing-wave acceleration. Superposition of the incident half and the reflected half of the laser pulse results in a standing wave pattern, and the modulated electric and magnetic fields in the standing wave can launch electrons in the forward and backward directions \cite{orban2015backward, ngirmang2016three, kemp2008hot}. Going to even shorter pulse duration matching the plasma wavelength ($t=\lambda_p/2c$), wakefields can be excited\cite{zaim2019few}. To have wakefield acceleration happen in laser-solid interaction, a few-cycle pulse driver ($\tau\sim T_0/2cos\theta$) is required\cite{zaim2019few} so that large-amplitude plasma waves can be generated along the density gradient up to the effective turning point $n_{cr}cos^2\theta$.

Most of the reported laser-solid experiments were set up in normal\cite{orban2015backward, feister2017relativistic, morrison2015backward} or oblique\cite{mordovanakis2009quasimonoenergetic, bocoum2016anticorrelated, tsymbalov2019well, thevenet2016vacuum} ($\sim45^{\circ}$) incidence. However, Naumova \textit{et al.} \cite{naumova2004attosecond} predicted the existence of attosecond electron bunches in a grazing incidence setup. Since then there have been extensive theoretical predictions and simulations for attosecond electron bunches using various geometries, such as droplet target\cite{liseykina2010relativistic, di2015relativistic}, nanofilm\cite{kulagin2007theoretical} and transversely-thin slice target\cite{ma2006dense}, no experimental indication has been reported yet. In this work, we confirm that generating attosecond electron bunches favors a larger angle of incidence (measured from target normal) in both experiments and simulations. We utilize \acs{PIC} simulation and particle tracking to investigate the formation mechanism of these attosecond electron bunches. Furthermore, we generate isolated attosecond electron bunches using single-cycle laser pulses in \acs{PIC} simulations. We study some key parameters that govern this process, such as preplasma density scale length, laser intensity, \acf{CEP} and focal-spot size.

\subsection{Experimental and computational setup}
\begin{figure}[ht]
\centering
\includegraphics[height=0.45\columnwidth]{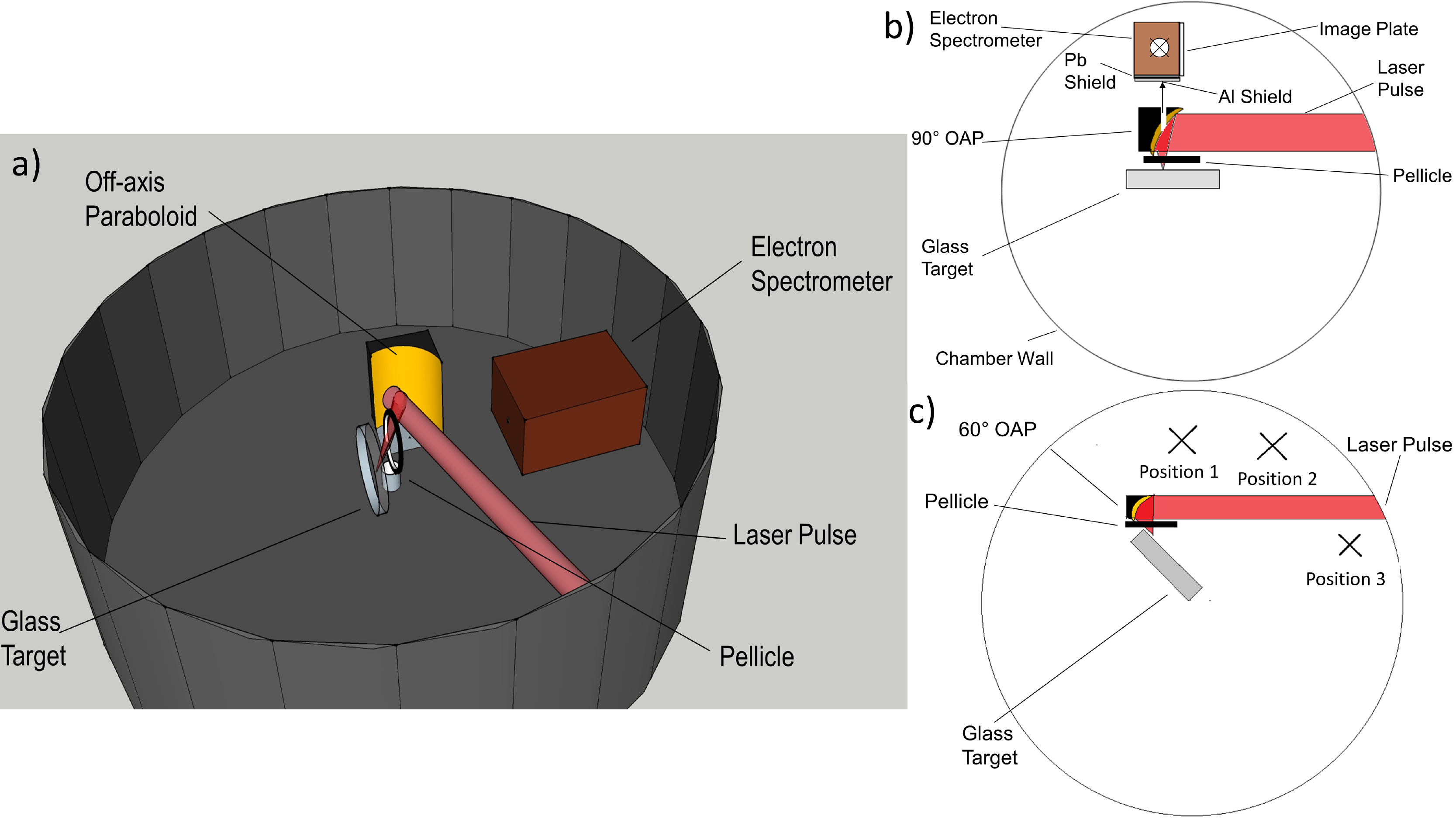}
\caption{Schematic of the experimental setup (a) and top view of the normal incidence geometry (b) as well as the grazing incidence geometry (c). Electron spectrometer: 1.15 kG magnetic spectrometer with a Fujifilm MS image plate covered with lead and aluminum shielding; pellicle: 2-inch diameter, 2 $\mu m$ thick nitrocellulose pellicle; target: 4-inch diameter, 6 mm thick glass; \acs{OAP}: off-axis paraboloid. The spectrometer is placed at three different positions in (c).}
\label{AttoSetup}
\end{figure}

The experiments were performed at the University of Michigan, using the $\lambda^3$ laser system with 30fs, 12mJ, $0.8\mu m$ pulses at a repetition rate of 480 Hz. The experimental setup is shown in Fig. \ref{AttoSetup}. A P-polarized laser pulse was focused from a $90^{\circ}$ ($60^{\circ}$) f/1 gold off-axis paraboloid onto the target at normal (grazing) incidence. Note that a hole was drilled in the center of the $90^{\circ}$ \acs{OAP} in order to capture signals on the spectrometer. The geometry of normal and grazing incidence cases are drawn in Fig. \ref{AttoSetup}b,c, respectively. In such a tight focus setup, the light rays cover a range of angles (approximately $\pm20^{\circ}$). The angle of the central ray in grazing incidence is estimated to be $\sim70^{\circ}$, which is as close to grazing as the beam could get without clipping. The focal spot size (\acs{FWHM}) was $1.5\mu m$, resulting in a peak intensity $I=\frac{0.94\cdot E_p}{\tau_p\cdot\pi\cdot w^2/2}=1.6\times 10^{19}\;W/cm^2$ and normalized vector potential $a_0\sim$ 2.6. The laser was focused through a thin pellicle to protect the \acs{OAP} from debris. A rotary target stage was used to provide degrees of freedom in rotation as well as movement radially and longitudinally. Most of the experiments were performed with glass targets unless specified to be copper. An external prepulse was available by placing a thin pellicle in the optical system to pick up $\sim8\%$ of the pulse energy. The prepulse intensity is around $5\times10^{16}W/ cm^{2}$. The generated electrons were diagnosed by a Fuji MS image plate after being deflected by a 1.15 KG magnetic field. The image plate efficiency was included using the published calibration data\cite{bonnet2013response}. Lead and aluminum shieldings were used to prevent noise from Bremsstrahlung radiation. A detailed description of the spectrometer can be found in reference\cite{mordovanakis2009quasimonoenergetic, mordovanakis2010temperature}.

\acs{PIC} simulations were performed using the OSIRIS\cite{OsirisRef, OsirisRef2} 4.4.4 framework in 2D3V Cartesian geometry. There are 100 macroparticles per cell and the grid size is $\lambda/32\times\lambda/32=0.025 \mu m\times0.025 \mu m$, where $\lambda=0.8 \mu m$ is the laser wavelength. The convergence of the simulation is checked using up to $\lambda/64\times\lambda/64$ grid size. The computational time-resolution is 0.027 fs or $\sim100$ step per laser cycle. The laser pulse is assumed to be Gaussian in both the longitudinal and transverse direction with a \acs{FWHM} pulse duration of $\tau=30 fs$. The laser pulse is continuously launched from the wall and focused down to beam waist $w0=\acs{FWHM}\times1.699/2=1.28 \mu m$, $a_0=2.5$. The simulation box size is $64 \mu m\times64\mu m$ in normal ($0^{\circ}$) incidence geometry, $48 \mu m\times48\mu m$ in oblique ($45^{\circ}$) incidence geometry and $72 \mu m\times28\mu m$ in grazing ($76^{\circ}$) incidence geometry. The reported angles are those of the central rays. The simulations were run for 270 fs, 210 fs, and 300 fs, respectively, until the interesting electrons left the simulation box. The initial plasma density profile is described by a uniform glass-solid-density ($5.01\times10^{22}cm^{-3}$) region plus an exponential tail, as is illustrated in Fig. \ref{AttoFig2}f. 

\subsection{Results}
\begin{figure}[ht]
\centering
\includegraphics[height=0.4\textheight]{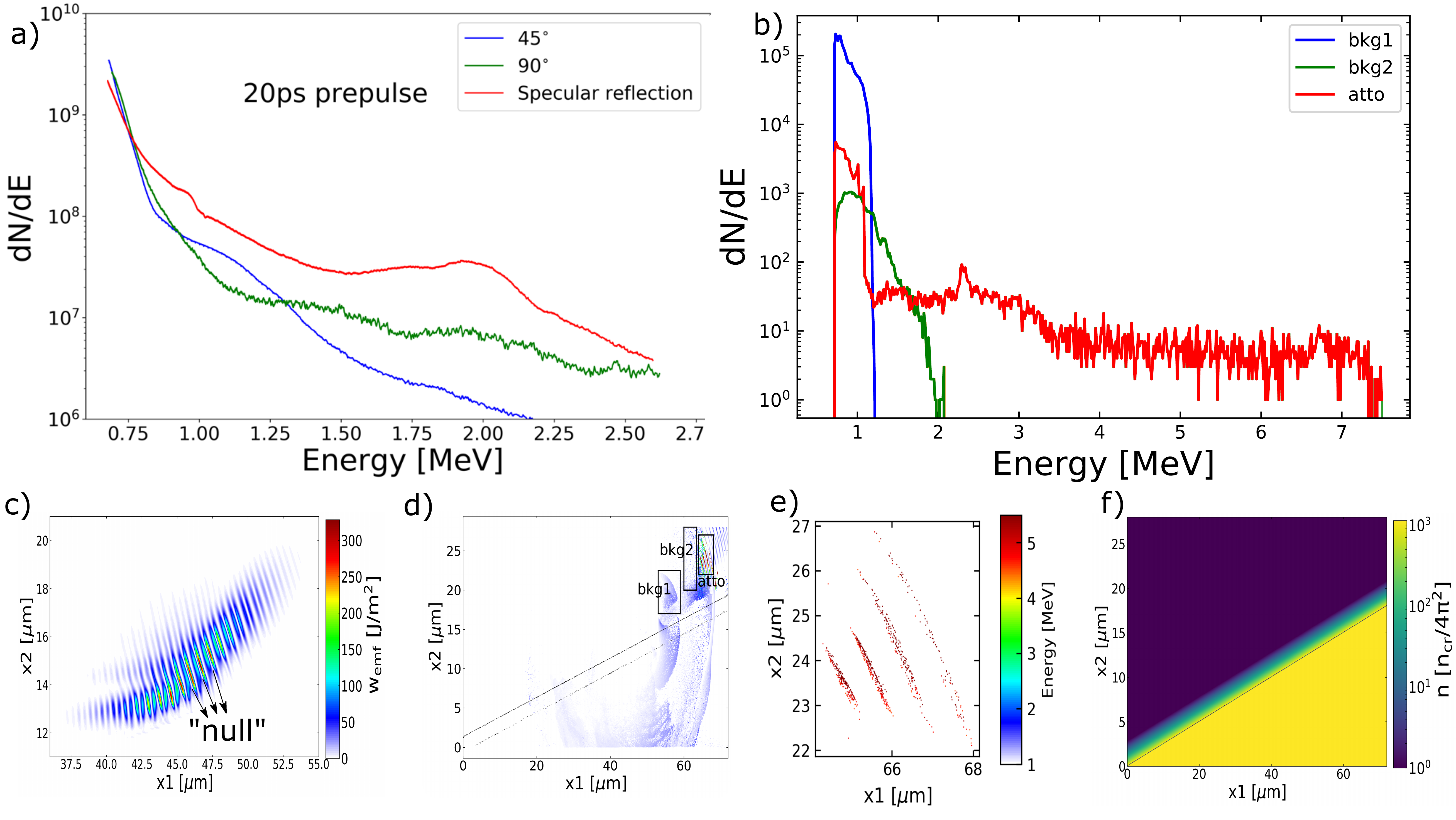}
\caption{Electron energy spectra from experiments (a) and simulations (b) in grazing incidence setup. $45^{\circ}$, $90^{\circ}$ and specular reflection in (a) correspond to position 1, 2 and 3 labeled in Fig. \ref{AttoSetup}c, respectively. (b): energy spectra of electrons labeled in (d). The bin size of the spectra is 72 bins/MeV. Both energy spectra in (a) and (b) have dn/dE in arbitrary units. (c): Electromagnetic energy density in the region of the reflected pulse at time t=210 fs when peak laser intensity interacts with the critical surface. (d): Total energy of particles in the simulation box beyond 1 MeV at time t=300 fs before the attosecond electron bunches leave the simulation box. (e): Particle energy of the attosecond electrons. It is the zoom-in of the square region "atto" in (d) with cutoff energy at 4 MeV.  (f) Initial charge density in unit $n_0=n_{cr}/4\pi^2$, where the scale-length is $L_s=0.5\lambda$.}
\label{AttoFig2}
\end{figure}

Fig. \ref{AttoFig2} demonstrates the similarities between the experimental and simulated electron energy spectra and the bunch duration measurement in simulation. The experiment was performed in a grazing incidence setup using a 20 ps prepulse and the measured electron energy spectra are shown in Fig. \ref{AttoFig2}a. The experimental setup is shown in Fig. \ref{AttoSetup} in the previous section. All spectra have been calibrated using published image plate efficiency \cite{bonnet2013response}. Fig. \ref{AttoFig2}b-f are from \acs{PIC} simulations, which were also conducted in grazing incidence geometry with preplasma density scale-length matches the experimental prepulse. This conversion from prepulse to preplasma scale-length is performed in the 1D hydrodynamic code (HYADES\cite{larsen1994hyades}) assuming isothermal expansion.

Comparing the experiment and simulation results, we notice a "bump" feature in the red curves: near 2 MeV in Fig. \ref{AttoFig2}a and left to 3 MeV in Fig. \ref{AttoFig2}b. This feature does not show up in other curves in blue or green. In the simulations, the "bump" component of the electron spectrum is always observed along the target surface and was always associated with the train of ultra-short duration bunches of energetic electrons. Since the red spectra in Fig. \ref{AttoFig2}b is taken from the attosecond electrons (bunch duration measured to be $\sim\lambda/8$ in Fig. \ref{AttoFig2}e), it suggests the existence of attosecond electron bunches in the red spectra in the experiment. The propagation direction of the attosecond electron bunches is found to be in the specular reflection direction through particle tracking, matching the experimental observation.

\begin{figure}[ht]
\centering
\includegraphics[height=0.6\columnwidth]{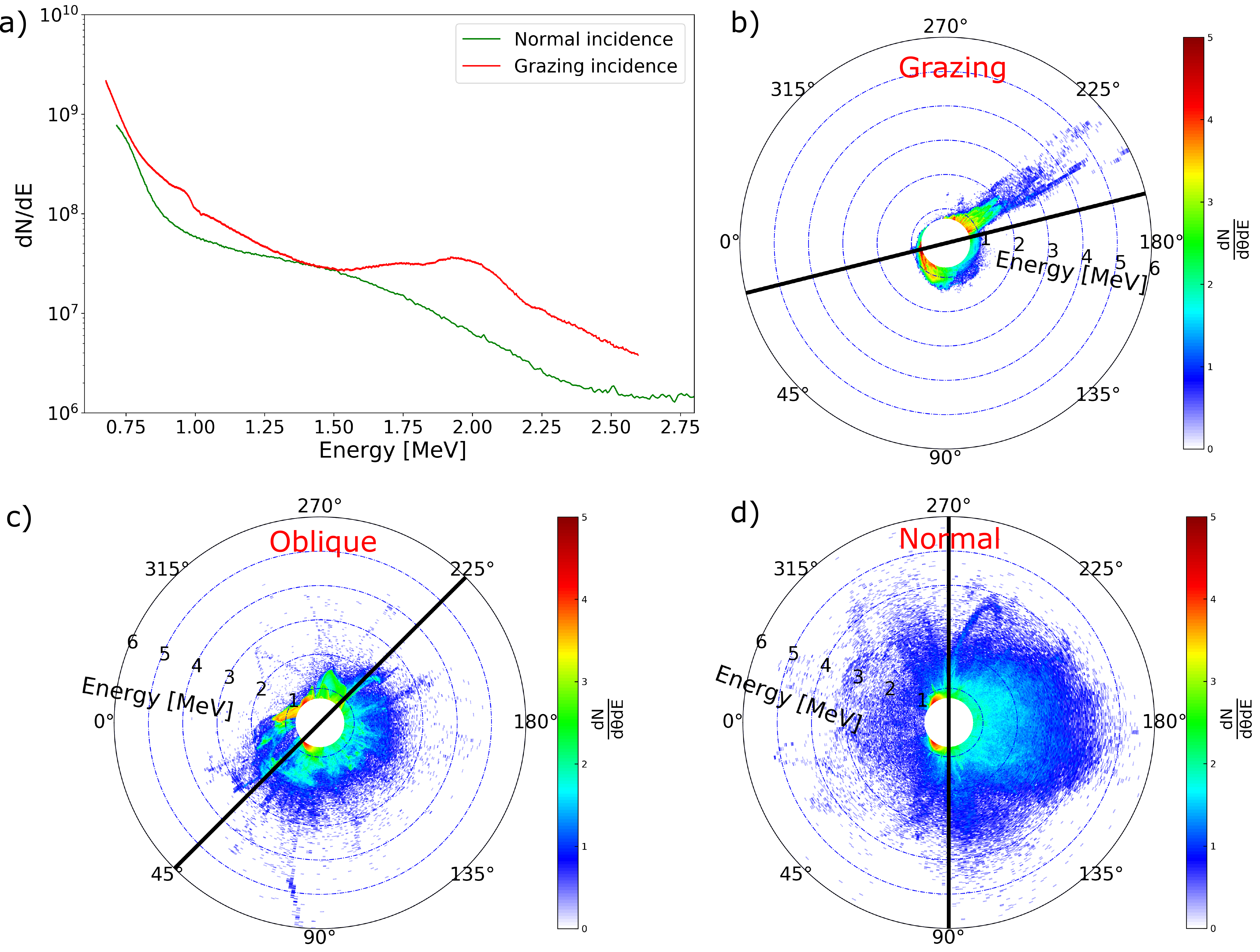}
\caption{(a): Experimental electron energy spectra from normal incidence and grazing incidence cases using a 20 ps prepulse. (b-d): Simulated angular energy distribution of electrons for grazing, oblique and normal incidence using scale-length $L_s=0.5\lambda$. The laser pulse comes into the simulation box from the left ($0^{\circ}$) and the target lies along the black line. The snapshots were taken when peak laser intensity interacts with the critical surface in each case.}
\label{AttoFig3}
\end{figure}

While Fig. \ref{AttoFig2} presents the direction of the emitted attosecond electron bunches, Fig. \ref{AttoFig3} investigates different incident angles of the laser pulses. The unique "bump" feature was observed only in the grazing incidence case but not in the normal incidence case, as is shown in Fig. \ref{AttoFig3}a. To understand it, we ran simulations at different angles of incidence. Fig. \ref{AttoFig3}b-d shows the angular distribution of all electrons when peak laser intensity interacts with the critical surface. This is also the time when the ultra-thin "null" in the electromagnetic energy density is formed in Fig. \ref{AttoFig2}c. Due to the self-intersection of electron trajectories, the electron concentration is abruptly peaked\cite{naumova2004attosecond} to initialize the bunching process. Fig. \ref{AttoFig3} shows that electrons are more spreading out in (c) and (d) than in (b), confirming that the bunching process favors grazing incidence. 

It is worth pointing out that these observations are consistent with respect to both experiments and simulations. From the 41 data shots we took at different geometries, 25 were taken at the grazing incidence geometry that encourages attosecond electron bunches where the laser pulse was at grazing incidence and the spectrometer was positioned in the laser reflection direction. We observed the "bump" feature in 12 of the 25 spectra. We observed no spectra with this feature in the other 16 shots taken at different geometries. Energetic attosecond electron bunches with noticeable coherence were also observed consistently in simulations as we scanned the preplasma electron density scale-length from $0.1\lambda$ to $2\lambda$, and detailed results will be presented in the following section.

\subsubsection{Preplasma effects}

\begin{figure}[H]
\centering
\includegraphics[height=0.5\columnwidth]{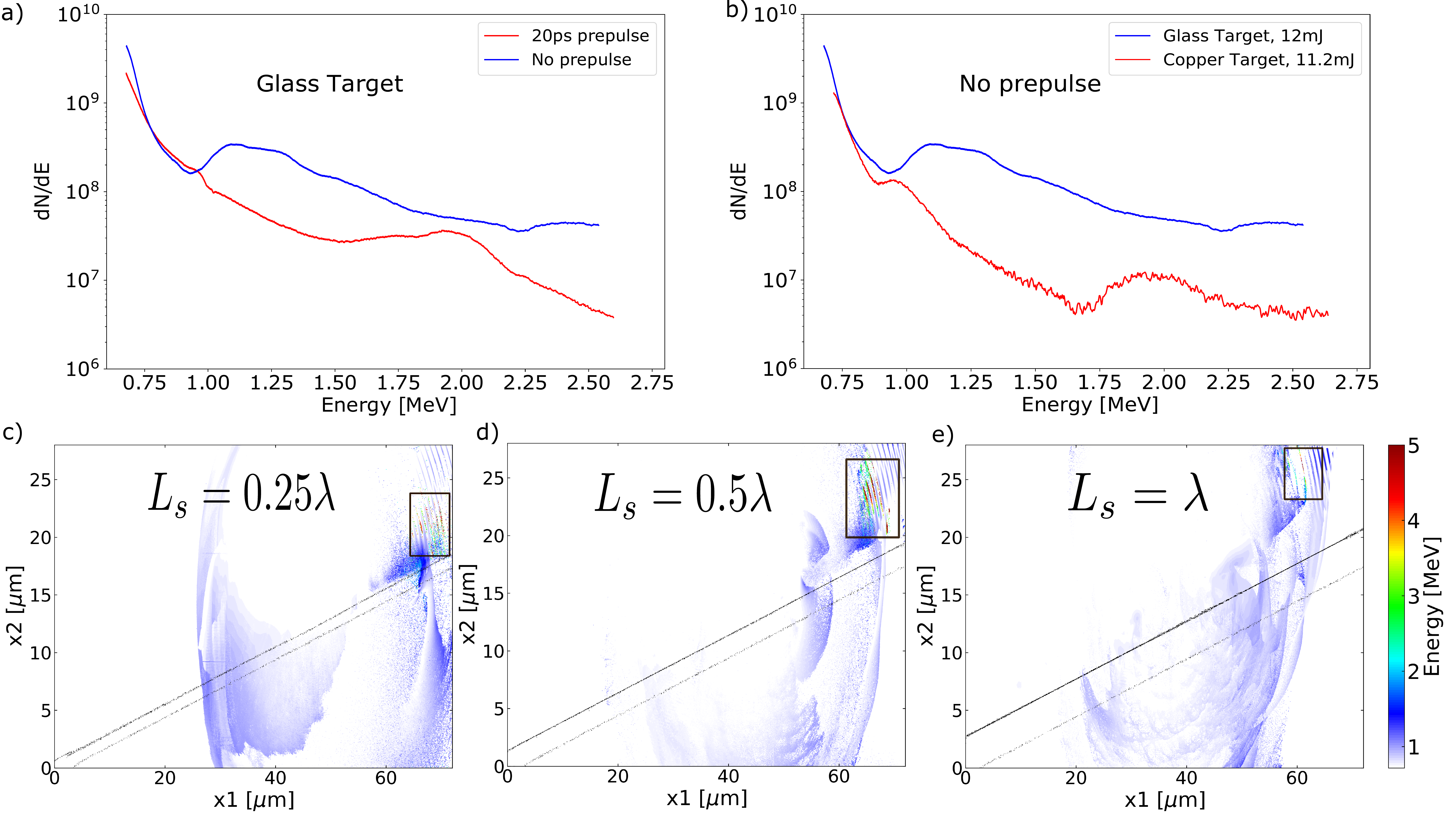}
\caption{(a): Experimental electron energy spectra with and without an external 20 ps prepulse using glass target. (b): Experimental electron energy spectra using glass and copper target without external prepulse. (c)-(e): Simulated spatial distribution of energetic electrons using preplasma scale-length $L_s=0.25\lambda(c),\; 0.5\lambda(d)\; and\; \lambda(e)$. Both experiments and simulations were performed at grazing incidence geometry.}
\label{AttoFig4}
\end{figure}

Preplasma density profile can affect the energy and propagation direction of the attosecond electron bunches. Fig. \ref{AttoFig4}a shows that including an additional 20 ps prepulse moves the "bump" feature towards the right on the spectra, suggesting bunching at higher energy. The same phenomenon is observed in Fig. \ref{AttoFig4}b as copper has a lower ionization threshold and thus a more developed preplasma profile with longer scale-length when the main pulse arrives. To explore this effect in parameter space, we tune the preplasma density scale-length from $0.1\lambda$ to $2\lambda$ in simulation. Energetic attosecond electron bunches with noticeable coherence were observed in moderate scale-length cases, as is highlighted in Fig. \ref{AttoFig4}c-e, but not in the two extreme cases where $L_s=0.1\lambda$ or $2\lambda$. Note that $L_s=0.5\lambda$ resulted in the most energetic attosecond electron bunches, matching the optimal condition for producing quasi-monoenergetic electron beams at $a_0\sim2$ in previous experiments\cite{mordovanakis2009quasimonoenergetic}. Besides, tuning the preplasma profile affects the bunch propagation direction. As the scale-length increases, the emission angle of the bunches becomes smaller as measured from the target normal. It is due to the critical surface moving away from the solid surface, as are the generated electron bunches.

\subsubsection{\acs{CEP} effects}

\begin{figure}[ht]
\centering
\includegraphics[width=0.95\columnwidth]{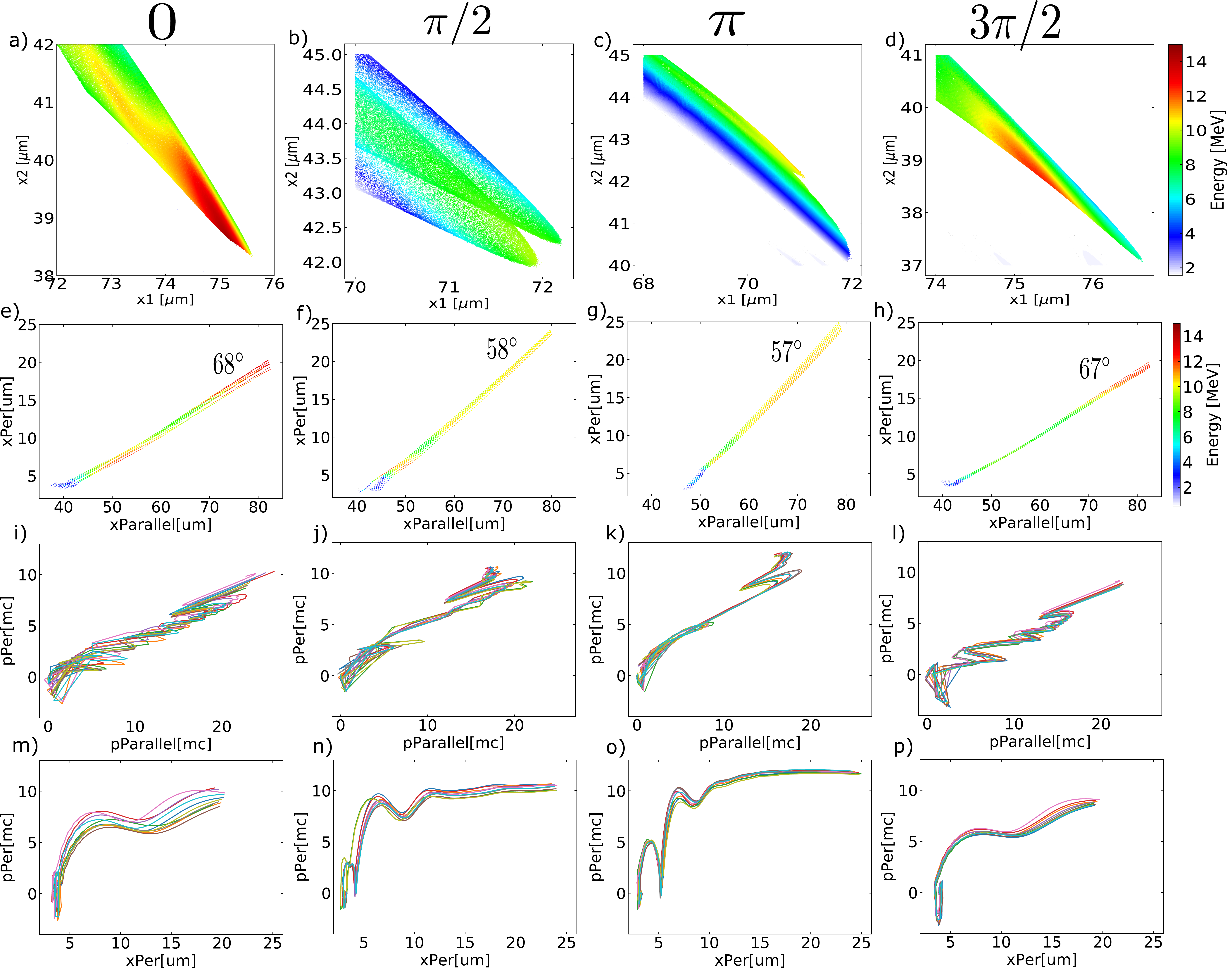}
\caption{Single attosecond electron bunch and particle tracking using \acs{CEP}$=\;0,\;\pi/2,\;\pi$, and $3\pi/2$. (a)-(d): Spatial energy distribution of bunched electrons at the last time-step. (e)-(h): Trajectory of the attosecond electrons with emission angle labeled. (i)-(l): Perpendicular momentum vs. parallel momentum, colors representing different particles. (m)-(p): Perpendicular momentum change in the perpendicular axis.}
\label{AttoFig5}
\end{figure}

To produce isolated attosecond electron bunches, we drove the interactions with single-cycle Gaussian pulses in the \acs{PIC} simulation, keeping the above grazing incidence geometry but at higher normalized vector potential $a_0=10$. This is to produce electron bunches at higher energy and more separable from the background while keeping a comparable driving laser pulse energy. It has been shown that \acs{CEP} plays a role in the electron acceleration process in \acs{LWFA} \cite{ouille2020relativistic} and in nanoplasma acceleration \cite{cardenas2019sub}. Fig. \ref{AttoFig5} demonstrates the generated single bunches using various carrier-envelope phases. It is observed that changing the \acs{CEP} can dramatically affect the energy and shape of the electron bunch. It can also change the direction of the generated bunch slightly, but not as effective as changing the preplasma density profile. To better understand the effect of \acs{CEP} on these electrons, we select ten macroparticles in the high energy part in Fig. \ref{AttoFig5}a-d and track their trajectory from birth. The tracking results are shown in Fig. \ref{AttoFig5}e-p in a rotated axis perpendicular/parallel to the target surface. Electrons in the \acs{CEP}$=\;\pi/2\;or\;\pi$ case in Fig.\ref{AttoFig5}j,k experience a momentum loss in the final acceleration stage. On the other side, electrons in the \acs{CEP}$=\;0\;or\;3\pi/2$ case in Fig.\ref{AttoFig5}i,l experience the momentum loss in earlier stages but find the correct phase with respect to the reflected laser field in later time to gain energy. This is in line with the difference in the final energy of the accelerated electrons in Fig. \ref{AttoFig5}a-d and e-h. Fig. \ref{AttoFig5}e-h also provide the emission angle of the bunched attosecond electrons: $\sim68^{\circ}$ in \acs{CEP}$=0\;or\;3\pi/2$ case and $\sim58^{\circ}$ in \acs{CEP}$=\pi/2\;or\;\pi$ case. The emission angle is defined as the angle between the electron bunch propagation trajectory and the target normal direction. While the different CEP results in different emission angle, they are all smaller than the incidence angle of $\sim76^{\circ}$. It could be explained by the magnetic field generated by the surface hot electrons. When a laser pulse hit the target at grazing angle, it produces hot electrons traveling along the target surface. A return current of cold electrons is therefore also formed and travels in the other direction. The currents yield a magnetic field where the emitted attosecond electrons feel the Lorentz force pointing away from the target. Therefore, the emission angle of the attosecond electrons is slightly smaller than the incident angle. A detailed investigation of such magnetic field produced from surface electrons is presented by Becker \textit{et al.}\cite{becker2019characterization} using a droplet target.

It is worth noting that electrons have negative initial perpendicular momentum in Fig. \ref{AttoFig5}m-o, moving towards the high-density target before the ejection. Electrons which have positive initial perpendicular momentum and travel away from the target cannot eject through the electromagnetic energy density gap until pulled back towards the target, as is shown in Fig. \ref{AttoFig5}p. However, despite losing the initial momentum, these electrons are not slowed down once they get ejected, as opposed to the momentum decrease of electrons in Fig. \ref{AttoFig5}n,o at $x_{per}\sim5\mu m$. Controlling \acs{CEP} offers the ability to inject electrons with various initial phases, which is important in the later stage of phase change and acceleration.

\subsubsection{Focal-spot size}

\begin{figure}[ht]
\centering
\includegraphics[width=0.9\columnwidth]{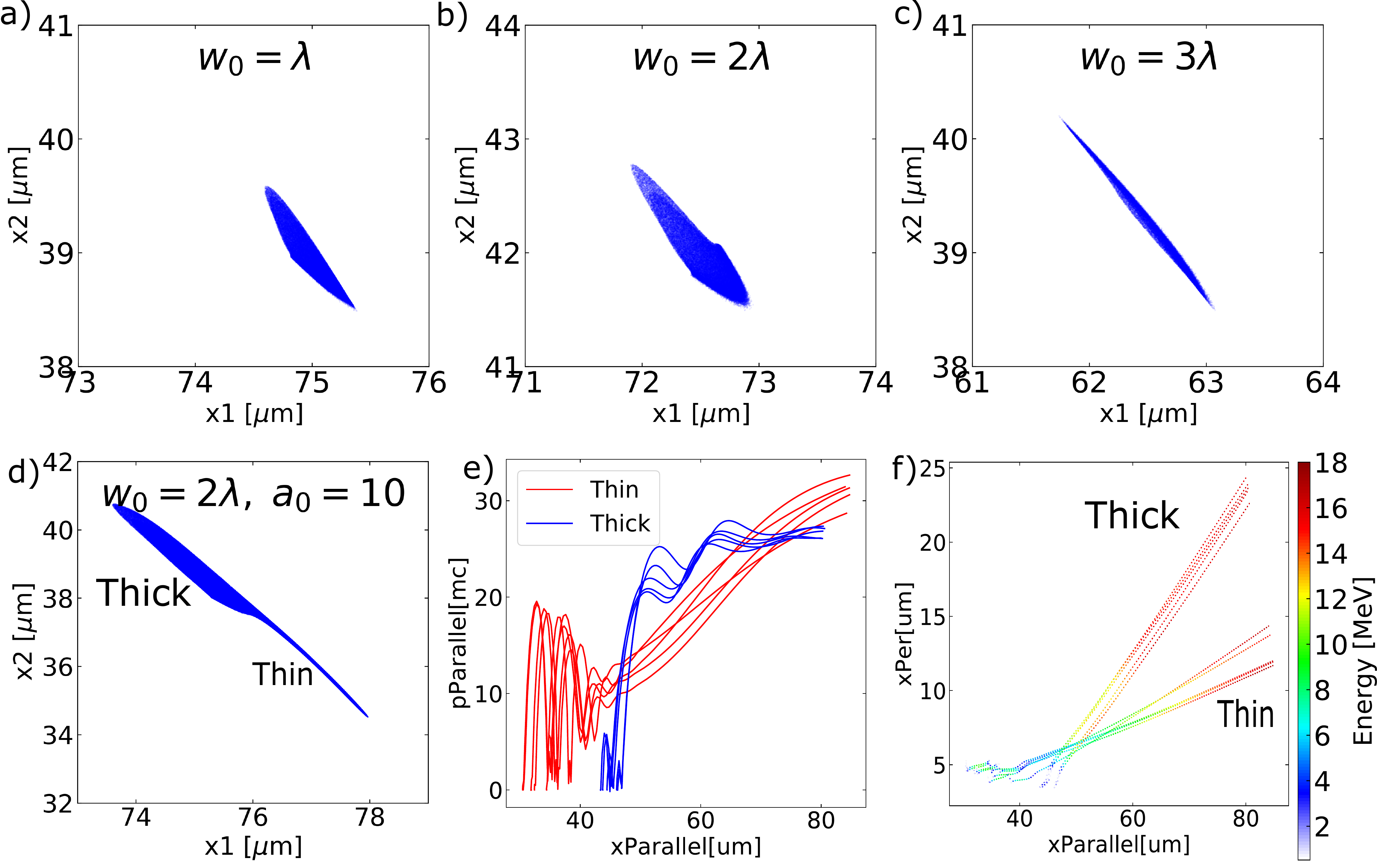}
\caption{Spatial profiles of attosecond electron bunch generated from different focal spot size, keeping the same pulse energy (a-c) or the same $a_0$ (a,d). Cutoff energy in (a)-(d) is set at the high energy edge on each spectra: 13 MeV, 9 MeV, 5 MeV and 18 MeV, respectively. (e) and (f) are the particle tracking results of the thick and thin part of the bunch in (d).}
\label{AttoFig7}
\end{figure}

Focusing the laser pulse to a larger focal spot can reduce the attosecond electron bunch duration. Fig. \ref{AttoFig7}a-c keeps the same pulse energy and varies the focal beam waist, achieving a thinner bunch at $w_0=3\lambda$. The numbers of electrons in these bunches are in the ratio 4:3:2. It is as expected that larger focal spots yield shorter bunches. In our grazing incidence setup using tight focus, the rays cover a range of angles $\alpha\sim1/(2\cdot f/\#)$. When focusing the beam to a small focal spot $w_0=\lambda$, this angle is around $20^{\circ}$. Given the incident angle of the central ray is $76^{\circ}$, the rays cover from $56^{\circ}$ to grazing. When the focal spot size is increased, this angle alpha decreases, i.e., $\alpha^{'} < \alpha$ and the rays cover from $\beta$ ($\beta=76^{\circ}-\alpha^{'}>56^{\circ}$) to grazing. Hence overall, the rays are incident at larger angles. Note that large incident angles are preferred to produce attosecond electron bunches, as is shown in Fig. \ref{AttoFig3}, and larger incident angles would lead to shorter electron bunch duration, as is shown in Eq. \ref{AttoBunchDuration}. A sketch of the ray angles in the focal spot geometry is attached below in Fig. \ref{AttoFigS1}.

\begin{figure}[ht]
\centering
\includegraphics[width=0.85\columnwidth]{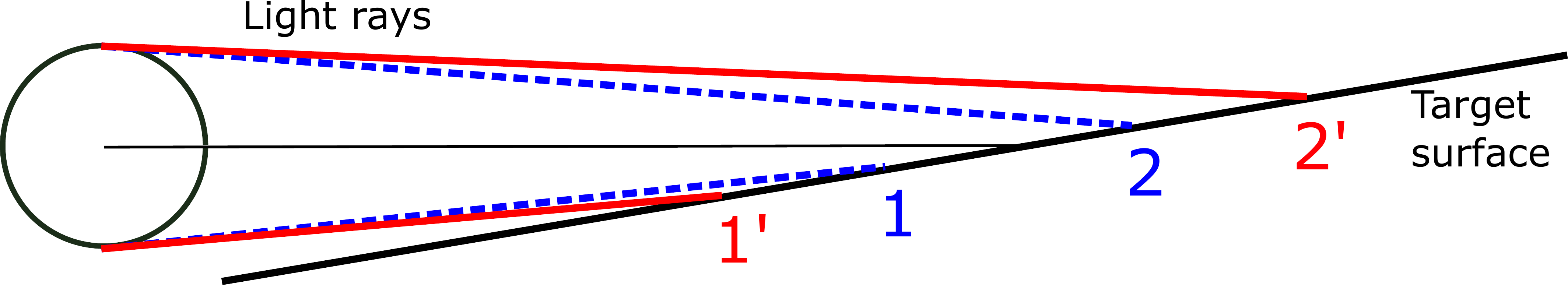}
\caption{Incidence angle increases with focal spot size. Blue and red rays represent small and large focal spot.}
\label{AttoFigS1}
\end{figure}

If we keep $a_0=10$ and increase the focal beam waist, an extra thin bunch is observed in addition to the thick bunch in Fig. \ref{AttoFig7}d. Particle tracking reveals that the thin bunch originates from earlier injection at $x_{parallel}\sim30\mu m$ in Fig. \ref{AttoFig7}f, while the thick part (as well as the bunch from the $w_0=\lambda$ case in Fig. \ref{AttoFig5}e) originate from $x_{parallel}>40\mu m$. Tracking the evolution of electron momentum along the target surface direction in Fig.\ref{AttoFig7}e, thin-bunch electrons (in red) oscillate within the laser field over a few cycles as it interacts with the dense plasma. They then get accelerated without the momentum loss that the thick-bunch electrons (in blue) experience, expected to reach even higher energy as they propagate.

\subsubsection{Energy-wise bunch characteristics}

\begin{figure}[ht]
\centering
\includegraphics[width=0.95\columnwidth]{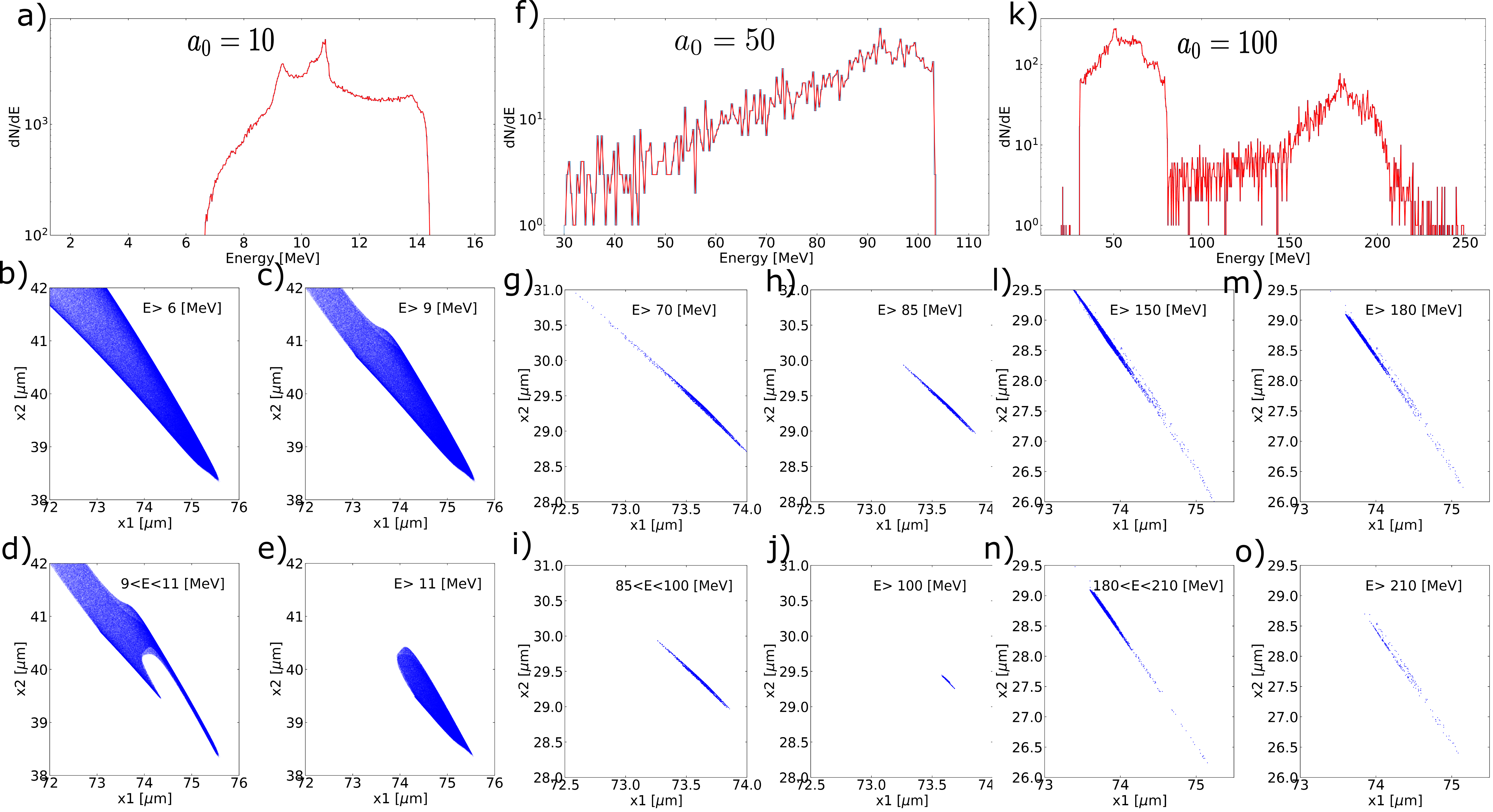}
\caption{Energy spectra and energy-wise bunch duration using $a_0=10$ in (a)-(e), using $a_0=50$ in (f)-(j) and using $a_0=100$ in (k)-(o).}
\label{AttoFig6}
\end{figure}

Duration of the isolated attosecond electron bunch and the electron concentration have been measured against energy. Fig. \ref{AttoFig6}a-e measure the bunch in Fig. \ref{AttoFig5}a ($a_0=10$, \acs{CEP}=0, $w_0=\lambda$). Fig. \ref{AttoFig6}f-o measure the bunch using all the same parameters but $a_0=50$ and $a_0=100$. Energy spectra in Fig. \ref{AttoFig6}a,f,k again show the "bump" feature, and the shape of these most concentrated electrons are shown in Fig. \ref{AttoFig6}d,i,n, respectively. The low energy part in the $a_0=100$ case represents background electrons that spatially overlap with the bunch at this time step, which goes away as the bunch propagates. Movies that track the electron bunches can be found in the Supplementary Video S1 in the associated publication \cite{lin2020towards}. Fig.\ref{AttoFig6}b-e, g-j, l-o present the spatial profile of the electron bunch at different cutoff energy, labeled respectively. Comparing them indicates that going to higher $a_0$ leads to a thinner bunch at higher energy. Note that the optimal scale-length for the energetic electron bunch decreases with $a_0$. We have scanned multiple scale-lengths and present $L_s=0.25\lambda$ for the $a_0=50$ case and $L_s=0.1\lambda$ for $a_0=100$. It is worth pointing out that these electron bunches hardly see any radiation reaction effects even when $a_0=100$. Classical radiation reaction effects become strong when the classical radiation reaction parameter $R_c\sim a_0\chi>0.01$ and dominate when $R_c>0.1$ \cite{blackburn2020radiation}, where $\chi$ is the relevant parameter for supercritical fields. To consider radiation reaction effects when $a_0=100$, one needs $\chi>0.01$. However, this is not available in our case even in the frame of the electrons because the beam energy is not high enough. The fact that the electron bunch is almost co-propagating with the reflected laser field makes $\chi$ even smaller as $\chi\sim\gamma|E|(1-\beta\cos{\theta})$, where $\theta$ is the angle between electron momentum and wave vector. In fact, these attosecond electron bunches produced under this geometry should see even weaker fields than the fields in the lab frame as $\theta\sim10^{\circ}$ and $\cos{\theta}\sim1$.

Another motivation to go to higher energy is to propagate the beam further as attosecond bunches before being dispersed to longer duration. The spatial dispersion between the fastest and slowest electron in a bunch is estimated using Eq. \ref{AttoEq1}, where $\gamma$ is the Lorentz factor.
\begin{equation}
    d L=0.5 *\left(\frac{1}{\gamma_{\text {low}}^{2}}-\frac{1}{\gamma_{\text {high}}^{2}}\right) \cdot L
\label{AttoEq1}
\end{equation}
For few-MeV electron bunches as in Fig. \ref{AttoFig2}e, the dispersion $dL/L$ is calculated to be $4\times10^{-3}$. The duration between the fastest and slowest electrons would exceed attosecond levels (beyond 1fs) at 0.1mm away from the source. We have also characterized the dispersion of the high-energy electron bunches in Fig. \ref{AttoFig6}. The estimated dispersion $dL/L$ is $4\times10^{-4}$, $1\times10^{-6}$ and $6\times10^{-7}$ for the bunch in Fig. \ref{AttoFig6}e,j,o, respectively. In other words, if the bunch in Fig. \ref{AttoFig6}0 propagates for one meter in free space, the spacing between its front and back edge will be 0.6 $\mu m$. The bunch can maintain duration below 1 fs until it propagates 0.45m. For direct measurements of the bunch duration, higher energy electrons with less dispersion are clearly favored. Such electron beams will also produce more optical transition radiation, which makes them easier to characterize.

\subsubsection{Tilted laser pulses}

\begin{figure}[ht]
\centering
\includegraphics[width=0.9\columnwidth, height=0.6\columnwidth]{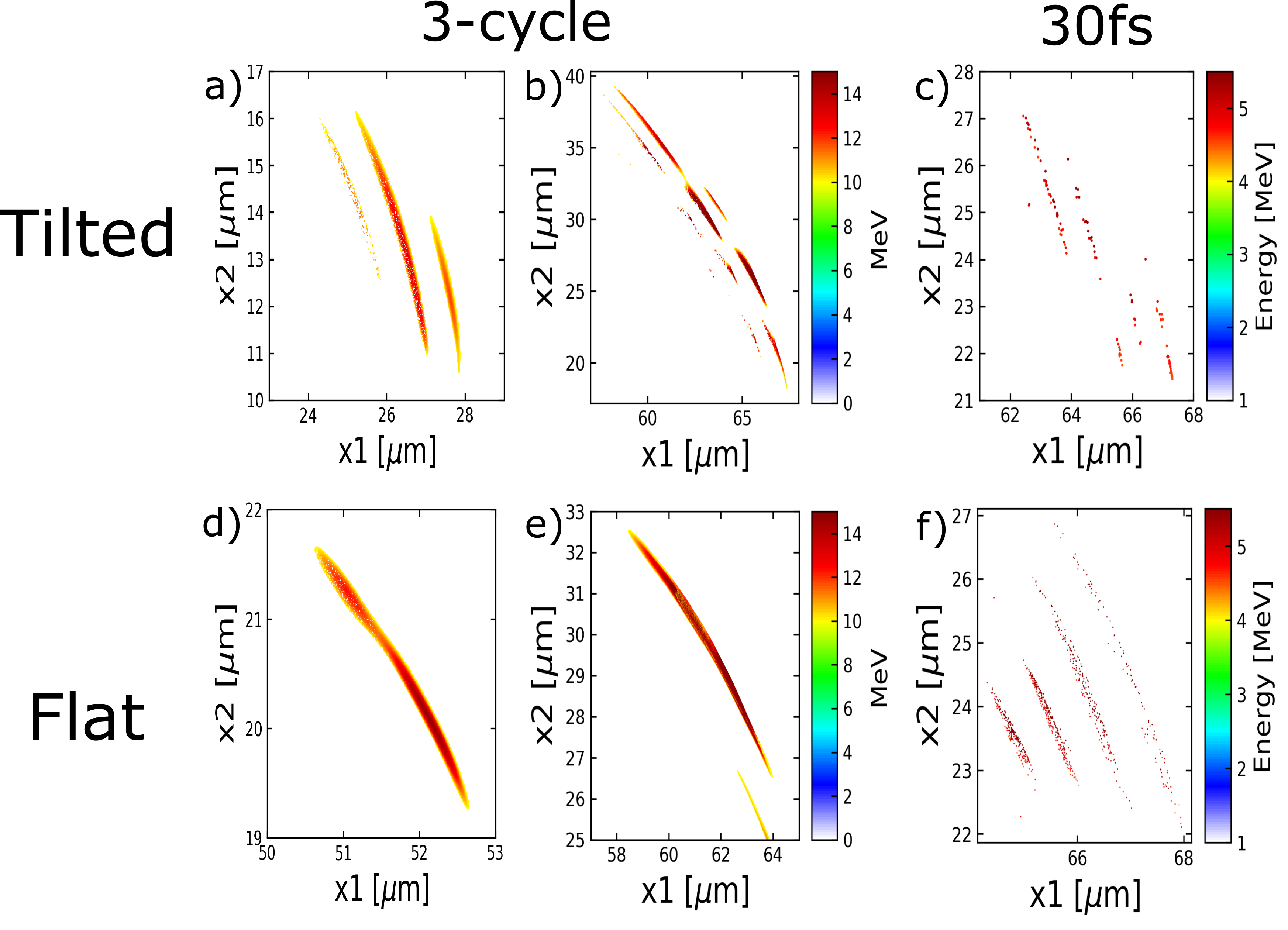}
\caption{Attosecond electron bunches generated using tilted pulses in (a)-(c) vs. flat pulses in (d)- (f).}
\label{AttoFig9}
\end{figure}

Apart from using single-cycle laser pulses, another approach to obtain isolated attosecond electron bunches is using tilted laser pulses. The idea is to spatially separate the electron bunches by rotating the wave vector within a pulse. This concept of spatial-temporal couplings in ultrashort pulses was first explored theoretically by Akturk \textit{et al.}\cite{akturk2010spatio} and was experimented to produce isolated attosecond pulses by Wheeler \textit{et al.}\cite{wheeler2012attosecond}. We show that this method also applies to the electron bunches as the tilted laser pulse interacts with solid-density plasmas. Fig .\ref{AttoFig9}a presents the spatial profile of energetic electrons ($>$10 MeV) produced by a tilted 3-cycle pulse, where a perpendicular spatial chirp $d\omega/dx_{2}$ is included at the focus. It can be achieved by introducing an angular dispersion in the laser beam, which allows different frequency components of the laser pulse to disperse along the perpendicular axis ($x_{2}$) and leads to the spatial chirp. As a result, the laser cycle period ($T_{l}=c/\omega$) varies along the perpendicular axis. This is equivalent to a perpendicularly varying spacing between laser field cycles in the spatial domain, which leads to a rotating wave vector. An intuitive illustration can be found in Fig. 1 in  Wheeler \textit{et al.}'s work\cite{wheeler2012attosecond}. The three cycles of the laser pulse go to focus at three different angles and produce three separate bunches, as is shown in Fig .\ref{AttoFig9}a. Compared to the central bunch generated during the second laser cycle, the first and third bunch are shorter in length as the interacting position varies over time and the phase of "null" EM energy density lasts longest amid the second cycle. Note that Fig .\ref{AttoFig9}a shows the electron bunches shortly after the laser pulse interacts with the dense plasma, while Fig .\ref{AttoFig9}b presents the profile of these electrons when they leave the simulation box. A movie that tracks the propagation of these three bunches in every half-cycle time-step can be found as Supplementary Video S2 online \cite{lin2020towards}. We also apply the tilted wavefront concept to a 30fs laser pulse, which is more readily accessible in high-power laser facilities. The resulted energetic electrons ($>$4 MeV) are presented in Fig .\ref{AttoFig9}c with observable separation as well. For reference, fig .\ref{AttoFig9}d-f show the energetic electron bunches from a flat wavefront without any spatial chirp using a 3-cycle pulse and a 30fs pulse, respectively. In conclusion, spatial-temporal couplings in ultrashort pulses can be inherited by the electron bunches through interaction with dense plasmas. However, maintaining these properties through propagation requires fine measures.

\subsection{Discussion}
Attosecond electron bunches are a unique product of ultra-short laser-solid interactions at grazing incidence geometry: "attosecond" comes from the ultra-thin "null" in the electromagnetic energy density shown in Fig. \ref{AttoFig2}c while "bunch" comes from the peak electron concentration shown in Fig. \ref{AttoFig3}b. Electrons oscillating in the fields at the correct phase can eject away from the target through the "null", which is the cause of the ultra-short duration of the bunched electrons. On the other side, the "bunch" is formed by the peaked electron density that these ejected electrons inherit\cite{naumova2004attosecond}, which prefers a large laser incidence angle.

The electrons are accelerated in the reflected laser pulse after being ejected from the target surface. This process is sensitive to the phase of the laser pulse as well as the initial phase of the electrons. As is introduced in Sec. \ref{sec:theorySingleElectron}, the momentum change of an electron in a plane electromagnetic wave can be described by: $p_{z}-p_{z 0}=\gamma-\gamma_{0}$ and $p_{\perp}-p_{\perp 0}=\mathbf{a}$, where $p_{z}$ is the longitudinal momentum and $p_{\perp}$ is the transverse momentum. Ejected electrons usually start with some non-zero momentum, as is shown in the particle tracking, and obtains longitudinal momentum by $\frac{a^{2}+2a\cdot p_{\perp0}}{2\left(\gamma_{0}-p_{z 0}\right)}$. There is a more generalized theory that describes the energy gain in three-dimensional space with dephasing\cite{arefiev2015novel}. In addition to the single electron model, the collective behavior of the plasma is nontrivial. Depending on the direction of the incident electric field in a single-cycle pulse, electrons move towards or away from the pulse to compress or stretch it. The vector potential and the electric field of the reflected pulse are\cite{naumova2004relativistic}:
\begin{equation}
    A_r=2\pi\int_{0}^{t^{\prime}} d \tau I_{y, e}(\tau)+I_{y, i} \xi_{r}, \;\;I_y=-e N_{0} V_{y}
    \label{AttoVectorPotential}
\end{equation}

\begin{equation}\label{AttoEfield}
    E_r=\frac{2 \pi N_{0} e\left(c \kappa_{y}+e A_{y}(\xi)\right) \mathcal{E}}{m_{e}^{2} c^{4}+\left(c \kappa_{y}+e A_{y}(\xi)\right)^{2}}+\frac{2 \pi N_{0} e}{c} V_{y}
\end{equation}
where $\xi$ is the initial phase of the laser pulse, $N_0$ is the plasma density and $V_y=c\sin(\theta)$ shifts an oblique incidence angle $\theta$ to normal incidence frame\cite{naumova2004relativistic}. Knowing the vector potential from Eq. \ref{AttoVectorPotential} we can determine the momentum gain $p_z$ in the modulated reflected fields:

\begin{equation}
    p_{z}=p_{z 0}+\frac{a_r^{2}+2a_r\cdot p_{\perp0}}{2\left(\sqrt{1+p_{z 0}^{2}+p_{\perp0}^{2}}-p_{z 0}\right)}, \;\;\;a_r=e A_r / m_{e} c^{2}
\end{equation}

This gives an estimation of the electron energy in the bunches. The electron bunch duration can also be predicted by estimating the "null" thickness in the electromagnetic energy density. Setting $E^2+B^2=0$, the phase of the reflected wave becomes $\xi_{r}=\int(1-a(2\pi\xi/\lambda)cot(\theta))d\xi$ , where $a(2\pi\xi/\lambda)$ is the vector potential of the incident wave. For energy density $0^+$ and $0^-$,
\begin{equation}
    \Delta\xi_{r}\sim[a(2\pi\xi^+/\lambda)-a(2\pi\xi^-/\lambda)]cot(\theta)
    \label{AttoBunchDuration}
\end{equation}
where $\xi^+$ and $\xi^-$ are the phases which correspond to density $0^+$ and $0^-$. This agrees with the simulation and experimental results that a larger incident angle is preferred. It also indicates that to get shorter electron bunch duration, a slowly-varying vector potential is preferred in phase space, or equivalently longer laser wavelength. Further experimental studies are expected to verify this prediction.

It should also be pointed out that the carrier-envelope phase of single-cycle pulses could affect the electric field strength directly. The intensity profile of a Gaussian pulse is:
% For a single-cycle Gaussian pulse with intensity profile 
\begin{equation}
    I(t)=\epsilon cn \cdot \frac{1}{T} \int_{t-\frac{T}{2}}^{t+\frac{T}{L}} |E(t')|^{2} dt'=\epsilon cn E_{0}^{2} \cdot \frac{1}{T} \int_{t-\frac{T}{2}}^{t+\frac{T}{L}} e^{-(t' / \tau)^{2}} \cos ^{2}(\omega t'+\phi) dt',
    \label{AttoCEP}
\end{equation}
In most cases when the pulse duration is much longer than one optical cycle, $t>T=2\pi/w_0$, we can apply the slowly varying envelope approximation (SVEA) and the intensity profile becomes: $I(t)=\epsilon cn E_{0}^{2}e^{-t^2/\tau^2} \cdot \frac{1}{T} \int_{t-\frac{T}{2}}^{t+\frac{T}{L}} \cos ^{2}(\omega t'+\phi) dt'=I_0e^{-t^2/\tau^2}$, where $I_0=\frac{1}{2}\epsilon cn E_{0}^{2}$ is the peak intensity. However, SVEA is no longer valid when the pulse duration is comparable to one optical cycle. In our case with a single cycle pulse, we have to integrate both the \acs{CEP} term and the exponential term in Eq. \ref{AttoCEP}.
The peak intensity scales with the electric field amplitude as $I_0\sim 0.441\epsilon cn E_{0}^{2}$ for \acs{CEP}$=0\;or\;\pi$ and $I_0\sim 0.454\epsilon cn E_{0}^{2}$ for \acs{CEP}$=\pi/2\;or\;3\pi/2$, which are different from the commonly used formula $I_0\sim0.5\epsilon cn E_{0}^{2}$ under SVEA. Thus for a fixed laser intensity, the carrier electric field is slightly stronger in the \acs{CEP}$=0\;or\;\pi$ cases. Note that this field is not exactly the electric field that drives the plasmas: the absolute phase of the carrier wave slides underneath the envelope as the beam focuses, especially in single-cycle pulses. It is due to the Gouy phase shift generated by isodiffracting ultra-braodband pulses through the beam waist\cite{feng2000spatiotemporal}. Besides the vacuum focus, plasmas act as another focusing lens as the ponderomotive force pushes the electrons outwards. The phase change due to the "plasma lens" is dependent on its thickness and refractive index, which are essentially determined by the preplasma density profile. 

Experimental measurements of ultra-short electron bunch duration may be accessible by measuring characteristics of \acf{COTR} produced by the electrons. When an electron bunch is incident on an interface between two media with different refractive indices, coherent electromagnetic fields are radiated and the electron bunch duration can be inferred. For example, Lundh \textit{et al.}\cite{lundh2011few} measured the \acs{COTR} from an electron beam generated in a colliding pulse injection regime, and compare the \acs{COTR} spectrum to analytic Gaussian \acs{COTR} spectra for different bunch duration to deduce the bunch duration is $\sim$few fs. In addition, Sears \textit{et al.}\cite{sears2008production} measured the duration of electrons from inverse free-electron-laser process to be $\sim410$ attoseconds. While using overdense plasmas in a grazing incidence setup, the electron bunch duration can potentially be shorter than a tenth of a femtosecond. Since shorter electron bunches produce transition radiation at shorter wavelengths, to measure the duration of such electron bunches requires diagnosing the \acs{COTR} further in the x-ray regime. These attosecond electron bunches can potentially be applied to probe the temporal evolution of dynamic systems such as magnetic field formation\cite{schumaker2013ultrafast} and attosecond electron microscopy and diffraction\cite{morimoto2018diffraction}. It would also be worth trying to drive the interactions with laser pulses that carry orbital angular momentum\cite{wang2019new, wang2019intense} to further manipulate the electron bunch characteristics.

To sum up, we show the experimental electron energy spectra from driving 30 fs, 800 nm laser pulses at grazing incidence onto a 6 mm thick glass target. The experimental energy spectra match the spectra of attosecond electron bunches observed in simulations. The duration of the bunches is measured to be $\sim$tenth of a micron ($\sim$100 attoseconds) in simulations. Direct experimental measurement of the bunch duration was not performed, although it can potentially be achieved by measuring characteristics of \acs{COTR} produced by the electrons into the x-ray regime. We find grazing incidence geometry is necessary to produce attosecond electron bunches in simulations and corresponding spectral features in experiments. We show that the generation of energetic attosecond electron bunches favors a larger incident angle, higher pulse energy, larger focal spot size, and moderately sharp preplasma density profile. We obtain isolated attosecond electron bunches using single-cycle pulses. Controlling \acs{CEP} offers the ability to inject electrons with various initial phases and to adjust the energy and shape of the bunch. Due to the Guoy phase shift, \acs{CEP} is changing as the single-cycle pulses focus through the plasmas. Preplasma density profile governs the propagation direction of the bunch and fine-tuning of the direction is accessible by tuning \acs{CEP}. Higher $a_0$ and larger focal spot size can result in an even shorter bunch duration with less dispersion after propagation. When operating with much higher pulse energy, a sharper preplasma profile is preferred to produce a cleaner bunch. Using tilted laser pulses can pass angular properties to electron bunches and spatially separate them. With a simplified analytic model, we predict the momentum gain of electrons and we predict that shorter electron bunch duration scales with larger incident angle and longer laser wavelength.

\clearpage
\section{Characteristic x-ray emission at different laser wavelengths}
% \section{Characteristic x-ray emission at 0.8 $\mu$m, 1.3 $\mu$m and 2 $\mu$m}
\label{sec:Xray}
\subsection{Introduction}
% !Work from the paper D:\CUOS\Thesis\prospectus\previousMaterials\LinTermPaperXrays.pdf

When a short-pulse laser interacts with overdense plasmas from solid targets, it accelerates electrons and generates radiation in the x-ray regime. For solid targets like plasma mirrors having short density scale lengths, \acf{HHG} is the most important mechanism for radiation generation. An experiment on \acs{HHG} has been discussed in Sec. \ref{sec:HHG}. For metallic targets with larger atomic numbers, other mechanisms are playing significant roles in producing radiation, such as characteristic x-ray emission and bremsstrahlung radiation. 

Bremsstrahlung radiation is a common form of broad-band radiation when collisional processes are involved \cite{jackson1999classical}. When a high-energy charged particle collides with another particle, it decelerates and loses energy via radiation emission due to relativistic effects. As introduced in the theoretical background in Sec. \ref{sec:theoryOverdense}, a significant feature of physical processes associated with overdense plasmas is the abundance of collisions. In short-pulse laser-solid interactions where electrons are accelerated during the $\sim$femtosecond pulse duration but ions are almost immobile, bremsstrahlung radiation mostly results from the scattering of electrons from ions. The total radiated power per unit area is given by:
\begin{equation}
    P_{B r}=1.54 \times 10^{-38} g_{f f} Z^{2} n_{i} n_{e}\left(\frac{T_{e}}{e V}\right)^{1 / 2} \frac{W}{m^{3}}
    \label{EqXrayBrem}
\end{equation}
where Z is the number of electrons in the ion, $g_{ff}$ the gaunt factor for quantum corrections, $T_e$ is the electron temperature, and $n_i$ and $n_e$ are the ion density and electron density, respectively. From Eq. \ref{EqXrayBrem}, the energy radiated via bremsstrahlung increases with the energy of the electron, and bremsstrahlung radiation becomes significant for relativistic particles.

Radiation can also be generated through inner-shell ionization by electron impact. When the laser field ionizes electrons from atoms, the emitted electrons can collide with other atoms and ionize them to release more electrons to trigger the avalanche. Details about the collisional ionization mechanism have been introduced in Sec. \ref{sec:theoryIonization}. Although most of the electrons to be ionized are in the outer shells of the atoms with lower binding energy, electrons in the inner shells can also be ionized, but at a much smaller probability. This process is known as inner-shell ionization. Note that the ionization leaves a vacancy in the inner-shell in the transition state, and an electron in the outer shell has to fall back to the inner shell to fill the vacancy. Due to the conservation of energy, the energy difference between the electron shells must be compensated, most probably through emitting radiation. Since the photon energy of the radiation is exactly the energy difference between the electron shells and purely dependent on the atom, it is called the characteristic x-ray emission. Under rare circumstances, the energy difference can be compensated by releasing an outer-shell electron via the Auger effect. The probability of inner-shell ionization by electron impact in different atoms has been studied extensively and is reviewed in Ref. \cite{bote2009cross, llovet2014cross}.

% And this paper Intro: Cross sections for ionization of K, L and M shells of atoms by impact of electrons and positrons with energies up to 1 GeV: Analytical formulas

X-rays produced via this mechanism have unique features compared to those from other mechanisms in short-pulse laser-solid interactions. The radiation spectrum has distinct peaks at the characteristic emission lines of the atom, unlike the broad-band spectrum from bremsstrahlung radiation. The peak energy of the produced characteristic x-rays is independent of the driving laser wavelength, 
unlike the harmonic dependence in \acs{HHG}. Furthermore, characteristic x-rays can be produced without strict vacuum condition \cite{hou2008generation, hou2008vacuum, hada2010effects}, which is usually required in \acs{HHG}. Because of these remarkable properties, producing characteristic x-ray emission using high-intensity lasers and solid targets has drawn increasing attention and has been demonstrated in many successful experiments \cite{hou2006hard, chen2008study, zamponi2009femtosecond, lee2015femtosecond, fourmaux2016laser, holtz2017towards, samsonova2019relativistic, reklaitis2019emission}. The source size, directional property, spatial coherence, and laser to x-ray conversion efficiency of the produced x-rays have been characterized and reported in Ref. \cite{boschetto2007spatial, hou2008directional, seely2011energetic, zhao2019characterization}. Owing to its high spatial coherence, an useful application of such x-ray sources is phase-contrast imaging for biological objects \cite{toth2007evaluation, chakera2008continuous, martin2021validation}. To better understand the characteristic x-ray emission from a laser-plasma point of view, the roles of some crucial laser-plasma conditions have been investigated, such as the laser wavelength \cite{weisshaupt2014high}, the laser pulse contrast ratio \cite{azamoum2018impact}, and the hot electron refluxing \cite{neumayer2010role}.  

In this section, we will present results of characteristic x-ray emission and bremsstrahlung radiation when short-pulse lasers with different wavelengths and pulse energies interact with the overdense plasma of various preplasma profiles from a molybdenum target. The study is performed both experimentally with hundreds of thousands of laser shots, and computationally with \acs{PIC} simulations scanning over the 4-dimensional parameter space consisting of laser wavelength, pulse energy, preplasma profile, and x-ray emission properties.

\subsection{Experimental setup}

The experiment was performed in the \acs{Lambda-cubed} at \acs{CUOS} at the University of Michigan. Various laser-plasma conditions were used in the experiment. Laser pulses at the near-infrared wavelength (0.8 $\mu m$, 6.1 mJ, 40 fs) and the mid-infrared wavelengths (1.3 $\mu m$, 0.4 mJ, 80 fs, and 2 $\mu m$, 0.2 mJ, 67 fs) were used. While the \acs{Lambda-cubed} outputs 0.8 $\mu m$ laser pulses, the 1.3 $\mu m$ and 2 $\mu m$ pulses are generated using an \acs{OPA}, as is introduced in Sec. \ref{sec:ExplaserSystem}. The pulse contrast ratio to the \acs{ASE} is $>10^8$. A tunable prepulse with intensity about two orders of magnitude lower than the main pulse intensity was introduced using a 2 $\mu m$ thick nitrocellulose pellicle. Details about the prepulse delay stage setup have been discussed in Sec. \ref{sec:ExplaserSystem} and shown in Fig. \ref{FigExpPrepulse}. The tunability of the prepulse covers 0 - 187 ps, which corresponds to a preplasma density scale length of 0.1 - 5.5 $\lambda$ from hydrodynamic simulations \cite{mordovanakis2009quasimonoenergetic}. Various laser pulse energies were also applied by adjusting the pump lasers.

\begin{figure}[ht]
\centering
\includegraphics[width=0.9\columnwidth]{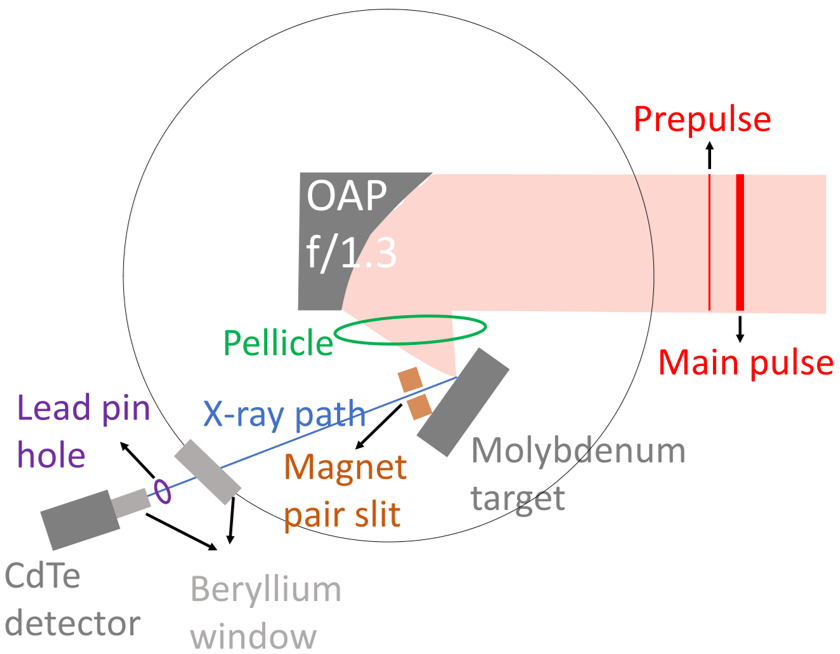}
\caption{Experimental setup for characteristic x-ray emission measurement. The laser pulses are focused onto the molybdenum target at an incident angle of $55^{\circ}$. The distance between the \acs{OAP} and focal spot on the target is 75 mm. The CdTe detector is located at 530 mm away from the target, at $65^{\circ}$ from the target normal.}
\label{xrayFigSetup}
\end{figure}

The laser pulses were directed into a vacuum chamber at a pressure of tens of mTorr, as is shown in Fig. \ref{xrayFigSetup}. More information on the experimental chamber operation can be found in Sec. \ref{sec:ExpTargetPrep}. The 0.8 $\mu m$, 1.3 $\mu m$, and 2 $\mu m$ pulses were focused by an f/1.3 \acs{OAP} to 2.1 $\mu m$ FWHM, 4.3 $\mu m$ FWHM, and 5 $\mu m$ FWHM, respectively. The peak laser intensity of the 2 $\mu m$ pulses and the 1.3 $\mu m$ pulses are on the order of $10^{16} W\cdot cm^{-2}$, while that of the 0.8 $\mu m$ pulses are on the order of $10^{18} W\cdot cm^{-2}$. Since the pulse energy is a control parameter in this study, a specific laser intensity or $a_0$ will be reported associated with each measurement. A 2 $\mu m$ thick nitrocellulose pellicle is inserted in the beam path to prevent the \acs{OAP} from the ablated debris from the target. The molybdenum target sits on a rotary stage for high repetition operation. Details on the target preparation and alignment have been described in Sec. \ref{sec:ExpTargetPrep}. 

A portion of the produced x-rays is measured, as is indicated in the blue path in Fig. \ref{xrayFigSetup}. A pair of magnet is placed next to the target to deflect high-energy electrons from the x-ray path. Beryllium windows are used at the chamber exit and at the detector entrance for high x-ray transmission efficiency. A lead pinhole is placed in front of the detector to avoid pile-up in the detection, and the size of the pinhole varies from $1.3\times10^4\;\mu m^2$ or $6.7\times10^4\;\mu m^2$.
To align the components along the x-ray path, a helium-neon laser is directed from the detector backward into the chamber to overlap with the focal spot on the target surface. The x-rays are captured by an Amptek XR-100 Cadmium Telluride (CdTe) single-photon detector. The detector is calibrated using an \acf{am-241} source, as is discussed in Sec. \ref{sec:ExpLPDiag}.

\subsection{Experimental results}
\begin{figure}[H]
\centering
\includegraphics[width=0.75\columnwidth]{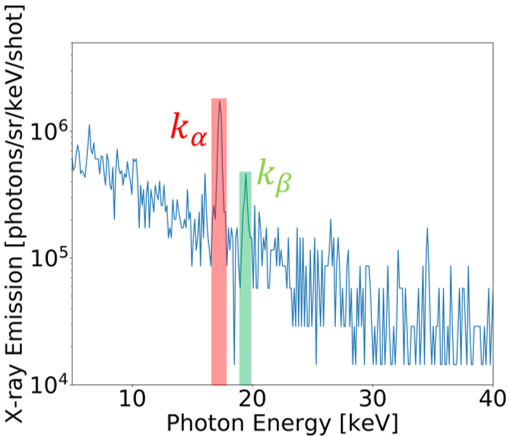}
\caption{An example of the measured x-ray spectrum from a molybdenum target. The spectrum was integrated over 10,000 laser shots.}
\label{xrayFigMolySpectrum}
\end{figure}
Fig. \ref{xrayFigMolySpectrum} shows the measured spectrum of the femtosecond x-ray source generated using 2 $\mu m$, 0.2 mJ, 67 fs, horizontally polarized laser pulses interacting with a molybdenum target. A prepulse arriving at 20 ps before the main pulse was used in this measurement. The flux of the $K_\alpha$ and the $K_\beta$ emission are $6.4\times10^{6}$ photons per $2\pi$ steradian per pulse and $1.8\times10^{6}$ photons per $2\pi$ steradian per pulse, respectively. The x-ray flux of the bremsstrahlung radiation is $6.2\times10^{7}$ photons per $2\pi$ steradian per pulse. 

The x-ray energy spectrum also tells the characteristic hot electron temperature. When short-pulse lasers interact with solids, some of the electrons are heated via the collisionless absorption mechanisms discussed in Sec. \ref{sec:theoryOverdense} to energies much higher than the initial background plasma temperature \cite{gibbon2005short}. Zulick \textit{et al.} \cite{zulick2014radiation} found an empirical relation to estimate the hot electron temperature from the bremsstrahlung spectra using \acf{MCNPX} simulations:
\begin{equation}
    T_h=0.73\times T_b^{1.09}
    \label{EqXrayZulick}
\end{equation}
where $T_b$ is the bremsstrahlung temperature fitted from the slope of the logarithm bremsstrahlung spectrum in Fig. \ref{xrayFigMolySpectrum}. The electron temperature associated with Fig. \ref{xrayFigMolySpectrum} is calculated to be 41.7 keV.

\begin{figure}[H]
\centering
\includegraphics[width=0.98\columnwidth]{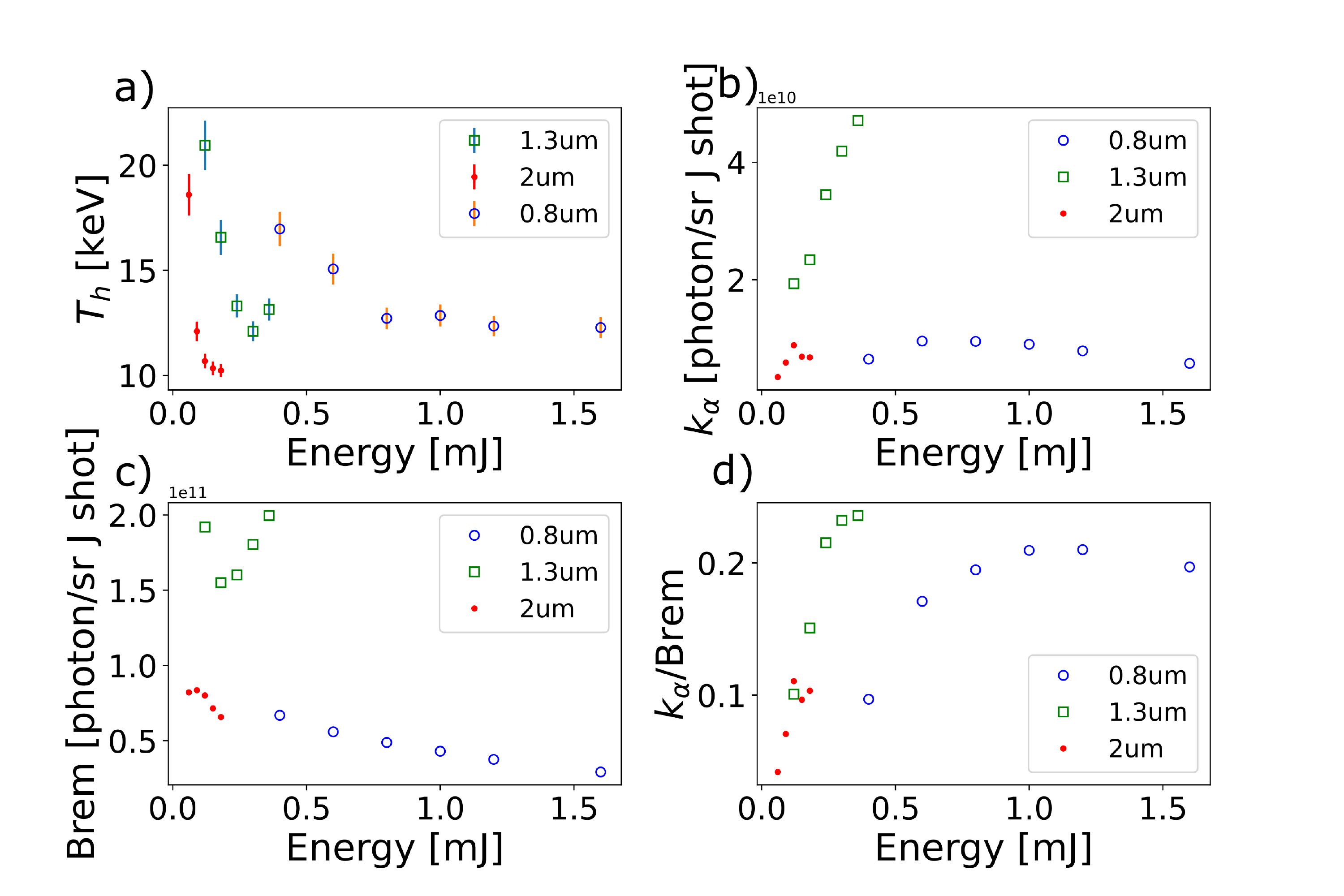}
\caption{Characteristic hot electron temperature (a), $K_\alpha$ emission flux per pulse energy (b), bremsstrahlung emission flux per pulse energy (c), and $K_\alpha$ to bremsstrahlung ratio (d) vs. laser pulse energy. The prepulse delay is kept at 20 ps for the 2 $\mu m$ pulses (red), the 1.3 $\mu m$ pulses (green), and the 0.8 $\mu m$ cases (blue).}
\label{xrayFiga0}
\end{figure}
The experiment was performed at various laser-plasma conditions for a parametric study. Fig. \ref{xrayFiga0} investigates the x-ray properties at various laser pulse energies. The $K_\alpha$ emission and the bremsstrahlung emission are normalized to the laser pulse energy, reported in the unit of photon per steradian per joule per pulse. The results from the 1.3 $\mu m$ pulses (green) are substantially different from the others. This is possibly due to a substantially thicker pinhole used for the x-ray measurements at 1.3 $\mu m$ cases, making the measurements much more sensitive to the pinhole placement if the x-rays do not exactly incident from the normal direction. Results from the 2 $\mu m$ pulses and the 0.8 $\mu m$ pulses are in line with the simulation results to be introduced in Fig. \ref{xrayFig4dka} and Fig. \ref{xrayFig4dbrem}. The $K_\alpha$ flux per pulse energy of the 2 $\mu m$ pulses (red) and the 0.8 $\mu m$ pulses (blue) behave non-monotonically vs. $a_0$, as is shown in Fig. \ref{xrayFiga0}b, agreeing with the simulation results in Fig. \ref{xrayFig4dka} where the normalized $K_\alpha$ and bremsstrahlung flux peak at some $a_0$. The measured normalized bremsstrahlung flux of the 2 $\mu m$ pulses (red) has a peak in Fig. \ref{xrayFiga0}c, but that of the 0.8 $\mu m$ pulses (blue) shows a monotonic relation as laser pulse energy increases. However, it has to be pointed out that the optimal laser pulse energy moves to the left from Fig. \ref{xrayFiga0}b to Fig. \ref{xrayFiga0}c. Analogously, the simulation results in Fig. \ref{xrayFig4dka} show that the optimal $a_0$ for the normalized $K_\alpha$ emission flux is 1 while the optimal $a_0$ for the normalized bremsstrahlung emission flux is 0.5. Therefore, it is possible that the blue curve in Fig. \ref{xrayFiga0}c has a peak to the left, and more experiments at smaller laser pulse energies are needed to draw any conclusion on the monotonicity of the normalized bremsstrahlung flux vs. pulse energy.

\begin{figure}[ht]
\centering
\includegraphics[width=0.98\columnwidth]{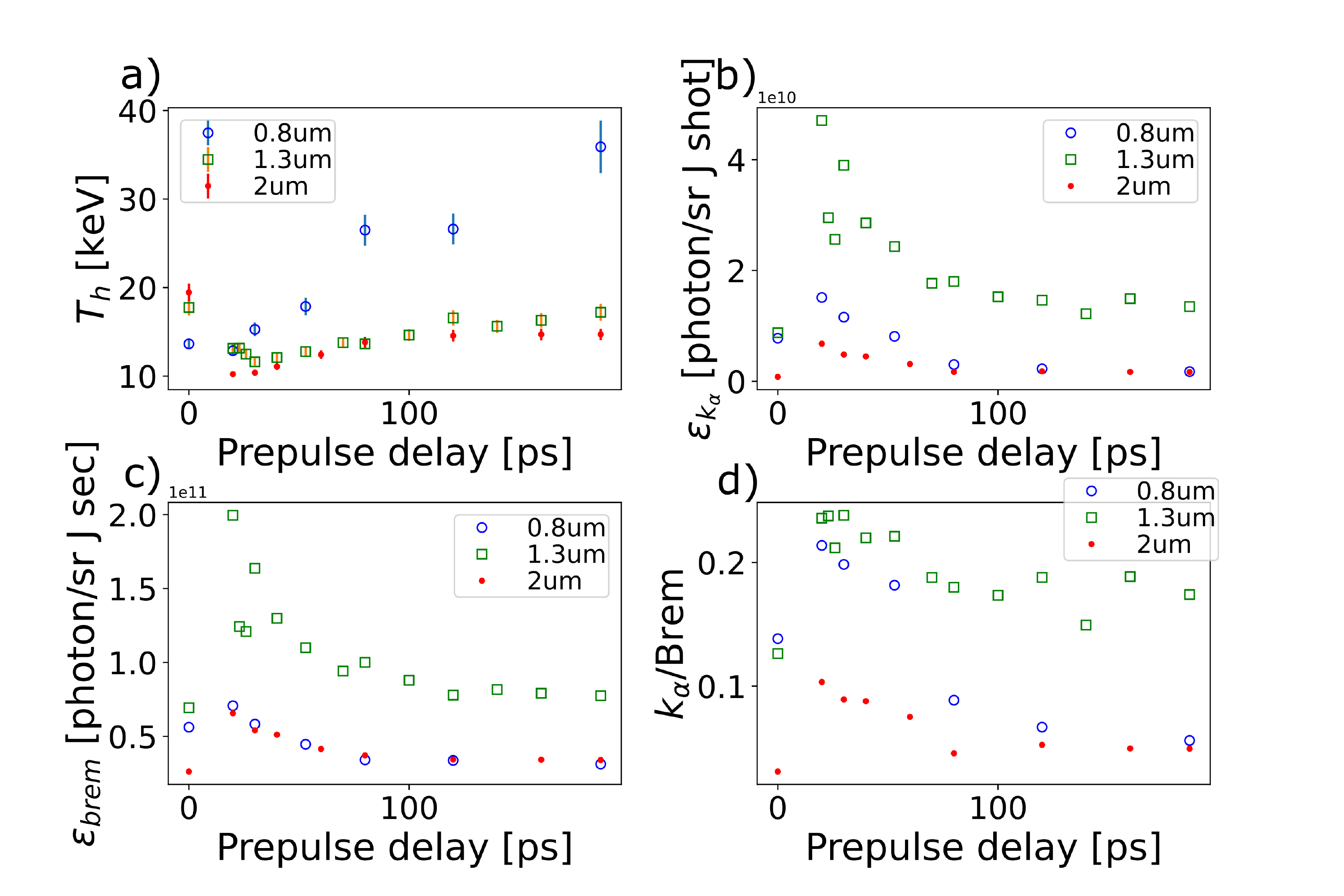}
\caption{Characteristic hot electron temperature (a), $K_\alpha$ emission flux per pulse energy (b), bremsstrahlung emission flux per pulse energy (c), and $K_\alpha$ to bremsstrahlung ratio (d) vs. prepulse delay. The pulse energy is 0.2 mJ, 0.4 mJ, and 0.6 mJ in the 2 $\mu m$ cases (red), 1.3 $\mu m$ cases (green) and 0.8 $\mu m$ cases (blue). The prepulse delay scans over 0 - 187 ps, which corresponds to a preplasma density scale length of 0.1 - 5.5 $\lambda$.}
\label{xrayFigPrepulse}
\end{figure}

Fig. \ref{xrayFigPrepulse} investigates the preplasma effect while the pulse energy was kept the same when tuning the prepulse. Results from 0.8 $\mu m$ pulses, the 1.3 $\mu m$ pulses, and the 2 $\mu m$ pulses are shown in blue, green, and red, respectively. The results from the 1.3 $\mu m$ pulses are somewhat anomalous, as is explained in the previous paragraph, while the trend of the 0.8 $\mu m$ pulses and the 2 $\mu m$ pulses agree with the simulation results to be shown in the next section in Fig. \ref{xrayFig4dka} and Fig. \ref{xrayFig4dbrem}. This is likely due to variability in beam quality and focal spot structure among the three wavelengths. It is observed that the measurement without a prepulse (0 ps) is distinct from the measurements with a prepulse in all cases. An optimal prepulse delay is suggested, and its value should be smaller than the shortest prepulse we used at 20 ps ($L_s=\sim0.5\lambda$). As the prepulse delay passes the optimal value and extends to 187 ps, the normalized$K_\alpha$ flux drops more than seven times smaller from its peak value while the normalized bremsstrahlung flux drops only $\sim$three-times. 

\subsection{\acs{PIC} simulations}
% PIC setup
PIC simulations were performed using the OSIRIS\cite{OsirisRef, OsirisRef2} 4.4.4 framework in 2D3V Cartesian geometry. Since the laser wavelength is a tuning parameter in this work, which affects many other parameters in the simulation setup like time-space resolution and critical density, these setup parameters will be described universally rather than in real units. An example of the OSIRIS input deck and the bash programming for parameter scan are appended in App. \ref{app:OSIRISinput} and App. \ref{app:bash}.

For the high-resolution simulations, there are 100 macroparticles per cell and the grid size is $\lambda/64\times\lambda/64$. For the simulations for parameter scan, there are 49 macroparticles per cell and the grid size is $\lambda/32\times\lambda/32$. The convergence of the simulation is checked using up to $\lambda/64\times\lambda/64$ grid size. The time step is chosen to resolve both the Courant criterion (see Sec. \ref{sec:ExpPIC}) and the plasma density. The laser pulse is assumed to be Gaussian in both the longitudinal and transverse direction with a FWHM pulse duration of $\tau=50 fs$. The laser pulse is continuously launched from the wall and focused down to beam waist $w_0=1.7\lambda$ ($FWHM=2\lambda$). The simulation box size is kept at $40 \mu m\times40\mu m$ throughout the parameter scan. The simulations ran until the interesting electrons left the simulation box, and the simulation time varies as the laser wavelength changes. The initial plasma density profile is described by a uniform solid density region plus an exponential preplasma. The solid density is kept the same at the molybdenum density: $1.46\times10^{23}cm^{-3}$, while the critical density varies as the laser wavelength changes according to Eq. \ref{EqTheoryCrit}.

Information of the electrons located within the solid density region are saved from \acs{PIC} simulations at every 15 fs, including their position and kinetic energy. Further analysis of such information yields an estimation of the hot electron temperature, the $K_\alpha$ emission, and the bremsstrahlung emission.

The hot electron temperature is fitted from the energy spectra of the electrons located within the solid density region, assuming a Boltzmann distribution starting from a cut-off energy $E_{os}$ \cite{liseykina2015collisionless}:
\begin{equation}
    \frac{dN}{dE}\sim exp(-E/T_h) \mbox{  if  } E>E_{os} = m_e c^2 [\sqrt{1+a_0^2/2}-1]
    \label{EqXrayBoltzmann}
\end{equation}
where the cut-off energy $E_{os}$ refers to the mean oscillation energy of the electrons in the laser field. Electrons with energies higher than $E_{os}$ are regarded as hot electrons, while electrons with lower energies are seen to be cold and will be included in the curve fitting. Even though Eq. \ref{EqXrayBoltzmann} is an exponential function, a linear curve fitting is applied by taking the logarithm of the spectra to reduce numerical errors. Note that the mathematical description of the electron energy spectra remains an open question. Another commonly used relation is $\frac{dN}{dE}\sim\sqrt{E} exp(-E/T_h)$, which assumes a Maxwellian distribution on the electron velocity rather than the electron energy. Liseykina \textit{et al.} \cite{liseykina2015collisionless} have an in-depth investigation on this topic. 
\begin{figure}[ht]
\centering
\includegraphics[
width=0.95\columnwidth
% , height=0.6\textheight
]{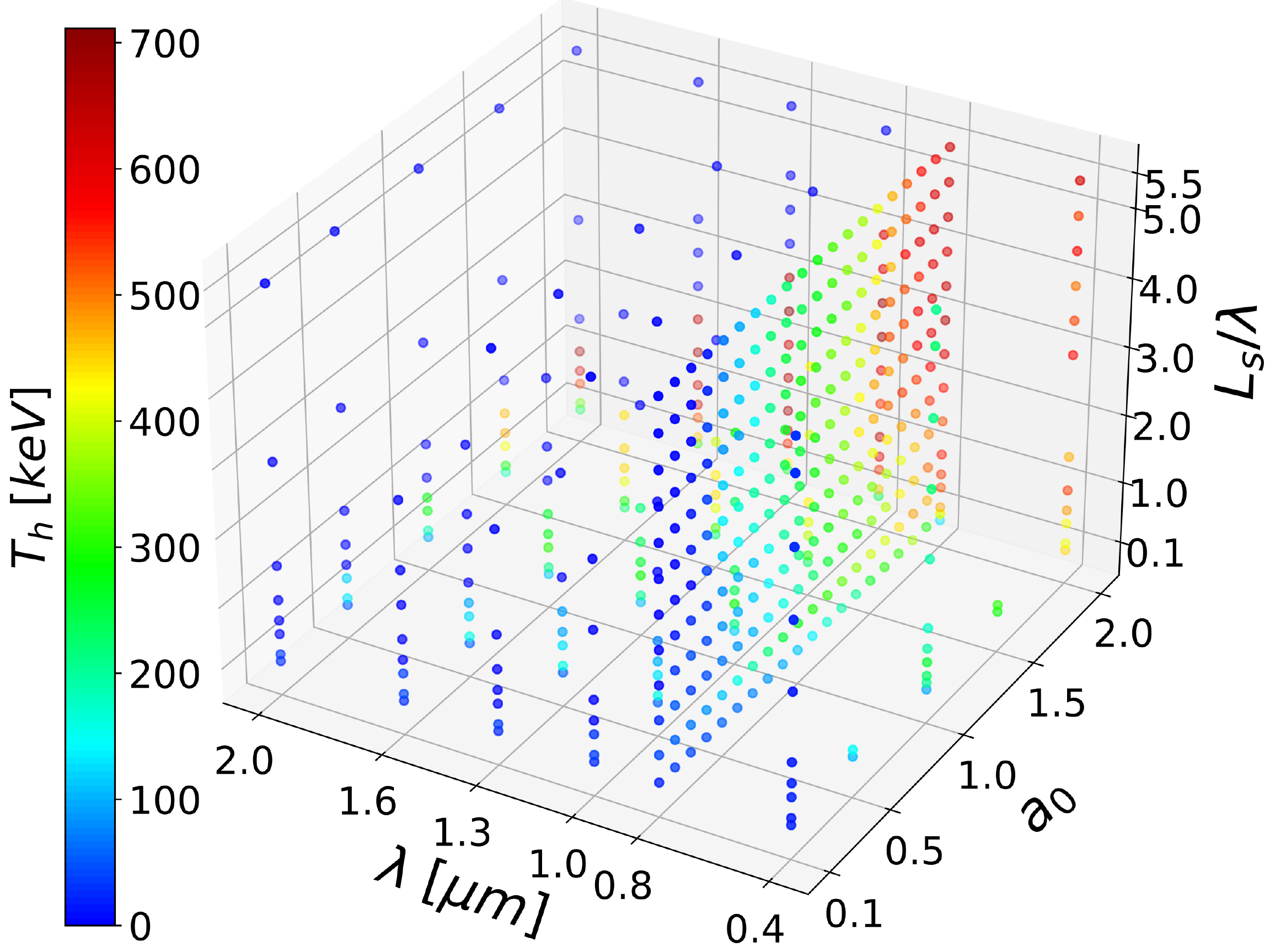}
\caption{Hot electron temperature (shown in color) from various laser-plasma conditions. Each dot represents a \acs{PIC} simulation where the laser wavelength $\lambda$ lies in between 0.4 $\mu m$ and 2 $\mu m$, the peak normalized vector potential $a_0$ lies in between 0.1 and 2, and the preplasma scale length $L_s$ lies in between 0.1 $\lambda$ and 5.5 $\lambda$.
}
\label{xrayFig4dTh}
\end{figure}

\begin{figure}[ht]
\centering
\includegraphics[
width=0.95\columnwidth, 
% height=0.4\textheight
]{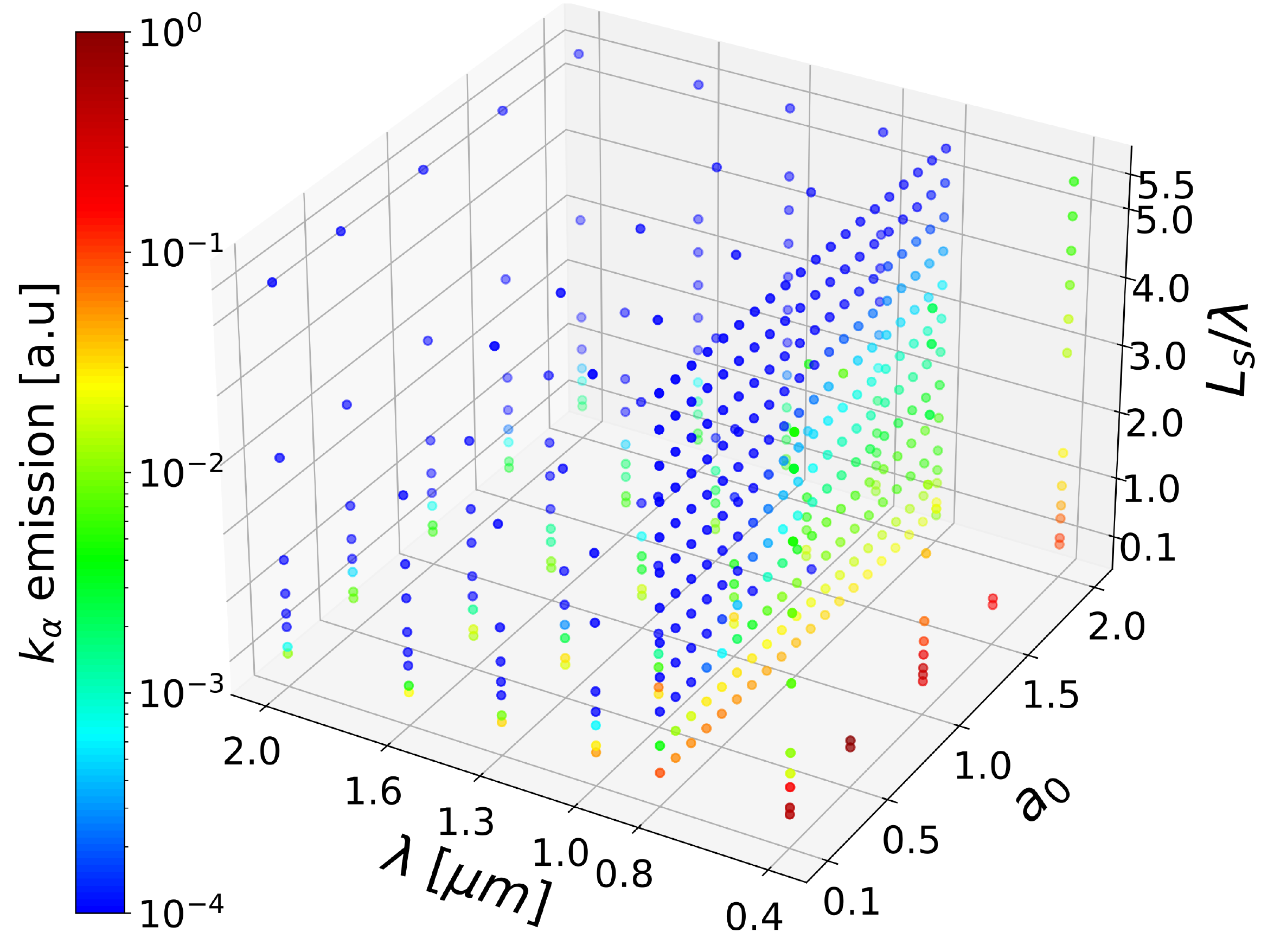}
\caption{Normalized $K_\alpha$ emission per pulse energy shown in color from various laser-plasma conditions. Each dot represents a \acs{PIC} simulation where the laser wavelength $\lambda$ lies in between 0.4 $\mu m$ and 2 $\mu m$, the peak normalized vector potential $a_0$ lies in between 0.1 and 2, and the preplasma scale length $L_s$ lies in between 0.1 $\lambda$ and 5.5 $\lambda$.}
\label{xrayFig4dka}
\end{figure}

The $K_\alpha$ emission (number of photons produced) is estimated given time t, spatial position $\vec{r}$ of the electron, and electron energy denoted by the Lorentz factor $\gamma$, assuming ions remain at rest:
\begin{equation}
    Yield  = \int dt \int d^3\vec{r} \int d\gamma f_e(\vec{r},t,\gamma) n_i(\vec{r},t)  \sigma_{\mathrm{K} \alpha}(\gamma) \sqrt{1 - 1/\gamma^2}\cdot c 
    \label{EqXrayKa}
\end{equation}
where $f_e(\vec{r},t,\gamma)$ is the electron distribution and $n_i(\vec{r},t)$ is the ion distribution and $\sqrt{1 - 1/\gamma^2} c$ is the velocity expressed in terms of $\gamma$. The spatial integral in Eq. \ref{EqXrayKa} is estimated by summing over all electrons in the solid-density region and assuming the ion and electron density is approximately uniform. The temporal integral in Eq. \ref{EqXrayKa} is estimated by summing over the entire simulation time. Instead of assuming $v\simeq c$, the particle velocity is calculated as $\sqrt{1 - 1/\gamma^2}\cdot c$ to take into account contributions from non-relativistic electrons. The cross-section for $K_\alpha$ emission $\sigma_{\mathrm{K} \alpha}$ of an electron is proportional to cross section for k-shell ionization $\sigma_{\mathrm{K}}$ \cite{llovet2014cross}: $\sigma_{\mathrm{K} \alpha} \propto \sigma_{\mathrm{K}}$. The analytical formula to calculate the cross-section for inner-shell ionization by electron impact is provided by Bote \textit{et al.} in Ref. \cite{bote2009cross} given the kinetic energy of the electron. Therefore, the $K_\alpha$ emission is estimated using Eq. \ref{EqXrayKa} and calculated regarding the laser pulse energy, then normalized to the largest $K_\alpha$ emission in the dataset, as is shown in Fig. \ref{xrayFig4dka}.

% The $K_\alpha$ emission is estimated by calculating the cross-section for $K_\alpha$ emission $\sigma_{\mathrm{K} \alpha}$ of each electron, which is proportional to cross section for k-shell ionization $\sigma_{\mathrm{K}}$ \cite{llovet2014cross}: $\sigma_{\mathrm{K} \alpha} \propto \sigma_{\mathrm{K}}$. The analytical formula to calculate the cross-section for inner-shell ionization by electron impact is provided by Bote \textit{et al.} in Ref. \cite{bote2009cross} given the kinetic energy of the electron. Therefore, the total $K_\alpha$ emission is estimated by integrating over all electrons in the solid-density region over the entire simulation time. 
% \begin{equation}
%     \sigma_{\mathrm{K} \alpha}=\frac{\Gamma_{\mathrm{K} \alpha}}{\Gamma_{\mathrm{K}}(\mathrm{R})} \omega_{\mathrm{K}} \sigma_{\mathrm{K}}
%     \label{EqXrayKa}
% \end{equation}
% where 

\begin{figure}[ht]
\centering
\includegraphics[
width=0.95\columnwidth, 
% height=0.4\textheight
]{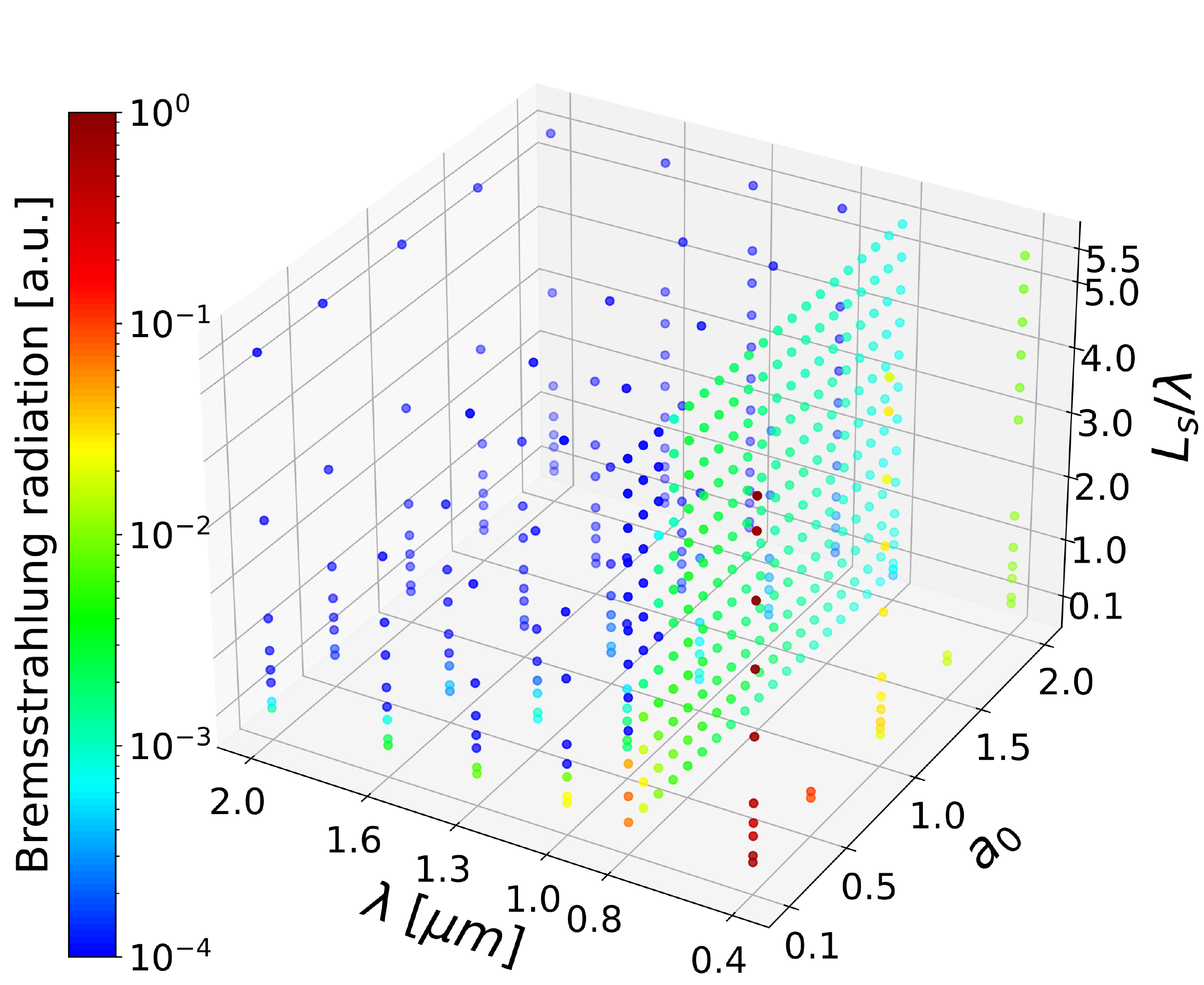}
\caption{Normalized bremsstrahlung radiation per pulse energy shown in color from various laser-plasma conditions. Each dot represents a \acs{PIC} simulation where the laser wavelength $\lambda$ lies in between 0.4 $\mu m$ and 2 $\mu m$, the peak normalized vector potential $a_0$ lies in between 0.1 and 2, and the preplasma scale length $L_s$ lies in between 0.1 $\lambda$ and 5.5 $\lambda$.}
\label{xrayFig4dbrem}
\end{figure}

The radiation power from bremsstrahlung is calculated using Eq. \ref{EqXrayBrem} given the fitted electron temperature in Fig. \ref{xrayFig4dbrem}. The bremsstrahlung radiation per pulse energy is calculated in $W/(m^3\cdot mJ)$ and then normalized to the largest bremsstrahlung radiation in the dataset. The code to estimate $K_\alpha$ emission, bremsstrahlung emission, and electron temperature from \acs{PIC} simulations are appended in App. \ref{app:Xray}.

The hot electron temperature, the normalized $K_\alpha$ emission (per pulse energy), and the normalized bremsstrahlung emission (per pulse energy) are plotted in Fig. \ref{xrayFig4dTh}, Fig. \ref{xrayFig4dka}, and Fig. \ref{xrayFig4dbrem}, with a multi-dimensional parameter scan across the laser wavelength, the peak normalized vector potential, and the preplasma density scale length. From Fig. \ref{xrayFig4dTh}, the hot electron temperature clearly favors larger $a_0$ but has a more complicated dependence on the other two parameters. The optimal normalized scale length $L_s/\lambda$ increases when the wavelength decreases, and the optimal absolute scale length $L_s$ lies in between 2 $\mu m$ and 3.25 $\mu m$ across the wavelength range. The highest electron temperature in Fig. \ref{xrayFig4dTh} occurs at $\lambda=1.3\;\mu m$ and $L_s=3.25\;\mu m$. Although more data points from more accurate \acs{PIC} simulations are needed to draw a conclusion on the exact optimal value, there does seem to be an optimal wavelength rather than a monotonic dependence. It can be explained by the fact that the laser pulse propagates deeper at a shorter laser wavelength (Eq. \ref{EqTheoryCrit}) while the ponderomotive force favors a longer wavelength (Eq. \ref{EqTheorysinglePondero4}).

The x-ray emission flux are normalized to the energy of the driving laser pulses $E\sim a_0^2$. From Fig. \ref{xrayFig4dka} and Fig. \ref{xrayFig4dbrem}, the laser wavelength is playing a much more significant role in radiation generation (both $K_\alpha$ and bremsstrahlung) than the preplasma density scale length and the $a_0$. As wavelength increases, the critical density decreases as $n_{cr}\propto \lambda^{-2}$ and the critical surface moves further away from the solid density surface, making it less efficient to couple the laser energy into the solid target. Therefore, the highest normalized emission flux occurs when the laser wavelength is shortest at 0.4 $\mu m$. The normalized $K_\alpha$ flux peaks at $a_0=0.5$ and $L_s=0.2\sim0.3\lambda$. This finding agrees with the observation in the experimental results in Fig. \ref{xrayFigPrepulse}b where the optimal prepulse is $<0.5\;\lambda$. It is also in line with the 0.8 $\mu m$ and 2 $\mu m$ experimental results in Fig. \ref{xrayFiga0}b where an optimal pulse energy exists. The normalized bremsstrahlung radiation peaks at $a_0=0.1$, which agrees with the experimental results in Fig. \ref{xrayFiga0}c where the bremsstrahlung radiation per pulse energy monotonically decreases with pulse energy. Similar to the experimental results in Fig. \ref{xrayFigPrepulse}c, the normalized bremsstrahlung in Fig. \ref{xrayFig4dbrem} favors a short plasma density scale length.

To proceed further with the parametric study, it is worth looking at parameters that are not directly controlled by the laser pulse or the plasma profile. The similarity parameter $S=\frac{n_e/n_{cr}}{a_0}$ is proposed by Gordienko \textit{et al.} \cite{gordienko2005scalings} for relativistic electrons in laser-plasma accelerators, where $a_0$ is the peak normalized vector potential, $n_e$ is the electron density and $n_{cr}$ is the critical density. Tab. \ref{tab:xRayS} summarizes three high-resolution \acs{PIC} simulations performed with various laser wavelength, peak normalized vector potential $a_0$, and normalized scale length $L_s/\lambda$. The similarity parameter is kept the same among these simulations by carefully choosing the tuning parameters, where the initial electron density is constrained by using the same absolute scale length $L_s$ = 0.8 $\mu m$.

\begin{table}[ht]
  \centering
    \begin{tabular}{|c|c|c|c|}
    \toprule
    Wavelength [$\mu m$] & 0.8   & 1.3   & 2 \\
    \midrule
    $n_{cr}\;[cm^{-3}]$ & 1.74$\times10^{21}$ & 6.60$\times10^{20}$ & 2.79$\times10^{20}$ \\
    \midrule
    $a_0$  & 0.4   & 1.06  & 2.5 \\
    \midrule
    Scale length [$\lambda$] & 1 & 0.62 & 0.4 \\
    \midrule
    Scale length [$\mu m$] & 0.8 & 0.8 & 0.8 \\
    \midrule
    S & 1 & 1 & 1 \\
    \midrule
    Normalized $K_\alpha$ emission $[\#/mJ]\cdot constant$ & 1 & 0.2 & 0.08 \\
    % \midrule
    % $\Delta K_\alpha/k_{\alpha, 0.8\mu m}$ & 0 & 39$\%$ & 7$\%$ \\
    \midrule
    Normalized bremsstrahlung $[W/(m^3\cdot mJ)]\cdot constant$ & 1 & 0.04 & 0.002 \\
    \midrule
    $T_h$ [keV] & 42.7  & 237   & 514 \\
    \bottomrule
    \end{tabular}%
  \caption{Comparison of $K_\alpha$ emission, bremsstrahlung radiation, and hot electron temperature from three high-resolution \acs{PIC} simulations at a fixed similarity parameter. The $K_\alpha$ emission per pulse energy and bremsstrahlung radiation per pulse energy are normalized to the largest values among the three, and their dimensions are listed in the seventh and eighth row.
%   In the sixth row, the constant is the scaling parameter that relates the cross-section for k-shell ionization to the cross-section for k-alpha emission.
    }
  \label{tab:xRayS}%
\end{table}%
It is observed that the bremsstrahlung emission and electron temperature vary drastically among the three cases. The vast variation agrees with the trend observed in Fig. \ref{xrayFig4dbrem} and Fig. \ref{xrayFig4dTh} where the bremsstrahlung emission is dominant by a short laser wavelength and the electron temperature favors a high $a_0$. The variation of the normalized $K_\alpha$ emission among the three cases is about an order of magnitude lower than that of the normalized bremsstrahlung radiation. Since the similarity parameter S scales the dynamics of relativistic electrons and given the fact that S works better for the $K_\alpha$ emission, it suggests that the most energetic (relativistic) electrons make substantial contributions to the $K_\alpha$ emission in our parameter space. This agrees with the fact that making an assumption on the particle velocity $v\simeq c$ in Eq. \ref{EqXrayKa} only increases the $K_\alpha$ emission by $\sim20\%$.

\subsection{Conclusion}

In summary, a parametric study is reported for the interdependent relationship between the tuning parameters (laser pulse wavelength, laser pulse energy, and preplasma density gradient) and the critical features in characteristic x-ray emission ($K_\alpha$ emission per pulse energy, bremsstrahlung emission per pulse energy, and electron temperature). Overall, the laser wavelength is the more dominant factor for the normalized $K_\alpha$ emission than the preplasma density gradient and the pulse energy are. The wavelength dependence is monotonic for the normalized $K_\alpha$ emission and for the normalized bremsstrahlung emission, and both favor short laser wavelength. It is because the critical surface is located closer to the solid density surface at short laser wavelength, making it easier for the energetic electrons to reach the atoms in the solid target. For the hot electron temperature, the wavelength dependence is no longer monotonic since the laser pulse propagates deeper at a shorter laser wavelength while the ponderomotive force favors a longer wavelength. Both experimental results and simulation results indicate some optimal preplasma profiles for the normalized $K_\alpha$ emission and for the normalized bremsstrahlung emission, respectively, and this optimal preplasma profile has a sharp density gradient ($L_s<0.5\lambda$). The dependence on $a_0$ is monotonic for electron temperature but not monotonic for the normalized x-ray emission, while the optimal $a_0$ is larger for the normalized $K_\alpha$ emission than for the normalized bremsstrahlung emission. Simulations using various control parameters but the same similarity parameter is performed to reveal the contribution of relativistic electrons to $K_\alpha$ emission.

Note that our results are contradictory to the results reported in \cite{weisshaupt2014high} regarding the wavelength dependence on $K_\alpha$ emission. However, there are a few differences in approaching the problem. First, we report on the normalized $K_\alpha$ flux ($K_\alpha$ emission flux per pulse energy) rather than the $K_\alpha$ emission flux reported in their work. Moreover, the calculation in \cite{weisshaupt2014high} does not include the collective behavior of plasmas so it is unlikely to be correct considering laser absorption and wave-particle interactions.
% while we address it using \acs{PIC} simulations.
% Inner shell ionization rate at high energy vs low energy

% !Work from this paper D:\CUOS\Thesis\prospectus\previousMaterials\LinTermPaperXrays.pdf

\chapter{Applications of Statistical Methods at High Repetition Rates}
\label{chap:ML}
\section[]{Focus optimization at relativistic intensity using genetic algorithms\footnote[3]{This section co-authored with Easter, J. H., Krushelnick, K., Mathis, M., Dong, J., Thomas, A. G. R., and Nees, J. (2018): Focus optimization at relativistic intensity with high numerical aperture and adaptive optics. Optics Communications, 421, 79-82.}}
\label{sec:FocusOpt}
% \footnotetext{This section co-authored with Easter, J. H., Krushelnick, K., Mathis, M., Dong, J., Thomas, A. G. R., and Nees, J. (2018): Focus optimization at relativistic intensity with high numerical aperture and adaptive optics. Optics Communications, 421, 79-82.}
\subsection{Introduction}
To achieve relativistic intensities in the ultrafast regime using
millijoule laser pulse energies, ultrashort pulses are typically focused
with high numerical aperture optics, such as \acf{OAP}s. However, wavefront aberrations resulting from these optics
can be significant. In addition, when approaching the diffraction limit,
aberrations caused by laser system distortions, thermal lensing, and self-phase modulation may also be important. Due to the difficulty of direct measurements of the laser focus at high intensity, wavefront correction and focus optimization are of significant importance. The analogous problem of correcting wavefront distortions is well known in the field of astronomy.
To correct for atmospheric aberrations, high-resolution telescopes
utilize adaptive optics to reshape incoming wave fronts \cite{beckers1993adaptive}. Relying on
reference beams to measure atmospheric distortions, adaptive optics
can be programmed to compensate for these and other distortions and
produce much higher quality images.

A typical adaptive optic is a deformable mirror (DM) with an array
of controllable actuators to deform the mirror surface. Such a mirror
may also be used to optimize the focal quality of a high-NA focusing
optic in a high-intensity, high-repetition-rate laser system. DMs together
with genetic algorithms (GAs) have already been utilized in several
high-power laser facilities. Traditional ways of determining the mirror
shape include direct measurement of the wavefront \cite{bahk2004generation}, optimization of \acf{SHG} in a nonlinear crystal \cite{albert2000generation} and in-situ optimization of surface \acs{SHG} \cite{planchon2006adaptive}. However, these methods all require attenuation of the beam to avoid damage to detection optics. An alternative method uses the \acf{SH} signal generated in a plasma by the
full power of the focused laser light to provide feedback for the optimization
process. It allows the DM to correct for additional wavefront
distortions that may not occur while the beam is attenuated (e.g. thermal
distortion of optics, thermal lensing, and self-focusing). It also avoids
possible wavefront aberrations introduced by the attenuation optics. In
addition, the optimization is no longer constrained to a specific focal
plane, which is a shared disadvantage of the three techniques mentioned
above. Since near-vacuum is maintained in between optimization and
experiment, shifting of optics due to pressure change in conventional
methods no longer exists. The use of optimization techniques to enhance
other phenomena generated from relativistic laser–matter interactions has
also been demonstrated recently at \acs{CUOS}, such as laser filamentation \cite{englesbe2016control}, laser-wakefield acceleration \cite{he2015coherent, he2015coherentPoP, lin2019adaptive}, THz generation \cite{hah2017enhancement}, and high-order harmonic generation \cite{beier2019relativistic}.

Harmonic generation occurs in gaseous media when high-intensity
laser ionizes atoms and plasma is formed. In addition to the possibility of
coherent X-ray generation and nonlinear Thomson scattering, frequency
doubling may also be observed \cite{lau2003nonlinear, bethune1981optical, mossberg1978optical}. \acs{SHG} in a plasma formed by a
high-NA focus is due to the formation of an electron density gradient
by the ponderomotive force. The electron gradient breaks the isotropy
of the gas that would normally prohibit the generation of even order
harmonics. The second order polarization in an underdense plasma is
given by \cite{bethune1981optical}:

\begin{equation}
\label{FocusOptEq1}
     P(2 \omega)=\chi\left[\frac{1}{2} \nabla E^{2}+\frac{2}{\epsilon_{p}} E\left(E \cdot \nabla \ln n_{e}\right)\right]
\end{equation}

where E is the electric field, $n_e$ is the electron density, $\chi$ is the
susceptibility and $\epsilon_p$ is the relative permittivity. The first term has zero curl and cannot radiate. Therefore, a better-corrected and smaller
focus will produce a stronger \acs{SH} signal both by producing
steeper electron density gradients and by delivering higher intensity.
Generation of \acs{SH} signal is a characteristic of laser–plasma
interactions at high intensities using both overdense and underdense
plasma targets \cite{beresna2009high}.

In this paper, we demonstrate a two-fold improvement in the focal quality at f/1.4 using a genetic algorithm.
% that in an f/1.4 focusing condition the beam focal quality can be improved two-fold using a genetic algorithm to optimize the wavefront. 
Experiments have been performed with laser pulses of both 800
nm and 2 $\mu$m wavelength, at $\sim$ mJ energy and with pulse duration of a few tens of femtoseconds. The relationship between the fundamental and
\acs{SH} signal is measured to be quadratic in both cases. The
phase distortion introduced by the back-filled gas in the experimental
chamber is determined to be negligible in both cases.

\subsection{Experimental methods}

\begin{figure}[ht]
\centering
\includegraphics[width=0.95\columnwidth
% , height=0.6\columnwidth
]{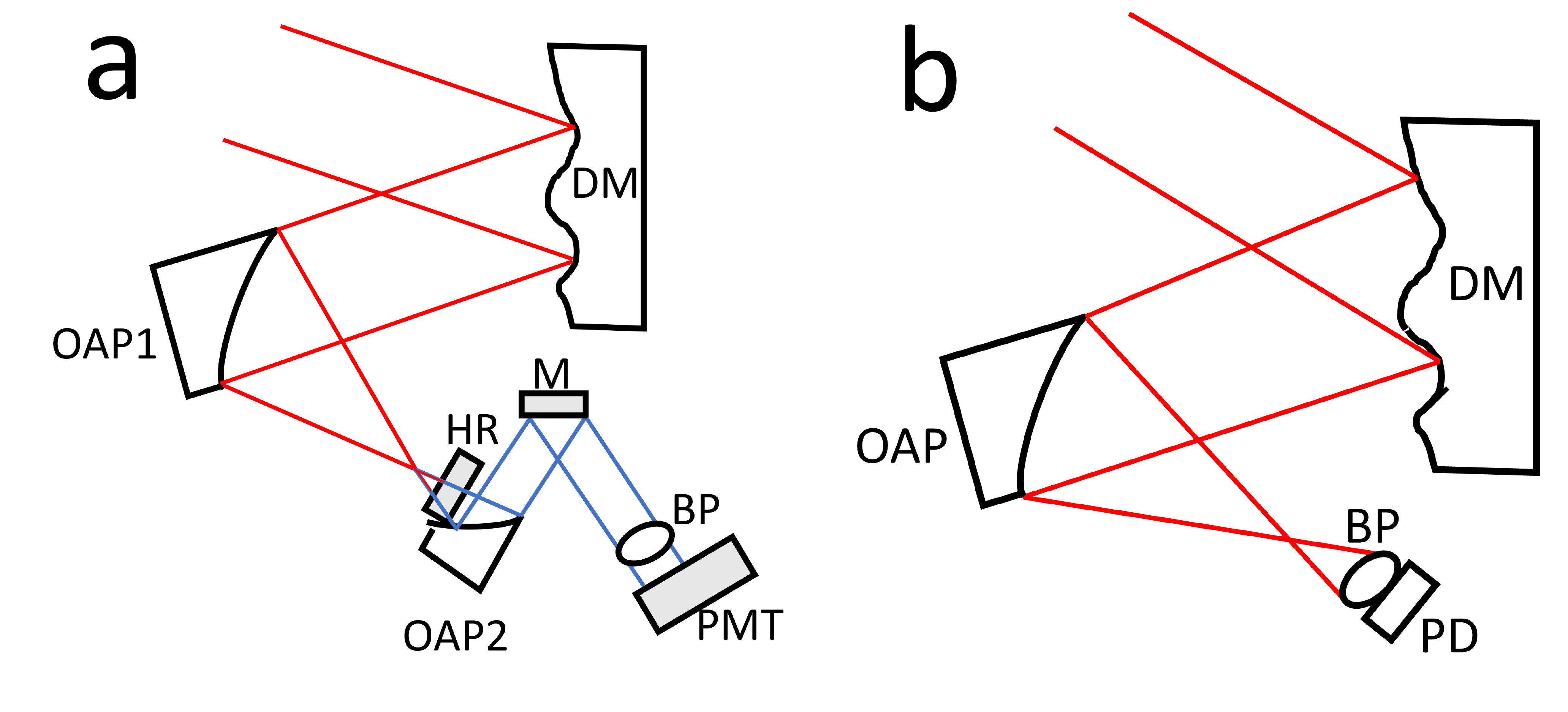}
\caption{(a) Diagram of the experimental setup for the 800 nm beam. DM: Xinetics 37 channel deformable mirror; \acs{OAP}1: 2-inch diameter f/1.4 off-axis paraboloid; HR: 800 nm high reflector; \acs{OAP}2: 1-inch diameter f/1 off-axis paraboloid; M: flat mirror; BP: 340 nm $\sim$ 460 nm bandpass filter; \acs{PMT}: photomultiplier tube. (b) Diagram of the experimental setup for the 2 $\mu$m beam. DM: Xinetics 37 channel deformable mirror; \acs{OAP}: 2-inch diameter f/1.3 off-axis paraboloid; BP: 900 nm $\sim$ 1050 nm bandpass filter; PD: photodiode detector.}
\label{FocusOptFig1}
\end{figure}

The experiments were performed using the \acs{Lambda-cubed} laser system at \acs{CUOS} at the University of Michigan. The experimental layout for the 800 nm
beam is shown in Fig. \ref{FocusOptFig1}a. The deformable mirror is controlled by 37 programmable piezoelectric actuators. The beam reflects off the DM at
an incident angle of $8^{\circ}$, and then propagates 2 m to a vacuum chamber
through an anti-reflection coated, 3 mm thick fused silica window. A 2-
inch diameter gold-coated f/1.4, $60^{\circ}$ \acs{OAP} focuses the 30 fs, 3 mJ pulses inside the chamber. Initial alignment of the paraboloid is performed
by maximizing the brightness of a visible spark generated in ambient
air with attenuation to low intensity. The chamber is then filled with
4 Torr of helium. A broadband dielectric mirror reflects most of the
fundamental light while passing the second harmonic. The beam is
recollimated by a $90^{\circ}$ f/1 \acs{OAP} and directed out of the chamber through a $M_gF_2$ window by a flat, silver mirror. A second filter (Hoya B390)
provides further discrimination against the fundamental. The signal is
then measured with a \acf{PMT}. The \acs{PMT} signal
is fed into a boxcar integrator and the boxcar output is read by the
control computer via a standard data-acquisition device. The computer
runs a genetic algorithm \cite{englesbe2016control, holland1992adaptation} to find the mirror configuration which produces the maximum second harmonic. Each sample used in the genetic algorithm optimization process is averaged over 25 laser shots.

The experimental setup for the focus optimization in the mid-infrared
regime is shown in Fig. \ref{FocusOptFig1}b. An optical parametric amplifier
\cite{xu2014nondegenerate} (OPA) with two BBO crystals is used to generate a beam of 1 mJ energy, 2 $\mu m$ wavelength, and $\sim$40 fs pulse duration. A mirror-based telescope
expands the beam to the size of the DM, which then directs the light into a vacuum chamber back-filled with 40 Torr of air. The chamber window
and initial alignment are all the same as the experiment above. A 50 mm
diameter f/1.4 \acs{OAP} focuses the beam, and a 20 mm diameter silicon
photodiode with a 0.90 $\mu m$ to 1.05 $\mu m$ bandpass filter is used to collect
the \acs{SH} signal. A lock-in amplifier and a control computer with data-acquisition
cards are used to integrate, amplify, and record the signal.
The same genetic algorithm is used to find mirror configurations that
optimize the \acs{SH} signal.

\subsection{Results}

\begin{figure}[ht]
\centering
\includegraphics[width=0.95\columnwidth
% , height=0.6\columnwidth
]{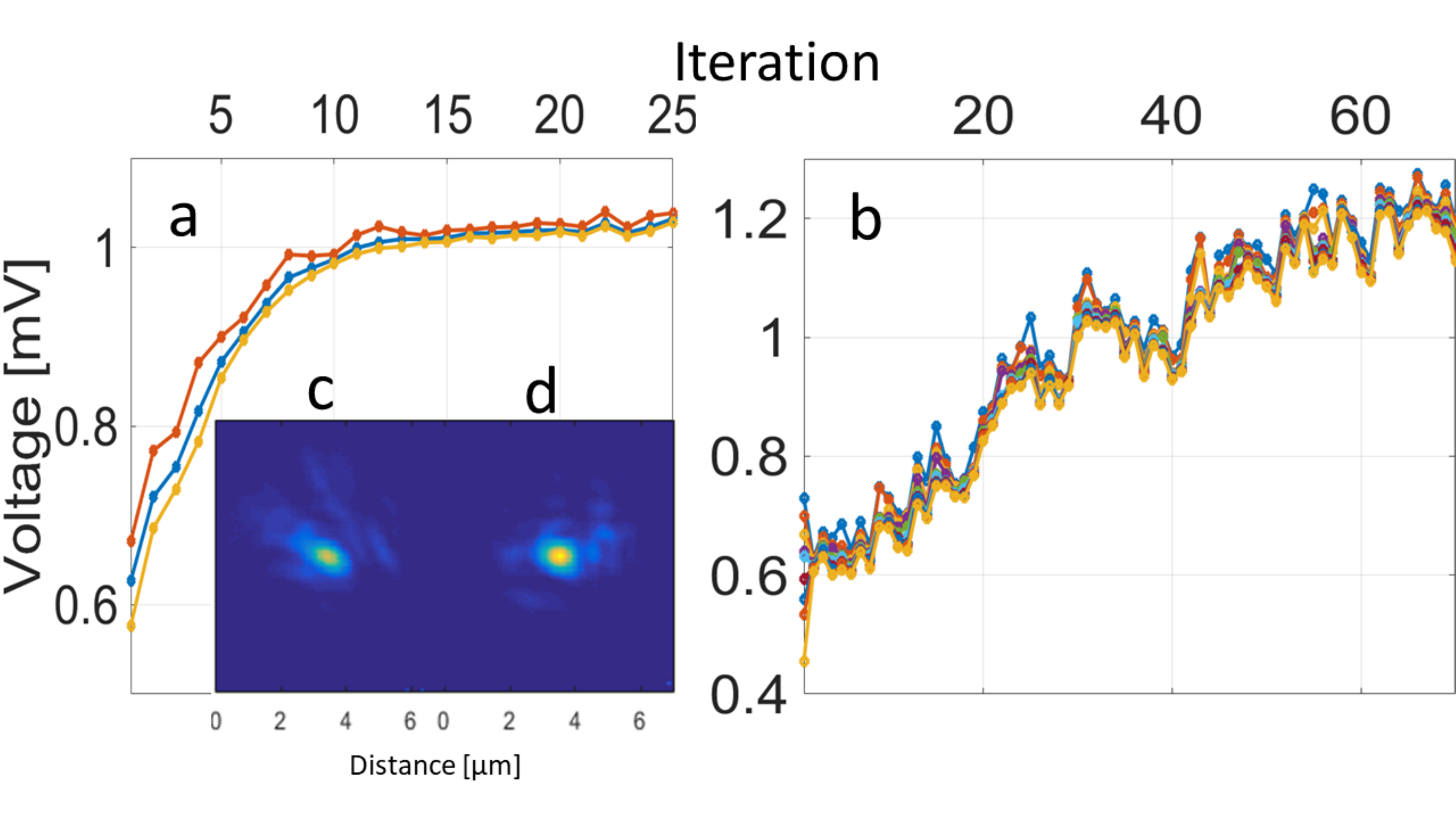}
\caption{Second harmonic signal of the 800 nm beam (a) and the 2 $\mu m$ beam. (b) Improvement charts against iteration number. Focal spot images are taken for deformable
mirror shape before (c) and after (d) correction for the 800 nm case.}
\label{FocusOptFig2}
\end{figure}

Generational improvement charts for the genetic algorithm are
shown in Fig. \ref{FocusOptFig2}. The algorithm takes the \acs{SH} signal as the figure of merit and produces 100 mirror figures for each generation from the ten best figure of merits in the previous generation. A mutation rate of $5\%$ is applied to introduce variation between generations. The algorithm starts from a
fixed figure, which has 30 Volts on all 37 actuators. Three best children
are plotted in the 800 nm case in Fig. \ref{FocusOptFig2}a while all ten children are plotted in the 2 $\mu m$ case in Fig. \ref{FocusOptFig2}b, indicating a higher noise level.

The improvement saturates after 20 generations in the 800 nm
experiment, taking 60 generations in the 2 $\mu m$ case. The \acs{SH} signal
is enhanced by $70\%$ and $100\%$, respectively. To further evaluate the
performance of the technique, images of the focus before and after
optimization are acquired with a $60\times$ microscope objective and CCD
camera at low power. A comparison of the focal quality for 800 nm is
also shown in Fig. \ref{FocusOptFig2}c,d. The focal spot size of the 800 nm beam
is 1.7 $\mu m$, suggesting a peak intensity of $1.1 \times 10^{18} w cm^{-2}$ and $a_0=0.7$.
Due to the limitation of the camera’s detection range, a focal spot
image is not taken for the 2 $\mu m$ beam but instead is approximated
to be $\frac{2\mu m}{0.8\mu m} = 2.5$ times larger than the spot size measured with
800 nm beam. The estimated laser intensity of the 2 $\mu m$ beam is thus
$5.9 \times 10^{16} w cm^{-2}$ and $a_0=0.4$. The corrected focus has a higher peak
intensity, better circularity, more energy above the noise level, and
a larger fraction of energy in the main spot. These features indicate
that the technique indeed does optimize for the highest focal intensity. A
calculation of Strehl ratio is performed based on the following definition:
the ratio between the peak intensity of an image divided by the peak
intensity of a diffraction-limited image with the same total flux \cite{born2013principles}.
The beam profile in the $\lambda^3$ laser system is measured to be close to Gaussian. The Strehl ratio is improved from 0.65 before optimization (Fig. \ref{FocusOptFig2}c)
to 0.95 afterwards (Fig. \ref{FocusOptFig2}d).

\begin{figure}[ht]
\centering
\includegraphics[width=0.95\columnwidth
% , height=0.6\columnwidth
]{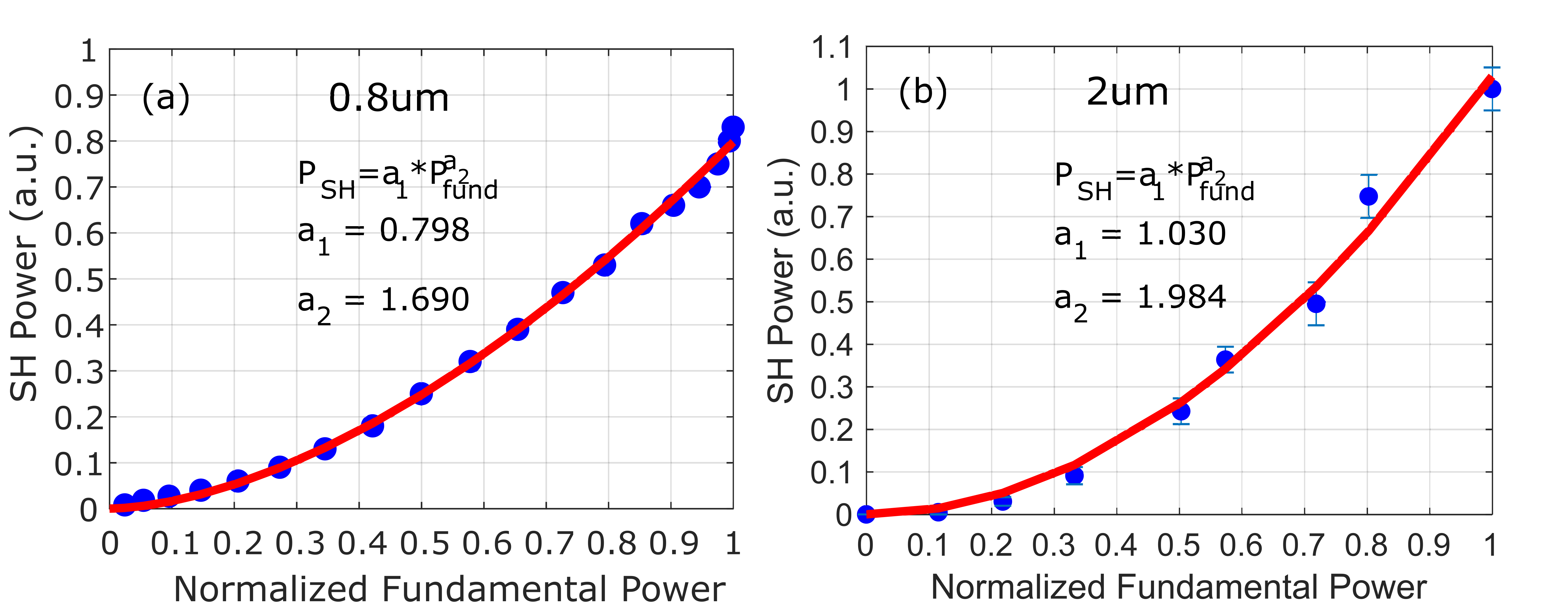}
\caption{Second harmonic signal vs. the fundamental laser power, for the $0.8\mu m$ case (a) and the $2\mu m$ case (b), respectively.}
\label{FocusOptFig3}
\end{figure}

Fig. \ref{FocusOptFig3} presents the data acquired to investigate the \acs{SH} scaling
laws for the two cases. As is shown in Eq. \ref{FocusOptEq1}, \acs{SH} signal is strongly
dependent on better focus for higher intensity and steeper electron
density gradient, both of which scale with laser energy. In the 800
nm case, a wave plate is rotated to control the input power, whereas,
in the $2\mu m$ case, calibrated neutral density filters are placed before
the paraboloid to control attenuation. This would vary the laser energy
without changing the pulse duration, and thus the \acs{SH} signal is only modified by the fundamental laser power. Note that the attenuation is performed only in the scaling measurement. The curves show a good fit
to the power relation that $P_{2\omega}\sim P_\omega$, which confirms that the figure of merit in the genetic optimization process is the second harmonic
signal. A larger uncertainty is observed in Fig. \ref{FocusOptFig3}b, showing higher
noise levels using longer wavelength and lower energy laser light. The
noise is mainly due to the fact that optical parametric amplification is a sensitive process to the input laser beam, suggesting that shot-to-shot fluctuations in pulse energy affect the efficiency of the optimization process.

Furthermore, it is worth pointing out that the exponent in the \acs{SHG} scaling does not necessarily equal to 2, which could be counter-intuitive. For example, \acs{SHG} scales with an exponent of $1.49\pm0.03$ in the nonlinear response of 50fs pulses to gold nanorods \cite{lien2017optical}. The unexpected scaling in the cited \cite{lien2017optical} study is explained as the effect of damping induced by a hot thermal distribution of single-particle excitations, and agrees well with the theoretically predicted value of 1.47 under this assumption. It would be interesting to investigate this scaling with more complete measurements as future work, which may reveal more about the underlying physics of \acs{SHG} in a rarefied gas.

\subsection{Discussion}
Since a rarefied gas is present in the chamber, it is important to
be sure that additional nonlinear effects (self-phase modulation, linear
plasma distortions, Etc.) are not significant. The phase shift can be
quantified by the B-integral:

\begin{equation}
\label{FocusOptEq2}
     B(z)=\frac{2 \pi}{\lambda} \int_{z_{0}}^{z} n_{2} I(z) d z
\end{equation}

where $n_2$ is the nonlinear index of refraction, and $I(z)$ is the spatially
dependent intensity. The beam profile can be approximated as Gaussian in the \acs{Lambda-cubed}. Therefore, the intensity is, with the maximum amount of accumulated phase occurs on axis at the peak of the pulse:

\begin{equation}
\label{FocusOptEq3}
     \left.I(z)\right|_{t=0, r=0}=I_{0} \frac{w_{0}^{2}}{w^{2}(z)}=I_{0} \frac{1}{1+z^{2} / z_{R}^{2}}
\end{equation}

Note that $I_{0}=4 E /\left(w_{0}^{2} \tau \sqrt{\pi^{3} \ln 2}\right)$ is the peak intensity in a Gaussian beam, and $z_R$ is the
Rayleigh range. Integrating the phase shift from the focusing optic at $z=z_0\gg z_R$ to the focal point at z=0:

\begin{equation}
\label{FocusOptEq4}
B(z)=\frac{4 E}{\lambda^{2} \tau} n_{2} \sqrt{\pi^{3} / \ln 2}
\end{equation}

In the case of the 800 nm, 3 mJ, 30 fs laser pulses, 4 Torr of
helium is used. The nonlinear refractive index of helium at atmosphere
is $3.5\times10^{21} w cm^{-2}$ \cite{bernhardt2008critical}, and it decreases linearly with pressure.
From Eq. \ref{FocusOptEq4}, the B-integral is approximately 0.008 rad, or $\lambda/800$. 
On the other hand, 40 Torr of air is backfilled when the
2 $\mu m$, 1 mJ, 30 fs pulses are used. The nonlinear refractive index of $N_2$, $O_2$, and Argon at $2\mu m$ wavelength are $7.3\times10^{-20} cm^{2} w^{-1}$, $8.2\times10^{-20} cm^{2} w^{-1}$, and $9.3\times10^{-20} cm^{2} w^{-1}$, respectively \cite{zahedpour2015measurement} . It yields a B-integral of 0.035 rad, or
$\lambda/180$. Since the intensity profile in the experiment is overestimated due to the Gaussian approximations, the obtained nonlinear phase shift is indeed an upper bound. Considering the chamber leaking, humid air can further
eliminate the phase shift. This can be inferred from the fact that water vapor has a lower refractive index than air \cite{owens1967optical, schiebener1990refractive} and presumably lower nonlinear refractive index, which accumulates less phase shift.

Linear plasma dispersion also contributes to the wavefront distortion.
Consider the phase difference of a wave in vacuum and in
oscillating plasma:

\begin{equation}
\label{FocusOptEq5}
\begin{aligned}
\Delta \phi &=\int \Delta k \cdot d z=\frac{z}{c}\left(\sqrt{\omega^{2}-\omega_{p}^{2}}-\omega\right) \cong \frac{z}{c} \cdot \frac{\omega_{p}}{\omega} \cdot\left(\omega-\frac{\omega_{p}}{2}-\omega\right)=\frac{z \omega_{p}^{2}}{2 c \omega}
\end{aligned}
\end{equation}

where the plasma frequency $w_p$ is much smaller than the laser frequency $w_l$ in low-pressure gas. The
atomic density at 4 Torr is $2.6\times10^{17} cm^{-3}$, which corresponds to a plasma
frequency of 29 rad/ps. The distance z in dispersive plasma starts from
the point where the pulses reach helium double ionization intensity at
$3\times10^{15} w cm^{-2}$ \cite{walker1994precision}. An f/1.4 optic focuses the 800nm, 3 mJ, 30 fs beam to this ionization
intensity at 91 $\mu m$ before the focus. The phase shift of an 800 nm
beam is calculated to be 0.05 rad, or $\lambda/116$ from Eq. \ref{FocusOptEq5}. Air at 40 Torr has an electron density of $2.6\times10^{18} cm^{-3}$, and has considerable ionization at intensities around
$3\times10^{14} w cm^{-2}$ \cite{liu2010tightly}. For the $2\mu m$, 1 mJ, 40 fs beam focused by an f/1.3 optic, the accumulated phase shift is approximately 0.9 rad or $\lambda/7$. This higher distortion is mainly due to the higher operating pressure,
but it is still smaller than the distortion one would expect from strongly
attenuating filters, which are necessary for other alternative methods to perform focus optimization. Furthermore, the second level ionization rate of air is lower than its first level ionization rate for orders of magnitudes at
intensities below $10^{15} w cm^{-2}$. Recall that our focal intensity is around $5.9\times10^{16} w cm^{-2}$, suggesting that for most of the propagation distance, the distortion
is dominated by single ionization. The accumulated phase shift
through singly ionized air is 0.5 rad, or $\lambda/13$. Note that both self-phase modulation and linear plasma distortion can be significantly reduced
by operating at lower pressure, where the nonlinear index of refraction
and plasma frequency are correspondingly lowered. Gas with a higher
ionization threshold can also decrease the effect of wavefront distortion.

\subsection{Conclusion}
A new method of optimizing the surface figure of a deformable mirror for high numerical aperture focusing to relativistic intensity using second harmonic signals generated in rarefied gas is presented. \acs{SH} signal is demonstrated to be a convenient and effective feedback for the genetic algorithm. Optimization at full intensity corrects for
all linear and nonlinear wavefront distortions present in the laser
and focusing systems without introducing additional distortions due to
attenuation. Another advantage of the method is that the optimization
is not constrained to a fixed plane or dependent on fine alignment of
diagnostics. This technique simplifies access to the relativistic intensity
regime, and should be applicable to higher intensity systems
and may employ other kinds of gas at lower gas pressures. For a 1 J, 40
fs, 800 nm laser beam with fast focus of f/1.4, the wavefront distortion
experienced in this focus optimization technique with helium pressure
down to 1 Torr is estimated to be $\lambda/125$. The B-integral would be more
significant, reaching f/10 but still small compared to the distortion from
thick optical filters. The benefits of applying full-power optimization
methods with adaptive optics may be even more significant in ultra-relativistic
experiments, owing to the additional difficulty in attenuating beams. For
example, in multi-joule systems, amplifiers must be disabled to achieve
sufficient attenuation. In addition, this technique also provides significant
experimental convenience and consistency: near-vacuum can be
maintained in between optimization and experiment, which eliminates
potential alignment errors caused by shifting of optics during pressure
change. The number of optics required to collect the second harmonic signal is limited, and they can be controlled remotely. However, to employ this technique as a robust pre-experiment routine in more relativistic high repetition rate laser facilities, there are still questions to be answered, such as a power scaling for the \acs{SHG} of these laser pulses in rarefied gas.

\clearpage
\section[]{Adaptive control of a laser-wakefield accelerator driven by mid-IR laser pulses\footnote[4]{This section co-authored with Ma, Y., Schwartz, R., Woodbury, D., Nees, J. A., Mathis, M., Thomas, A. G. R., Krushelnick, K., and Milchberg, H. (2019): Adaptive control of laser-wakefield accelerators driven by mid-IR laser pulses. Optics express, 27(8), 10912-10923.}}
\label{sec:MIRLWFA}
% \footnotetext{This section co-authored with Ma, Y., Schwartz, R., Woodbury, D., Nees, J. A., Mathis, M., Thomas, A. G. R., Krushelnick, K., and Milchberg, H. (2019): Adaptive control of laser-wakefield accelerators driven by mid-IR laser pulses. Optics express, 27(8), 10912-10923.}
\subsection{Introduction}

Coherent control of dynamic processes through systematic optimization of the phase of laser has been applied to a variety of systems, such as quantum dots \cite{bonadeo1998coherent, nowack2007coherent}, qubits \cite{nakamura1999coherent}, two-photon transitions \cite{meshulach1998coherent}, photocurrent generation \cite{hache1997observation} and chemical reactions \cite{assion1998control}. In the field of intense laser-matter interactions, deformable mirrors (DMs) controlled by genetic algorithms (GAs) taking in feedback measurements have already been utilized widely in high-power laser facilities. These adaptive optical systems have been implemented to control THz generation \cite{hah2017enhancement}, multi-filament configuration \cite{englesbe2016control}, high order harmonic generation \cite{wang2018enhanced} and optimization of the focal spot \cite{albert2000generation, bahk2004generation, lin2018focus}. Plasma waves produced from the interaction process can also be controlled via this phase shaping technique, suggesting that a particular laser wavefront can steer the plasma wave to a final state using an optimal electric field structure \cite{he2015coherent}. Plasma waves produced by high-power lasers, in particular relativistic electrons from \acf{LWFAs}, have been studied extensively \cite{tajima1979laser, mangles2004monoenergetic, geddes2004high, faure2004laser, esarey2009physics, leemans2006gev, liu2011all} as it has extremely large accelerating gradients and consequently the short accelerating distance compared to conventional accelerators. There remain issues with beam pointing, stability control, energy spread and dark current for the use of such beams. In \acs{LWFA}, plasma electrons are expelled from the relativistic laser pulse and form a cavity, or a void of electrons, behind the pulse. The cavity's spatial extent is close to a plasma wavelength in length, a laser focal spot size in width and close to the speed of light in phase velocity. Background electrons can get captured in the cavity and get accelerated to high energy. However, when the laser power is high enough ($P > P_{cr}$), relativistic self-focusing modifies the laser wavefront to overcome diffraction limit and focuses the laser pulse to higher intensity or guides the beam, depending on the ratio $P/P_{cr}$. If the pulse length is long relative to the plasma wavelength, it also overlaps multiple plasma buckets and the laser pulse can modulate and break up into a train of short pulses with pulse length around the plasma wavelength. Operating in this \acf{SM-LWFA} regime with higher density, a large amplitude wakefield approaching the wave-breaking limit is generated to trap background electrons, but the acceleration length is limited. \acs{SM-LWFA} has recently aroused interest in betatron radiation using picosecond duration laser pulses at large laser facilities \cite{albert2018betatron}. 

It is worth pointing out that both the critical power (Eq. \ref{EqTheoryUnderCriticalPower}) and the ponderomotive force (Eq. \ref{EqTheorysinglePondero4}) scale with the square of the laser wavelength. For laser wavelengths at $\lambda=0.8\mu m$ and $3.9\mu m$, the critical densities are $2\times10^{21}cm^{-3}$ and $7\times10^{19}cm^{-3}$, respectively. Assuming an electron density of $3\times10^{19}cm^{-3}$, the critical powers are 987 GW and 42 GW, respectively. Over the past decade, remarkable progress has been made in the generation \cite{andriukaitis201190, liang2017high} and application of mid-infrared (MIR) laser pulses, showing their superiority in generating high order harmonics \cite{popmintchev2012bright, mitrofanov2018high, beier2019relativistic}, electromagnetic pulses \cite{zheltikov2018intensity}, filaments \cite{burger2018intense} and x-rays\cite{weisshaupt2014high}. For instance, it has been found that the characteristic $k_\alpha$ flux of hard x-rays from $3.9\mu m$ laser driver is much greater than that from the 800 nm driver \cite{weisshaupt2014high}. Towards reaching longer wavelengths in \acs{LWFA} has also drawn attention, not only for its lower critical power threshold and higher $a_0$ but also for being less difficult in achieving near-critical density interactions, which enables the generation of MeV-scale electrons with moderate laser intensity. These MeV-scale electron sources from high repetition rate laser systems have demonstrated their use in electron radiography \cite{mangles2006table, bussolino2013electron}. \acs{SM-LWFA} at near-critical density can be approached using cryogenically cooled, high-density gas jets with 800nm Ti:Sapphire lasers \cite{goers2015multi, salehi2017mev, salehi2019characterization}, and more recently using moderate density gas jets with a mid-infrared laser at $\lambda=3.9\mu m$\cite{woodbury2018laser}.

It is natural to consider coherent control of the \acs{LWFA} dynamics with mid-\acf{IR} lasers. In this work, we present the first experiment to optimize the quality of the electron beam from mid-\acs{IR} ($\lambda=3.9\mu m$) light interacting with near-critical density plasma. Beam charge, energy spectrum, beam pointing and fluctuation have been improved by controlling the laser wavefront via an evolutionary algorithm. Wavefront reconstruction and \acs{PIC} simulations illustrate that changes on laser wavefront lead to different laser focusing and self-guiding in plasma. Filamentation has been observed in the case of a flat laser wavefront, and can be corrected by the adaptive control system for better electron acceleration. This work also demonstrates the ability to have regular deformable mirrors with 4 $\mu m$ full stroke to properly function in a mid-\acs{IR} laser system, and the ability to reconstruct wavefronts without the presence of a mid-\acs{IR} wavefront sensor.

% \subsection{Results}
\subsection{Experimental setup}

\begin{figure}[ht]
\centering
\includegraphics[width=0.75\columnwidth,height=0.25\textheight]{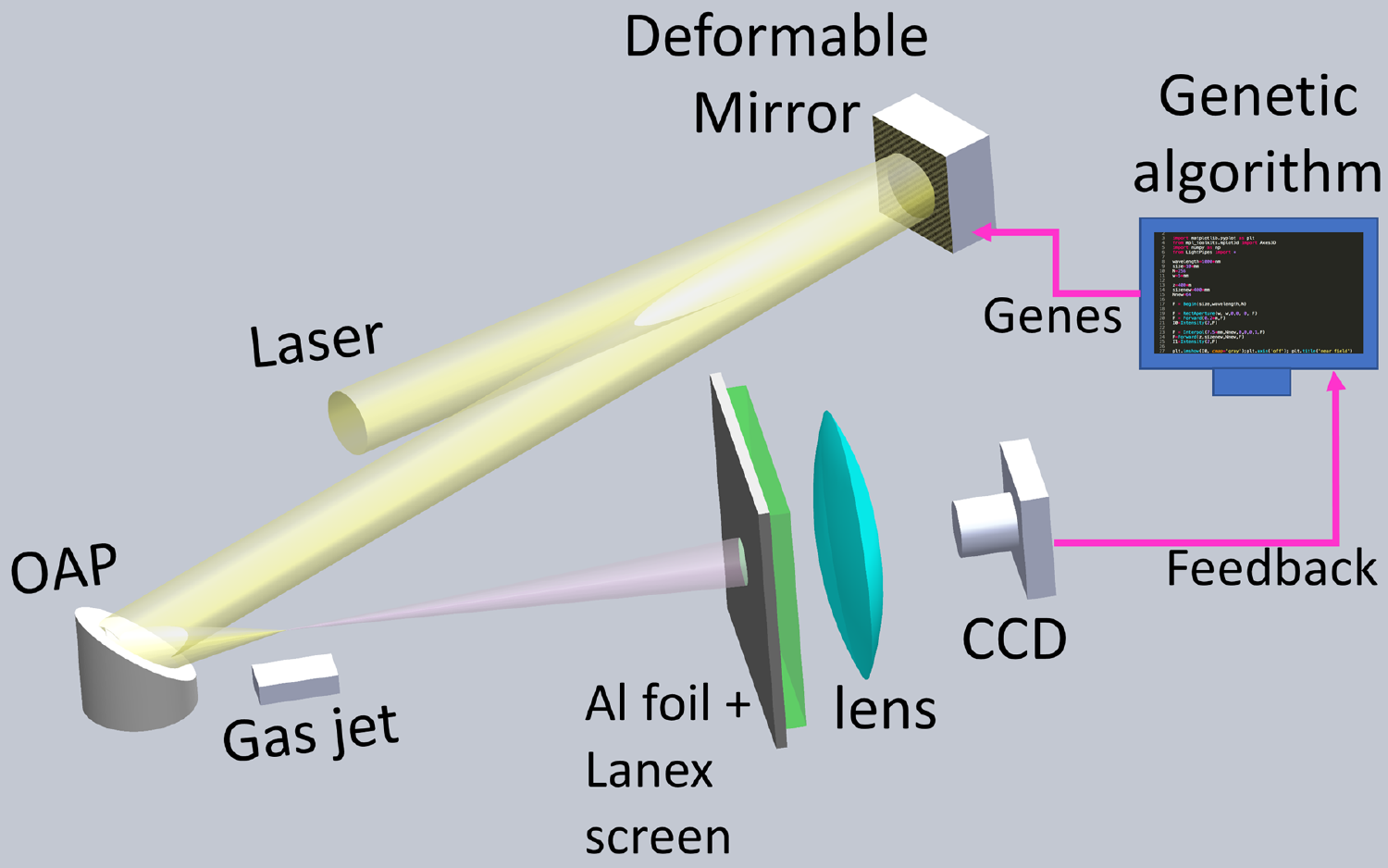}
\caption{Schematic of the setup: Deformable mirror: AOA Xinetics 37-channel 2 inch; OAP: f/2.7; Gas jet: 150$\mu m$ orifice diameter nozzle; CCD: The Imaging Source DMK41BU02.H Charged Particle Device (CCD) camera; Lanex: LANEX Regular screen.}
\label{MIRLWFAsetUp}
\end{figure}

The experiment was conducted at the University of Maryland, using a hybrid optical parametric amplifier/optical parametric chirped-pulse amplifier (OPA/OPCPA) laser system which generates $25\pm1$ mJ, 87 fs, $3.9\mu m$ pulses at a repetition of 20 Hz \cite{andriukaitis201190}. A deformable mirror (DM) controlled by the evolutionary algorithm was used to adjust the laser wavefront based on the diagnostic feedback. The experiment setup is shown in Fig. \ref{MIRLWFAsetUp}. An f/2.7 focus was achieved from the paraboloid and the 37.5mm diameter laser beam was focused to $15 \mu m$ at the beam waist, measured using a knife-edge scan. The pulse energy on target was measured to be 15 mJ, resulting a peak intensity $I=\frac{0.94\cdot E_p}{\tau_p\cdot\pi\cdot w^2/2}=4.6\times 10^{16}\;W/cm^2$ and $a_0\sim$ 0.7. Plasma density up to $3\times 10^{19}cm^{-3}$ (or $40\%$ of the critical density at $\lambda=3.9 \mu m$) can be easily reached by the gas jet without cryogenic cooling. The jet was mounted onto a 3-D translation stage to adjust the position of the laser focus throughout the hydrogen gas target. A LANEX regular screen with a shield of 100 $\mu m$ thick aluminum was placed 9 cm from the jet and imaged onto the CCD camera. The camera was synchronized to the 3.9 $\mu m$ pulse and integrated over 2 ms. In each iteration, the genetic algorithm analyzed 50 electron beam profiles corresponding to 50 deformable mirror surfaces, and the median of 10 shots was used to evaluate a figure of merit function for each mirror surface. While the system repetition rate was limited to 1 Hz due to radiation safety requirements, this was still adequate for averaging over the shot-to-shot fluctuation while keeping the data acquisition period reasonable. The starting point of the optimization process was chosen at the condition of minimum phase changes, where the DM was initialized to a flat mirror surface. It took the evolutionary algorithm $\sim30$ iterations to find an improved mirror surface where the generation curve approached convergence, as is shown in Fig. \ref{MIRLWFAimproveChart}.

\subsection{Results}
\subsubsection{Optimizing the total electron beam charge}

\begin{figure}[ht]
\centering
\begin{subfigure}{.5\textwidth}
  \centering
  \includegraphics[width=.85\linewidth,height=0.17\textheight]{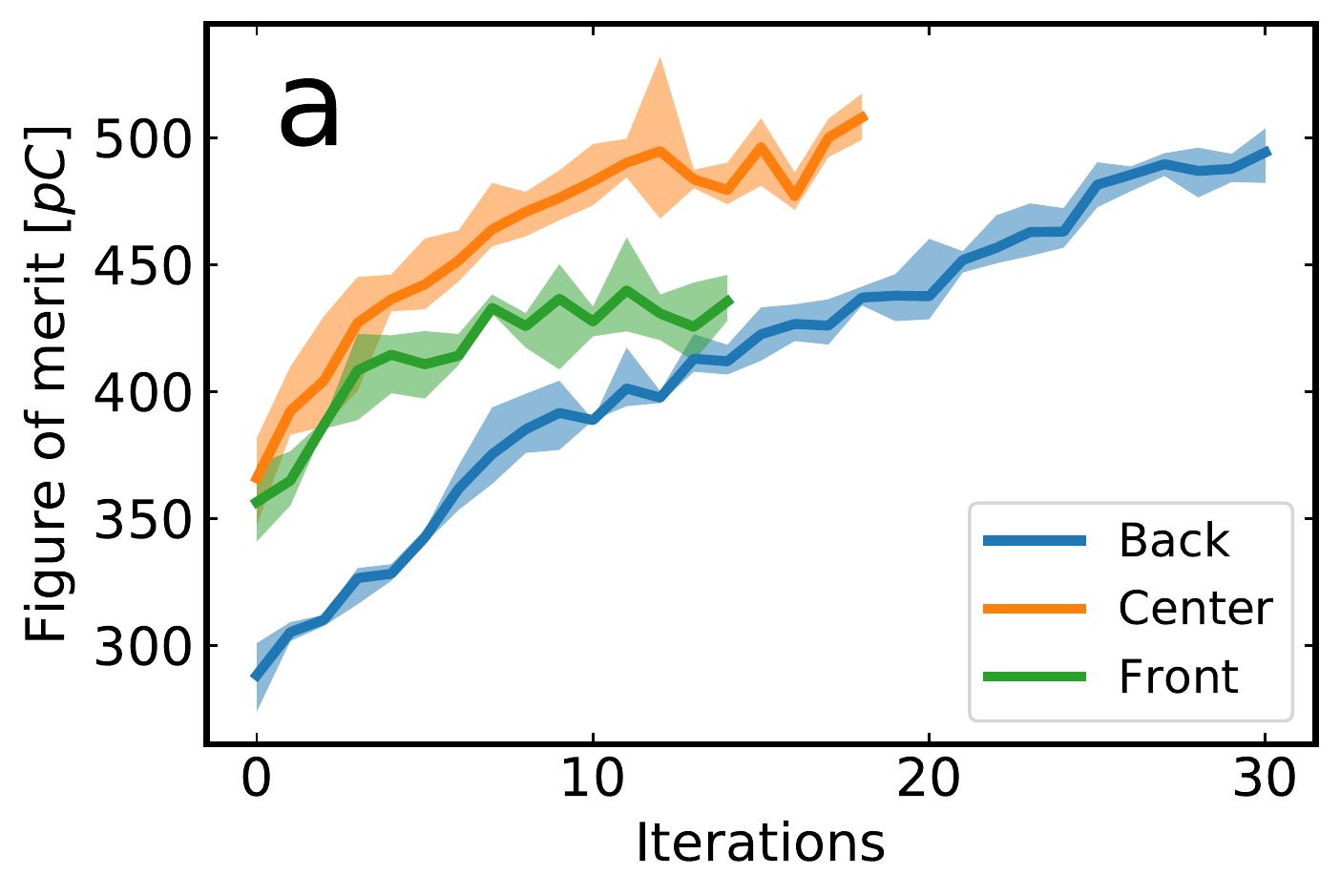}
%   \caption{}
%   \label{charge}
\end{subfigure}%
\begin{subfigure}{.5\textwidth}
  \centering
  \includegraphics[width=.85\linewidth,height=0.17\textheight]{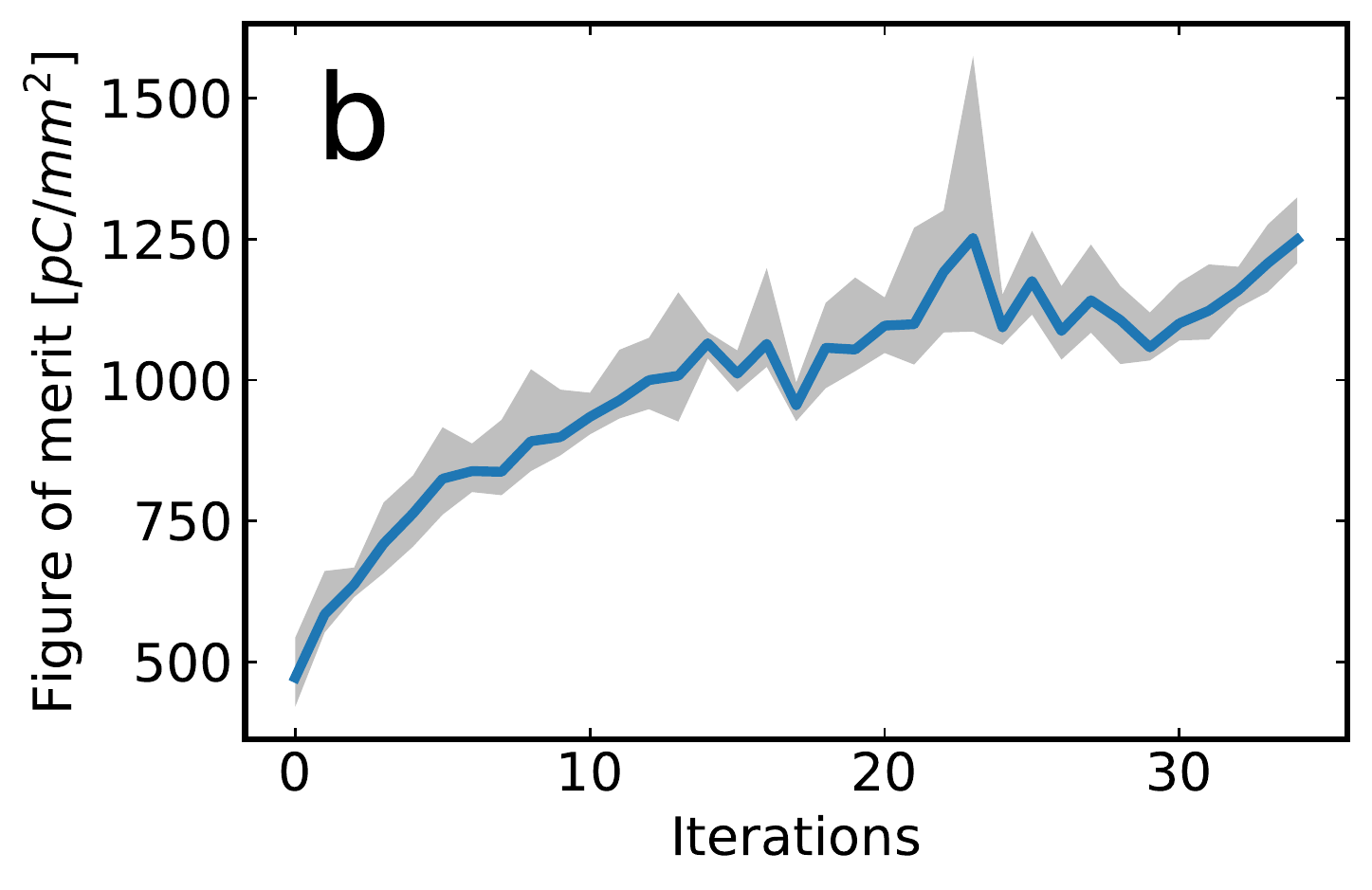}
%   \caption{}
%   \label{MIRLWFAimageMoment}
\end{subfigure}
\caption{Improvement charts using different figure of merit (FOM) functions: (a) total charge collected in the region of interest on CCD image after background subtraction; (b) the fitness function defined in Eq. (\ref{MIRLWFAimageMomentEq}) with n=2. Both optimizations started from initializing the deformable mirror to a flat surface. The shaded area refers to the variation of 5 best genes in each iteration. The number of iteration was limited by the experimental time considering the system repetition rate was as low as 1 Hz. The figure of merit values were calibrated to real units taking into account the geometry and the efficiency of the optics, the LANEX screen \cite{glinec2006absolute} and the \acs{CCD} camera.}
\label{MIRLWFAimproveChart}
\end{figure}

\begin{figure}[ht]
\centering
\includegraphics[width=.9\linewidth,height=0.4\textheight]{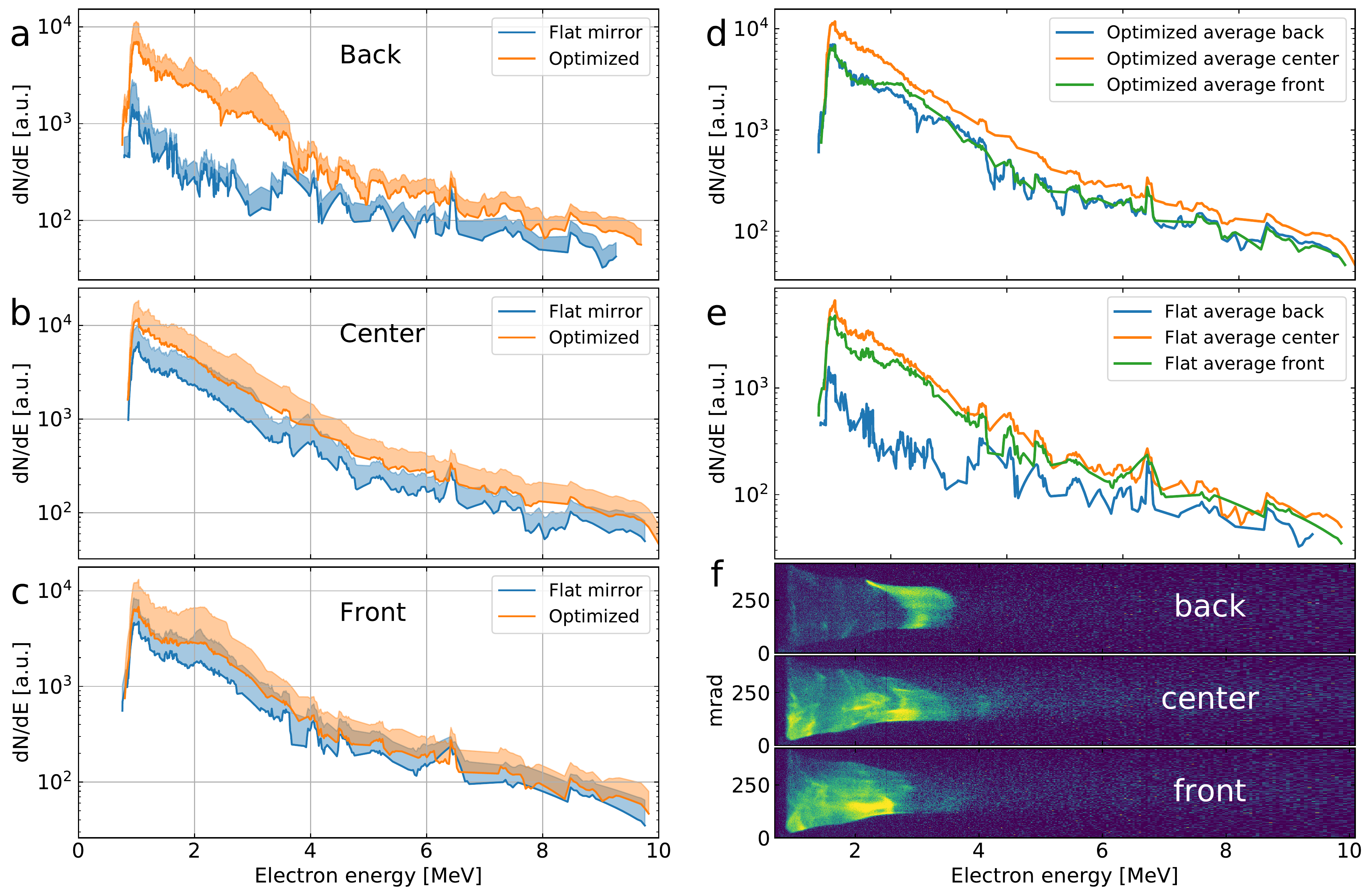}
\caption{Comparison of electron beam energy spectra: (a)-(c) before and after optimization; (d),(e) at front, center and back of the gas jet. 20 consecutive images were taken in each case. The shot-to-shot variation is shown in (a)-(c) while the statistic mean was shown in (d),(e). Examples of raw spectra with non-thermal peak features are shown in (f), in which the optimal laser wavefront found by the genetic algorithm was focused at back, center and front of the gas jet. Note that peaks do not occur on all shots.}
\label{MIRLWFAtotalCharge}
\end{figure}

Fig. \ref{MIRLWFAimproveChart}(a) illustrates the improving curves using total charge as FOM with laser focus at different positions on the Gaussian gas density profile (\acs{FWHM} $\sim 250 - 1000 \mu m$ \cite{woodbury2018laser}). Note that the low energy electrons ($<$500 keV) were filtered by the aluminum foil and the total charge collected was $\sim$450 pC. The gas jet was moved in 10-$\mu m$-step along the laser direction. A Nomarski interferometer with 515 nm probe light indicated that the plasma density was $37\%$, $35\%$ and $29\%$ of critical density at the front, center, and back side of the gas jet. 

Electron energy spectra were compared at different focusing positions before and after optimization in Fig. \ref{MIRLWFAtotalCharge}(a)-\ref{MIRLWFAtotalCharge}(e),  while examples of raw energy spectra are shown in Fig. \ref{MIRLWFAtotalCharge}(f). A high-energy bump around 3 MeV shows up in some individual shots for the back focus, and gets lower and weaker as the focus moves towards the front of the density ramp. Improvements are observed at all three focal positions and focusing at the center ($0.35n_{cr}$) gives the best energy spectrum.  Focusing on the back ($0.29n_{cr}$) gives very high beam charges, shown in Fig.\ref{MIRLWFAimproveChart}(a), but worse energy spectra. It could be caused by the electron beam missing the slit of the spectrometer, due to inferior beam collimation and pointing stability. The trend in Fig. \ref{MIRLWFAtotalCharge} indicates the existence of an optimal plasma density for electron acceleration in the mid-\acs{IR} regime, as was mentioned by Woodbury \cite{woodbury2018laser}. 

\subsubsection{Optimizing the electron beam profile}

To further improve the quality of the electron beam profile, the image moment function, as is defined in Eq. (\ref{MIRLWFAimageMomentEq}), was applied to the genetic algorithm.

\begin{equation}
\label{MIRLWFAimageMomentEq}
     FOM = \sum\limits_{(i,j)}^{ r_{ij}\neq r_0} \frac{I_{ij}}{|r_{ij}-r_0|^n}
\end{equation}

\begin{figure}[H]
\centering
\begin{subfigure}{.9\textwidth}
  \centering
  \includegraphics[width=.85\linewidth,height=0.15\textheight]{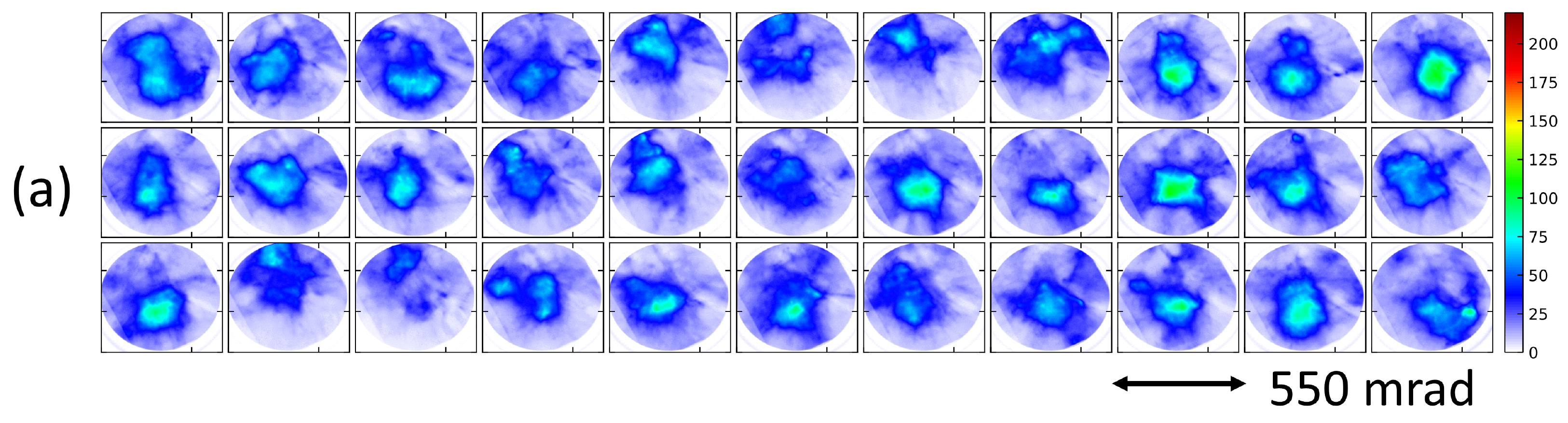}
\end{subfigure}\\
\begin{subfigure}{.9\textwidth}
  \centering
  \includegraphics[width=.85\linewidth,height=0.15\textheight]{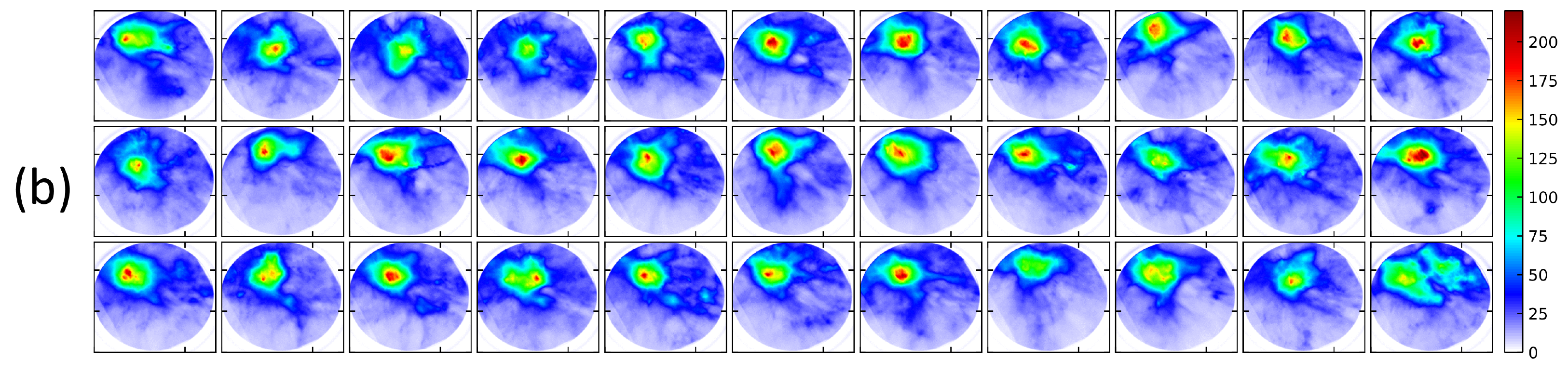}
  %\caption{}
%   \label{}
\end{subfigure}\\
\begin{subfigure}{.4\textwidth}
  \centering
  \includegraphics[width=.85\linewidth,height=0.16\textheight]{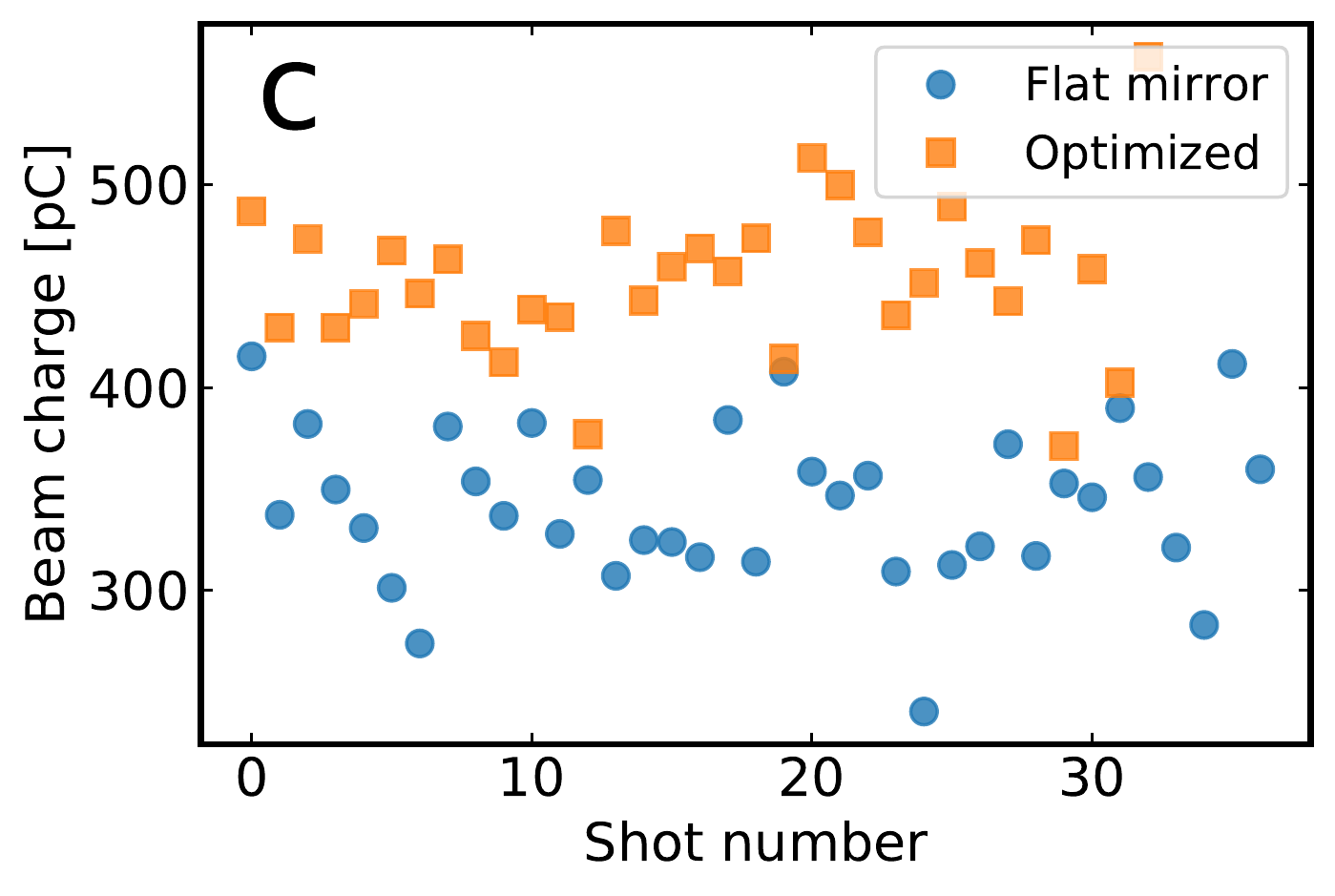}
  %\caption{}
%   \label{}
\end{subfigure}
\begin{subfigure}{.4\textwidth}
  \centering
  \includegraphics[width=.85\linewidth,height=0.16\textheight]{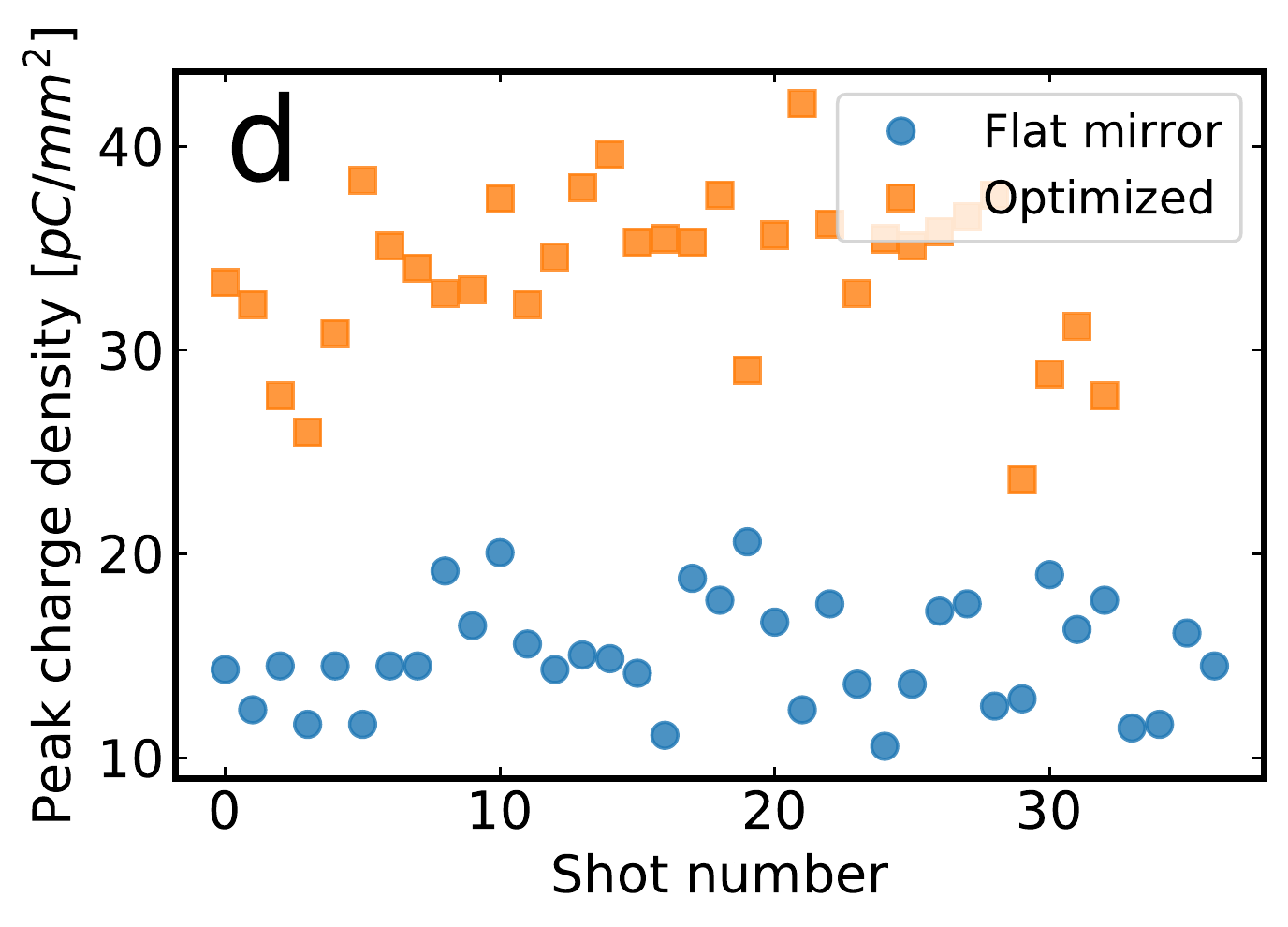}
  %\caption{}
%   \label{}
\end{subfigure}\\
\begin{subfigure}{.4\textwidth}
  \centering
  \includegraphics[width=.85\linewidth,height=0.16\textheight]{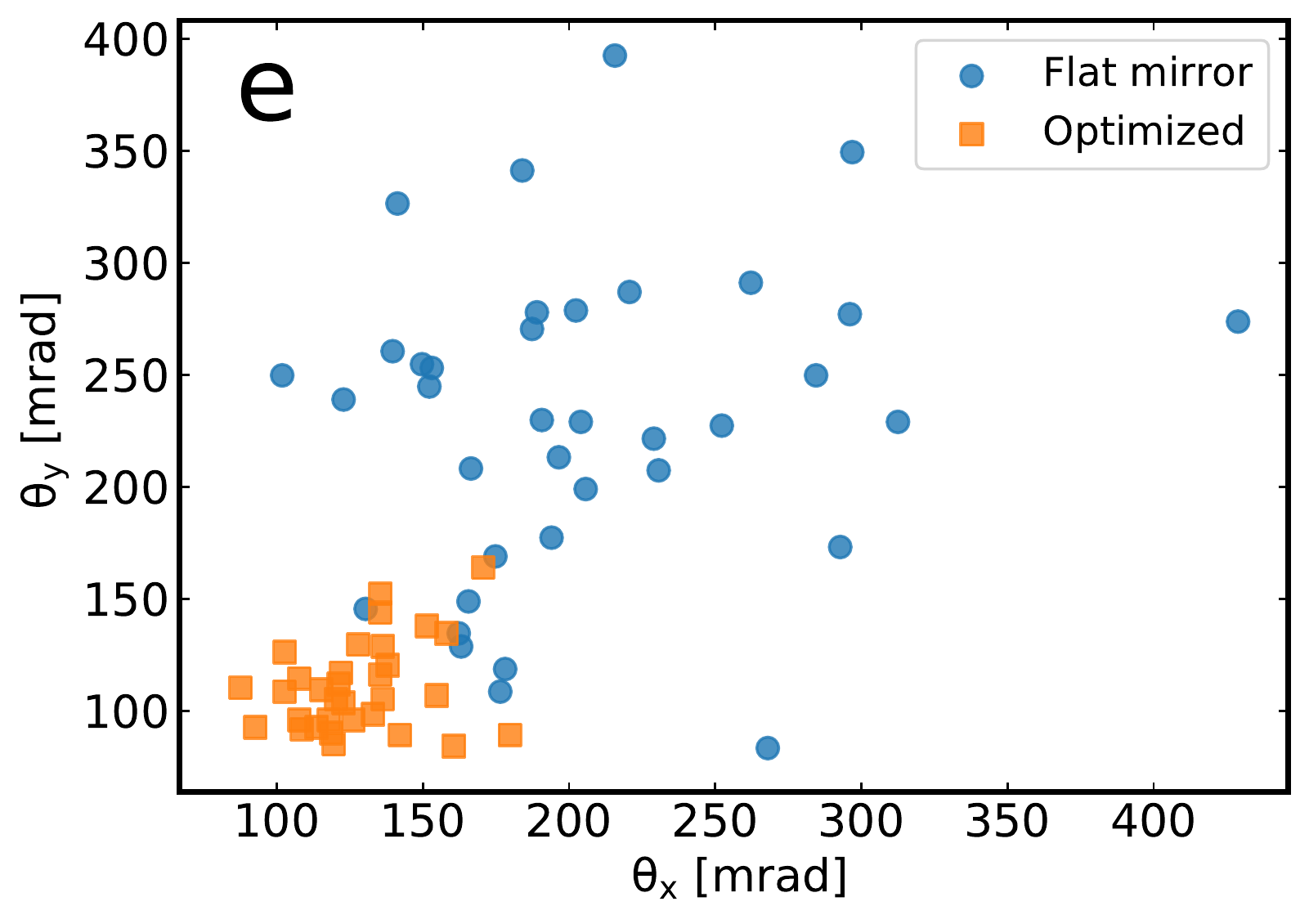}
  %\caption{}
%   \label{}
\end{subfigure}
\begin{subfigure}{.4\textwidth}
  \centering
  \includegraphics[width=.9\linewidth,height=0.16\textheight]{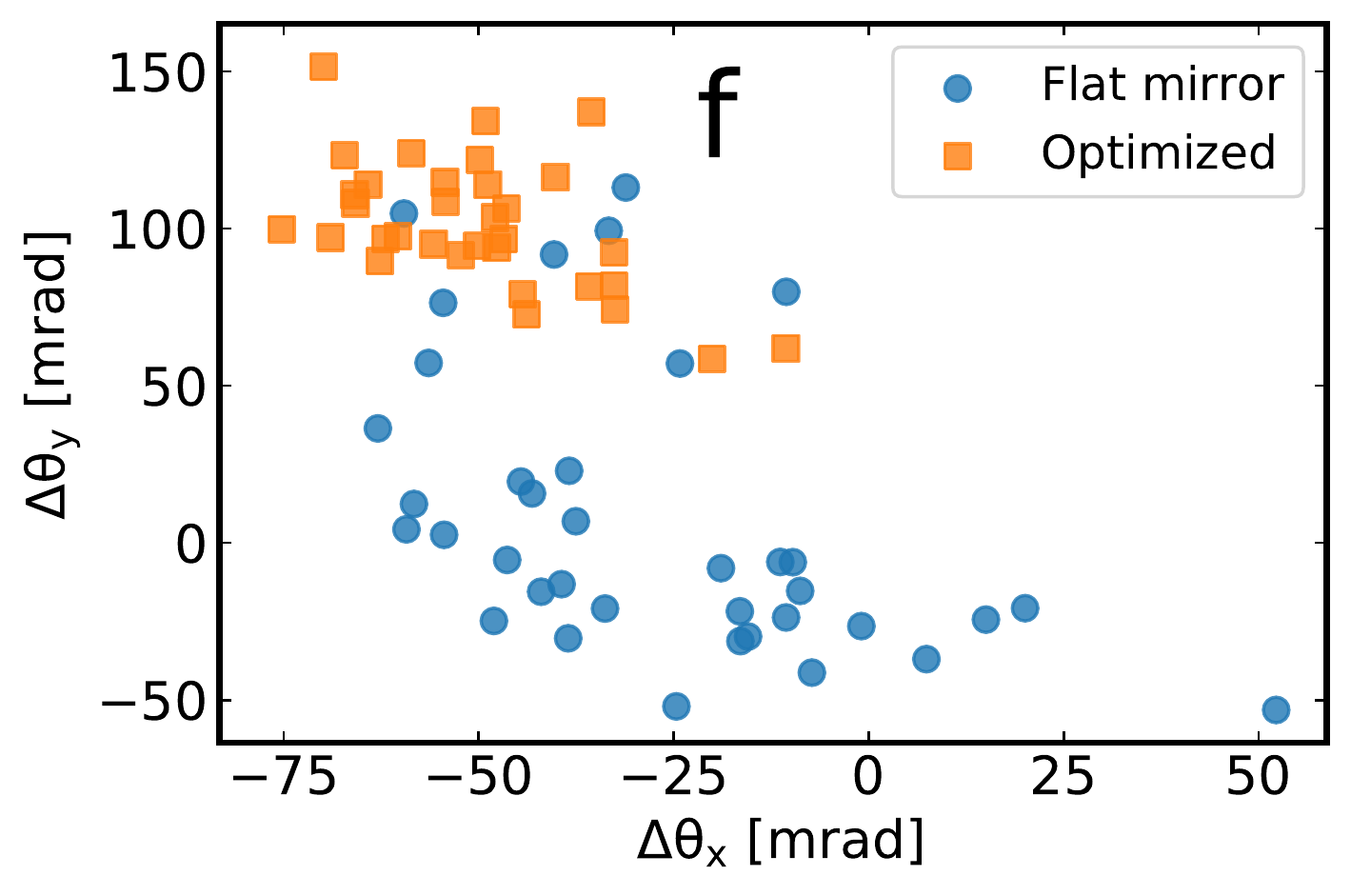}
  %\caption{}
%   \label{}
\end{subfigure}\\
\caption{Electron beam profile optimization using the image moment as figure of merit, defined in Eq. (\ref{MIRLWFAimageMomentEq}). (a) and (b): 30 consecutive raw images before and after optimization. The circular edge, due to a collimation tube in front of the LANEX, corresponds to a solid angle of 550 mrad. (c) - (f) are the visualization of beam quality in terms of total beam charge, peak charge density, divergence angle and beam pointing, respectively. Each dot represents one shot.}
\label{MIRLWFAbeamProfile}
\end{figure}

Where $I_{ij}$ is the intensity collected at pixel position $r_{ij}$ on the camera. The beam center $r_{0}$ is determined in each shot from the center of mass calculation after background subtraction. It quantifies not only the total charge but also beam collimation and pointing. The improvement chart is shown in Fig. \ref{MIRLWFAimproveChart}(b) and the optimization performance is shown in Fig. \ref{MIRLWFAbeamProfile}. The genetic algorithm started from initializing the DM to a flat surface. It is observed that the raw images in Fig. \ref{MIRLWFAbeamProfile}(a) are divergent and have significant pointing fluctuations while the ones in Fig. \ref{MIRLWFAbeamProfile}(b) are collimated and directional. Detailed analysis can be found in Fig. \ref{MIRLWFAbeamProfile}(c)-\ref{MIRLWFAbeamProfile}(f). Averaging over 30 shots, the optimization was able to increase the total beam charge by $\sim40\%$ and the peak charge density to threefold in 35 iterations. In Fig. \ref{MIRLWFAbeamProfile}(e) the beam divergence $\theta_x$ and $\theta_y$ in transverse directions were reduced from $206\pm64$ mrad and $228\pm69$ mrad to $128\pm21$ mrad and $110\pm20$ mrad, respectively. The pointing instabilities, defined as the standard deviations of beam pointing $\delta\theta_x$ and $\delta\theta_y$ in Fig. \ref{MIRLWFAbeamProfile}(f), were reduced from 25.2 mrad and 45.7 mrad to 14.5 mrad and 20.6 mrad.

\subsubsection{Wavefront reconstruction}

The laser wavefront was measured $ex\;situ$ by applying the recorded voltage distribution on the deformable mirror and subsequently measuring the wavefront with visible light. An imaging system involving the deformable mirror and a Shack-Hartman wavefront sensor was set up using a helium-neon laser after the experiment. The FrontSurfer wavefront analyzer (Version 1.4.7, OKO Technologies), consisting of a high-precision lenslet array and a CMOS UI-2210M CCD camera, can describe the wavefront in Zernike polynomials up to $200^{th}$ order: 
\begin{equation}
\label{MIRLWFAZernikes}
     \Delta \phi = \sum\limits_{j=1}^{200} A_j Z_j
\end{equation}
where $A_j$ and $Z_j$ represent the $j^{th}$ coefficient and base of the Zernike polynomials, respectively. Knowing the voltages on the deformable mirror, the coefficients could be obtained from the influence matrix $C_{ij}$:
\begin{equation}
\label{MIRLWFAvoltage}
     A_j = \sum\limits_{i=1}^{37} C_{ij} V_i
\end{equation}

\begin{figure}[ht]
\centering
\begin{subfigure}{0.3\textwidth}
  \centering
  \includegraphics[width=.95\linewidth]{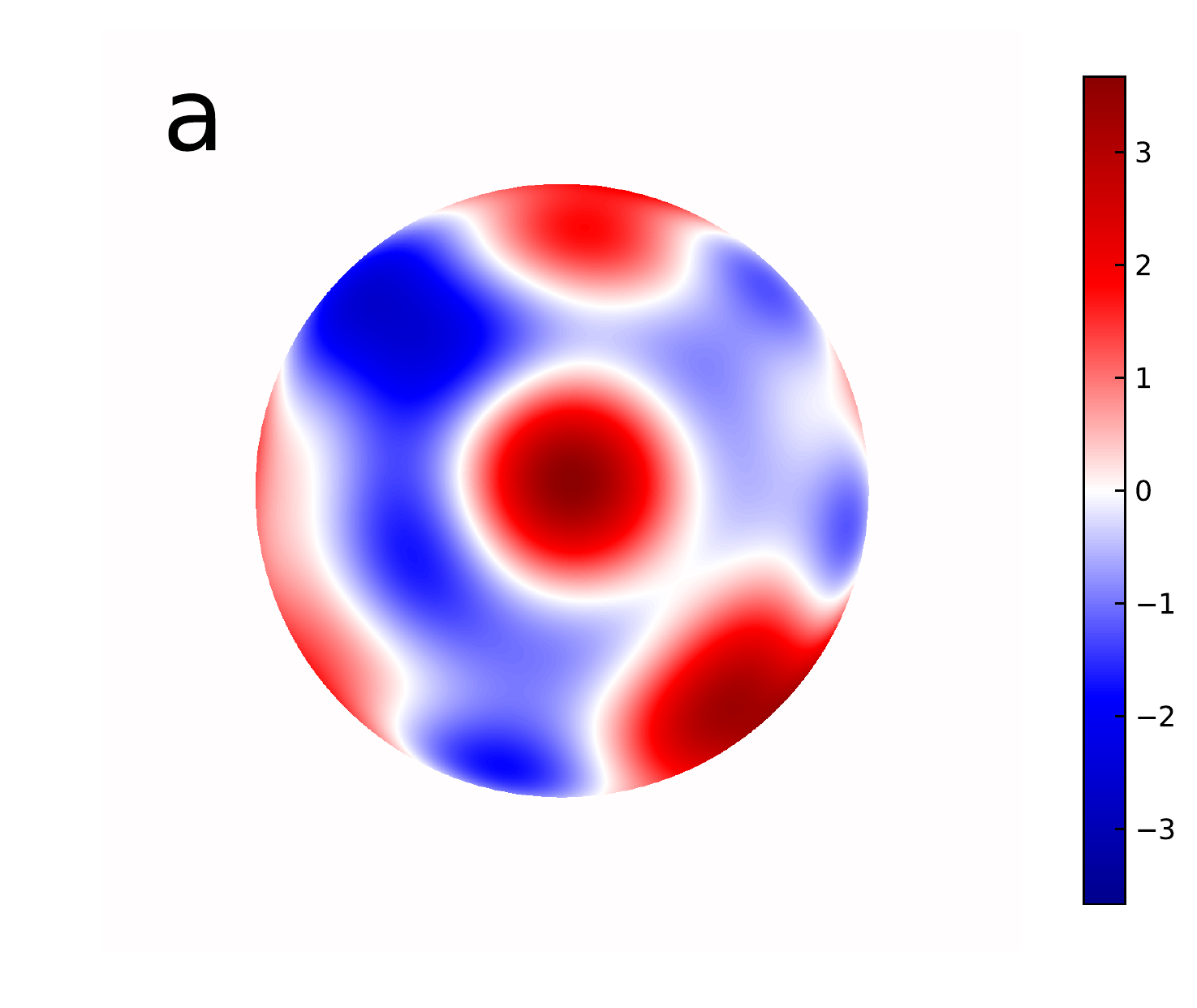}
  %\caption{}
%   \label{centerDM}
\end{subfigure}
\begin{subfigure}{0.3\textwidth}
  \centering
  \includegraphics[width=.95\linewidth]{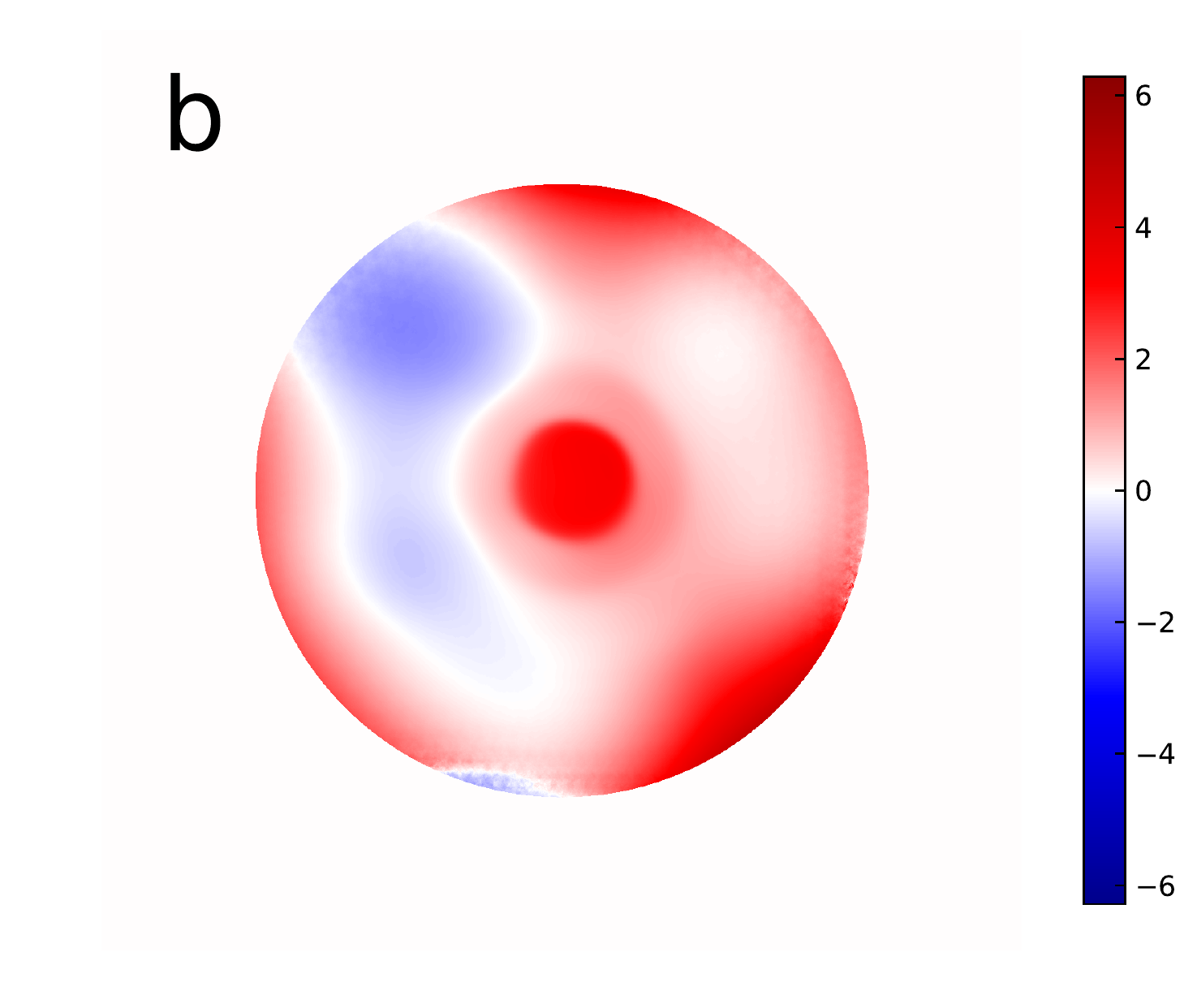}
  %\caption{}
%   \label{centerProp}
\end{subfigure}
\begin{subfigure}{0.3\textwidth}
  \centering
  \includegraphics[width=.95\linewidth]{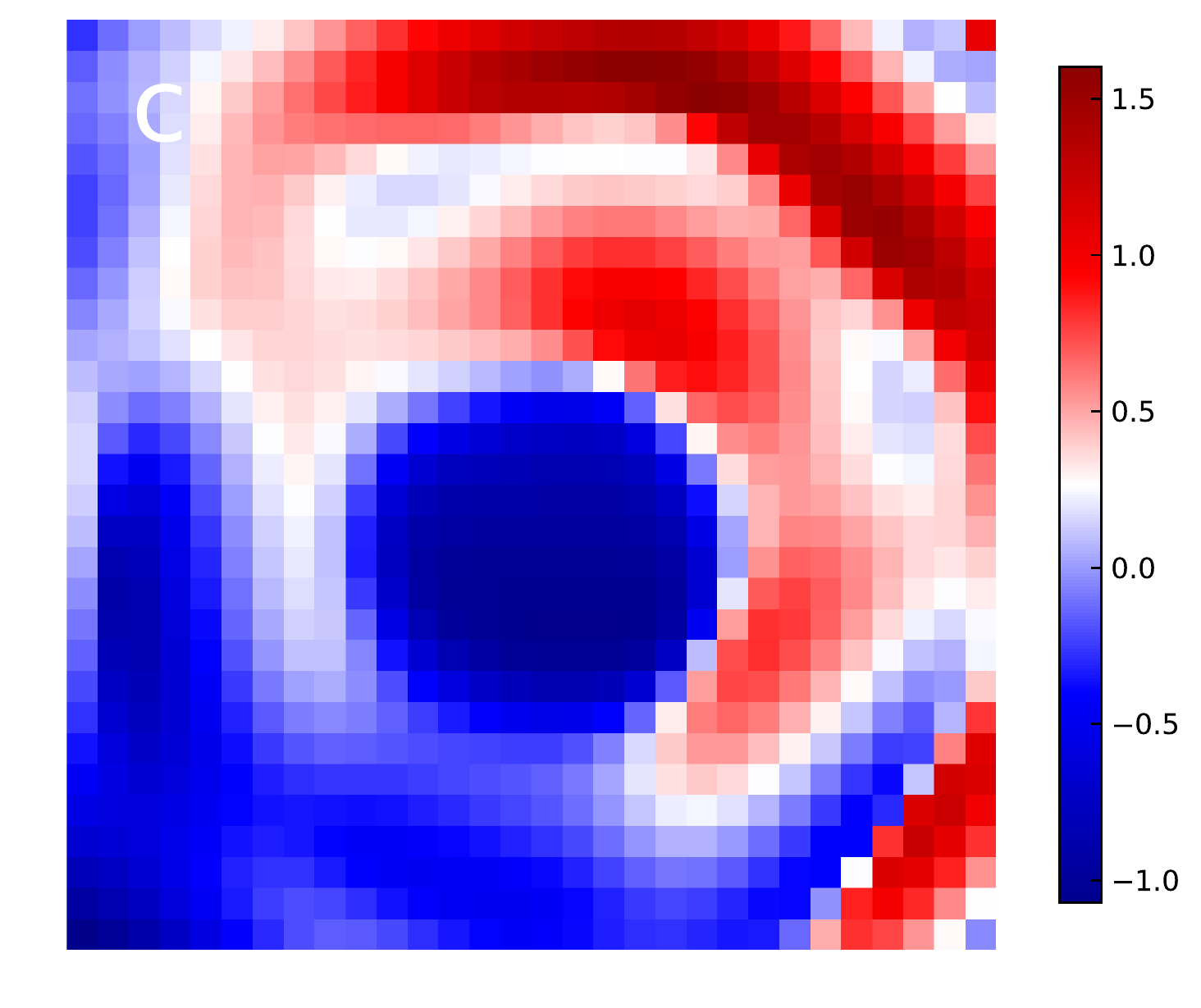}
  %\caption{}
%   \label{center500}
\end{subfigure}\\
\begin{subfigure}{0.3\textwidth}
  \centering
  \includegraphics[width=.95\linewidth]{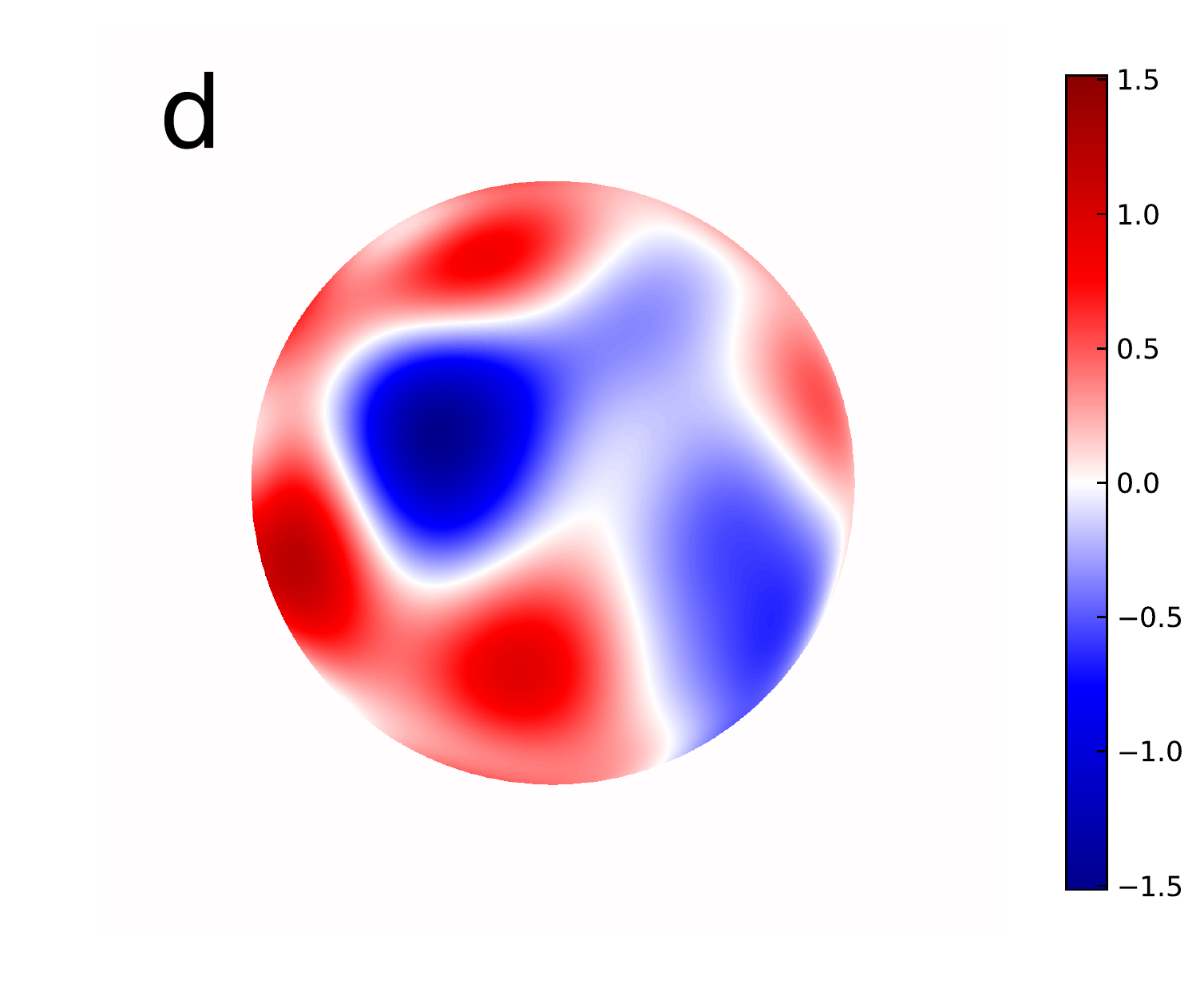}
  %\caption{}
%   \label{backDM}
\end{subfigure}
\begin{subfigure}{0.3\textwidth}
  \centering
  \includegraphics[width=.95\linewidth]{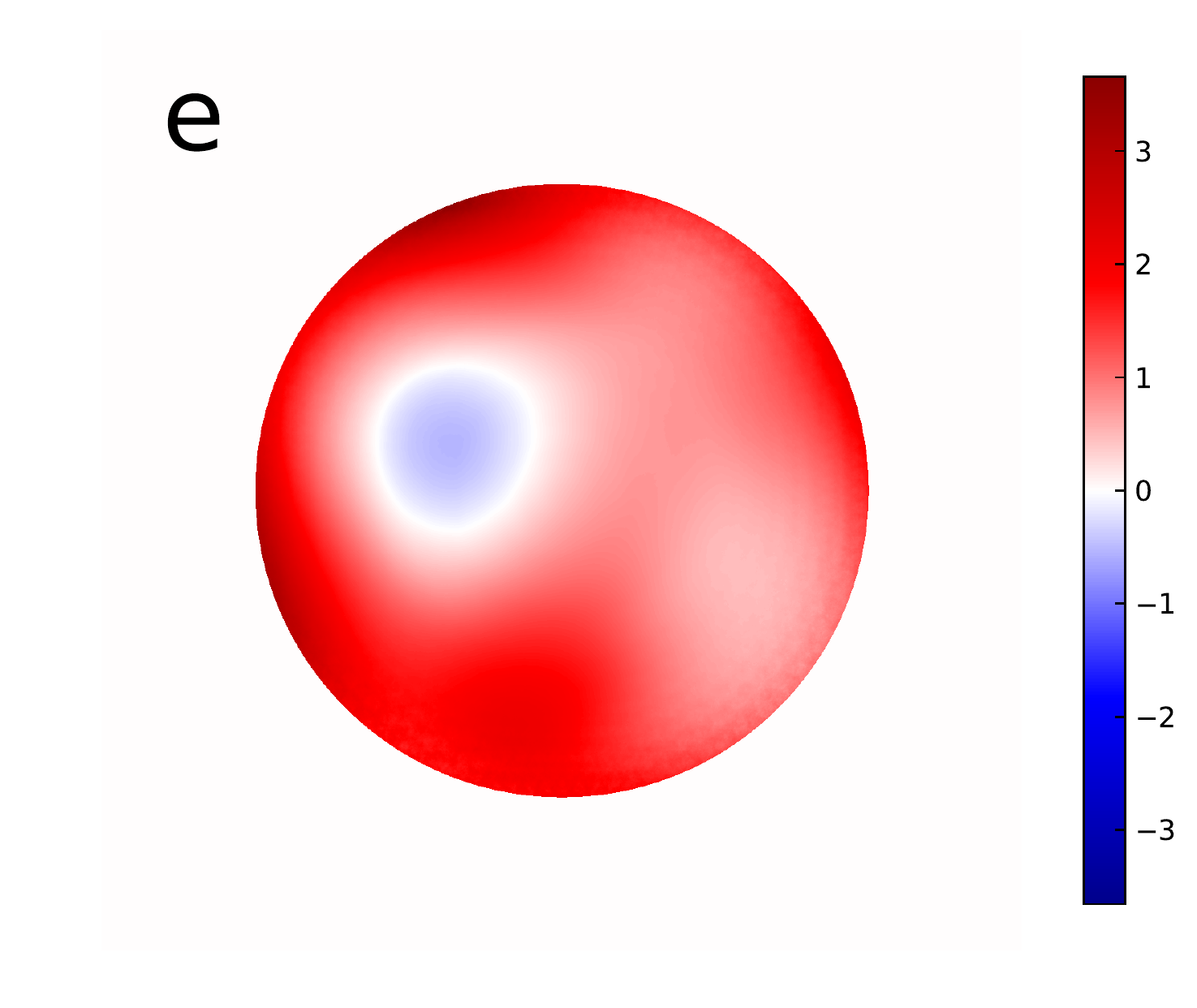}
  %\caption{}
%   \label{backProp}
\end{subfigure}
\begin{subfigure}{0.3\textwidth}
  \centering
  \includegraphics[width=.95\linewidth]{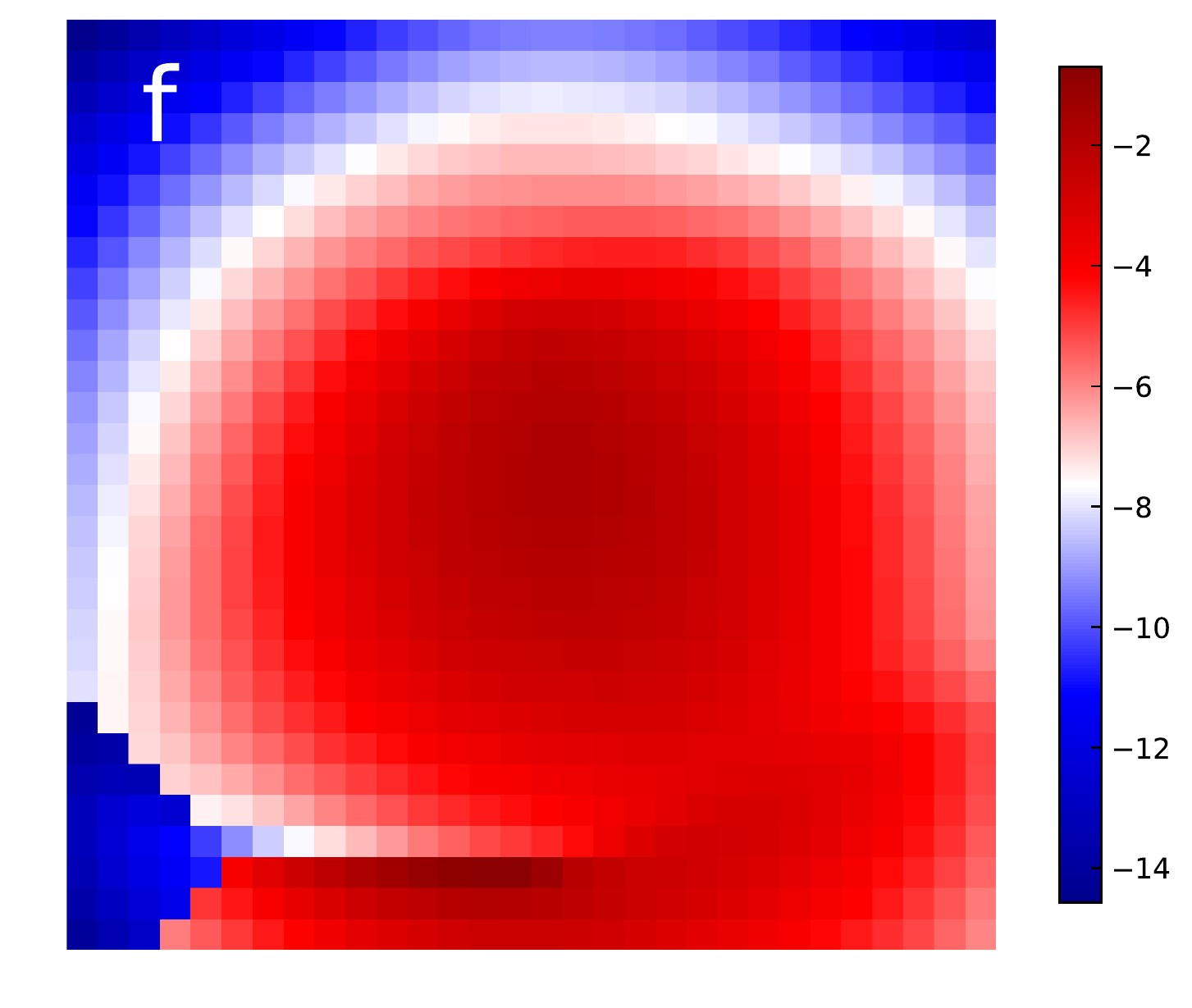}
  %\caption{}
%   \label{back500}
\end{subfigure}\\
\begin{subfigure}{0.3\textwidth}
  \centering
  \includegraphics[width=.95\linewidth]{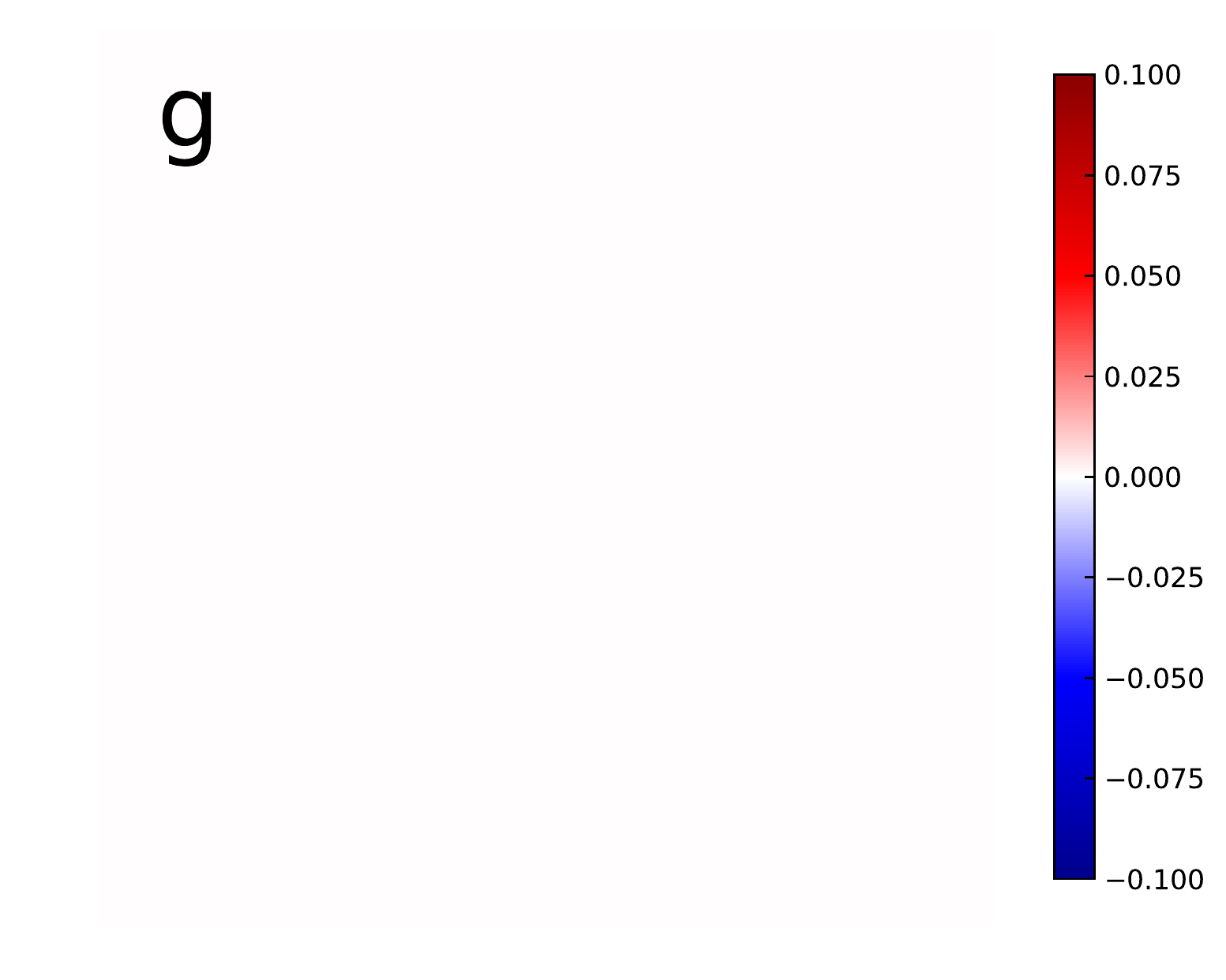}
  %\caption{}
%   \label{frontDM}
\end{subfigure}
\begin{subfigure}{0.3\textwidth}
  \centering
  \includegraphics[width=.95\linewidth]{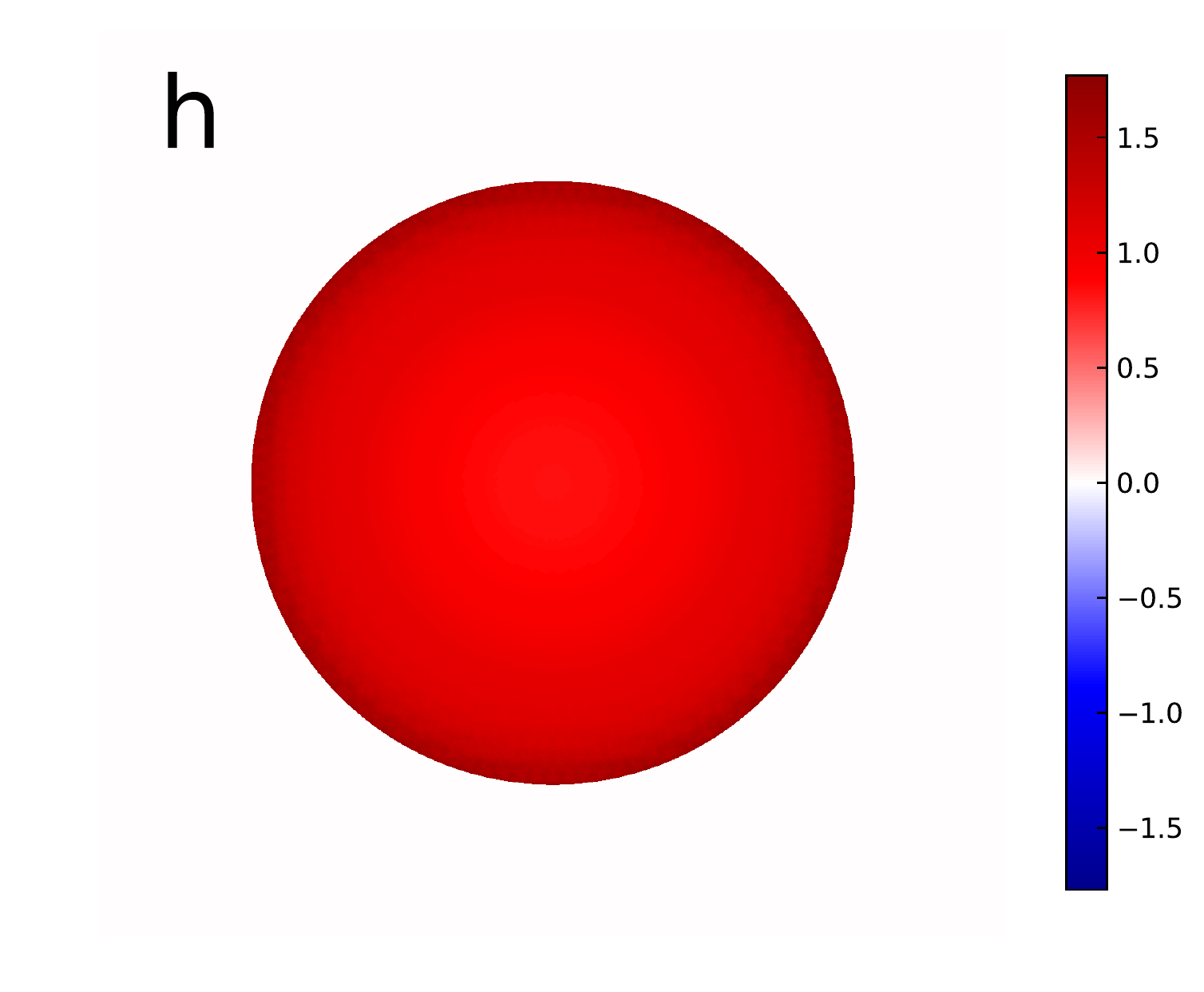}
  %\caption{}
%   \label{frontProp}
\end{subfigure}
\begin{subfigure}{0.3\textwidth}
  \centering
  \includegraphics[width=.95\linewidth]{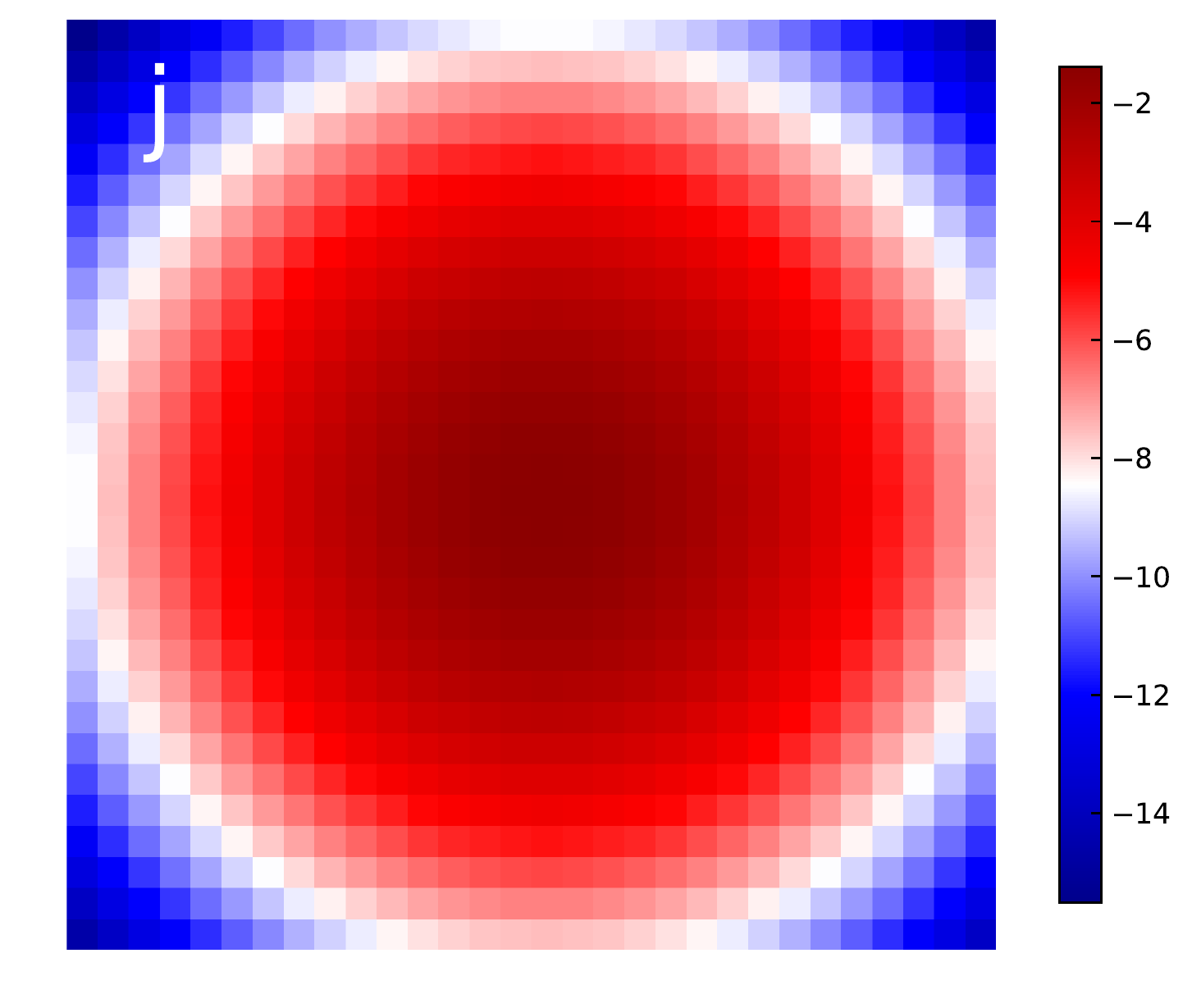}
  %\caption{}
%   \label{front500}
\end{subfigure}\\
\caption{Reconstruct the laser wavefront propagation for three cases. (a-c): wavefront for the optimal electron beam,  (d-f): wavefront from a flat mirror surface, and (g-i): perfect Gaussian wavefront. Wavefronts leaving the DM, propagating 4.5m and focused by the OAP are shown in the first, second and third column. First 50 Zernike coefficients are included in the reconstruction.
%\yongm{[Comment: can we use a circular mask for h as well?]}
}
\label{MIRLWFAwavefront}
\end{figure}

The influence matrix method is useful for real-time analysis of the wavefront, but here only the optimized wavefront is to be analyzed. Instead of measuring each element in the influence matrix, an alternative way to restore the wavefront change due to the DM ($\phi_{DM}$) is to directly apply the voltage recorded in the experiment and measure the mirror surface. A reference mirror surface with a known wavefront would be necessary to reconstruct the 3.9$\mu m$ laser wavefront. Analogous to the setup in Fig. \ref{MIRLWFAsetUp}, the laser beam was attenuated and focused onto an AGS crystal to generate second harmonic (SH) signal. The genetic algorithm was run to improve the SH signal to threefold till convergence, which suggested the highest peak intensity available \cite{albert2000generation}. It corresponds to the smallest focal spot and the flattest wavefront available. A knife-edge scan showed the focal spot was decreased from $\sim25\mu m$ to $\sim15\mu m$ after the second harmonic optimization. This voltage map on the DM was recorded and afterwards applied to the wavefront analysis system with visible light to obtain the phase $\phi_{SHG}$. Assuming the laser wavefront going into the DM was $\phi_{laser}$, the wavefront after the SHG optimization would be almost Gaussian: 
\begin{equation}
\label{MIRLWFAshg}
     \phi = \phi_{laser} + 2\times \phi_{SHG} \simeq \phi_{Gaussian}
\end{equation}

On the other hand when the DM was set to optimize the electron beam, the :
\begin{equation}
\label{MIRLWFAopt}
     \phi_{opt} = \phi_{laser} + 2\times \phi_{DM} 
\end{equation}

Subtracting Eq. (\ref{MIRLWFAshg}) from Eq. (\ref{MIRLWFAopt}) would give the laser wavefront leaving the deformable mirror during the experiment, $2\times (\phi_{DM} - \phi_{SHG})$, as is shown in Fig. \ref{MIRLWFAwavefront}(a). Fresnel diffraction was taken into account to propagate the wavefront 4.5 meters to the OAP, as is shown in Fig. \ref{MIRLWFAwavefront}(b), and the phase change was calculated with LightPipes \cite{LightPipes} using direct integration approach. Fig. \ref{MIRLWFAwavefront}(c) shows the wavefront at 500 $\mu m$ before the geometric focus, which was used in the \acs{PIC} simulation. Fig. \ref{MIRLWFAwavefront}(d)-\ref{MIRLWFAwavefront}(f) present the laser wavefront before the evolutionary algorithm was run where the DM was initialized to a flat surface. The propagation of a perfect Gaussian beam is included in Fig. \ref{MIRLWFAwavefront}(g)-\ref{MIRLWFAwavefront}(i) for comparison. Note that the validity of this whole reconstruction process is dependent on a list of factors, including the stability of the voltage on the DM actuators, the accuracy of measurement using the visible wavefront sensor, and mostly the flatness of the wavefront after the second harmonic optimization, or the validity of Eq. (\ref{MIRLWFAshg}).

\subsubsection{Particle-in-cell simulations}

\begin{figure}[ht]
\centering
\includegraphics[width=0.85\columnwidth,height=0.4\textheight]{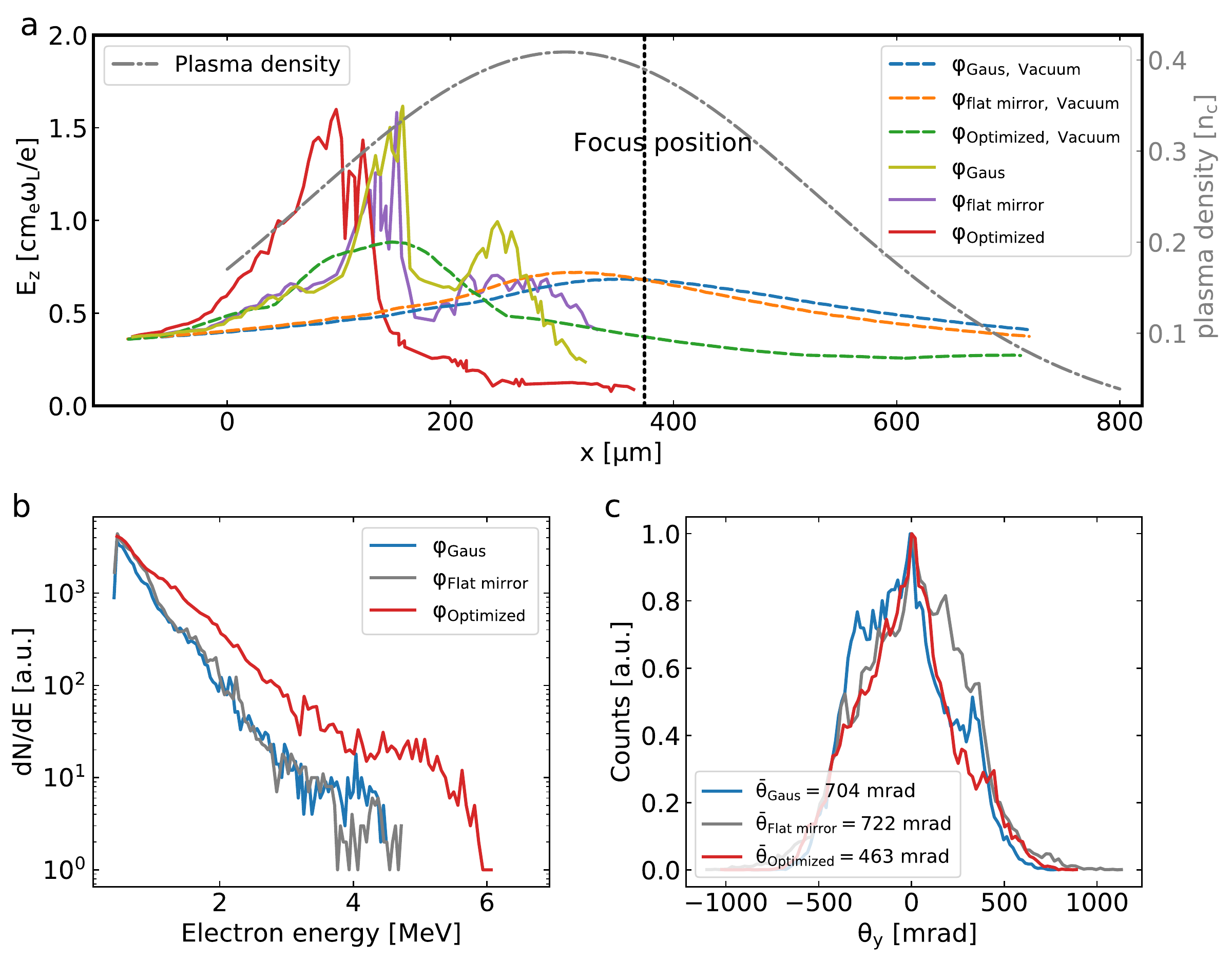}
\caption{Laser field evolution and electron beam qualities with different laser wavefronts in \acs{PIC} simulations. (a) Evolution of peak laser field strength with different wavefronts in vacuum and plasma, respectively. (b) and (c) Electron spectra and angular distributions at the end of the simulation ($t = 2$ ps) with different laser wavefronts and the same plasma profile showing in (a).
}
\label{MIRLWFAsimWavefront}
\end{figure}

The effect of wavefront changes on wakefield acceleration was further investigated with two-dimensional \acs{PIC} simulations in the EPOCH framework \cite{arber2015contemporary}.
The simulation box with moving window is 200 $\mu \mathrm{m} \times$ 160 $\mu$m with grid size of $1/32$ and $1/16~\lambda_L$ in $x$ and $y$, where  $\lambda_L = 3.9~\mu$m is the laser wavelength, $x$ is the laser propagation direction, $y$ is the transverse direction and $z$ is the laser polarization direction. There are 64 macro-particles per cell. The laser pulse is Gaussian in both transverse and longitudinal directions with a \acs{FWHM} pulse duration $\tau = 100$ fs, a $1/e^2$ spot size $w_0 = 13 \mu$m and a normalized vector potential $a_0 = 0.7$. The plasma density distribution along the laser propagation direction was fitted from  interferometric measurements which indicate a Gaussian distribution with a peak density of $0.4~n_c$ and a \acs{FWHM} of 505 $\mu$m, as shown in Fig. \ref{MIRLWFAsimWavefront}(a). The focus position of the laser pulse was initially set at 374 $\mu$m, which corresponds to the ``Center'' case, for a perfect Gaussian laser beam. In \acs{PIC} simulations, we compared three different cases, namely ``Optimized'', ``Flat mirror'' and ``Gaussian'', with wavefronts shown in Fig. \ref{MIRLWFAwavefront}(c), \ref{MIRLWFAwavefront}(f) and \ref{MIRLWFAwavefront}(j), respectively. 

\begin{figure}[H]
\centering
\includegraphics[width=0.55\columnwidth,height=0.32\textheight]{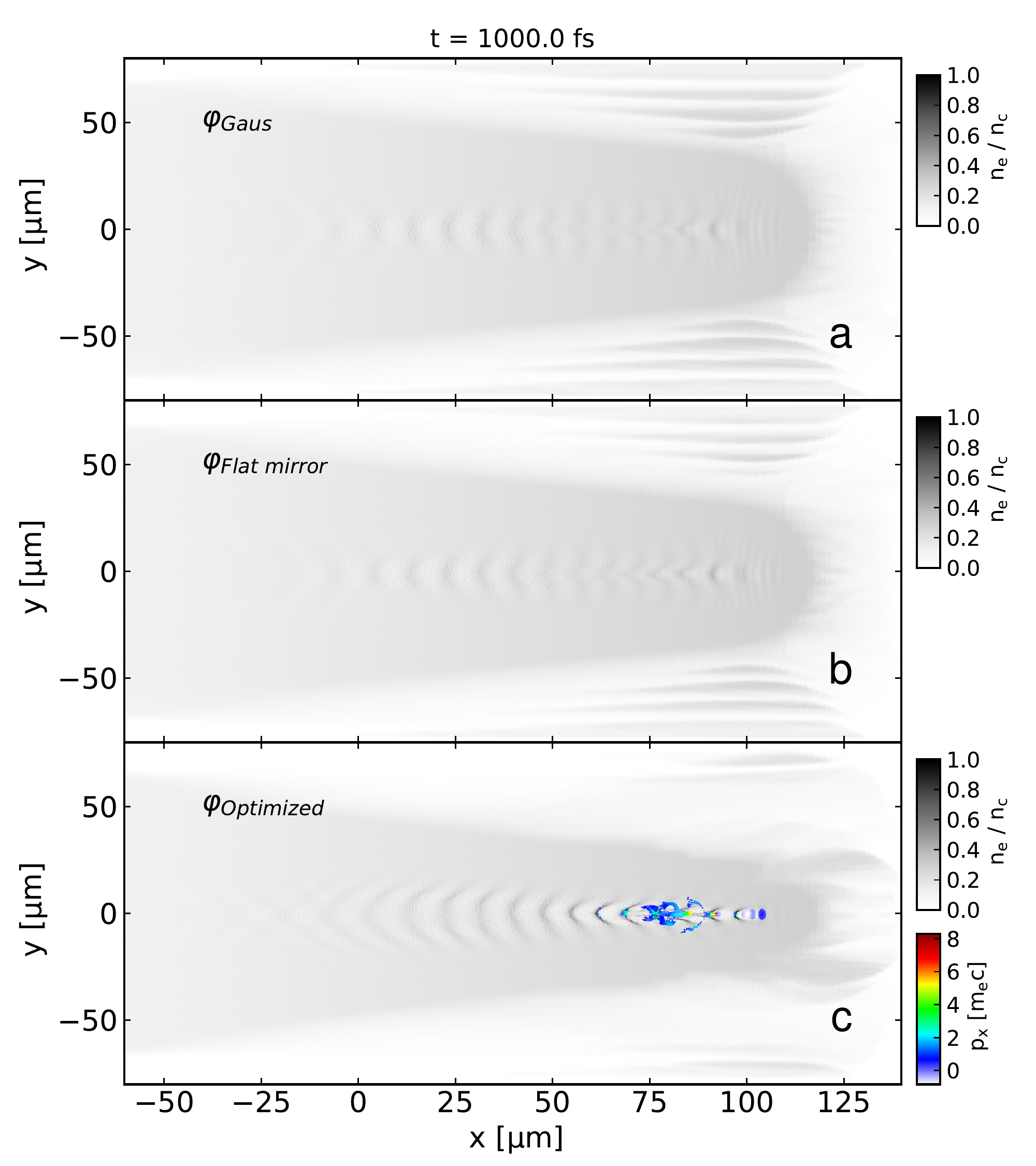}
\caption{
Snapshot for plasma density and electron beam distribution with (a) Gaussian, (b) flat mirror and (c) optimized wavefront at the same time, $t = 1$ ps. Self-injection has occurred with optimized wavefront in (c) while not in the other two cases.
}
\label{MIRLWFAsim_1ps}
\end{figure}

\begin{figure}[H]
\centering
\includegraphics[width=0.85\columnwidth, height=0.6\textheight]{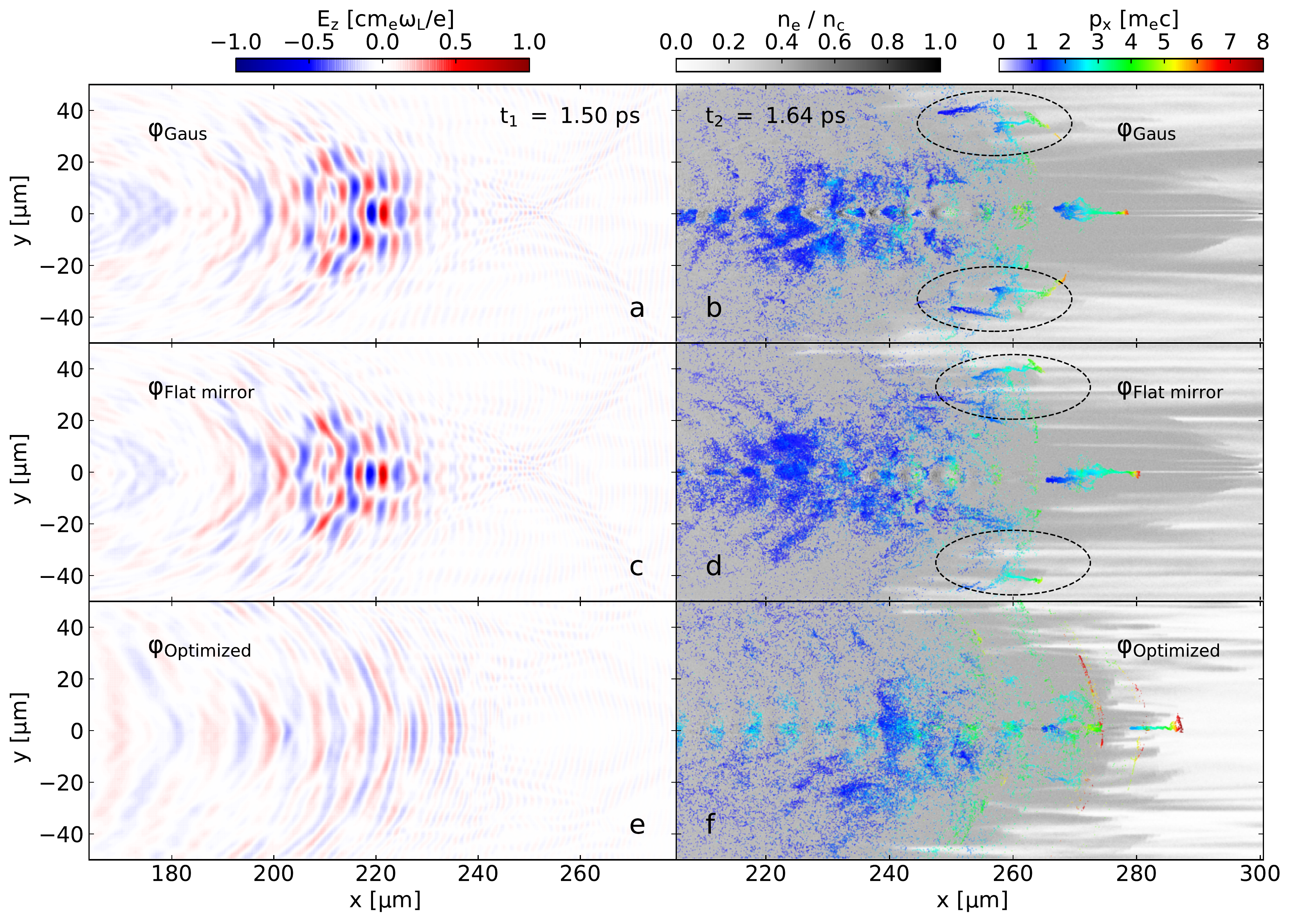}
\caption{
Snapshots of \acs{PIC} simulations with different wavefronts.
Laser field distribution at $t = 1.5$ ps for (a) Gaussian, (c) Flat mirror and (e) Optimized wavefront, respectively. Plasma density distribution at $t = 1.64$ ps for (b) Gaussian, (d) Flat mirror and (f) Optimized wavefront, respectively. Spatial distribution of accelerated electron macro-particles are overlaid on the plasma density distribution where color scale represents longitudinal momentum of the particles. 
}
\label{MIRLWFAfilament}
\end{figure}

The propagation of laser pulses with different wavefronts in both vacuum and plasma were examined. As shown in Fig. \ref{MIRLWFAsimWavefront}(a), the peak laser field strength reaches its maximum much earlier in the optimized wavefront case than in the other cases. The \acs{LWFA} process starts as the laser field reaches its maximum during the self-focusing, and almost the whole acceleration happens within the density up-ramp region. The laser pulse with optimized wavefront initiates the acceleration earlier, as is shown in Fig.~\ref{MIRLWFAsim_1ps}, and thus experiences a lower plasma density. Since the maximum energy gain in \acs{LWFA} \cite{lu2007generating} scales as $\Delta E_{max} \propto n_p^{-2/3}$, the relative lower plasma density for the ``Optimized'' case would result in higher final energy gain. This has been confirmed by the electron spectra at the end of the simulations, as shown in Fig.~\ref{MIRLWFAsimWavefront}(b). The electron spectra from \acs{PIC} simulations agree with the experimental results in Fig. \ref{MIRLWFAtotalCharge} qualitatively. 

Moreover, the laser pulses in the ``Gaussian'' and ``Flat mirror'' cases suffer more from the transverse self-modulation, leading to the self-filamentation shown in Fig.~\ref{MIRLWFAfilament}(a) and (c). These filaments are intense enough to drive \acs{LWFA} on their own, which eventually lead to the wing structure of the electron beam in Fig.~\ref{MIRLWFAfilament}(b) and (d). Consequenctly, the electron beam collimation in these two cases is worse than that in the ``Optimized'' case, as is presented in Fig.~\ref{MIRLWFAsimWavefront}(c), which also agrees with the experimental results.

\subsection{Discussion}

Understanding the phase front condition of the laser to the plasma is crucial to the success of wakefield acceleration. The non-Gaussian features of laser wavefront in experiments can strongly affect the acceleration mechanism and betatron sources \cite{ferri2016effect}. It is, therefore, of great interest to control the phase front in \acs{LWFA}. Optical steering of the electron beam direction \cite{popp2010all}, enhancement of betatron radiation \cite{yu2018enhanced} and spectral control of the x-rays produced in the process \cite{mangles2009controlling} have been achieved by modifying the laser wavefront. Here we have demonstrated the ability to coherently control the relativistic electron beam from wakefield acceleration by mid-\acs{IR} laser pulses in near-critical density plasma. Electron total charge, energy spectrum, beam pointing and fluctuation are improved and the effect of wavefront changes on the acceleration process are studied with \acs{PIC} simulation. The optimal wavefront initiates the acceleration earlier on the density up-ramp and thus experiences a lower plasma density, which leads to higher energy gain during the interaction. It also sees less filamentation from the transverse self-modulation, which would be responsible for the wing structure and divergence of the electron beam. With this improved wavefront, better electron beam collimation and energy spectra are observed in both experiment and simulation. The computer modeling is based on wavefront reconstruction using the voltage applied to the deformable mirror, in the absence of a mid-\acs{IR} wavefront sensor. 

Improvement in electron beam quality is independent of the improvement in laser focus since the highest intensity laser focus produced an order of magnitude lower electron charge. This behavior, together with the intensity wings from optimized wavefront, has been observed in previous work \cite{he2015coherent} with $\lambda=800nm$ as well. Analogously, looking at the x-rays producing by the wakefield acceleration, a wavefront with coma aberration generates more high-energy photons than a flat wavefront \cite{mangles2009controlling} does. Modifying the phase of the light can cause strong optical nonlinear effects in the plasma interactions, which can affect the plasma wave dynamics in a complex but deterministic manner.

This work opens a new window to the study of coherent control of relativistic mid-\acs{IR} laser-plasma interactions. It is worth noting that the full stroke of the deformable mirror surface is 4 $\mu m$, or a wavelength of the mid-\acs{IR} driver. Namely, without upgrading the DMs to deeper stroke or the wavefront sensors to a longer wavelength range, current adaptive optical systems are capable of conducting experiments using mid-\acs{IR} lasers. Our work shows the potential for the use of long-wavelength lasers in \acs{LWFA} in near-critical density plasma which would be difficult to achieve using near-\acs{IR} lasers. Recently, pulse shaping implemented into the system algorithm \cite{streeter2018temporal} has been validated using near-\acs{IR} lasers and can be extended to the mid-\acs{IR} critical-density regime. 
% Future work could also include a theoretical validation of the wavefront reconstruction approach by means of the \acf{QOCT} \cite{werschnik2007quantum}.

\clearpage
\section{Beyond optimization - supervised learning applications in a laser-wakefield accelerator}
\label{sec:MLLWFA}
% \begin{abstract}
% We explore the applications of a variety of machine learning techniques in relativistic laser-plasma experiments beyond optimization purposes. With the trained supervised learning models, the beam charge of electrons produced in a laser wakefield accelerator is predicted given the laser wavefront change caused by a deformable mirror. Feature importance analysis using the trained models shows that specific aberrations in the laser wavefront are favored in generating higher beam charges, which reveals more information than the genetic algorithms and the statistical correlation do. The predictive models enable operations beyond merely searching for an optimal beam charge. The quality of the measured data is characterized and anomaly detection is demonstrated. The model robustness against measurement errors is examined by applying a range of virtual measurement error bars to the experimental data. This work demonstrates a route to machine learning applications in the highly nonlinear problem of relativistic laser-plasma interaction for in-depth data analysis to assist physics interpretation.
% \end{abstract}

% \maketitle

\subsection{Introduction}
High-repetition-rate laser systems have been widely used with evolutionary algorithms to solve optimization problems in the field of relativistic laser-plasma interactions, including laser wakefield acceleration \cite{he2015coherent, dann2019laser, lin2019adaptive}, ion acceleration \cite{smith2020optimizing, noaman2018controlling}, x-ray production \cite{streeter2018temporal}, terahertz generation \cite{hah2017enhancement}, laser filamentation \cite{englesbe2016control, lefebvre2018phase, finney2021filament}, and laser focus optimization \cite{nayuki2005production, lin2018focus}. A detailed review of high-repetition-rate laser-plasma experiments has been given in Sec. \ref{sec:IntroHigh}. However, evolutionary algorithms usually provide little information other than a local optimum, which can be difficult to interpret. Instead, machine learning (ML) methods can generate predictive models that reveal more information in the dataset to help understand the physical processes. 

The broader discipline of plasma physics has adopted various machine learning methods in recent years. For instance, supervised learning regression algorithms have been applied to \acf{ICF} experiments with growing interests, such as Deep Jointly-Informed Neural Networks \cite{gaffney2019making, hsu2020analysis, humbird2019parameter, maris2019finding} and Gaussian Process regressor \cite{hatfield2019using}. Another popular machine learning technique called Random Forest has found success in magnetic confinement fusion experiments for both classification and regression problems \cite{rea2018disruption, piccione2020physics}. In space physics, Gaussian Processes are used to classify solar wind plasmas into categories \cite{camporeale2017classification},  and deep neural networks are used to predict solar flares from sunspot data \cite{jiao2020solar, chen2019identifying}. Beyond supervised learning, plasma physicists have utilized in other powerful and increasingly popular machine learning methods, such as transfer learning \cite{humbird2019transfer} and reinforcement learning \cite{witman2019sim, kain2020sample}. The laser-plasma community is starting to embrace machine learning techniques as well. Artificial neural networks are employed to analyze features in high-order-harmonic spectra \cite{gonoskov2019employing} and laser-induced-breakdown spectra \cite{li2020laser}. Our work explores the capability of machine learning techniques in the field of laser-wakefield acceleration using all the supervised learning methods mentioned above.

\acf{LWFAs}, first proposed by Tajima and Dawson \cite{tajima1979laser}, provide a possible alternative to conventional particle accelerators at a substantially smaller size and cost. Taking advantage of the electric fields in laser-produced plasmas, \acs{LWFA} can reach acceleration gradients of tens of GeV/m, which are many orders of magnitude greater than those produced in conventional accelerators. Extensive experiments have been performed to understand \acs{LWFA} mechanisms and to generate energetic electron beams \cite{malka2002electron, mangles2004monoenergetic, geddes2004high, faure2004laser}, and the highest electron energy achieved so far is 7.8 GeV \cite{gonsalves2019petawatt}. While the highest energy laser facilities usually fire a few shots a day, there have been rapid developments in high repetition rate laser systems with lower peak power \cite{guenot2017relativistic, salehi2017mev, prencipe2017targets, roso2018high, salehi2019high, feister2019development}. One of the rationales to increase the repetition rate is that many applications demand higher repetition rates but at only moderate beam energies. MeV-level electrons from \acs{LWFAs} above 1 Hz have been used in transmission electron radiography \cite{bussolino2013electron}, picosecond electron diffraction \cite{he2016capturing}, and generating $\gamma$-ray sources through bremsstrahlung conversion for imaging \cite{dopp2016bremsstrahlung}. In addition, having higher repetition-rate allows meeting the statistical requirements to have better control over experiments \cite{he2015coherent, dann2019laser, lin2019adaptive, kim2017stable, liu2014adaptive, tsai2018control, shalloo2020automation}. 

In this work, we demonstrate the use of machine learning in \acs{LWFA} beyond optimization purposes. We build four regression models using supervised learning algorithms: Random Forest, Neural Network, Deep Jointly-Informed Neural Network, and Gaussian Process. The models are trained to predict the electron beam charges in an \acs{LWFA} given the laser wavefront modification caused by a deformable mirror, while Random Forest turns out to be the best considering the model performance and the computational cost. The trained models help examine the quality of the measurement by evaluating the model performance on every measured data point. Three of the models show similar performance, providing a potential way for anomaly detection without repeated measurements. To investigate if the ML models can make accurate predictions when the measured data have uncertainty, we characterize the model robustness against a range of virtual error bars assigned to the data. Gaussian Process and Random Forest are found to be more resistant to measurement uncertainties. We rank the important Zernike terms that lead to high electron beam charge according to the ML models, compare them to the results from the genetic algorithms and from the statistical correlation. The feature importance from the ML models is found to reveal more information than the latter methods do.

% In this work, we have predicted the electron beam charges in LWFA given the laser wavefront modification caused by a deformable mirror using four supervised learning regression methods: Random Forest, Neural Network, Deep Jointly-Informed Neural Network, and Gaussian Process. We benchmark these models and demonstrate applications beyond optimization. We show that generating higher beam charges favors specific wavefront aberrations, which is revealed by ranking the feature importance in the form of the Zernike decomposition. We analyze the experimental data quality by evaluating the model performance on every measured data point. We also characterize the model robustness against a range of virtual measurement error bars assigned to the experimental data.

\subsection{Data and Methods}
% \subsubsection{Laser-wakefield acceleration}
% \subsubsubsection{Physics mechanisms and precise control}
% \paragraph{Physics mechanisms and precise control}
\subsubsection{\acs{LWFA} precise control}
The physics mechanism of \acs{LWFA} has been introduced in Sec. \ref{sec:theoryUnderdense}. Over the past decade, advances in theoretical analysis \cite{esarey2009physics, lu2007generating} and experimental demonstrations \cite{gonsalves2019petawatt, wen2019polarized} of \acs{LWFA} has pushed our understanding of its mechanisms to the forefront. The critical questions to address in the next decade within the \acs{LWFA} community \cite{milchberg2019workshop} have moved towards precise control of \acs{LWFA} experiments to serve as bright sources of relativistic particles \cite{gonsalves2019petawatt, guenot2017relativistic} and high energy photons \cite{albert2016applications}. While unprecedented stability of laser-plasma accelerators with many hours of operation has been demonstrated in recent studies \cite{maier2020decoding, rovige2021optimization}, challenges remain to precisely control the highly nonlinear physical processes in \acs{LWFA}. As an effective tool for data analysis in nonlinear and high-dimensional problems, machine learning is gaining attention \cite{he2015coherent, dann2019laser, shalloo2020automation}.

\subsubsection{Experimental}
The experiments were conducted using the Lambda Cubed laser system in the Gerard Mourou Center for Ultrafast Optical Science at the University of Michigan, which produces 35 fs, 20 mJ laser pulses at 800 nm wavelength. 
The laser pulses were focused by an f/2 off-axis paraboloidal mirror to a vacuum spot size of 2.5 $\mu m$ FWHM, resulting a peak intensity of $3\sim4\times10^{18} W\cdot cm^{-2}$. The experiments were operated at a plasma density roughly a Gaussian with a 120 $\mu m$ FWHM and a peak of $6.5\times10^{18} cm^{-3}$. The laser focus was generally placed on the down-ramp of the target density profile. The target is free-flowing argon gas from a capillary with inner-diameter of 100 $\mu m$, and the optimal backing pressure was in the range of $21\sim23$ psi. Details of the experimental setup and electron acceleration mechanisms are described in reference \cite{he2013high}.  The laser wavefront change was induced by a deformable mirror (AOA Xinetics 37-channel 2 inch diameter) and recorded as the 37 actuator voltages that control the mirror surface. The electron beam was captured via a scintillator screen that was imaged by an electron-multiplying charge-coupled device (CCD) camera (Andor Luca-R, 14-bit).
% This particular system runs at a very high repetition rate relative to the systems used for similar experiments, where repetition rates generally top out on the order of 1 Hz. These high repetition rates lead to a greater ability to accumulate larger amounts of data, which enables us to employ optimization algorithms to improve experimental results. 
The \acs{LWFA} electron beam data were recorded from four separate experimental days when running optimization algorithms.
% We are confident that the data accumulated across the four experiment days can be merged and treated as one set of data. 
The optics that deliver the laser to the experimental chamber were unchanged throughout the four days. The target is placed with respect to the laser focus in the same position, and the alignment procedures were routinely carried out the same way each day. An example of the raw electron beam image is shown below. 
% Hero: Typically one provides empirical evidence for this, e.g. a histogram comparison from day to day. Otherwise, I suggest that you take this sentence out of the paper as it actually  raises a red flag (If you need to state that you are confident without evidence, then the interpretation is that you are actually not so confident)

\begin{figure}[ht]
\centering
\includegraphics[width=0.6\columnwidth]{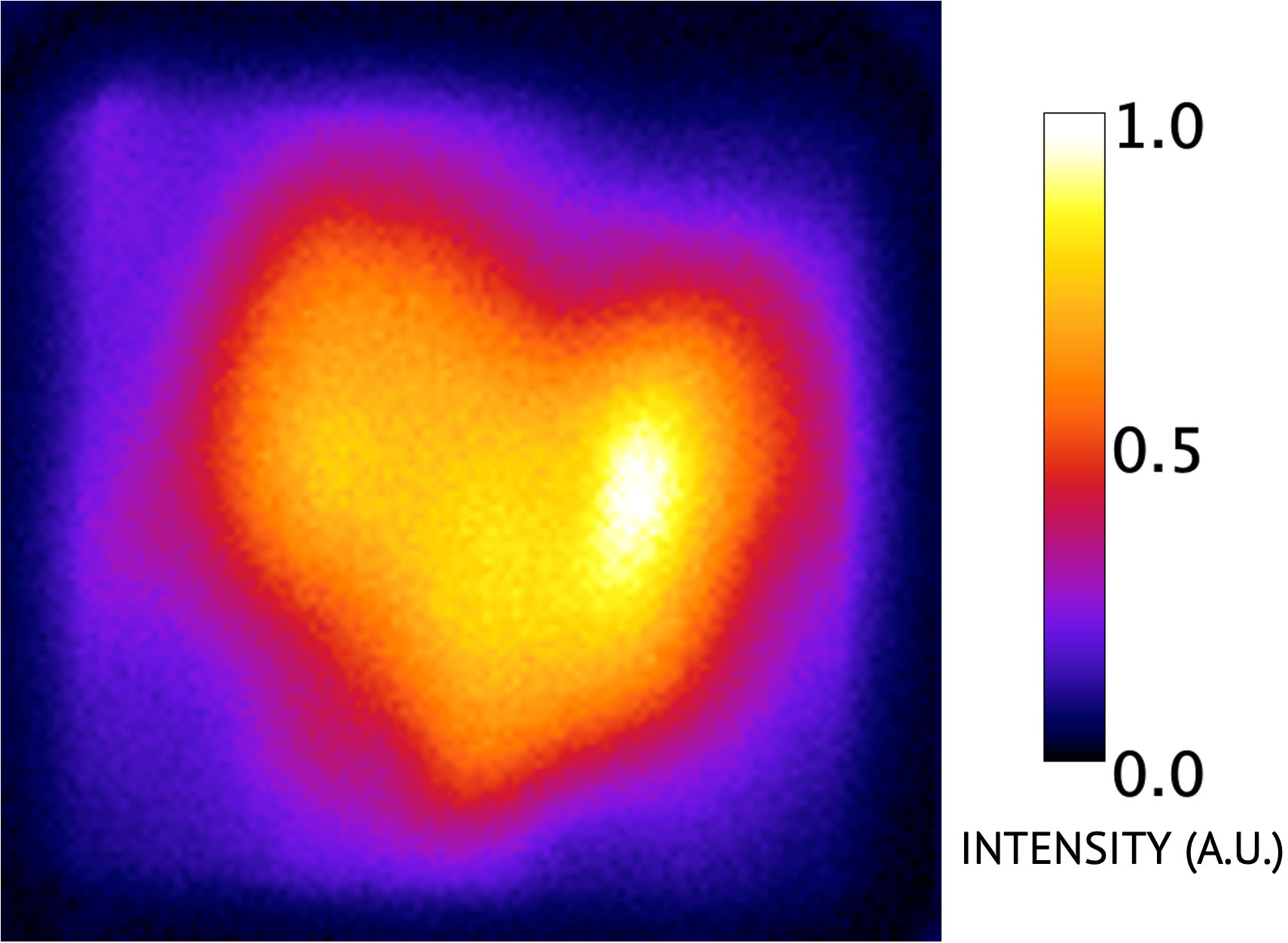}
\caption{A sample image of the measured electron beam from the \acs{LWFA}. Color scale indicates the intensity of the electron signal.}
\label{MLLWFAbeamImage}
\end{figure}

\subsubsection{Data pre-processing and correlation}
% \subsubsubsection{Pre-processing}
The dataset is pre-processed before the regression modeling. The dataset contains the information of the optical wavefront change of the driving laser caused by a deformable mirror, as well as the electron beam charge of the accelerated electrons. The wavefront changed by the deformable mirror has 37 dimensions in space and can be described mathematically by a polynomial, known as the Zernike polynomial \cite{born2013principles}, to reduce the dimensionality. In this case, the coefficients of the first five layers (15 terms) in the Zernike polynomial can accurately reproduce the wavefront. The 15-dimensional vectors consisting of the Zernike coefficients are used as the input to our supervised learning models, while the electron beam charges integrated from the beam image are the output, normalized to the range [0, 1]. In the context of machine learning, the input is called a feature and the output is called a label. We have 208 data samples in total, while each data point was averaged over $\sim120$ laser shots. The dataset is split into two subsets: $80\%$ of the data points are used to train the models while $20\%$ are for testing. The feature matrix in the training set is sphered so that its rows have zero sample mean and unity sample variance. The feature matrix in the test set is updated accordingly with the same transformation.

% \subsubsubsection{Correlation}
The statistical correlation coefficient measures the dependence between two variables, and its value falls in the range of [-1,1]. The absolute value represents the strength of the dependence while the sign represents the direction. We calculate the correlation between each feature (Zernike coefficients) and the electron beam charge in the test dataset, as is illustrated in Tab. \ref{MLLWFAcorrelation}. Among all the Zernike coefficients, while $z_{10}$ has the largest magnitude of correlation. The correlation matrix will be compared to the machine learning model predictions in Tab. \ref{MLLWFAfeatureImportanceTest}, \ref{MLLWFAfeatureImportanceAll} in Section. \ref{subsec:MLLWFAresult}.

\begin{table}[H]
  \centering
%   \resizebox{\columnwidth}{!}{
    \begin{tabular}{|r|r|r|r|r|r|r|r|r|r|r|r|r|r|r|r|}
    \hline
    $z_0$ & $z_1$ & $z_2$ & $z_3$ & $z_4$ & $z_5$ & $z_6$ & $z_7$ \\
    \hline
    0.51 & -0.33 & -0.28 & 0.29 & 0.14 & 0.21 & -0.15 & 0.37\\
    \hline\hline
    $z_8$ & $z_9$ & $z_{10}$ & $z_{11}$ & $z_{12}$ & $z_{13}$ & $z_{14}$ & charge \\  
    \hline
    -0.42 & 0.20 & -0.55 & 0.35 & 0.47 & -0.42 & -0.019 & 1 \\
    \hline
    \end{tabular}%
    % }
  \caption{Statistical correlation between Zernike coefficients ($z_0$ - $z_{14}$) and the electron beam charge.}
  \label{MLLWFAcorrelation}
\end{table}%

\subsubsection{Machine learning methods}
Four supervised learning regression methods are used to predict the electron beam charges based on laser wavefront changes: \acf{RF}, \acf{GP}, \acf{DNN}, and \acf{DJINN}. Supervised learning is a branch of machine learning, which learns a function that maps an input (feature) to an output (label) based on example input-output pairs in a training sample. In each of the supervised learning methods, the model is trained on the training dataset recursively until it can accurately predict the labels using the features. The model performance is then characterized by the test dataset. Details about the supervised learning methods have been introduced in Sec. \ref{sec:ExpAdaptiveOS}.

In the \acs{RF} method, the algorithm is implemented using the \textit{RandomForestRegressor} library in Scikit-learn \cite{pedregosa2011scikit}. The hyper-parameters to tune are the number of trees, the maximum depth in a tree, and the maximum number of features when splitting. In the \acs{DNN} method, a fully-connected five-layer DNN is built using the \textit{Tensorflow.Keras} library \cite{chollet2015keras} based on Google’s deep learning software TensorFlow \cite{abadi2016tensorflow}. When constructing the network, we use the rectified linear unit (ReLU) function and the Sigmoid function as the activation functions for different layers. The activation function for the first, the fourth, and the fifth layer is the rectified linear unit (ReLU) function, and that for the second and the third layer is the Sigmoid function. 
The cost function is the mean squared error loss governed by the Adam optimizer \cite{kingma2014adam} to update the network weights. A $L_2$ norm regularization is added to the loss function to reduce overfitting. The main tuning parameters are the number of layers, the number of neurons in each layer, the epoch size, and the initialization of the weight matrix. The \acs{DJINN} regression source code is accessible at the LLNL/DJINN github directory. The main tuning parameters in the \acs{DJINN} method are the maximum depth of trees, and the number of trees (nets) in ensemble. In the \acs{GP} method, the algorithm is implemented using the $Sklearn.gaussian\_process$ library in Scikit-learn \cite{pedregosa2011scikit} with a combination of Matern kernel and Rational Quadratic kernel. The hyper-parameters to tune are the smoothness, the length-scale, and the scale mixture parameter in the kernels.

\subsection{Results}
\label{subsec:MLLWFAresult}

\begin{figure}[H]
    \centering
    \includegraphics[width=0.9\columnwidth]{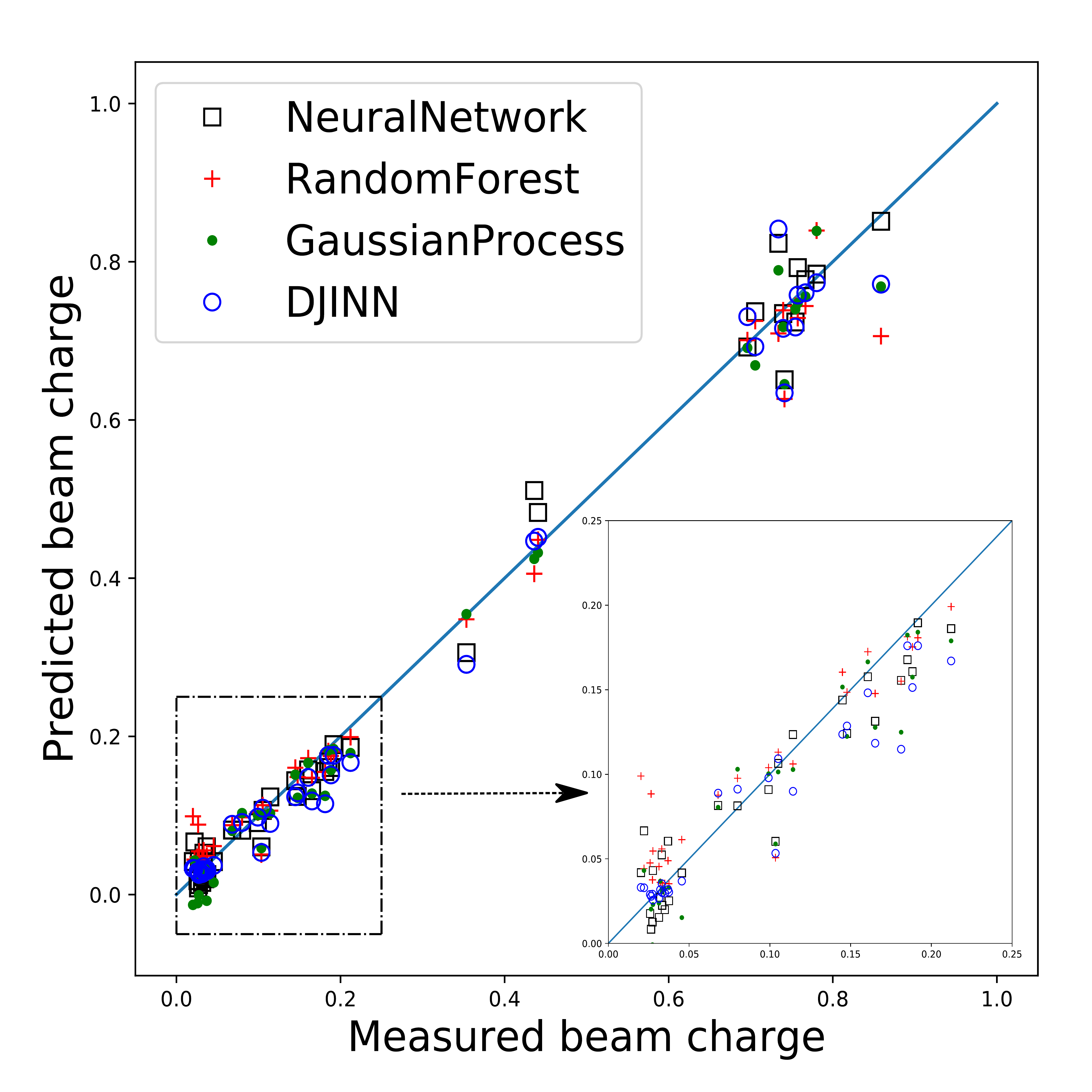}
    \caption{Predicted electron beam charges using DNN, RF, GP, and DJINN vs. measured electron beam charge in the test dataset.}
    \label{MLLWFApredVSreal}
\end{figure}

\begin{table}[ht]
\centering
\begin{tabular}{ |c|l|c|c|c| } 
 \hline
 Model &  MSE & MAE & $R^2$ & ExVar\\
\hline\hline
 Random Forest & 0.00132 & 0.0268 & 0.986 & 0.987 \\ 
\hline
  Neural network &0.00162 &0.0292  &0.983  &0.984 \\ 
 \hline
  DJINN & 0.00154 & 0.02741  & 0.98403  & 0.98404 \\ 
 \hline
  Gaussian Process & 0.00185 & 0.0305 & 0.981 & 0.981\\ 
 \hline
\end{tabular}
\caption{Evaluation matrix: the mean-square-error, the mean-absolute-error, $R^2$, and the explained variance of the predictions in test dataset using four models.}
\label{MLLWFAall_results}
\end{table}

All codes are written in Python. After training the models, we predict the electron beam charge using the laser wavefront change in the test dataset. Predicted electron beam charges using the above models are shown in Fig. \ref{MLLWFApredVSreal} against measured electron beam charges. A reference line at $45^{\circ}$ is included, and data points closer to the reference line are considered better predictions. The bottom left corner of the plot is magnified and shown to the right. Detailed statistical evaluations are summarized in Tab. \ref{MLLWFAall_results}, in which we report the mean-square-error (MSE), mean-absolute-error (MAE), R-squared ($R^2$), and explained variance score (ExVar) based on the predicted charge and the measured charge. MSE measures the average squared difference between the predictions and the real values, which contains information of both variance and bias. It is the most popular metric when evaluating machine learning models and we use MSE as the target for the hyperparameter tuning process. The problematic aspect of MSE is that it can be sensitive to outliers, which MAE handles better by measuring the absolute error instead of the squared error. $R^2$ is the proportion of variance of the measured value from the prediction. It tells how likely a new sample (out of the dataset) can be predicted by the model. Explained variance considers bias on top of $R^2$. It is the same as $R^2$ if the mean of error is 0. In general, one would like to have MSE and MAE close to 0 while $R^2$ and ExVar close to unity. 

% \jinpu{use scientific notation?}
All four models demonstrate similar statistics in Tab. \ref{MLLWFAall_results}, though RF performs slightly better and GP gives the largest MSE and MAE scores while the smallest $R^2$ and ExVar scores. It also predicts negative values when the electron beam charges are small. However, it does not necessarily mean that Random Forest is the best model and Gaussian Process is the worst. The results in the evaluation matrix are sensitive to the way we split the training set and test set. We will show in the next section that training and evaluating the model on different data points yield different results. We will also present more analyses, such as the model consistency against measurement errors and overfitting-related issues.

\subsubsection{Data quality}
\begin{figure}[H]
    \centering
    \includegraphics[width=0.9\columnwidth]{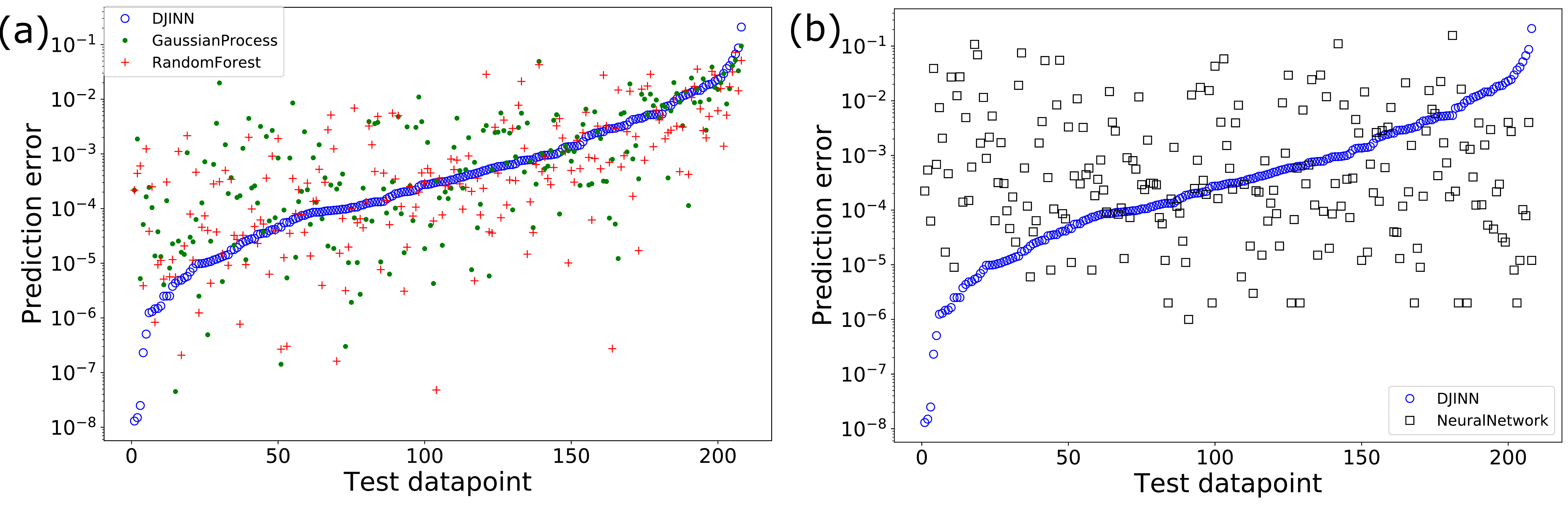}
    \caption{Prediction error when testing ML models with every data point. The test data point is presented in an order where the DJINN prediction error is monotonically increasing. (a) compares the results from DJINN, GP and RF, and (b) compares the results from DJINN and DNN.}
    \label{MLLWFADataquality}
\end{figure}
Experimental measurements in \acs{LWFA} can suffer from a lack of reproducibility and may have outliers in the dataset. Possible sources include shot to shot fluctuations in the beam pointing and pulse energy in high power laser systems, as well as the irreproducibility in the plasma density profile from gas jets. A natural question to ask is how much we can trust each of the measured data points. In this section, we train machine learning models to justify the quality of our measured electron beam charge. Instead of splitting the dataset into $80\%$ for training and $20\%$ for testing, we test only one data point while all the other data points are used to train the model. We then compare the predicted beam charge to the measured electron beam charge of this particular data point and calculate their difference. This process is looped over the entire dataset. Fig. \ref{MLLWFADataquality}a plots the prediction error ($\sigma=|y_{predicted}-y_{measured}|$) at each data point from GP, RF, and DJINN. The prediction errors from DNN are shown in Fig. \ref{MLLWFADataquality}b. Fig. \ref{MLLWFADataquality} is presented in a monotonic order of the DJINN prediction error. It is observed that the three models in Fig. \ref{MLLWFADataquality}a have a similar trend while the prediction errors from DNN in Fig. \ref{MLLWFADataquality}b behave differently. We interpret three messages from these two plots. 1. Prediction errors vary across seven orders of magnitude, i.e., some data points can be accurately predicted ($\sigma\sim1e-8$) while some data points can hardly be predicted ($\sigma\sim0.1$). Therefore, selecting different data points into the training or test set can lead to different evaluation matrices from the one in Tab. \ref{MLLWFAall_results}. 2. This huge variation can be caused by either the inconsistency of the model across data points, or this specific data being very different from all the other data points in the dataset that are used to train the model. We are satisfied with the reliability of the models, as is characterized in Tab. \ref{MLLWFAall_results}. In addition, the similarity among the three models' performance in Fig. \ref{MLLWFADataquality}a suggests that the models are less likely to be inconsistent at the same time since they behave in a similar manner regarding the data points. Thus it provides a potential characterization of the quality of each data point, i,e., a data point that has a very large prediction error in all three models can be considered as having poor data quality and may be dropped as an outlier. 3. DNN performs differently from the other models, suggesting that it overfits the data set and it is less reliable in this scenario. It matches the learning curves in Fig. \ref{MLLWFAlearningcurve}, and a detailed discussion on the overfitting issue can be found later.
% \jinpu{How do we justify the validity of NN despite of overfitting? How reliable are the numbers in the first table?}

Evaluating the prediction error on every data point can assist anomaly detection. For example, if we drive the same laser pulse into the laser plasma accelerator twice, we might observe different electron beam charges due to a lack of reproducibility in \acs{LWFA}. A typical solution would be to calculate the mean value and the error bar. However, it could include misinformation if one of the measurement is an outlier due to technical glitch and should be dropped. By performing the above analysis, we would be able to tell which one of the measured beam charges is more reliable. Moreover, it can help identify outliers not only in repeated measurements but anywhere in the parameter space that the experiment scans across. The data points in the top-right corner in Fig. \ref{MLLWFADataquality}a are examples of possibly poor data quality, where all three models have large prediction errors.

\subsubsection{Robustness against measurement errors}
\label{subsec:MLLWFAVirtualMeasurementError}
\begin{figure}[H]
\centering
\includegraphics[width=0.85\columnwidth]{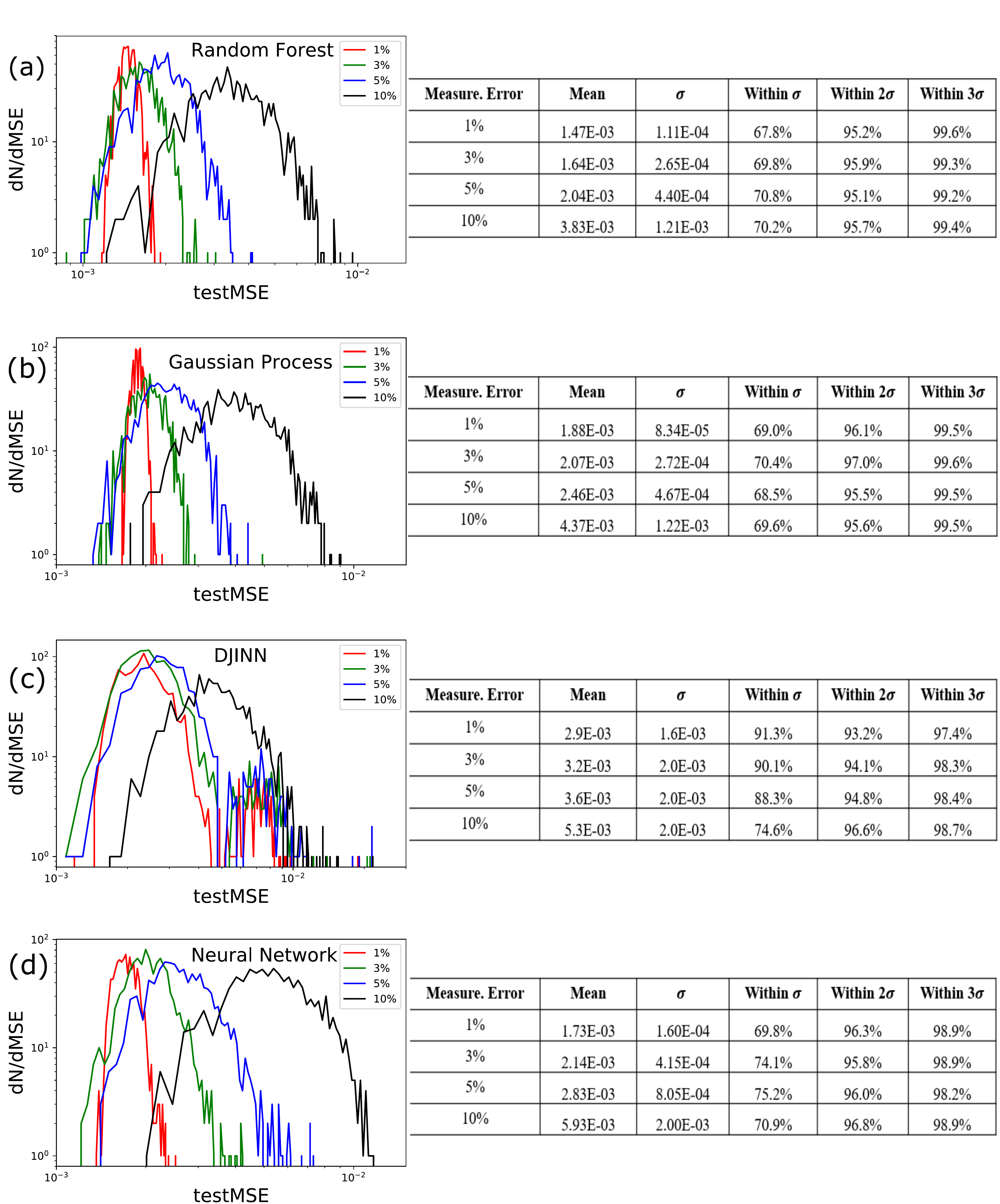}
\caption{Model performance against virtual measurement errors. The figures on the left show the distribution of test MSE using RF (a), GP (b), DJINN (c), and DNN (d). The color indicates the amount of virtual measurement error applied. The tables on the right list the mean value, standard deviation ($\sigma$), and the percentage of points that fall within one, two, or three standard deviations around the mean value.}
\label{MLLWFAnoisyData}
\end{figure}

Another fundamental question to ask about the data is if they are sufficient for the machine learning models to make accurate predictions. It is crucial for experimental data considering all experimental measurements are associated with some degrees of uncertainties. In this section, we investigate the performance of these models against measurement errors of the electron beam charges. Since measurement errors were not recorded during our experiments, we include various virtual error bars to every measured electron beam charge. At each measurement, the true value is assumed to lie in the range of measured value $\cdot(1\pm X\%)$, where $X=1,3,5,10)$, and 1000 points are drawn randomly from a normal distribution within this range. Therefore we get 1000 copies of the original dataset with the same wavefront but different electron beam charges. The reason to have 1000 datasets is to generate enough statistics to justify the model performance against unsure measurements. Results are presented in Fig . \ref{MLLWFAnoisyData}.

Fig. \ref{MLLWFAnoisyData}a shows the distribution of the test MSE using RF. Each colored line is generated from 1000 MSEs. During the training process, the model configuration was kept the same among the 1000 datasets but the weight learning was updated in each dataset. Measurement errors that define the range of the dataset fluctuation are $1\%,\;3\%,\;5\%$ and $10\%$, while the corresponding test MSE distributions are plotted in red, green, blue, and black, respectively. Statistical analysis is summarized in the adjacent table to the right, illustrating the mean value and standard deviation ($\sigma$) of the 1000 test MSEs as well as the distribution within one, two, or three standard deviations around the mean value. Fig. \ref{MLLWFAnoisyData}b-d present the results using GP, DJINN, and DNN. The four models share some common performances. The mean test MSE value increases with the measurement error, which means it is more likely to make a less accurate prediction when the measurement itself is less accurate, as expected. The standard deviation also increases with the measurement error, suggesting a less consistent or less precise model prediction at larger measurement errors. There are noticeable differences in the last three table columns of the four models. Remember that the virtual measurement errors are drawn from a perfect normal distribution, where the percentage of values that lie within one, two, or three standard deviations around the mean value are $68.3\%$, $95.5\%$, and $99.7\%$, respectively. As is shown in the tables, RF and GP retain almost normal distributions in the sample prediction, DNN gives normal-like distributions, and the results from DJINN are far from normal distributions. 
% \jinpu{What does this mean?}

\subsubsection{Learning curve}
\begin{figure}[H]
\centering
\includegraphics[width=0.9\columnwidth]{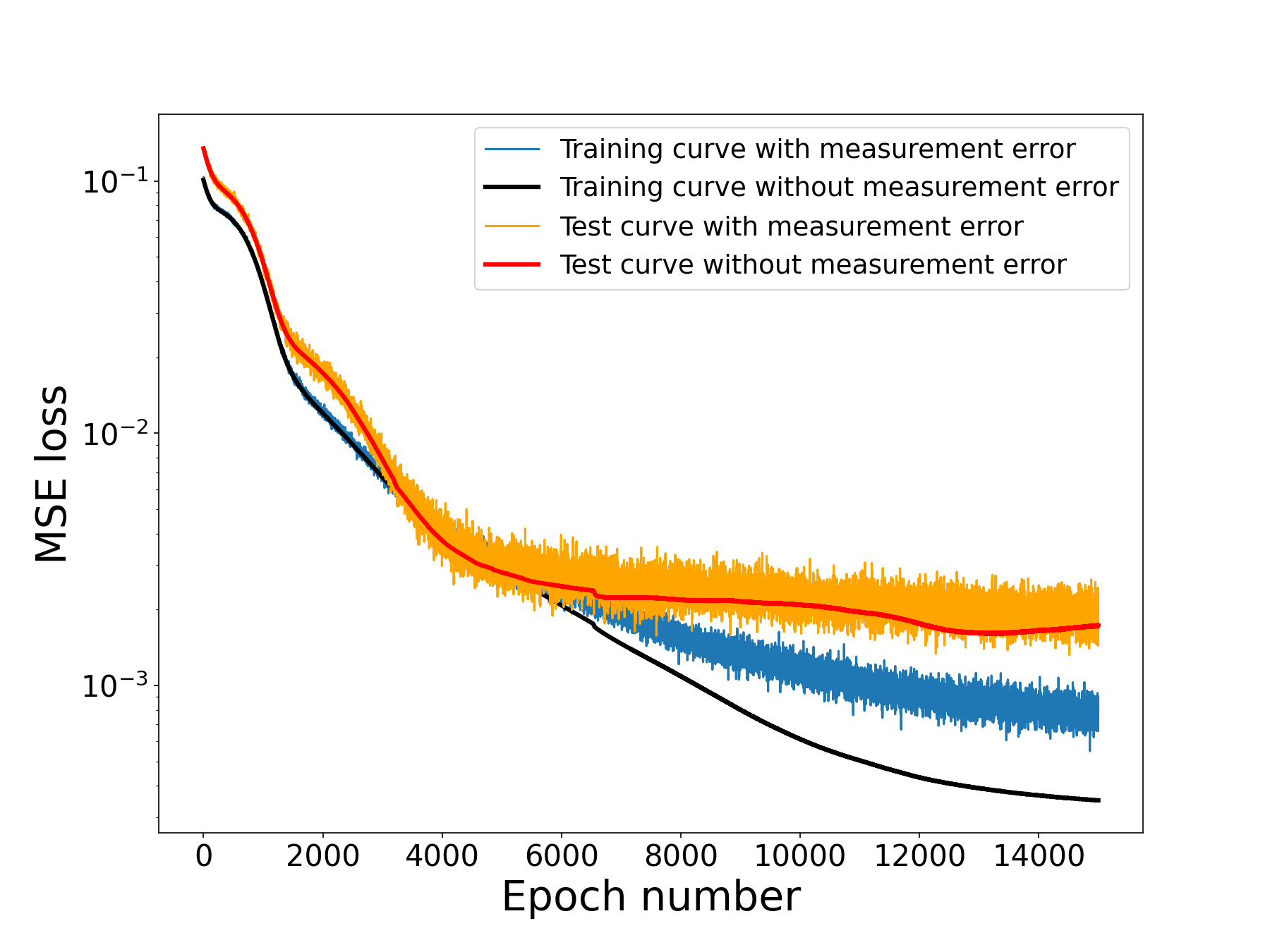}
% \vspace*{-3mm}
\caption{Learning curves in the neural network model: including measurement errors decreases overfitting. Training and test MSE without measurement errors are plotted in black and red, while training and test MSE with measurement errors are plotted in blue and yellow, respectively.}
\label{MLLWFAlearningcurve}
\end{figure}

Overfitting occurs in machine learning when a model has learned the training data so well that it also learns the statistical noise or random fluctuations in the data. The learning curve is an intuitive tool to visualize the degree of overfitting. Fig. \ref{MLLWFAlearningcurve} shows the learning curves in our Neural Network model, which plots MSE at each epoch for both training data (black) and test data (red). The training curve tells how well the model learns, while the test curve tells how well the model generalizes. Since the red curve does not decrease as much as the black curve does, the model overfits. In other words, the overfitted model performs worse outside the training dataset. Fig. \ref{MLLWFAlearningcurve} also plots the learning curves considering measurement errors in blue and orange to better reproduce experimental conditions. Measurement errors are included in a similar way to the ones in the previous section, where the true value is assumed to lie in the range of measured value$\cdot(1\pm 3\%)$. Instead of generating 1000 copies of datasets at a time, here we generate only one dataset with measurement errors and update the measurement errors at every epoch in the learning process. Obtained learning curves are noisier but demonstrate less overfitting, as is shown in Fig. \ref{MLLWFAlearningcurve}. After applying measurement errors, the training curve moves higher, but the test curve does not shift much. Namely, the model finds it harder to learn, but it is still able to make equally accurate predictions. Therefore, including virtual measurement errors is beneficial as it not only represents practical experimental conditions better but also decreases overfitting. Note that we have also plotted the learning curve with and without measurement errors using the DJINN model, and they almost overlap with each other. It is not surprising since the DJINN model does not overfit as much as the neural network model does.      

\subsubsection{Feature importance}

\begin{table}[ht]
\centering
\begin{tabular}{ |c|c|c|c|c| } 
 \hline
 Model & Most important & $2^{nd}$  & $3^{rd}$ & $4^{th}$\\
\hline\hline
 Random Forest & $z_{0}$ & $z_{6}$ & $z_{1}$ & $z_{10}$ \\ 
\hline
 Gaussian Process &  $z_{1}$ & $z_{10}$ & $z_{0}$ & $z_{6}$ \\ 
\hline
  Neural Network & $z_{0}$ & $z_{10}$ & $z_{1}$ & $z_{3}$\\ 
 \hline
  DJINN & $z_{1}$ & $z_{10}$  & $z_{0}$ & $z_{6}$ \\
 \hline
  Correlation & $z_{0}$ & $z_{10}$ & $z_{1}$ & $z_{13}$ \\
 \hline
\end{tabular}
\caption{Feature importance according to the test dataset and correlation raking. The Zernike coefficients (features) are ranked by their importance to producing high beam charge in the columns.}
\label{MLLWFAfeatureImportanceTest}
\end{table}

\begin{table}[ht]
\centering
\begin{tabular}{ |c|c|c|c|c| } 
 \hline
 Model & Most important & $2^{nd}$  & $3^{rd}$ & $4^{th}$\\
\hline\hline
 Random Forest & $z_{0}$ & $z_{6}$ & $z_{1}$ & $z_{12}$ \\ 
\hline
 Gaussian Process &  $z_{10}$ & $z_{1}$ & $z_{0}$ & $z_{6}$ \\ 
\hline
  Neural Network & $z_{0}$ & $z_{10}$  &  $z_{1}$ & $z_{8}$\\ 
 \hline
  DJINN & $z_{1}$ & $z_{10}$ & $z_{0}$ & $z_{6}$ \\
 \hline
 Correlation & $z_{10}$ & $z_{0}$ & $z_{12}$ & $z_{13}$ \\
 \hline
 GA subset 1 & $z_{0}$ & $z_{13}$ & $z_{9}$ & $z_{1}$ \\
 \hline
 GA subset 2 & $z_{0}$ & $z_{2}$ & $z_{8}$ & $z_{4}$ \\
 \hline
 GA subset 3 & $z_{0}$ & $z_{10}$ & $z_{14}$ & $z_{1}$ \\
 \hline
 GA subset 4 & $z_{0}$ & $z_{8}$ & $z_{12}$ & $z_{6}$ \\
 \hline
\end{tabular}
\caption{Feature importance according to the entire dataset, correlation ranking, and wavefront of the optimized electron beam using genetic algorithms. The Zernike coefficients (features) are ranked by their importance to producing high beam charge in the columns. The genetic algorithm ran $\sim$50 iterations in each subset.}
\label{MLLWFAfeatureImportanceAll}
\end{table}
We have trained models to predict the electron beam charge upon laser wavefront modification represented by the first 15 Zernike coefficients. It is natural to ask how sensitive the beam charge is to these features. We evaluate the feature importance using our four models and compare them to the correlation ranking, summarized in Tab. \ref{MLLWFAfeatureImportanceTest} and \ref{MLLWFAfeatureImportanceAll}. In each row, we list the four most important features decided by that model, while the numbers are the orders of the Zernike coefficients. The importance of a feature is measured by calculating the increase in the model's prediction error (MSE) after setting the feature values to a constant. A feature is considered important if the prediction error increases significantly, and less important when the prediction error does not change much. When evaluating each feature, the values of this feature of all test data are set to their mean value. Note that model training is performed prior to this process and the training data are not modified. It is debatable whether the feature importance should be computed on the test data or the training data. The former tells how much the model relies on each feature for making predictions, while the latter tells how much the feature contributes to the performance of the model on unseen data \cite{molnar2020interpretable}. We show the feature importance computed from both ways. In Tab. \ref{MLLWFAfeatureImportanceTest}, we split the dataset to a training set ($80\%$) and a test set ($20\%$) and evaluate the feature importance on the test set. In Tab. \ref{MLLWFAfeatureImportanceAll}, we train the model using the entire dataset and measure the feature importance also on the entire dataset. For comparison, Tab. \ref{MLLWFAfeatureImportanceAll} also includes the laser wavefront that optimized the electron beam charge using genetic algorithms, where each subset contains $\sim$50 data points.

Depending on the model and the evaluation data, the feature importance rankings in Tab. \ref{MLLWFAfeatureImportanceTest} and \ref{MLLWFAfeatureImportanceAll} are slightly different. Overall, the $0^{th}$, $1^{st}$, $6^{th}$, and $10^{th}$ Zernike terms are generally believed to be the more important ones according to the trained models, physically representing the piston, tilt, vertical trefoil, and oblique quadrafoil in wavefront aberration, respectively. It suggests that controlling these features would be more effective at producing high electron beam charges in this scenario. Similar to the machine learning models, the statistical correlation and the genetic algorithm optimization emphasize on the low-order ($0^{th}$ and $1^{st}$) and high-order ($\geq10^{th}$) Zernike terms. However, they ca not recognize the $6^{th}$ Zernike term that repeatedly shows up in the feature importance ranking among the ML models except the DNN. The fact that the DNN is more overfitted and behaves differently from other models in Fig. \ref{MLLWFADataquality} adds validity to this finding on the $6^{th}$ Zernike term. It indicates that machine learning can reveal information that can not be observed in genetic algorithms or statistical correlation who do not involve configuring models.

% \begin{table}[ht]
% \resizebox{\columnwidth}{!}{
% \begin{tabular}{ |c|c|c|c|c| } 
%  \hline
%  Model & $1^{st}$ & $2^{nd}$  & $3^{rd}$ & $4^{th}$\\
% \hline\hline
%  Random Forest & 0 (0.0188) & 6 (0.00365) & 1 (0.00335) & 12 (0.00224) \\ 
% \hline
%  Gaussian Process &  10 (0.436) &  1 (0.0412) & 0 (0.0347) & 6 (0.0341)\\ 
% \hline
%   Neural Network & 0 (0.092) & 10 (0.069)  &  1 (0.067)& 14 (0.048)\\ 
%  \hline
%   DJINN & 1 (0.0566) & 10 (0.0444) & 0 (0.0247) & 6 (0.0139) \\
%  \hline
%  Correlation & 10 & 0 & 12 & 13 \\
%  \hline
% \end{tabular}
% }
% \caption{Evaluate feature (Zernike coefficient) importance on entire dataset. The values in range $[0,1]$ inside parentheses is the change of MSE.}
% \label{featureImportanceAll}
% \end{table}

% \begin{table}[ht]
% \resizebox{\columnwidth}{!}{
% \begin{tabular}{ |c|c|c|c|c| } 
%  \hline
%  Model & $1^{st}$ & $2^{nd}$  & $3^{rd}$ & $4^{th}$\\
% \hline\hline
%  Random Forest & 0 (0.0176) & 6 (0.00645) & 1 (0.00474) & 10(0.0035) \\ 
% \hline
%  Gaussian Process &  1 (0.0573) &  10 (0.0557) & 0 (0.0442) & 6 (0.0335)\\ 
% \hline
%   Neural Network & 0 (0.0570) & 10 (0.049765) & 1 (0.049759) & 8 (0.0474)\\ 
%  \hline
%   DJINN & 1 (0.0884)  & 10 (0.0789)  & 0 (0.0334) & 6 (0.00406) \\
%  \hline
%   Correlation & 0 & 10 & 1 & 13 \\
%  \hline
% \end{tabular}
% }
% \caption{Evaluate feature (Zernike coefficient) importance on the test dataset. The values in range $[0,1]$ inside parentheses is the change of MSE.}
% \label{featureImportanceTest}
% \end{table}

\subsection{Discussion}
% \section{Physics Discussion}
% \Revision{\section{Physics Discussion}}
When tuning the wavefront of the laser beam that drives an laser-wakefield accelerator, the highest electron beam charge or the best electron beam profile are not necessarily associated with a flat laser wavefront, even though a flat wavefront produces the highest vacuum focal intensity. This has been shown in previous experiments using genetic algorithms \cite{he2015coherent, lin2019adaptive}. As the laser pulse propagates in the plasma, its wavefront can be altered by nonlinear interactions such as self-modulation, self-focusing, Raman scattering and even the ionization dynamics before it reaches the vacuum focus. Moreover, a wavefront with aberration can move the focal point along the plasma density gradient and thus affects the energy gain in the produced electrons, which is dependent on the plasma density. Such a non-flat optimal wavefront has also been observed in other relativistic laser-plasma experiments when changing the wavefront, for example filament-induced breakdown spectroscopy studies \cite{fibich2004control, dergachev2014plasma, finney2021filament}, which are sensitive to the laser beam ellipticity and the ionization in the plasma.

There has been growing interest in the relativistic laser-plasma community to determine what value machine learning can provide. Machine learning methods are not expected to offer some generalized predictive models that save experimentalists from carrying out every experiment. This is due to the lack of reproducibility in high-power laser-plasma experiments, where laser systems usually suffer from shot-to-shot fluctuations and a given plasma density profile is hard to duplicate. However, using machine learning techniques can help us better understand the experiment performance and improve the design of next-step experiments. Tab. \ref{MLLWFAfeatureImportanceAll} has shown that machine learning can reveal deeper information that are buried in the statistical correlation or the optimization results from genetic algorithms. Moreover, it enables deeper physical interpretation of the data since the predictive accuracy of the regression models is determined by the data quality. For example, by ranking the feature importance, we are able to identify the Zernike terms that are most sensitive to noise. Some of these turned out to be the high order terms (vertical trefoil and oblique quadrafoil) that located at the edge of the wavefront. The "importance" of these high order terms can be explained from an experimental point of view as follows. Deformable mirrors are manufactured to have actuators forming into a hexagon matrix, while a wavefront is usually defined in a circular or rectangular shape, leading to lack of information on the edge. If the wavefront is measured directly with wavefront sensors, it is usually necessary to manually draw a circle that covers most lighted pixels on the detector as the region of interest. As a result, uncertainty arises at the pixels on the edges of the region of interest. If the wavefront is reconstructed using the actuator displacement on the deformable mirror surface, the phase in these unknown edges also needs to be defined manually. Another possible source of noise is the imperfect overlap of the laser and the deformable mirror surface: either the laser beam clipped off the mirror edge, or it did not fulfill the whole mirror surface.
% \qian{Also discuss the overfitting in NN model in this paragraph. (somthing like NN model can have overfitting when we don't have a large number of data to train it. In other words, when we only have a limit number of data, Random forest or DJINN has a better performance compare to NN.) }

Including virtual measurement uncertainties can be useful even if the experimental data come with some uncertainties. It may be able to narrow down the range of uncertainty when the measurement uncertainty is large. For instance, if a data point in the dataset has some considerable uncertainty and we would like to know the true measurement value or at least narrow down its range, machine learning regression methods can provide us a possible solution. We can start by randomly sampling N points within the range of uncertainty of this measurement value. The next step is to make N copies of the measurement dataset and replace the point of large uncertainty with one of our sampled points in each dataset. Therefore, we obtain N similar datasets with difference only at one point. We then train and test the machine learning models on each dataset. Those datasets that lead to less accurate predictions (large test MSE) are less likely to contain true measurement at the uncertain point if the models make sense.

It is also worth discussing the physics interpretation induced from the kernel functions in the Gaussian Process method. The common way of kernel selecting is either to have expert knowledge about the dataset or to compare candidate kernels for the best performance. In this project, we have tried different kernels and decide that a combination of Matern kernel and Rational Quadratic kernel works best. We can thus infer some knowledge about our dataset based on the rationale of these kernels. Matern is a generalization of the popular Gaussian radial basis function (RBF) with tunable smoothness \cite{rasmussen2003gaussian}. The smoothness of the model can be controlled by a parameter $\mu$ while the $\mu$ value in our case is as low as 0.3, suggesting that the resulting function is far from smooth. The other kernel that fits our model, the Rational Quadratic kernel, is a sum of RBF kernels with different length scales. Note that the length scale decides the safe distance to extrapolate when modeling, and the discontinuity can be handled with a short length scale. According to the optimized parameters in these two kernels, our model function is neither smooth nor continuous, which is not surprising as the high-dimensional dataset was taken from a highly-nonlinear physical process.
% \newline
% An ideal thing would be to take the wavefront to some PIC simulation and verify the result. However, we didn't do that because: 1. we didn't record the reference wavefront in the laser beam. 2. Simulation might not have all the physics considered in this wavefront-dependent process. \jinpu{Do we need to mention this?}

% \Revision{\section{Conclusion}}
% \section{Conclusion}
In summary, we demonstrate several applications of machine learning in relativistic laser-plasma experiments beyond optimization purposes. We have built four supervised learning regression models to predict the electron beam charge using the laser wavefront change in an \acs{LWFA} experiment. All four models present similar statistics in the evaluation matrix, although Random Forest performs slightly better and Gaussian Process performs slightly worse. To justify the data quality affected by the irreproducibility in experiments, we characterize the model prediction on every single data point. Three of the models show similar performance, providing a potential way of recognizing outliers without repeated measurements. The Deep Neural Network is easy to overfit our dataset and thus not the best candidate for analyzing data quality. We include virtual measurement errors to the measured electron beam charges, where Gaussian Process and Random Forest are found to be less sensitive to measurement fluctuations. Having virtual measurement errors is beneficial as it not only represents experimental conditions better but also decreases overfitting. The significance of the Zernike coefficients in terms of generating high electron beam charge is analyzed using the trained models, which reveals more information than the genetic algorithms and the statistical correlation can provide. The Deep Neural Network requires the most computational cost, followed by DJINN, while Gaussian Process and Random Forest consume the least. Therefore, Random Forest is recommended when working with datasets from similar relativistic laser-plasma experiments.

\chapter{Conclusions and Outlook}
\label{chap:Conclusion}
\subsection{Summary}
In this dissertation, six experiments are presented on electron acceleration and
radiation generation using relativistic laser-plasma interactions at high repetition rates. This work demonstrates radiation produced via surface \acs{HHG} and via characteristic x-ray emission, and relativistic electrons accelerated via \acs{LWFA} and from solid-density plasmas. The main objects of this work are to investigate the impact of long-wavelength laser pulses in the interactions, and to explore the application of statistical methods in the experiments given the high-repetition-rate capability.

The three experiments presented in Chap. \ref{chap:Solid} were performed with bulk solid targets at the \acs{Lambda-cubed} at \acs{CUOS}. In Sec. \ref{sec:HHG}, \acs{HHG} spectra and harmonic divergence were measured experimentally when 2 $\mu m$ laser pulses interacted with silicon and glass targets. The harmonic efficiency scales with harmonic order in a power law as $I(\omega)\propto (\omega/\omega_L)^{-2.752}$, which is close to the frequently quoted value of -8/3 predicted by the \acs{ROM} model. The scaling law of the (third) harmonic efficiency vs. laser intensity is also studied in a power-fit $I_{3\omega}\propto I_L^n$, which suggests a nonlinear scaling (n$\sim$2) predicted by the \acs{ROM} model. The intensities of harmonics polarized in horizontal and vertical directions are characterized when the driving laser pulses are polarized in horizontal, vertical, left-circular, and right-circular directions. For linearly-polarized driving pulses, the measurements do not display the distinct feature in even and odd harmonics predicted by the unique selection rule of the \acs{ROM} model. For circularly-polarized laser pulses, both even and odd harmonics were observed, which is in line with the \acs{ROM} selection rule. Overall, generating horizontally-polarized harmonics with P-polarized interactions yields the highest laser-to-harmonic efficiency.

In Sec. \ref{sec:Atto}, MeV-level attosecond electron bunches are studied when driving a laser pulse onto a glass target at grazing incidence. The experimental energy spectra match the spectra of attosecond electron bunches observed in \acs{PIC} simulations. The duration of the bunches is measured to be $\sim$tenth of a micron ($\sim$100 attoseconds) in simulations. Direct experimental measurement of the bunch duration was not performed, although it can potentially be achieved by measuring characteristics of \acs{COTR} produced by the electrons into the x-ray regime. It is found that the generation of energetic attosecond electron bunches favors larger incident angle, higher pulse energy, larger focal spot size, and moderately sharp preplasma density profile. Single-cycle pulses are used to obtain isolated attosecond electron bunches in \acs{PIC} simulations. Controlling \acs{CEP} offers the ability to inject electrons with various initial phases and to adjust the energy and shape of the bunch. Due to the Guoy phase shift, \acs{CEP} changes as the single-cycle pulses get focused and propagate through the plasmas. The propagation direction of the electron bunch is dominated by the preplasma density profile, while fine-tuning is accessible by varying the \acs{CEP}. Higher $a_0$ and larger focal spot size can result in even shorter bunch duration with less dispersion after propagation. When operating with much higher pulse energy, a sharper preplasma profile is preferred to produce a cleaner bunch. Using tilted laser pulses can pass angular properties to electron bunches and spatially separate them. A simplified analytic model is used to predict the momentum gain of electrons, and short electron bunch duration is found to favor large incident angle and long laser wavelength.

In Sec. \ref{sec:Xray}, the key features in characteristic x-ray emission ($k_\alpha$ emission efficiency, bremsstrahlung emission efficiency, and electron temperature) are investigated at various tuning parameters in the laser-plasma experiment (laser pulse wavelength, laser pulse energy, and preplasma density gradient). The parametric study is performed both experimentally and computationally with \acs{PIC} simulations. Among the three controlling parameters, the laser wavelength is the most dominant one regarding the $k_\alpha$ emission efficiency. The wavelength dependence is monotonic for both $k_\alpha$ efficiency and for bremsstrahlung efficiency, while both favor short laser wavelengths. Such a dependence of the x-ray efficiency is determined by the fact that the critical surface is located closer to the solid density surface at a short laser wavelength, making it easier for the energetic electrons to reach the atoms in the solid target. The hot electron temperature, however, no longer has a monotonic dependence on laser wavelength since the laser pulse propagates deeper at a shorter laser wavelength while the ponderomotive force favors a longer wavelength. The other two tuning parameters have smaller impact on the $k_\alpha$ emission efficiency but are crucial for the bremsstrahlung efficiency and the hot electron temperature. The electron temperature favors a larger $a_0$ but sees an optimal preplasma density scale length, while the bremsstrahlung efficiency sees an optimal preplasma density scale length and an optimal $a_0$. Moreover, simulations using various control parameters but the same similarity parameter is performed to reveal the contribution of relativistic electrons to $k_\alpha$ emission.

The three experiments presented in Chap. \ref{chap:ML} were performed with gaseous plasma targets assisted by statistical methods. In Sec. \ref{sec:FocusOpt}, a deformable mirror and a genetic algorithm are used to optimize the high numerical
aperture focal spot of a 800 nm, 30 fs, 3 mJ pulse and of a 2 $\mu m$, 67 fs, 1.6 mJ pulse. The focus optimizations are performed at relativistic intensity without attenuation. The laser pulses were directed into rarefied gas to produce second harmonic signal, which was found to be a convenient and effective feedback for the
genetic algorithm. This technique provides significant experimental convenience compared to other methods that require attenuating the laser or breaking the vacuum condition in between the focus optimization and the laser-plasma experiments.

In Sec. \ref{sec:MIRLWFA}, the first experiment to optimize the quality of the electron beam from mid-IR ($\lambda=3.9\mu m$) laser pulses interacting with near-critical density plasmas is presented. Electron beam charge, energy spectrum, beam pointing and fluctuation have been improved by manipulating the laser wavefront via an evolutionary algorithm and a deformable mirror. Wavefront reconstruction and PIC simulations illustrate that changes on laser wavefront lead to different laser focusing and self-guiding in plasma. Filamentation has been observed in the case of a flat laser wavefront, and can be corrected by the adaptive control system for better electron acceleration. This work also demonstrates the ability to have regular deformable mirrors with 4 $\mu m$ full stroke to properly function in a mid-IR laser system, and the ability to reconstruct wavefronts without the presence of a mid-IR wavefront sensor.

In sec. \ref{sec:MLLWFA}, applications beyond optimization purposes of machine learning in \acs{LWFA} are demonstrated. Four supervised learning regression models are built to predict the electron beam charge using the laser wavefront change caused by a deformable mirror. The quality of the measurement is examined by evaluating the model performance on every measured data point, showing a potential way for anomaly detection without repeated measurements. To investigate if the ML models can make accurate predictions when the measured data have uncertainty, the model robustness is characterized against a range of virtual errors assigned to the data. Feature importance analysis using the trained models shows that specific aberrations in the laser wavefront are favored in generating higher beam charges, which reveals more information than the genetic algorithms and the statistical correlation do. Overall, Random Forest works best for this experimental dataset considering the model performance and the computational cost.

\subsection{Future work}
% Future work, reinforcement learning for 1 hz lasers, transfer learning for single shot lasers
% reinforcement learning - see prospectus slides and Ted norris course project

% hhg - increase intensity limit

% midIR electrons from solid target atto?
The reported results in this dissertation may guide several follow-up experiments. In Sec. \ref{sec:Atto}, the duration of the ultrashort electron bunches is measured only in \acs{PIC} simulations. Future experiments could measure the bunch duration by diagnosing the \acs{COTR} of the produced electron bunches. The simplified model suggests that a longer wavelength laser would produce electron bunches of shorter duration. It would be interesting to verify that experimentally using mid-infrared laser pulses. In Sec. \ref{sec:Xray}, the parametric study on characteristic x-ray emission can be further investigated both experimentally and computationally. Future experiments could endeavor to employ more laser wavelengths to enrich the validity of the parametric study, such as 400 nm pulses from frequency doubling the 800 nm pulses and 1.3 $\mu m$ pulses from the idler of the 2 $\mu m$ \acs{OPA}. Future simulation work should take into account the ionization physics to differentiate the difference in driving laser wavelengths. The current OSIRIS framework only provides a field ionization module, which is not the main source of ionization in solid-density plasma. Collisional ionization is to be included in future \acs{PIC} simulations. In Sec. \ref{sec:FocusOpt}, two-fold optimization has been presented using the second harmonic signal of the laser pulses when focusing in a rarefied gas. Despite the exceptional convenience, this method is not as efficient as other methods that require attenuating the laser or breaking the vacuum condition \cite{he2014laser}. To better understand the physics process that leads to the \acs{SHG} in such conditions, a power-law scaling for the \acs{SHG} at various laser intensities is to be investigated with more complete measurements. In Sec. \ref{sec:MIRLWFA}, the optimization of the \acs{LWFA} electrons driven by a mid-\acs{IR} pulse is achieved by a deformable mirror and a genetic algorithm. Future experiments could extend it to optimization using temporal control of the pulse with a Dazzler.

% ML: CNN+RL, transfer learn
Machine learning as a tool for relativistic laser-plasma experiments has been preliminarily explored in Sec. \ref{sec:MLLWFA}, and more ambitious applications are expected in the future. In this study, in order to avoid the complexity of image processing, the laser wavefront is converted to the Zernike polynomials and the electron beam charge is used instead of the electron beam profile. Future work could employ a \acf{CNN} to address this issue. Instead of quantifying the image into a number (\acs{FOM}) in traditional methods or analyzing the raw image in neural networks, a \acs{CNN} can extract much more information from the image, such as recognizing a specific pattern. Its ability to deal with complex images should find wide applications in diagnosing the particles and radiation produced in relativistic laser-plasma experiments. The concept of \acf{RL} is another category of machine learning to be explored to assist high-intensity laser facilities. While most of the previous optimization experiments were performed with high-repetition-rate lasers $\sim$kHz, a number of laser systems with lower repetition rates $\sim$Hz but much higher intensities are emerging, for example. the ZEUS laser system at \acs{CUOS}. To run optimization experiments on these laser systems, algorithms with higher efficiency are needed. Receiving growing interests in the past few years in the machine learning community, \acs{RL} should also be able to integrate with high-intensity laser via adaptive optical systems. There is potential for \acs{RL} to take over evolutionary algorithms to enable more powerful and more efficient real-time interpretation and optimization. For lasers with the highest power and record intensities, machine learning could also find usefulness despite the fact that such lasers usually fire only a few shots per day. Future work could utilize transfer learning to narrow the gap between expensive experimental results and low-cost simulation results, which allows to construct a predictive model using the latter and to transfer the learned knowledge to a model constructed by the former.

%% Appendix %%
\startappendices

\appendix{Example OSIRIS input deck}
\label{app:OSIRISinput}
\begin{minted}
[
xleftmargin=\parindent,
breaklines,
frame=lines,
framesep=2mm,
baselinestretch=1.2,
fontsize=\footnotesize,
linenos
]
{fortran}

! 2um solid target exp - Molybdenum
! scale length = 0.5*lambda
! laser wavelength = 2um, pulse duration = 50fs, spot size = 2*lambda FWHM diameter
! In Vacuum
! written on: 2/23/2021 by Jinpu Lin @linjinp@umich.edu
! ---------------------------------------------------
! Units normalised to laser units
! t0 to To, 6.67 fs, which is 2um/3e8m/s
! x0 to lambda0, 2 um
! omega0 = 2pi, 6.283
! Grid size [1/64, 1/64] lambda0
! 20ps prepulse - 0.5lambda
! ---------------------------------------------------

simulation
{
n0 = 7.06e18,       ! [cm^-3], n0 = nc/4pi^2
}

!----------the node configuration for this simulation----------
node_conf 
{
  node_number(1:2) = 4,4,
  if_periodic(1:2) = .false., .false., 
}


!----------spatial grid----------
grid 
{
 nx_p(1:2) = 1280, 1280,      ! (20) * 64 = 1280             
 coordinates = "cartesian",
}


!----------time step and global data dump timestep number----------
time_step 
{
  dt     =  0.0034739, ! dt = 0.995 * sqrt[1/(c^2/dx^2 + c^2/dy^2)] / T0 ; dt<0.5/sqrt(20716), 20716*no in density profile, choose the smaller dt
  ndump  = 1079,        ! 1079 iterations * 0.0034739 * T0 = 25 fs. The time step for output is 25 fs, pulse duration is 50fs
}

!----------restart information----------
restart 
{
  ndump_fac = 0,
  if_restart = .false.,
}

!----------spatial limits of the simulations----------
!(note that this includes information about
! the motion of the simulation box)
space 
{
  xmin(1:2) =   0.000d0,  0d0, 
  xmax(1:2) =    20d0,   20d0,  !20*2um=40um in x1 and x2
  if_move= .false., .false.,  
}

!----------time limits ----------
time 
{
  tmin = 0.0d0, tmax  = 53, ! total simulation time is 53*T0=350fs
}

el_mag_fld
{

}

!----------boundary conditions for em-fields ----------
emf_bound 
{
    type(1:2,1) = "open", "open",
    type(1:2,2) = "open", "open",
}


!----------diagnostic for electromagnetic fields---------- 
diag_emf 
{
  ndump_fac = 1,
  reports   = "e1", "e2", "ene_emf",
}

!----------number of particle species----------
!----------number of particle species----------
particles 
{  
  num_species    = 2,
  num_neutral    = 0,
  interpolation  = "cubic",
}



species
{
    name           = "preplasma",         
    num_par_max    = 1.0d9,
    rqm            = -1.000,          ! Means 'electron'
    num_par_x(1:2) = 10, 10,            
    add_tag        = .true.,
}

!----------inital proper velocities-----------------
udist {
  !uth(1:3) = 0.014, 0.014, 0.014, !100eV. 1 keV calculated as sqrt(1/rqm)*uth_e, where uth_e = sqrt(T/m_ec^2)
  uth(1:3) = 0.0014, 0.0014, 0.0014,    !1eV
  ufl(1:3) = 0.0d0 , 0.0d0 , 0.0d0 ,
}
profile
{
    density        = 1, 

    profile_type(1:2) = "math func" ,

    math_func_expr = "39.48*(x2+sqrt(2)*0.4*log(20716/39.48)>=x1)*exp(((x1-x2)/sqrt(2))/0.4)", !L1=0.4*log(20716/39.48) is distance between critical sueface and solid surface. ncr=39.48*n0.

}

spe_bound
{
    type(1:2,1) = "open", "open",
    type(1:2,2) = "open", "open",
}

diag_species 
{
  ndump_fac = 1,
  reports = "charge",

  ndump_fac_pha = 1,
  ndump_fac_raw = 1,
  !raw_math_expr = "t > 22", !select particles at time > 30*T0, right before focus
  !raw_math_expr = "step(g-1.03)*step(t-22)",!>15kev, gamma>1.03
  raw_math_expr = "step(g-1.03)*step(x1-x2)",!all electrons in target higher than k_alpha energy

  ps_gammamin = 1.0, 
  ps_gammamax = 20.0,
  ps_ngamma = 1000,


  phasespaces = "g", 
}


species
{
    name           = "eTarget",         !ionized electrons 
    num_par_max    = 1.0d9,
    rqm            = -1.000,          ! Means 'electron'
    num_par_x(1:2) = 10, 10,            
    add_tag        = .true.,
}

!----------inital proper velocities-----------------
udist {
  uth(1:3) = 0.00626, 0.00626, 0.00626,  !20 eV. 
  ufl(1:3) = 0.0d0 , 0.0d0 , 0.0d0 ,
}

profile
{

    density        = 1, 

    profile_type(1:2) = "math func" ,

    math_func_expr = "20716*(x1-sqrt(2)*0.4*log(20716/39.48)>x2)",

}


spe_bound 
{ 
    type(1:2,1) = "open", "open",
    type(1:2,2) = "open", "open",
}

diag_species 
{

  ndump_fac = 1,
  reports = "charge",

  ndump_fac_pha = 1,
  ndump_fac_raw = 1,
  !raw_math_expr = "t > 22", !select particles at time > 30*T0, right before focus
  !raw_math_expr = "step(g-1.4)*step(t-22)",!>200kev, gamma>1.4
  raw_math_expr = "step(g-1.03)*step(x1-x2)",!all electrons in target higher than k_alpha energy

  ps_gammamin = 1.0, 
  ps_gammamax = 20.0,
  ps_ngamma = 1000,

  phasespaces = "g", 
}


zpulse_wall
{ 
  a0             = 2.5,
  omega0         = 6.283, 
  pol            = 0, 
  propagation    = "forward",
  tenv_type       = "gaussian",
  tenv_duration   = 18.5, ! fwhm = 50fs
  tenv_range      = 74, ! total length of the laser pulse. ~ 4*fwhm

  per_type       = "gaussian",
  per_center     = 10,
  per_w0(1:1)     = 1.7, !with a spot size (radius) of ?*lambda. fwhm = 2*lambda. w0=fwhm*1.699/2. keep this the same for all wavelengths
  per_focus       = 10, !position of focus. depth. focused at critical density 
}


! --------------------- end of osiris input file ---------------

\end{minted}

\appendix{Iterative bash submission for parameter scan}
\label{app:bash}
\begin{minted}
[
xleftmargin=\parindent,
breaklines,
frame=lines,
framesep=2mm,
baselinestretch=1.2,
fontsize=\footnotesize,
linenos
]
{bash}
#!/bin/bash

# Dec 2020, by Jinpu Lin
#This script will run all of my simplified simulation files
#Need an "inputDeck" for the software (OSIRIS, etc), and a "runscript.sh" for the computing system, both under the same root path
#Edit file and save: sed 's/"p"/0/g' file.txt > fileNew.txt. "g" stands for global
# remove file: rm

#Loop over a0 
for i in {1..17}
do
	#Loop over scale length
	for j in {1..55}
	# for i in {1..56..5}
	# for j in 0.1 0.5 1
	do
		((x=$i/10))
		((y=$i-10*$x))
		# sed s/a0value/printf %.1f "$((10**3 * $i/10))e-3"/ inputDeck>temp
		sed s/a0value/$x.$y/g inputDeck>temp
		# echo $x
		# echo $y
		((c=$j/10))
		((d=$j-10*$c))
		sed s/scaleLength/$c.$d/g temp>a_${x}.${y}_L_${c}.${d}
		# sed s/scaleLength/$c.$d/ temp>a_$x.$y_L_$c.$d
		sed s/deckname/a_${x}.${y}_L_${c}.${d}/g runscript.sh>runscripta_${x}.${y}_L_${c}.${d}.sh
		sbatch runscripta_${x}.${y}_L_${c}.${d}.sh
		rm temp
		rm runscripta_${x}.${y}_L_${c}.${d}.sh
		# rm a_${x}.${y}_L_${c}.${d}
	done
done

clear

\end{minted}

\appendix{Characteristic x-ray emission analysis algorithm}
\label{app:Xray}
\begin{minted}
[
xleftmargin=\parindent,
breaklines,
frame=lines,
framesep=2mm,
baselinestretch=1.2,
fontsize=\footnotesize,
linenos
]
{python}

# -*- coding: utf-8 -*-
"""
Created on Wed Feb  3 20:10:43 2021

@author: linji
"""

import os
import numpy as np
import h5py as h5
import scipy.constants as const
from scipy.optimize import curve_fit
import math

pathRoot = '/Users/jimlin/Desktop/umich/CUOS/OSIRIS/dataFile/solidXrays/scan3param/1.6um/all/'
pathResult = '/Users/jimlin/Desktop/umich/CUOS/OSIRIS/results/solidXrays/scan3param/1.6um/postProcess/'


# calculate kalpha emission caused by electron impact of all energies
def IonizeXsec(ene):
    ene=ene*1000 #input energy in kev, converts into ev
    z=42;en=1.997e4;a00=5.29177210903e-11 #Bohr radius
    u=ene/en
    if ene<16*en:
        a1=1.281e-3;a2=4.105e-5;a3=-1.410e-3;a4=1.450e-3;a5=-1.642e-3
        xsec=4*math.pi*a00**2*(u-1)*u**(-2)*(a1+a2*u+a3/(1+u)+a4/(1+u)**3+a5/(1+u)**5)**2
    else:
        bm=2.692e-1;anj=2.647e-8;g1=7.942e-1;g2=5.981;g3=-2.59;g4=1.369
        beta=math.sqrt(ene*(ene+2*const.m_e*const.c**2))/(ene+const.m_e*const.c**2)
        X=math.sqrt(ene*(ene+2*const.m_e*const.c**2))/(const.m_e*const.c**2)
        pwba=4*math.pi*a00**2*anj*beta**(-2)*((math.log(X**2)-beta**2)*(1+g1/X)+g2+g3*(1-beta**2)**0.25+g4/X)
        xsec=pwba*ene/(ene+bm*en)
    xsec=xsec*1e25
    return xsec
    
def getXrays(pathJob, a0, ls):
    path_rawI = pathJob + '/MS/RAW/eTarget/'
    path_raw = pathJob + '/MS/RAW/preplasma/'
    path_gammaI = pathJob + '/MS/PHA/gamma/eTarget/'
    path_gamma = pathJob + '/MS/PHA/gamma/preplasma/'
    x0 = 1.6
    n0 = 1.103e19
    ns = 5800
    
    kalpha=[];
    for i in range(0, 22):
        f_raw=h5.File(path_raw+"RAW-preplasma-%06d.h5"%i, "r")
        ene = f_raw['ene'][()]; #local kinetic energy
        x1 = f_raw['x1'][()];
        x2 = f_raw['x2'][()];
    
        f_rawI=h5.File(path_rawI+"RAW-eTarget-%06d.h5"%i, "r")
        eneI = f_rawI['ene'][()]; #local kinetic energy
        x1I = f_rawI['x1'][()];
        x2I = f_rawI['x2'][()];
        x1T=np.hstack([x1,x1I])
        x2T=np.hstack([x2,x2I])
        eneT=np.hstack([ene,eneI])
        x1T = x1T * x0;
        x2T = x2T * x0;
        
        enetarget=eneT[(x1T-x0*np.sqrt(2)*ls*np.log(ns/39.48)>x2T)]

        xsec=enetarget+1; #define an array of the same size as eneT
        kyield=enetarget+1; #define an array of the same size as eneT
        for j in range(0,enetarget.size):
            xsec[j]=IonizeXsec(enetarget[j]*1e3)
            kyield[j]=xsec[j]*np.sqrt(1-(enetarget[j]+1)**(-2)) #x-ray yield ~ corss section * particle velocity
            
        kyieldArray = np.array(kyield)
        kyieldArray[np.isneginf(kyieldArray)] = 0
        kalpha.append(sum(kyieldArray))
        
        
    # fit Te
    allTemp=[];
    allTempErr=[];
    allBrem=[];
    def log_func(E,a,T):
        return (a-E/T)
    
    for i in range(14, 22):
        f_gamma=h5.File(path_gamma+"gamma-preplasma-%06d.h5"%i, "r")
        f_gammaI=h5.File(path_gammaI+"gamma-eTarget-%06d.h5"%i, "r") 
        gamma = f_gamma['gamma'][()];
        xstart = f_gamma['AXIS']['AXIS1'][(0)]
        xend   = f_gamma['AXIS']['AXIS1'][(1)]
        gammaI = f_gammaI['gamma'][()];
        gammaT = gamma + gammaI
        x = np.linspace(xstart, xend,  np.size(gammaT), endpoint=True)
        E=(x-1)*511; #keV
        y=abs(gammaT)
        for j in range(0, len(y)):
            if y[j]==0:
                y[j]=1e-10
        
        f_raw=h5.File(path_raw+"RAW-preplasma-%06d.h5"%i, "r")
        ene = f_raw['ene'][()]*1e3; #local kinetic energy in kev
        
        low=((1+a0**2/2)**0.5-1)*511
        idxLow = (np.abs(E - low)).argmin() + 1 #avoid extreme indices like 0
        Elow=E[idxLow]
        yLow=y[idxLow]
        yHigh=yLow*1e-2
        idxHigh=(np.abs(y - yHigh)).argmin()
        Ehigh=E[idxHigh]
        lny=np.log(y)
        
        if ene.max()>low and lny[(E<Ehigh)&(E>Elow)].size>3: #avoid cases where idxHigh-idxLow <= 2
            popt, pcov = curve_fit(log_func, E[(E<Ehigh)&(E>Elow)], lny[(E<Ehigh)&(E>Elow)], p0=(5, 100))
#            for small a0 and sharp gradient, temperature can be << 100kev
            if popt[1]<0:
                popt, pcov = curve_fit(log_func, E[(E<Ehigh)&(E>Elow)], lny[(E<Ehigh)&(E>Elow)], p0=(4, 20))
            perr = np.sqrt(np.diag(pcov))

            allTemp.append(popt[1])
            allTempErr.append(perr[1])
            Z = 42; ne = n0*4*math.pi**2;
            Te = popt[1];
            brem = 1.54*1e-38*Z**2*ne**2*(Te*1e3)**0.5 # is the total bremsstrahlung power per unit volume
            allBrem.append(brem)
    
    if not kalpha: #if list is empty
        totalKalpha=0
    else:
        totalKalpha=sum(kalpha)
        
    if not allTemp: #if list is empty
        avgTe=0
    else:
        avgTe=np.mean(allTemp)
        
    if not allTempErr: #if list is empty
        avgTeErr=0
    else:
        avgTeErr=np.mean(allTempErr)
        
    if not allBrem: #if list is empty
        totalBrem=0
    else:
        totalBrem=sum(allBrem)
        
    xrayDiag = [totalKalpha, avgTe, avgTeErr, totalBrem]
    xrayDiagNorm = [totalKalpha/a0**2, avgTe, avgTeErr, totalBrem/a0**2]

        
    return xrayDiag, xrayDiagNorm


resultJob = []
resultJobNorm = []
jobList = os.listdir(pathRoot)

for x in jobList:
    if not x.startswith('.'):
        jobContent=os.listdir(pathRoot+x)
        print(pathRoot+x)
        for xx in jobContent:
            if xx.startswith('a'):
                a0str=(xx[2:5:1])
                a0=float(a0str)
                lsstr=(xx[8:11:1])
                ls=float(lsstr)
                
                xrayEmission = getXrays(pathRoot+x, a0, ls)[0]
                resultJob.append(os.listdir(pathRoot+x)[0])
                resultJob.append(xrayEmission)
                
                xrayEmissionNorm = getXrays(pathRoot+x, a0, ls)[1]
                resultJobNorm.append(os.listdir(pathRoot+x)[0])
                resultJobNorm.append(xrayEmissionNorm)

np.savetxt(pathResult+'XrayDiag.txt',resultJob, fmt='%s')
np.savetxt(pathResult+'XrayDiagNorm.txt',resultJobNorm, fmt='%s')




\end{minted}

%% Bibliography %%
\startbibliography
 \begin{singlespace} % Bibliography must be single spaced
  \bibliography{References}% Use the BibTeX file ``References.bib''.
 \end{singlespace}

% \bibliographystyleMyPaper{unsrt}
% \bibliographyMyPaper{mypaper}
% An external Abstract that can be printed at the end of the document, 
% for separate submission to Rackham.
%\startextabstractpage
%{High-resolution Experiments and Computations on Mixing of Turbulent Buoyant Round Free Jets in Uniform and Stratified Environments}{Sunming Qin}{Chairs: Prof. Annalisa Manera \& Dr. Victor Petrov}
%\input{Abstract/Abstract}
%\label{ExtAbstract}

\end{document}